\documentclass[letterpaper, 11pt]{article}

\usepackage{mathtools}
\usepackage{lmodern}
\usepackage{array}
\usepackage{multirow}
\usepackage[T1]{fontenc}
\usepackage[utf8]{inputenc}
\usepackage{helvet}
\usepackage{enumitem}
\usepackage{amsmath,amsfonts,amssymb,bm}
\usepackage{titlesec}
\usepackage{pgfgantt,rotating}
\usepackage{float}
\usepackage{graphicx}
\usepackage{sidecap}
\usepackage{wrapfig}
\usepackage{bold-extra}
\usepackage{authblk}
\usepackage{appendix}
\usepackage{amsfonts}
\usepackage{amsmath}
\usepackage{upgreek}
\usepackage{newtxmath}
\usepackage[utf8]{inputenc}
\usepackage{color}
\usepackage{rotating}
\usepackage{cite}
\usepackage{bm}

\usepackage[margin=1in]{geometry}

\usepackage[linesnumbered,ruled]{algorithm2e}
\usepackage{algpseudocode}

\newcommand{\beq}{\begin{equation}}
\newcommand{\eeq}{\end{equation}}
\newcommand{\bgqar}{\begin{eqnarray}}
\newcommand{\enqar}{\end{eqnarray}}
\newcommand{\bgqarn}{\begin{eqnarray*}}
\newcommand{\enqarn}{\end{eqnarray*}}
\newcommand{\bgary}{\begin{array}}
\newcommand{\enary}{\end{array}}

\newcommand{\etal}{{\it et al. }}

\graphicspath{{figures/}}

\title{Data-driven Framework for Forward and Inverse Problems in Guided Waves-Based Structural Health Monitoring Under Varying Environmental and Operating Conditions}

\author{Yiming Fan}
\author{Fotis Kopsaftopoulos\footnote{Corresponding author.}}

\affil{\small Intelligent Structural Systems Laboratory (ISSL) \\ Department of Mechanical, Aerospace and Nuclear Engineering \\ Rensselaer Polytechnic Institute, Troy, NY, USA \\ Email: \{fany5,kopsaf\}@rpi.edu}

\date{\today}

\titleformat{\paragraph}
{\normalfont\normalsize\bfseries}{\theparagraph}{1em}{}
\titlespacing*{\paragraph}
{0pt}{3.25ex plus 1ex minus .2ex}{1.5ex plus .2ex}

\begin{document}

\maketitle


\begin{abstract}

Recently, guided waves-based techniques have garnered increased attention from researchers in the field of Structural Health Monitoring (SHM) for damage detection and quantification. Extracting features that are sensitive to changes in structural conditions has become a critical step. Guided waves, being one of the most widely used techniques, are highly responsive to structural changes caused by various types of damage. However, due to disturbances from environmental and operational conditions (EOCs) and waveforms reflected from boundaries and damage in complex systems, manually extracting features related to the system state is challenging.
To address this, various neural networks, capable of automatically deriving features, have been developed and applied to system identification tasks. Among these, Convolutional Autoencoders (CAEs) have proven effective as damage detectors, projecting data efficiently into a latent space. However, existing CAE-based models for damage diagnosis often overlook the influence of EOCs, which can significantly impact model performance.
In this work, a scheme that provides accurate damage level classification while estimating EOCs, such as loading conditions, making the model adaptable to varying external factors is introduced. By restructuring and combining the CAE with feedforward neural networks (FFNNs), the proposed scheme also enables signal reconstruction, a potentially valuable feature when data is limited. The method has been validated on an aluminum plate with various damage levels under different loading conditions, achieving near-perfect state classification and low-error signal reconstruction. The framework has also been tested on incomplete datasets to verify its robustness.
\end{abstract}
\newpage\pagebreak 
\tableofcontents 
\section{Introduction} \label{sec:intro}
Structural Health Monitoring (SHM) models are designed to provide early warnings of damage, assess structural performance, and support decision-making and design improvements while reducing costs. These models are widely applied in tasks such as damage detection, quantification, and localization \cite{Amer-Kopsaftopoulos19a,Amer-Kopsaftopoulos20,Janapati-etal16}. Structural materials, including metals and complex composites, are prone to various damage types, such as delamination, cracks, and corrosion, which can lead to significant degradation and even failure \cite{farrar2007introduction,diamanti2010structural}. Therefore, techniques that are adaptable to multiple materials and environmental conditions are essential. Ultrasonic Lamb waves have gained considerable attention in SHM due to their ability to propagate over long distances and their high sensitivity to structural changes \cite{2016-9,2017-5}. Typically, guided wave-based SHM techniques utilize sensor arrays, such as piezoelectric lead zirconate titanates (PZTs), to generate excitations and collect data for monitoring \cite{ahmed2022statistical}. 

Existing SHM approaches can be broadly classified into two main categories: physics-based and data-driven techniques. Physics-based techniques typically involve the use of analytical or numerical models to predict the behavior of structures under varying conditions. This approach provides a physical understanding of structural behaviors, enabling the estimation of causal factors such as damage severity \cite{badihi2022comprehensive}. Moreover, these models can integrate multiphysics aspects, such as thermal and environmental factors, allowing for more realistic simulations \cite{shen2016combined,han2009finite}. However, the performance and reliability of these models depend heavily on the accuracy of the assumptions made during development, particularly regarding initial and boundary conditions. Such assumptions can limit the generalizability of the models, especially when applied to different structures or environmental conditions \cite{doebling1996damage,friswell1995finite}. Additionally, certain models in this category, such as finite element methods (FEM), can be computationally intensive, requiring significant resources for large or complex structures, which poses challenges for real-time prediction \cite{goyal2016vibration}. 

Data-driven techniques, in contrast to physics-based methods, focus on extracting patterns, relationships, and insights directly from data, rather than relying heavily on explicit physical models. These techniques are particularly valuable for SHM tasks due to their ability to capture complex, nonlinear structural behaviors and provide real-time monitoring results, which are often challenging for conventional physics-based models. Typically, data-driven approaches involve a feature extraction step that is crucial for accurate classification. The model's classification performance is closely tied to the quality and sensitivity of the extracted features. Since the phase and group velocities of Lamb waves depend on structural properties, there is potential to extract meaningful features from the collected waveforms. However, wave distortion during propagation, coupled with reflections from boundaries and damage, can complicate direct feature extraction from the raw signals \cite{gorgin2020environmental,giurgiutiu2005tuned}.

Various methods have been proposed to capture features that are sensitive to damage type, severity, and location, depending on the specific application and available data. For guided wave-based approaches, features are typically defined in the time or frequency domain, or both. In the time domain, features like time of flight (ToF)—the time taken for guided waves to travel between two points—are sensitive to wave speed, which is determined by material and structural conditions \cite{ostachowicz2011guided}.
Frequency-based features, on the other hand, focus on analyzing the spectral content of guided wave signals [18]. Techniques such as Fourier Transform (FT) or Wavelet Transform are commonly employed to extract frequency-related information, as changes in the frequency content can indicate the presence of damage and help in determining its type. Fast Fourier Transform (FFT), an extension of FT, accelerates the transformation process and is widely used for guided wave signal analysis \cite{mitra2016guided}. However, these approaches often lack generality and robustness across different structures and conditions, as they are typically limited to specific material types and geometric properties.

The Damage Index (DI) offers an alternative approach that does not rely on material properties or geometric information due to its formulation \cite{ihn2008pitch, giurgiutiu2011piezoelectric}. While DI can take various forms, it is typically derived by comparing signals from an unknown state to baseline signals collected under healthy conditions. Its value changes in response to factors such as time delay of specific wave packets, signal amplitude, phase, and energy content \cite{jin2019monitoring, nasrollahi2018multimodal}. DI is a versatile tool for both damage detection and quantification \cite{Janapati-etal16,Soman-etal18, fan2022damage}. 
Ahmad \etal proposed a probabilistic framework that integrates DI with one-dimensional and two-dimensional Gaussian Process Regression Models (GPRMs) for predicting damage size in active sensing-based systems \cite{amer2023gaussian}. Although DIs reduce the dependency on material properties, they may still lack the sensitivity needed to address different environmental and operational conditions (EOCs). Furthermore, these features can lead to complex and computationally intensive models due to their nonlinear characteristics \cite{kopsaftopoulos2018stochastic, spiridonakos2014non}.

As an alternative to manually derived features, artificial neural networks (ANNs), another class of data-driven approaches, can automatically extract features. Various ANN models have been employed in SHM applications \cite{worden2007application, humer2022damage, posilovic2022deep}, with convolutional neural networks (CNNs) being particularly well-suited for time series data due to their ability to extract temporal features. By incorporating convolutional and downsampling layers, CNNs can retain essential information while reducing computational complexity, making them a promising tool for system identification tasks. For example, Avci et al. developed a real-time damage detection method using CNNs to extract damage-sensitive features from raw acceleration signals \cite{avci2017structural}. Pandey \etal applied 1D CNNs for damage detection on a thin aluminum plate using Lamb waves, and further explained the classifier's prediction performance using Interpretable Model-Agnostic Explanations (LIME) \cite{pandey2022explainable}.

%
\begin{figure}[!t]
\centering
\includegraphics[scale=0.46]{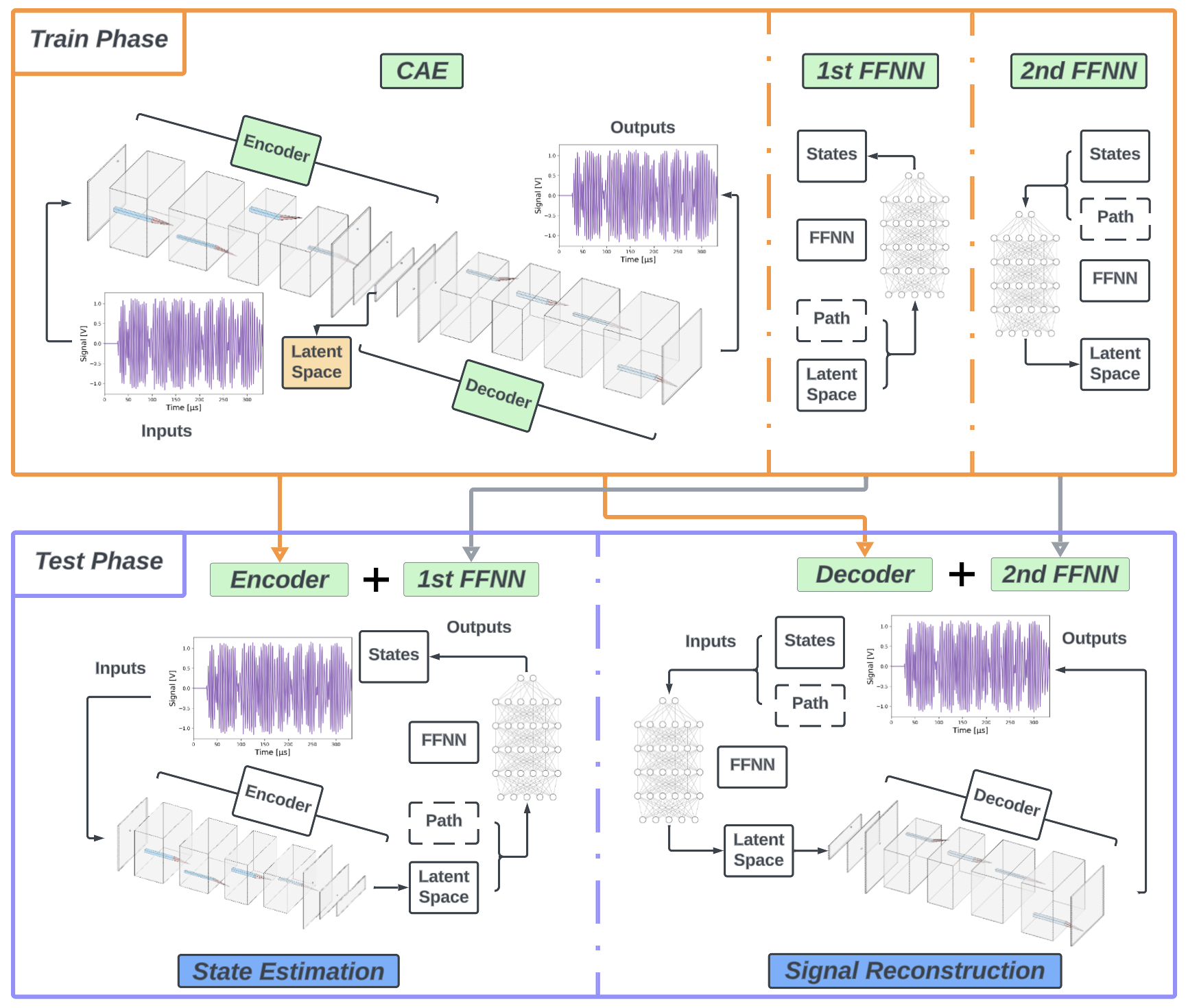}
\caption{The flowchart that demonstrates the main steps of guided-wave based multi-fidelity GPRM training procedure.}.
\label{fig:flowchart} 
\vspace{10pt}
\end{figure}

Autoencoder (AE), a popular variant of artificial neural networks developed recently, is capable of compressing and reconstructing data to learn efficient representations of inputs. Various forms of AEs, including convolutional autoencoders (CAEs), variational autoencoders (VAEs), and long short-term memory autoencoders (LSTM-AEs), have been applied to anomaly detection tasks. Among these variations, the VAE provides opportunities for statistical estimations by modeling the distribution of variables in the latent space \cite{coracca2023unsupervised,ma2020structural,shu2023unsupervised,zhang2022unsupervised}. LSTM-AEs are particularly suitable for time series data, although they may require extensive training time when handling high-dimensional datasets \cite{rizvi2023anomaly}. CAEs demonstrate high reconstruction accuracy while maintaining reasonable computational requirements. Lee et al. developed a fatigue damage detector by comparing the root mean square error (RMSE) of dominant features from the CAE's latent space against a predetermined threshold \cite{lee2022automated}.
Rautela \etal evaluated fault alarm performance based on CAEs in comparison to conventional dimensionality reduction methods such as principal component analysis (PCA) and independent component analysis (ICA) on guided wave datasets \cite{rautela2022delamination}. Additionally, higher-order temporal–frequency representations transformed from time series data have been utilized as inputs for CAEs.
For example, Guo \etal created an indicator using CAEs by comparing the similarities between features learned from baseline and new samples \cite{guo2022unsupervised}.
While these damage detectors can accurately identify anomalies by setting thresholds, many existing AE-based models fall short of providing precise estimates of damage levels, which limits their effectiveness for comprehensive structural assessments. Furthermore, EOCs such as temperature, moisture, applied loads, and boundary conditions can adversely affect the performance and robustness of guided wave-based models by altering wave propagation, particularly in complex structures \cite{sohn2007effects}. Although some methods exist to mitigate the effects of varying EOCs, most of them focus on single factor and are challenging to extend when multiple EOC factors coexist and change simultaneously \cite{dworakowski2016multi, fendzi2016data}.

In this work, a ML-based framework employing CAE and FFNN for structural state estimation with two major functionalities: system identification under varying EOCs and signal reconstruction based on the structural state, is proposed. During the training phase, the CAE is trained to produce accurate reconstructed signals and latent space representations, while FFNNs are trained to establish precise mappings between state index vectors and latent variables extracted from the CAE's bottleneck layer. In the inspection phase, the framework includes two branches: the first addresses the inverse problem, namely damage detection and quantification under varying EOCs using unknown signals, while the second addresses the forward problem, which involves generating signals based on specific states. These branches are realized through the partitioning and recombination of the trained networks. The first branch serves as a state estimation method, whereas the second can augment the dataset when data collection is time-consuming and restricted. The main contributions of this study are as follows: 
(i) the introduction of a unified framework to address both the forward and inverse problems, specifically state estimation under varying EOCs and signal generation at target states; (ii) simultaneous prediction of structural states and operating conditions, with the approach easily adaptable to multiple EOCs; (iii) investigation of different model structures and tensor configurations for optimal performance; and (iv) analysis of time-varying and time-invariant latent spaces for enhanced model interpretability.



\section{Methodology}

This section introduces a multifunctional CAE-based framework that can be adapted to various input matrix construction methods. As illustrated in Figure \ref{fig:flowchart}, the framework consists of two primary components: the training phase and the testing phase. During the training phase, three neural networks (NNs) are utilized—one convolutional autoencoder (CAE) and two feedforward neural networks (FFNNs)—each serving distinct purposes. The CAE is responsible for data reconstruction and latent space extraction. By minimizing the difference between the output signals and the original inputs, the CAE ensures that the latent space retains the majority of the data's information. The two FFNNs are designed to establish mappings between latent space vectors and state vectors, with or without path information, depending on the structure of the CAE. Since these two FFNNs have opposite inputs and outputs, they exhibit complementary structures, allowing us to study just one of them to determine the configurations of both. In the testing phase, the trained NNs are organized and combined in various sequences to achieve different objectives. For the first primary function—state estimation—the outputs from the encoder are fed into the first FFNN, generating state vectors based on the provided signals. For the second function—signal generation—the second FFNN is connected to the decoder, producing signals from the state vectors. The framework's generality and stability stem from its inherent adaptability to different types of EOC factors. Further details, including the mathematical formulations, will be presented in the subsequent sections.

\subsection{Data Representation}

Model identification is based on noise-corrupted output data records corresponding to a sample of the admissible structure and operating conditions. The data records are of length $N$, each one corresponding to a specific value of the state vector $\bm{k}$, which, without loss of generality, is assumed to be two-dimensional. A sample of $M_1$ values is used for the first measurable variable $k^1$ (first element of vector $\bm{k}$), while a sample of $M_2$ values is used for the second externally measurable variable $k^2$ (second element of vector $\bm{k}$). 

Each experiment is characterized
by a specific element of $\bm{k}$, say $\bm{k} = [k_i^1,k_j^2]$. This vector is, for simplicity of notation, also designated as the duplet $k_{i,j}=(k_i^1,k_j^2)$ (the first term designating the value of $k^1$ and the second that of $k^2$). As a result, a total of $\sum_{i=1}^{M_1} \sum_{j=1}^{M_2} n_{i,j}$ experiments ($n_{i,j}$ is the number of trials repeated at state $k_{i,j}$ to decrease noise effects) are performed, with the complete
series covering the required range of each scalar parameter, say $[k_{min}^1, k_{max}^1]$ and $[k_{min}^2, k_{max}^2]$, via the
discretizations $k^1=k_1^1,k_2^1,...,k_{M_1}^1$ and $k^2=k_1^2,k_2^2,...,k_{M_2}^2$.

Data records from different structure and operating points corresponding to various values of the operating parameter vector can be written as:
\begin{equation}
 \bm{Y}\triangleq \{y_{\bm{k}}[t]|\bm{k}\triangleq[k^1,k^2]^T, t=1,...,N, k^1\in{k_1^1,k_2^1,...,k_{M_1}^1}, k^2\in{k_1^2,k_2^2,...,k_{M_2}^2}\}
 \label{eq:ae_eqn1}
\end{equation}

In this expression, $t$ designates normalized discrete time (the corresponding analog time being $t\cdot T$ with $T$ standing for the sampling period), $y_{\bm{k}}[t]$ the noise-corrupted output signals corresponding to $\bm{k}$. Each data record can then be expressed as $\bm{y}_{\bm{k}}\in\mathbb{R}^{N}$.

\subsection{Latent Space Representation}

In contrast to conventional feature extraction techniques, artificial neural networks (ANNs) can automatically extract nonlinear features, eliminating the need for tedious computations related to state-sensitive features. Within this framework, the convolutional autoencoder (CAE) is selected as the feature extraction tool, as it allows for extensive compression of inputs through the encoder and achieves low reconstruction error. While this study focuses on damage level and external load as the primary variables, it is straightforward to incorporate additional factors into this framework, following the same construction approach, thereby enhancing its adaptability to various working conditions and scenarios. 

As data $\bm{y}_{\bm{k}}\in\mathbb{R}^{N}$ passing through the encoder portion, it will be compressed by the convolutional layers as well as the followed downsampling and dense layers. Maximum compression occurs in the bottleneck layer which also serves as the connector of the encoder and the decoder. The latent space vector, also called feature space, is derived from this layer. Denoting as $\bm{z}_{\bm{k}}\in\mathbb{R}^{D}$, where $D$ is the pre-defined latent space width, the latent vector can then be regarded as a compressed yet informative representation of the original data $\bm{y}_{\bm{k}}$. To express the latent space variable under different states clearly, 
$\bm{z}_{m}(k^1,k^2)$
will be used later in this paper, where $m$ indicates the $m$-th latent space variable.


\subsection{State Estimation}

As shown in Figure \ref{fig:flowchart}, the test phase contains two parts. The trained NNs can be arranged in specific sequences to achieve different objectives. For the purpose of state estimation, the signals are initially projected through the trained encoder, resulting in a compressed vector. This compressed data is then input into the first trained FFNN to generate the corresponding state vector estimates. This process can be expressed mathematically as follows:

\begin{equation}
 \widehat{\bm{k}}=\varphi_{1}(\bm{z})= \varphi_{1}(\alpha_{yz}(\bm{W}_{yz}\bm{y}+\bm{b}_{yz}))
 \label{eq:ae_eqn2}
\end{equation}
where $\varphi_{1}$ is the 1st FFNN function in the state estimation branch, $\alpha_{yz}$ is the nonlinear function for the encoder, $\bm{W}_{yz}$ and $\bm{b}_{yz}$ are weights and bias for encoder, respectively.

\subsection{Signal Reconstruction}

For signal reconstruction propose, the inputs are the states with or without the paths information mentioned in Section \ref{sec:data_process}. Whether the paths information is needed depends on the model types that will be discussed later. The second FFNN aims to find the corresponding latent vectors, which will be further streched by the decoder to obtain the reconstructed signal. This process can be expressed as:
\begin{equation}
 \widehat{\bm{y}}=\alpha_{zy}(\bm{W}_{zy}\bm{z}+\bm{b}_{zy})=\alpha_{zy}(\bm{W}_{zy}{\varphi_{2}(\bm{k})}+\bm{b}_{zy})
 \label{eq:ae_eqn3}
\end{equation}

where $\varphi_{2}$ is the 2nd FFNN function in the signal reconstruction branch, $\alpha_{zy}$ is the nonlinear function for the decoder, $\bm{W}_{zy}$ and $\bm{b}_{zy}$ are weights and bias for decoder, respectively.

To assess the signal reconstruction performance, RSS/SSS($\%$) is used as the criteria to quantify how well the reconstructed signals match with the original ones. Given the amplitude of original signal and the reconstructed signal at time $t$ as $y[t]$ and $\widehat{y}[t]$ respectively, the formulation of RSS/SSS($\%$) is:

\begin{equation}
 RSS/SSS(\%) =
    \frac{\sum_{t=1}^{N}(y[t]-\widehat{y}[t])^2}{\sum_{t=1}^{N}{y[t]}^2}
 \label{eq:rss}
\end{equation}

\section{Experimental Setup and Data Processing}
\begin{figure}[h!]
\centering
\includegraphics[scale=0.24]{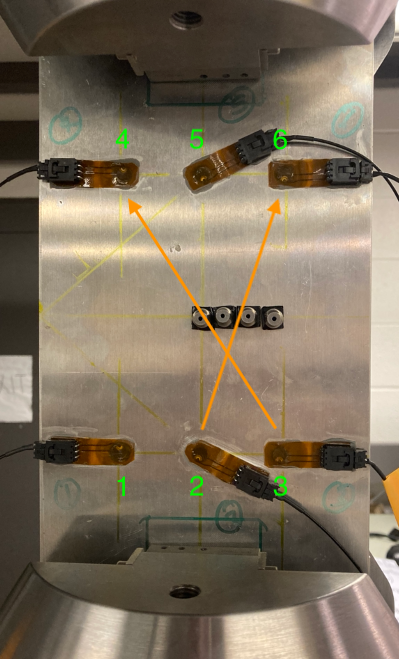}
\caption{The Al coupon used in this study with attached weights to simulate damages. Different loading conditions can be applied by attaching the coupon onto a tensile testing machine (Instron, Inc). Sensors 1-3 serve as activators while sensors 4-6 serve as receivers. The yellow arrows indicate the two paths analyzed in this test case.}
\label{fig:Al_coupon} 
\end{figure}

\begin{figure}[t!]
    \begin{picture}(500,410)
    \put(10,250){\includegraphics[width=0.45\textwidth]{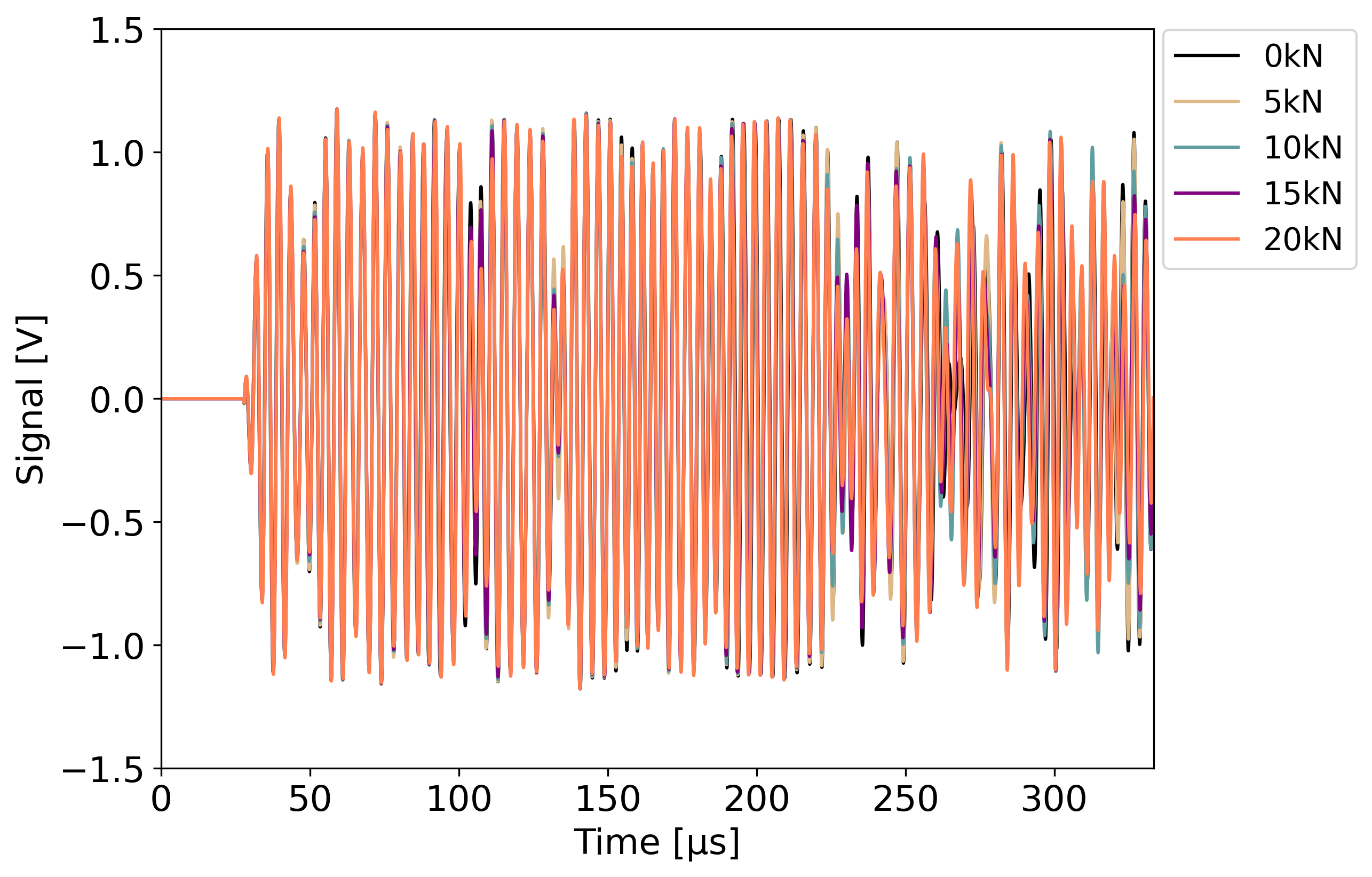}}
    \put(224,252){\includegraphics[width=0.45\textwidth]{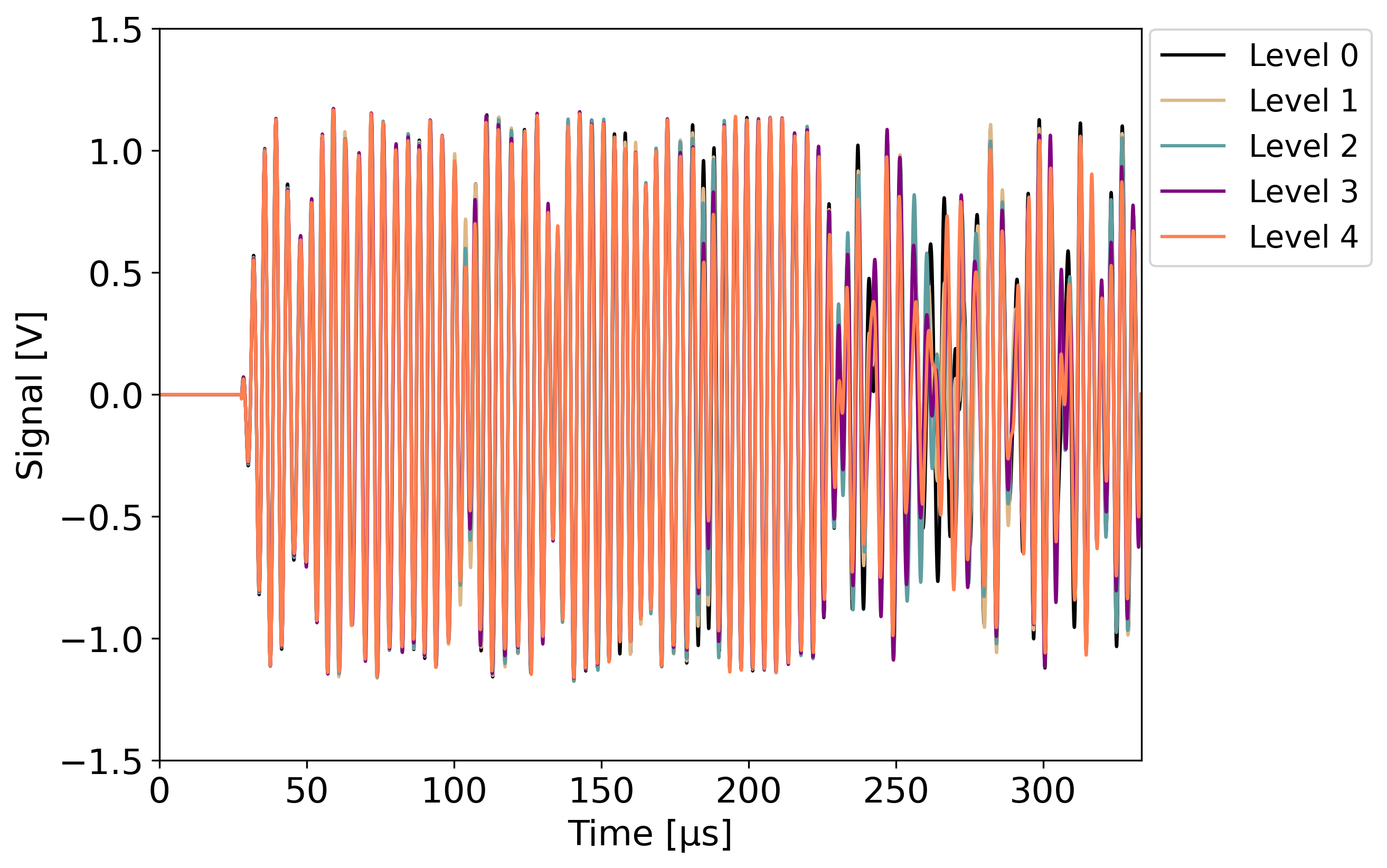}}
    \put(10,112){\includegraphics[width=0.45\textwidth]{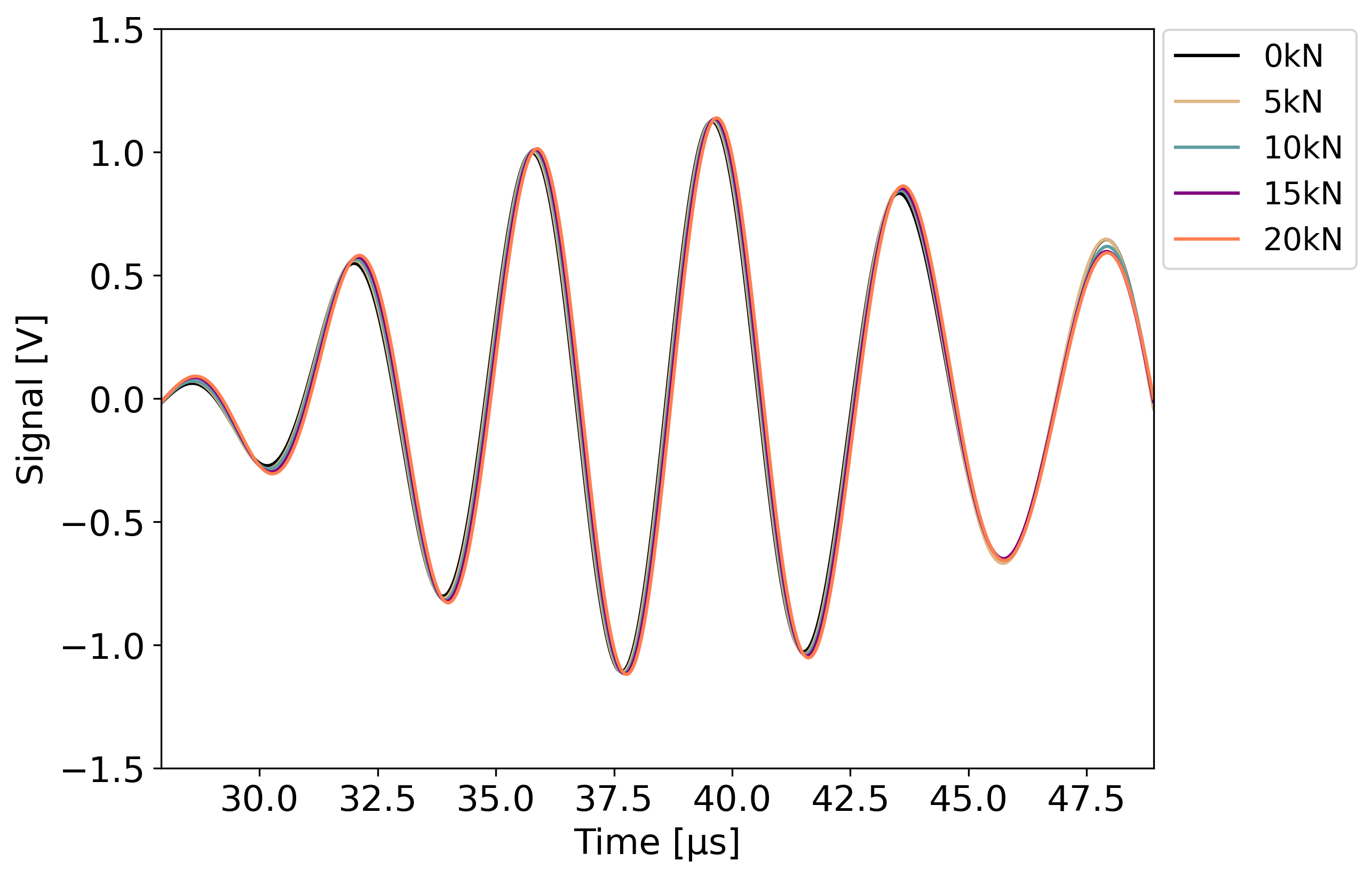}}
    \put(224,114){\includegraphics[width=0.45\textwidth]{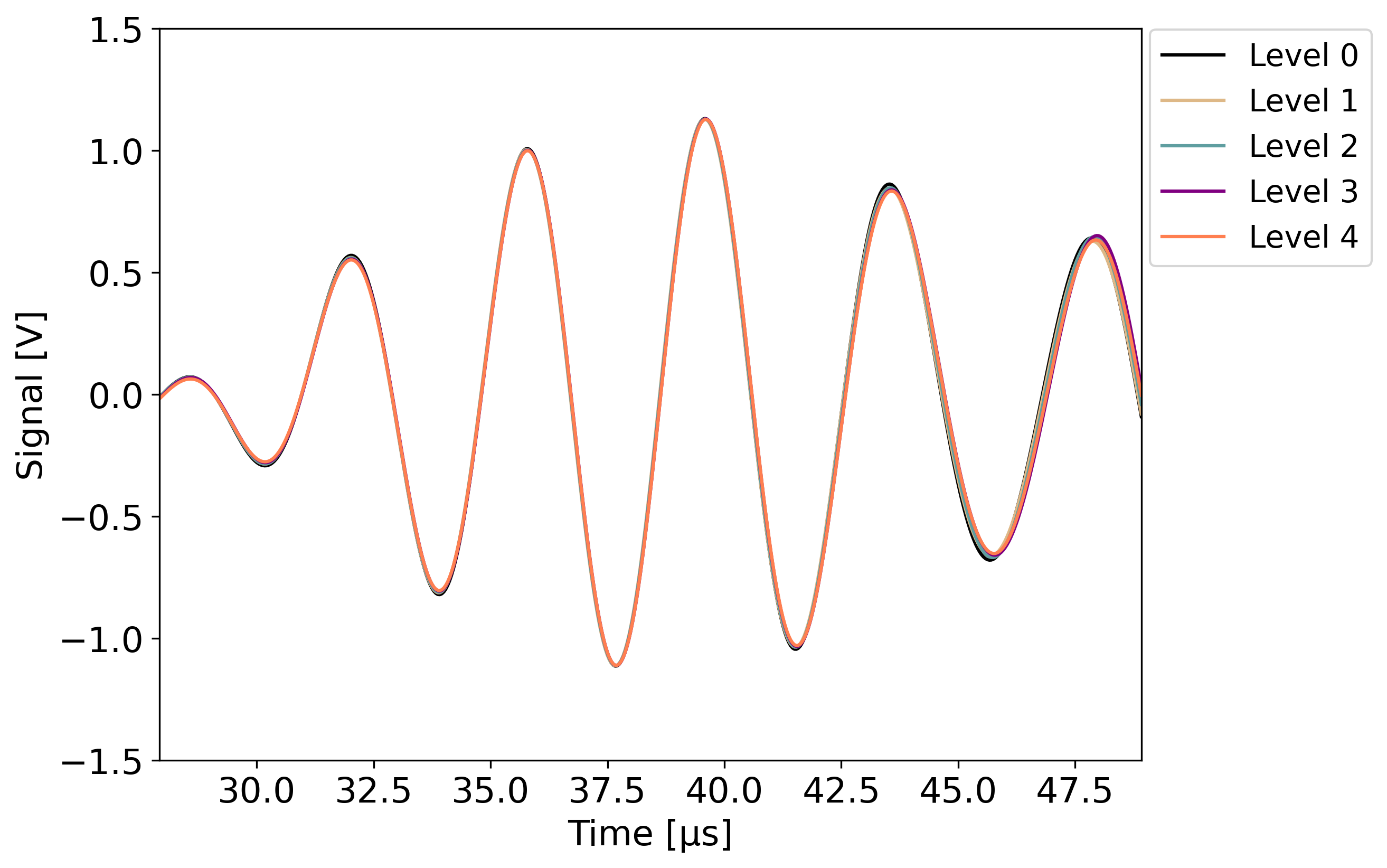}}
    \put(10,-26){\includegraphics[width=0.45\textwidth]{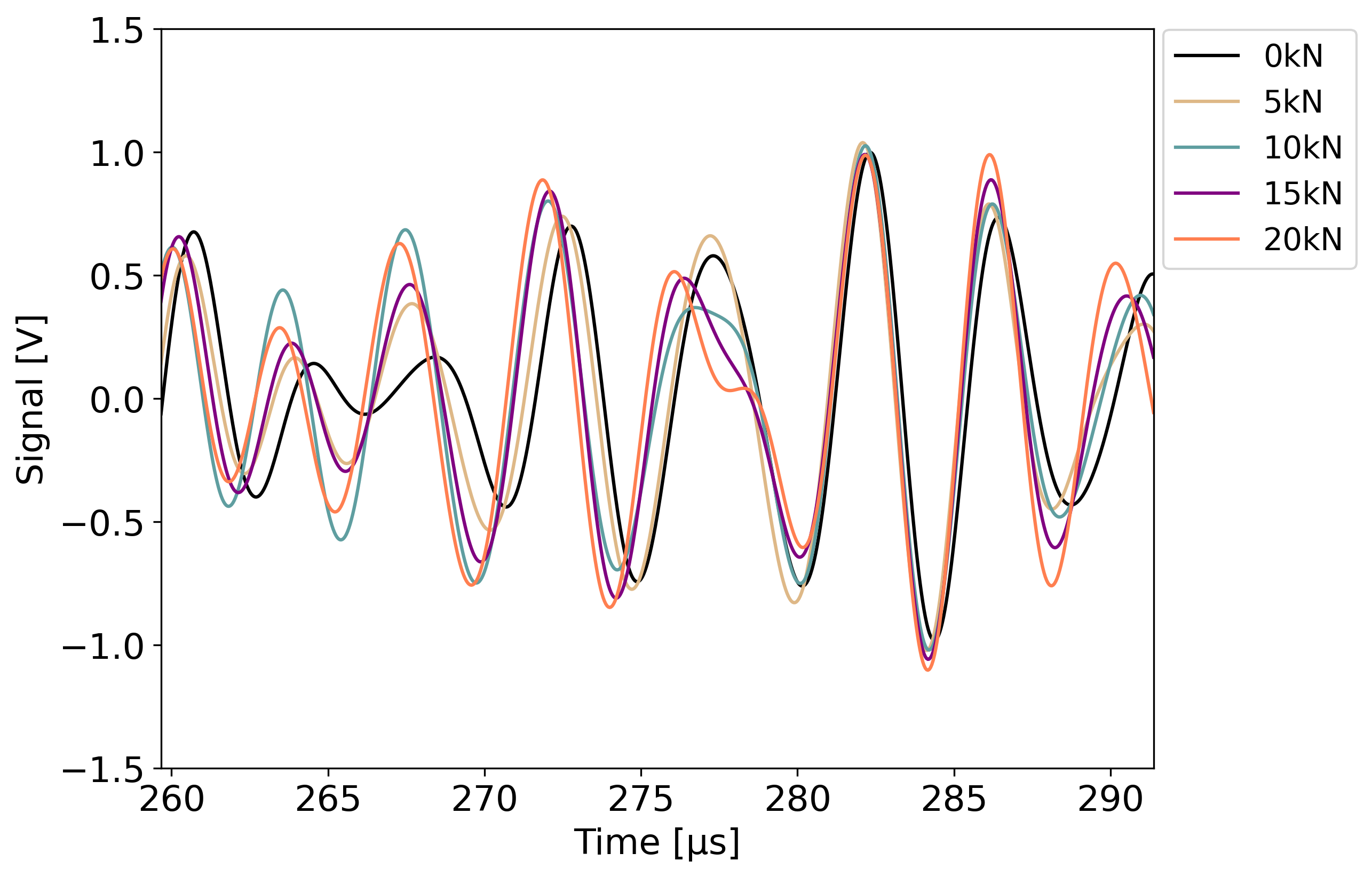}}
    \put(224,-24){\includegraphics[width=0.45\textwidth]{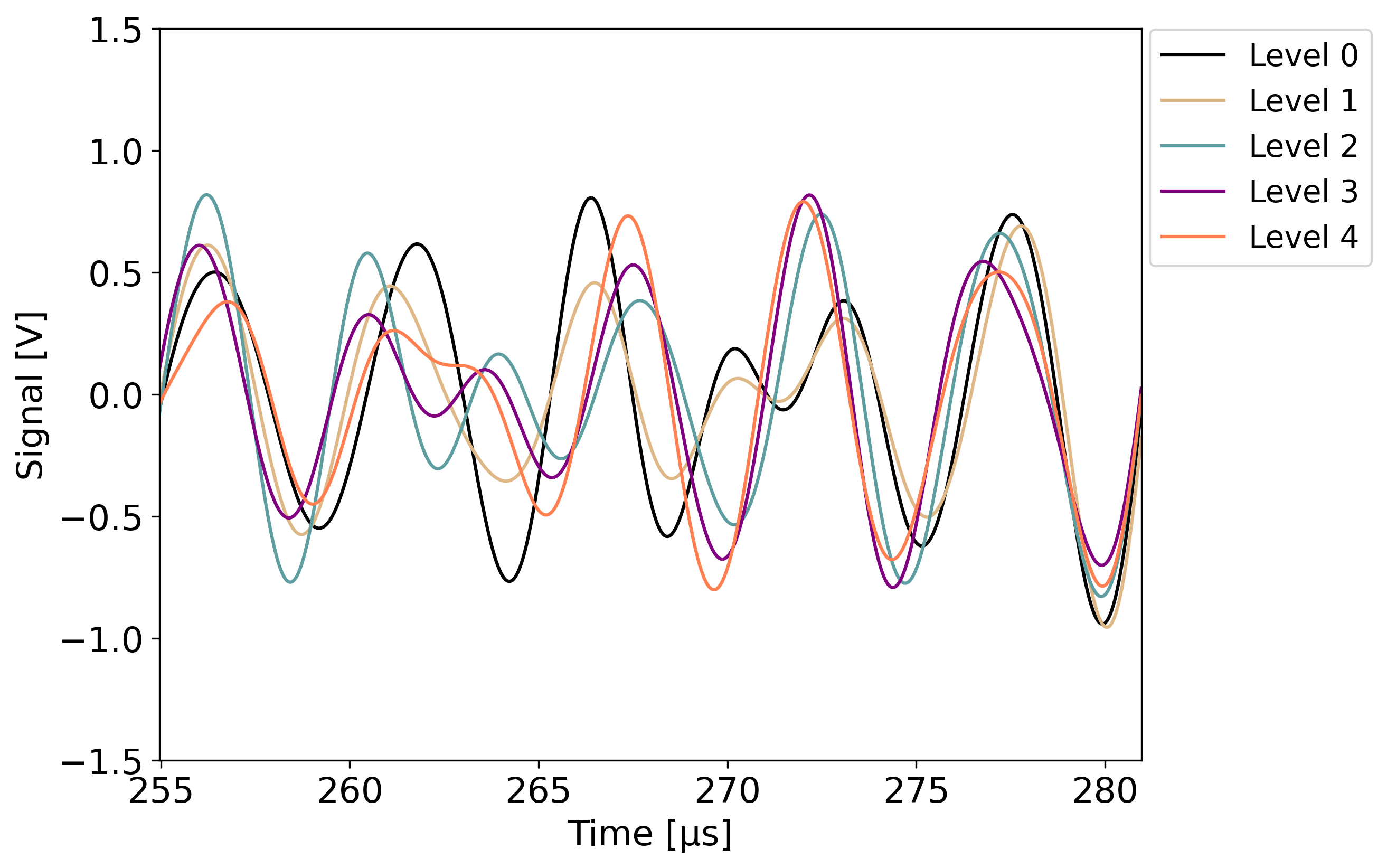}}
    \put(200,280){\color{black} \large {\fontfamily{phv}\selectfont \textbf{a}}}
    \put(410,280){\large {\fontfamily{phv}\selectfont \textbf{d}}}
   \put(200,140){\large {\fontfamily{phv}\selectfont \textbf{b}}} 
   \put(410,140){\large {\fontfamily{phv}\selectfont \textbf{e}}}
   \put(200,0){\large {\fontfamily{phv}\selectfont \textbf{c}}} 
   \put(410,0){\large {\fontfamily{phv}\selectfont \textbf{f}}} 
    \end{picture} 
    \vspace{10pt}
    \caption{Sample signals collected in this study. Panel a: full signals at damage level 2 under all loads from path 2-6; panel b: the first wave packets at damage level 2 under all loads from path 2-6; panel c: partial signals at damage level 2 under all loads from path 2-6; panel d: full signals under 5kN at all damage levels from path 3-4; panel e: the first wave packets under 5kN at all damage levels from path 3-4; panel f: partial signals under 5kN at all damage levels from path 3-4.}
\label{fig:signals_samples} 
\end{figure}
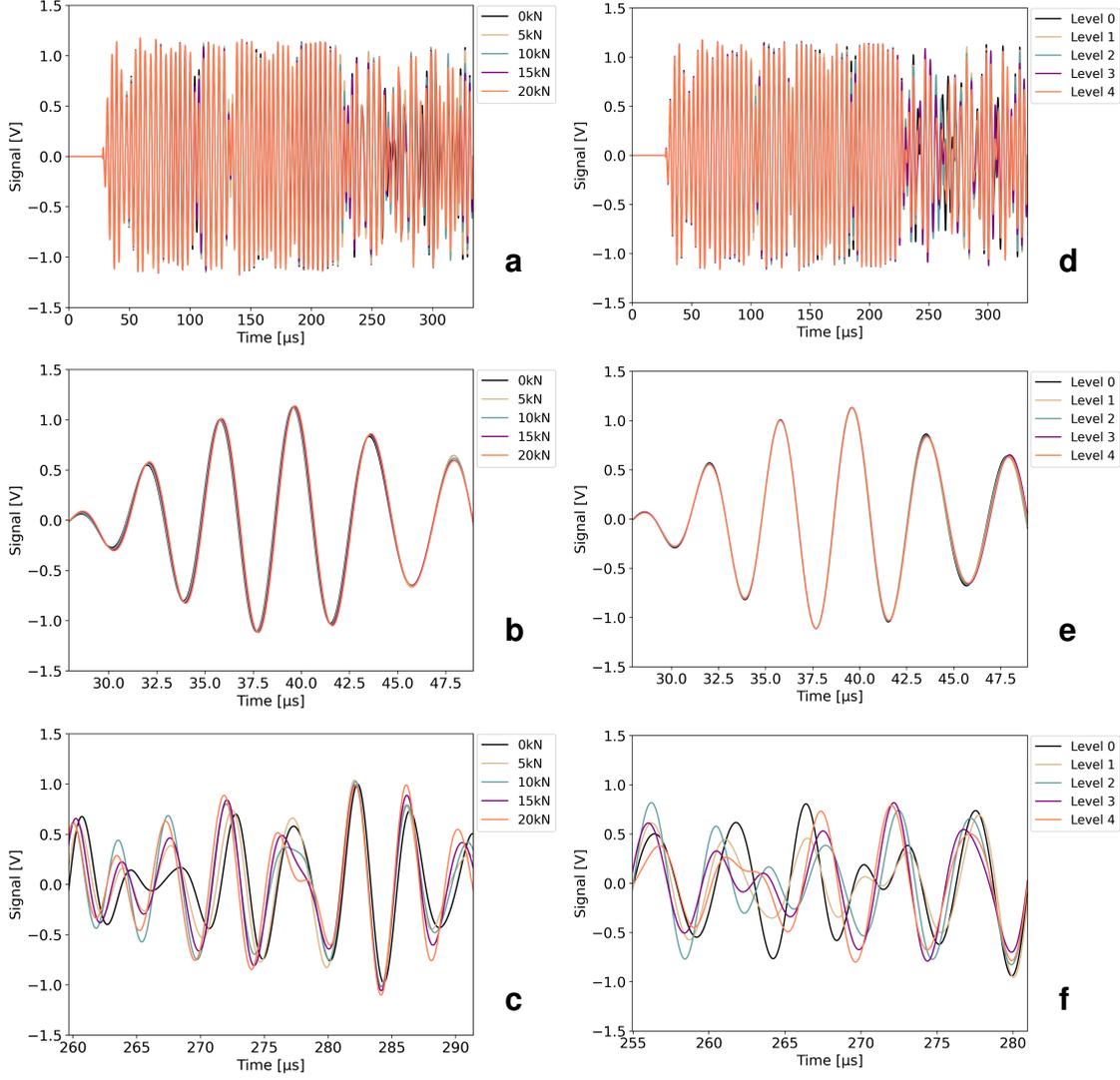

\begin{figure}[b!]
\centering
\includegraphics[scale=0.4]{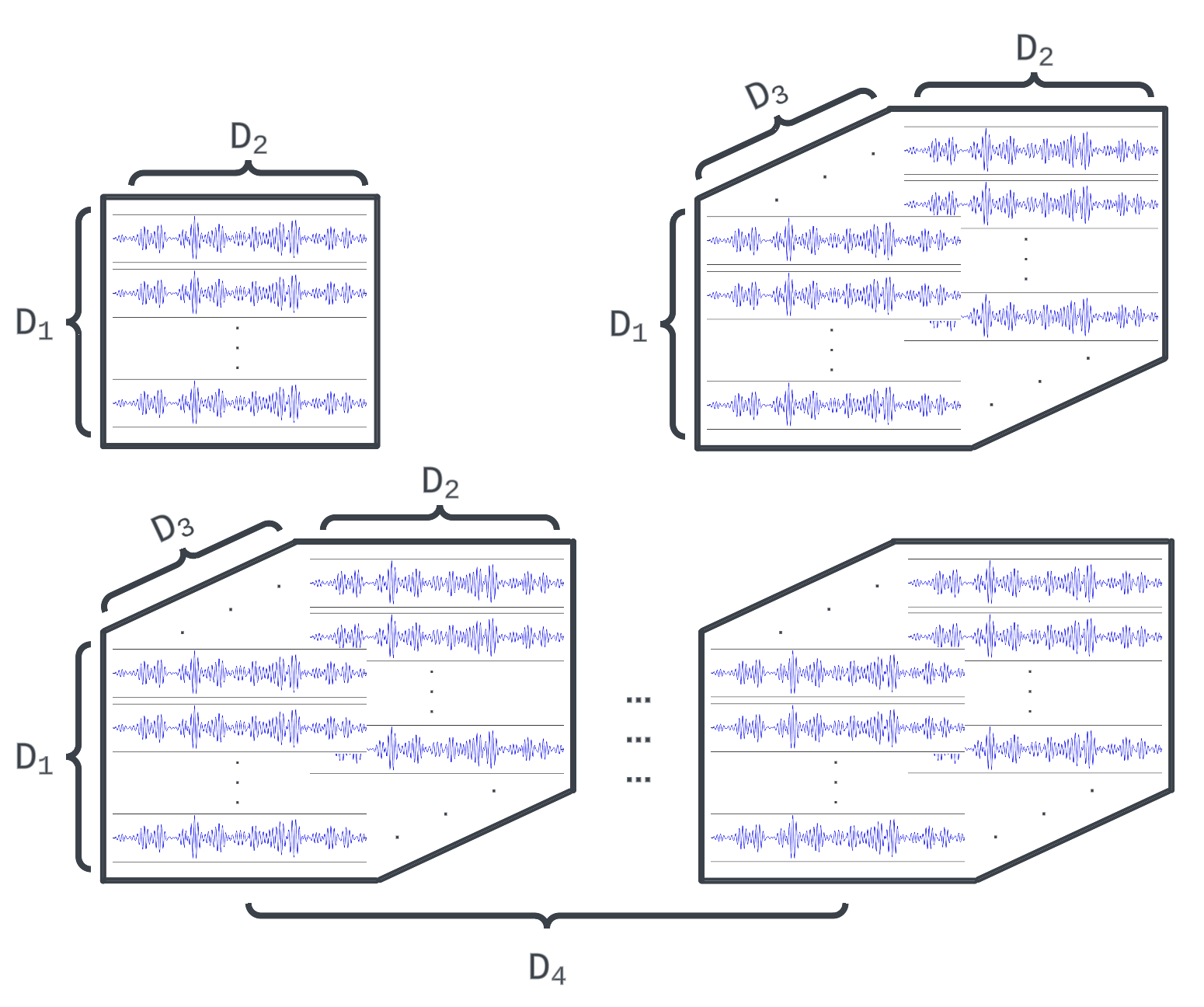}
\put(-200,145){\large {\fontfamily{phv}\selectfont \textbf{a}}}
\put(-20,145){\large {\fontfamily{phv}\selectfont \textbf{b}}} 
\put(-20,30){\large {\fontfamily{phv}\selectfont \textbf{c}}} 
\caption{Three ways tested in this work to construct the input tensor.}
\label{fig:const} 
\end{figure}
%
%
\subsection{{Experimental Setup}} \label{sec:exp_setup}

To assess the proposed framework, experiment was conducted on a 152.4 × 304.8 mm (6 × 12 in) 6061 Aluminum coupon (0.093 in thick) in a Lamb wave-based active sensing scope. As shown in Figure \ref{fig:Al_coupon}, six PZT-5A sensors that form a spatially distributed array were instrumented on the coupon surface at certain locations serving as actuators and receivers depending on their positions. To guarantee stable functionalities, the adhesive was cured under vacuum at room temperature for one day. In this work, sensors in the first row (indices 1-3) served as actuators in turn to provide 5-peak tone bursts sine wave signals. Meanwhile, sensors in the second row (indices 4-6) were used as receivers for data collection. 

To simulate the scenarios with different damage levels, an increasing number of weights have been attached starting at the middle spot in a sequence. The four three-gram weights can imitate up to five different damage conditions including the healthy case. In addition, external loads were provided by inserting the Al plate onto a tensile testing machine (Instron, Inc). Five different loads, i.e., 0, 5, 10, 15 and 20 kN, with an increment of 5kN were applied to simulate different EOCs. This setting gave a total of 25 states with $M_1=M_2=5$.

Under each state, the actuators functioned one at a time in a round-robin fashion while the receivers worked simultaneously. A total of 20 waveforms were collected from each path excluding at 20kN loading condition where only 2 trials were completed in order to figure out how lacking of data would affect the model performance:
\begin{equation}
n_{i,j}=
\begin{cases}
  2, & \text{if}\ k_j=20kN \\
  20, & \text{otherwise}
\end{cases}
\end{equation}

There were therefore 3690 pieces of time series collected from the nine sensor paths in total. Each time series was gathered within a time interval of 333.33$\mu$s containing 8,000 data points with a 24 MHz sampling rate. All the data was digitized and collected using a ScanGenie III data acquisition system (Acellent Technologies, Inc). In normal states, 8 out of the 20 trials were used as training set and the rest 12 trials were used as testing set. In terms of 20kN case, 1 trial was used for training and the other one  was for testing. Figure \ref{fig:signals_samples} shows some sample signals collected. Panels a to c include full and partial signals of damage level 2 under all available loads from path 2-6 while panels d to f show the signals under 5kN at all available damage levels from path 3-4. It can be observed that the first wave packets under various damage conditions and EOCs as shown in panels b and e have only slight differences which can be hard to capture. While the time series have much more distinct patterns at other periods as in panels c and f. For this reason, full signals were considered in this work to create the state classifiers. The NNs were trained and tested on a workstation with Intel Xeon Gold 6148 CPU with 64 GB RAM running at 2.4GHz .

\subsection{{Data Processing}} \label{sec:data_process}

    
%

The data processing method is relatively trivial in this study, making it easy to be extended and generalized in other conditions. First, data was down-sampled from $N=8,000$ to $N=800$ per time series while decreasing the sample frequency from 24 to 2.4 MHz—only one-tenth of the original rate. This down-sampling helps mitigate noise and decreases the number of model parameters that need to be trained. Second, the signals were standardized prior to being used as inputs to enhance model stability and performance, following common practices in data-driven methods, as described by the following equation:

\begin{equation}
{y'}[t] = \frac{{y}[t]-{\mu}}{\sigma}
\end{equation}
where $\mathbf{\mu}$ and $\sigma$ are the mean and standard deviation of the time series, respectively. 

The last step to process the data was to construct the input tensors. As discussed, three different ways of construction were considered as shown in Figure \ref{fig:const}. The first way is a common one that is widely used in related work, in which signals under different states are stacked along a single dimension $D_1$ as shown in Figure \ref{fig:const} panel a. With this construction, the tensor can be easily enlarged when novel data is available, simplifying both training and testing processes. The first dimension herein integrates all available states including both damage level $k^1$ and external loads $k^2$ as well as the path index $p$ and hence has the length of $P\sum_{i=1}^{M_1} \sum_{j=1}^{M_2} n_{i,j}$, where $P$ is total number of paths. When put into the model, despite the tensor’s flexibility in the first dimension, it can only be convolved along the second dimension $D_2$ with length $N$ in CAE using 1D kernels. In terms of a large data set where EOCs are complex, value of $D_1$ might be excessively large, making the parameter tuning during the training process time-consuming. Under this concern, more construction methods were considered.

The second way can be regarded as an extension of the first one, where the path information is extracted into an individual dimension $D_3$. This setting gives the possibility to decrease the length of $D_1$ while other dimensions can be further compressed during the convolution process. To be specific, $D_1 = \sum_{i=1}^{M_1} \sum_{j=1}^{M_2} n_{i,j}$, $D_2$ still equals to the length of each time series $N$ while $D_3$ has a length of $P$. By using 2D kernels, $D_2$ and $D_3$ can be convolved simultaneously in a more efficient manner comparing to the previous case.

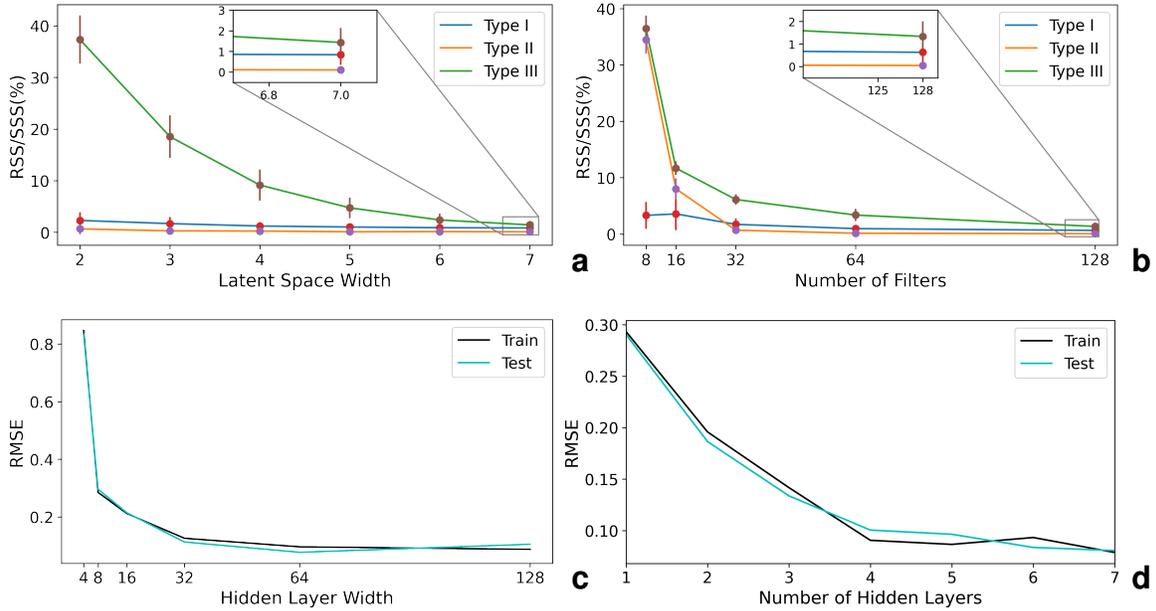
\begin{figure}[!t]
    \begin{picture}(500,240)
    \put(10,120){\includegraphics[width=0.45\textwidth]{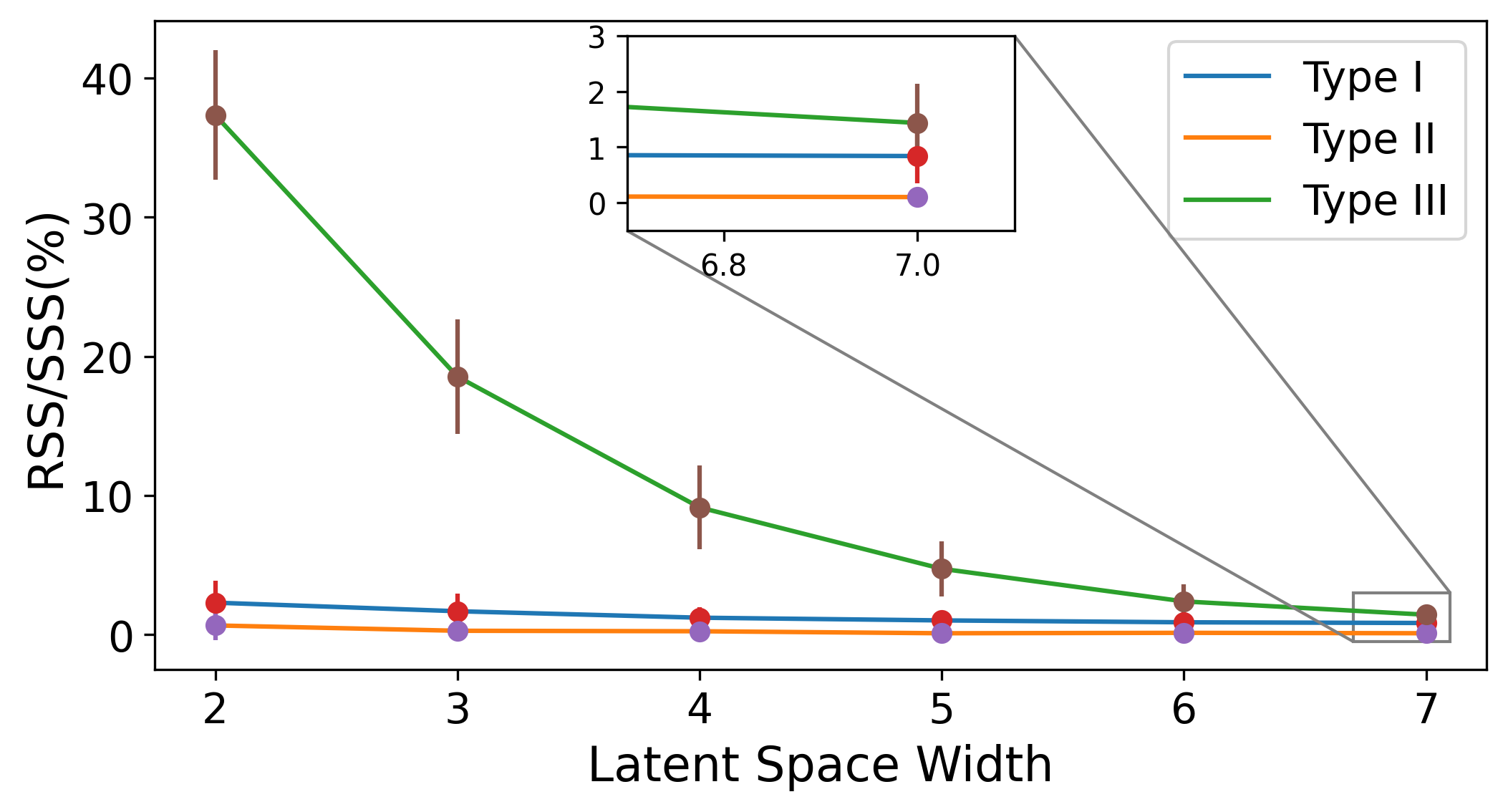}}
    \put(224,120){\includegraphics[width=0.45\textwidth]{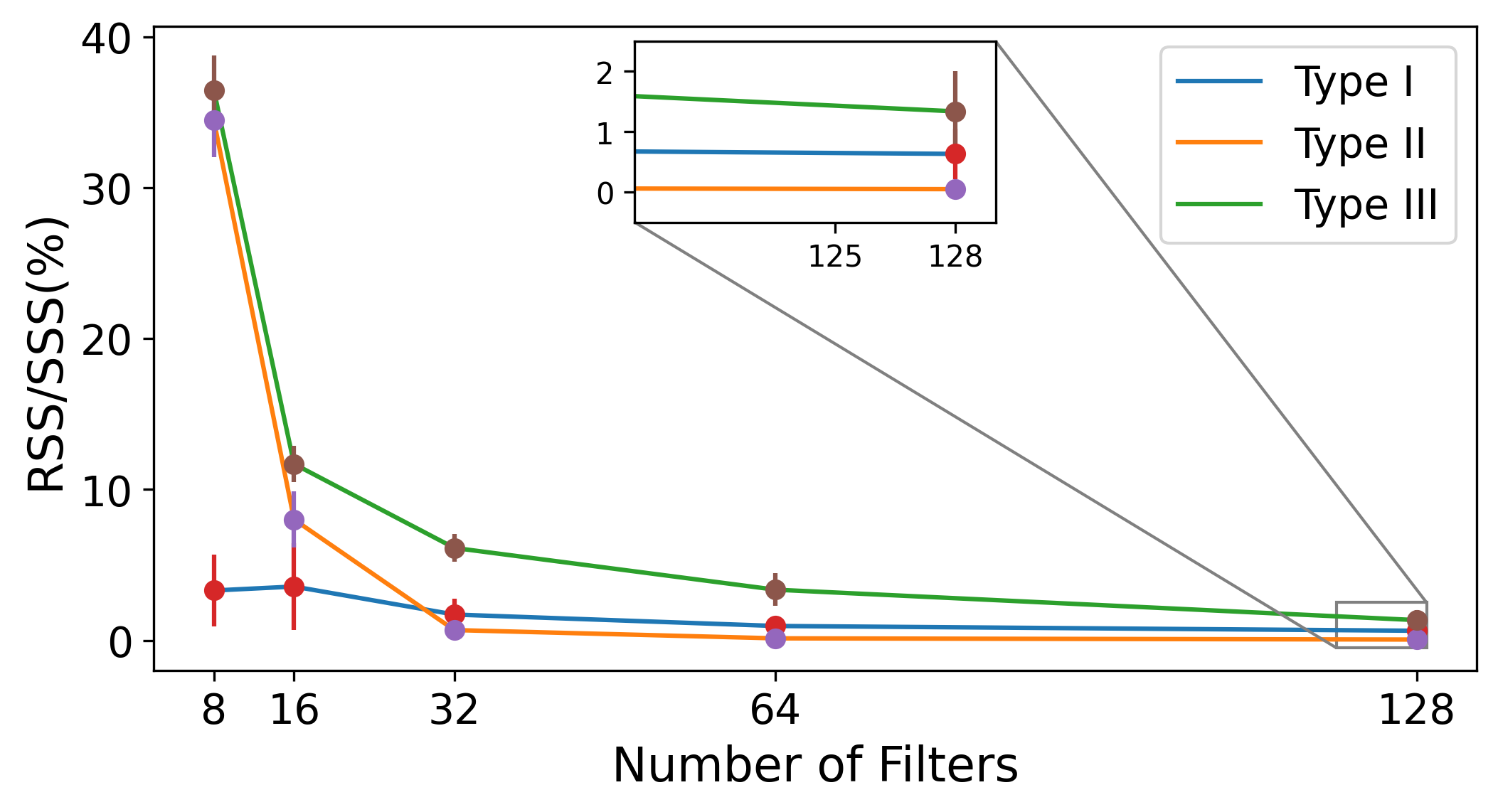}}
    \put(10,0){\includegraphics[width=0.45\textwidth]{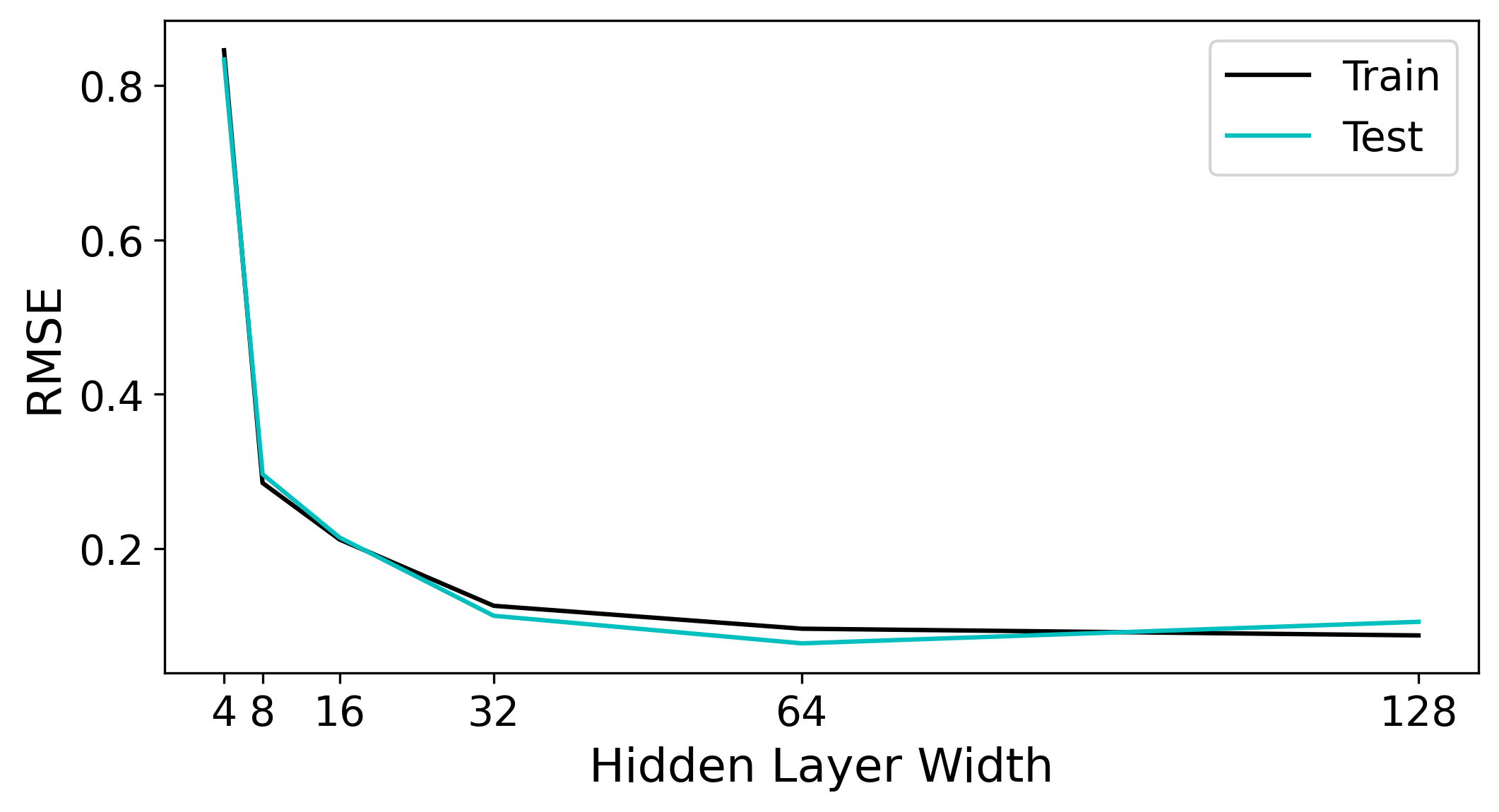}}
    \put(220,0){\includegraphics[width=0.46\textwidth]{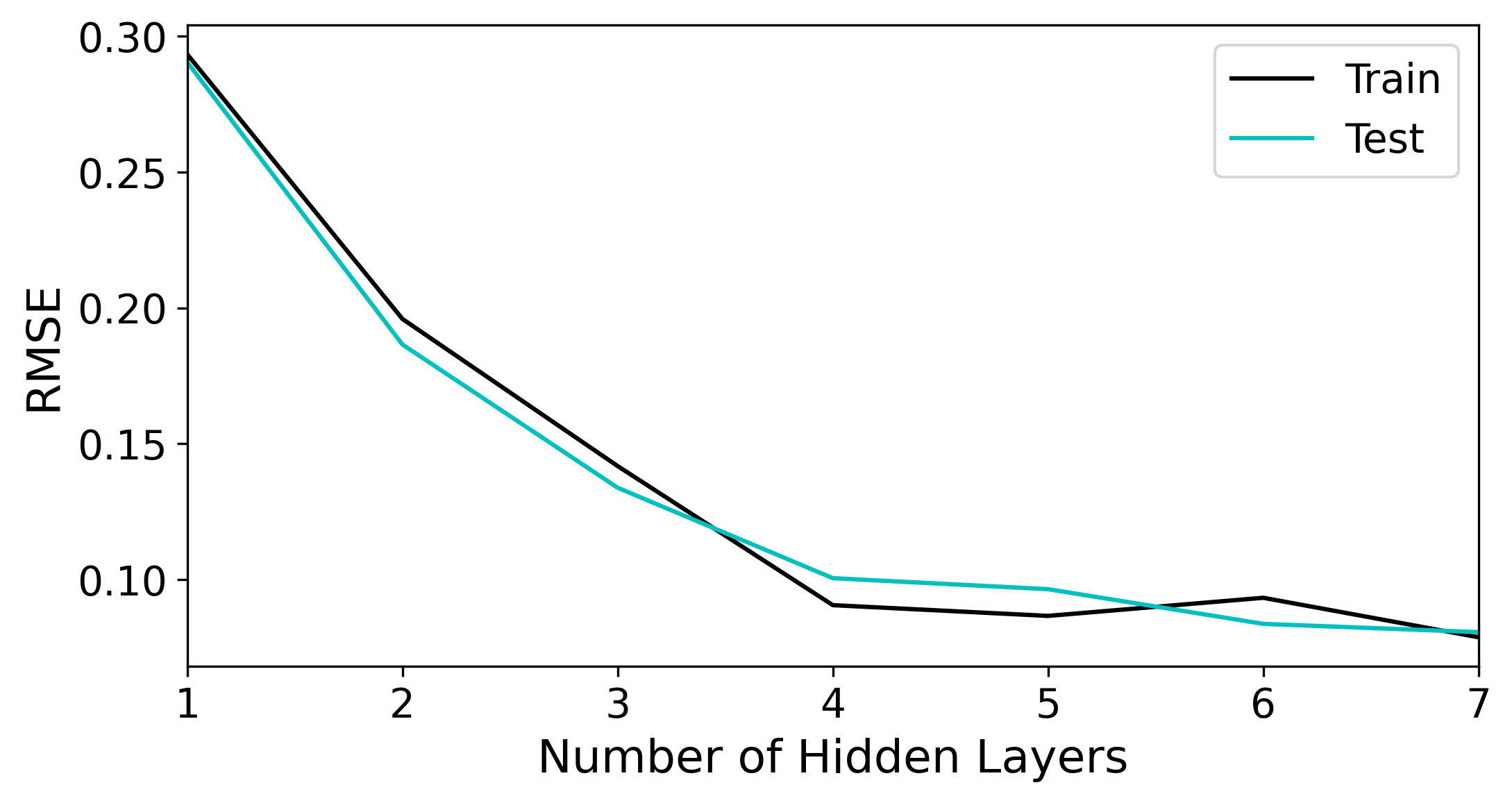}}
    \put(438,130){\color{black} \large {\fontfamily{phv}\selectfont \textbf{b}}}
    \put(226,130){\large {\fontfamily{phv}\selectfont \textbf{a}}}
   \put(438,10){\large {\fontfamily{phv}\selectfont \textbf{d}}} 
   \put(226,10){\large {\fontfamily{phv}\selectfont \textbf{c}}} 
    \end{picture} 
    \caption{Hyperparameter configuration of the networks in this study. Panel a and b: Error bar plots of the reconstruction results for CAE configuration. Panel c and d: RMSE w.r.t number of hidden layer width and hidden layers, respectively, for FFNN configuration.}
\label{fig:cae_config} 
\end{figure}

For the third option, as shown in Figure \ref{fig:const} panel c, the dimension representing the paths in the second method is further refactored into two dimensions, i.e., $D_3$ and $D_4$, aiming to further compress the path information. 2D convolution is then applied with respect to these two dimensions. This option might be advantageous when the path number is large in situations such as complex sensor array applications. These three methods will lead to CAEs with three different architectures. The performance of the three types of models, i.e., Type I, Type II and Type III, will be assessed and compared in Section \ref{Sec:results}.


\begin{table}[b!]
\centering
\caption{\uppercase{Type I CAE architecture.}} \label{tab:cae_compare1} 
{\small
\begin{tabular}{lllr}
\hline
  Layer & Output Shape & Parameters   \\
\hline
  \textbf{Encoder:}  &    &     \\
  Input Layer  &  800 $\times$ 1  &  0 \\
  Conv1D & 800 $\times$ 64  & 256 \\
  MaxPooling1D & 400 $\times$ 64  & 0 \\
  Conv1D & 400 $\times$ 32  & 6,176 \\
  MaxPooling1D & 200 $\times$ 32  & 0 \\
  Flatten & 6,400 & 0 \\
  Dense & 32 & 204,832 \\
  Dense & 7 & 231 \\
\hline
  \textbf{Decoder:} & & \\
  Dense & 32 & 256 \\
  Dense & 6400 & 211,200 \\
  Reshape & 200 $\times$ 32 & 0 \\
  Upsampling1D & 400 $\times$ 32 & 0 \\
  Conv1D & 400 $\times$ 64 & 6,208 \\
  Upsampling1D & 800 $\times$ 64 & 0 \\
  Conv1D & 800 $\times$ 1 & 193 \\
\hline
 Encoder Parameters & - & 211,495 \\
 Decoder Parameters & - & 217,857 \\
\hline
 Training Time & 15.8316s/epoch & \\
\hline

\end{tabular}
}
\end{table}

\begin{table}
\parbox[t!]{.5\linewidth}{
\small
\caption{\uppercase{Type II CAE architecture.}} \label{tab:cae_compare2} 
\begin{tabular}{ccc}
\hline
  Layer & Output Shape & Parameters   \\
\hline
  \textbf{Encoder:}  &    &     \\
  Input Layer  &  800 $\times$ 9 $\times$ 1  &  0 \\
  Conv2D & 800 $\times$ 9 $\times$ 64  & 640 \\
  MaxPooling2D & 400 $\times$ 3 $\times$ 64  & 0 \\
  Conv2D & 400 $\times$ 3 $\times$ 32  & 18,464 \\
  MaxPooling2D & 200 $\times$ 1 $\times$ 32  & 0 \\
  Flatten & 6,400 & 0 \\
  Dense & 32 & 204,832 \\
  Dense & 7 & 231 \\
\hline
  \textbf{Decoder:} & & \\
  Dense & 32 & 256 \\
  Dense & 6400 & 211,200 \\
  Reshape & 200 $\times$ 1 $\times$ 32 & 0 \\
  Upsampling2D & 400 $\times$ 3 $\times$ 32 & 0 \\
  Conv2D & 400 $\times$ 3 $\times$ 64 & 18,496 \\
  Upsampling2D & 800 $\times$ 9 $\times$ 64 & 0 \\
  Conv2D & 800 $\times$ 9 $\times$ 1 & 577 \\
\hline
 Encoder Parameters & - & 224,167 \\
 Decoder Parameters & - & 230,529 \\
\hline
 Training Time & 2.9310s/epoch & \\
\hline

\end{tabular}
}
\hfill
\parbox[t!]{.5\linewidth}{
\caption{\uppercase{Type III CAE architecture.}} \label{tab:cae_compare3}
\begin{tabular}{ccc}
\hline
  Layer & Output Shape & Parameters   \\
\hline
  \textbf{Encoder:}  &    &     \\
  Input Layer  &  800 $\times$ 3 $\times$ 3 $\times$ 1  &  0 \\
  Conv2D & 800 $\times$ 3 $\times$ 3 $\times$ 64  & 640 \\
  MaxPooling2D & 800 $\times$ 1 $\times$ 1 $\times$ 64  & 0 \\
  Conv2D & 800 $\times$ 1 $\times$ 1 $\times$ 32  & 18,464 \\
  Reshape & 800 $\times$ 32 & 0 \\
  Dense & 800 $\times$ 7 & 231 \\
\hline
  \textbf{Decoder:} & & \\
  Dense & 800 $\times$ 32 & 256 \\
  Reshape & 800 $\times$ 1 $\times$ 1 $\times$ 32 & 0 \\
  Conv2D & 800 $\times$ 1 $\times$ 1 $\times$ 64 & 18,496 \\
  Upsampling2D & 800 $\times$ 3 $\times$ 3 $\times$ 64 & 0 \\
  Conv2D & 800 $\times$ 3 $\times$ 3 $\times$ 1 & 577 \\
\hline
 Encoder Parameters & - & 19,335 \\
 Decoder Parameters & - & 19,329 \\
\hline
 Training Time & 3.4680s/epoch & \\
\hline

\end{tabular}
}
\end{table}

\section{Results and Discussion} \label{Sec:results}

\begin{figure}[b!]
    \begin{picture}(500,390)
    \put(10,200){\includegraphics[width=0.45\textwidth]{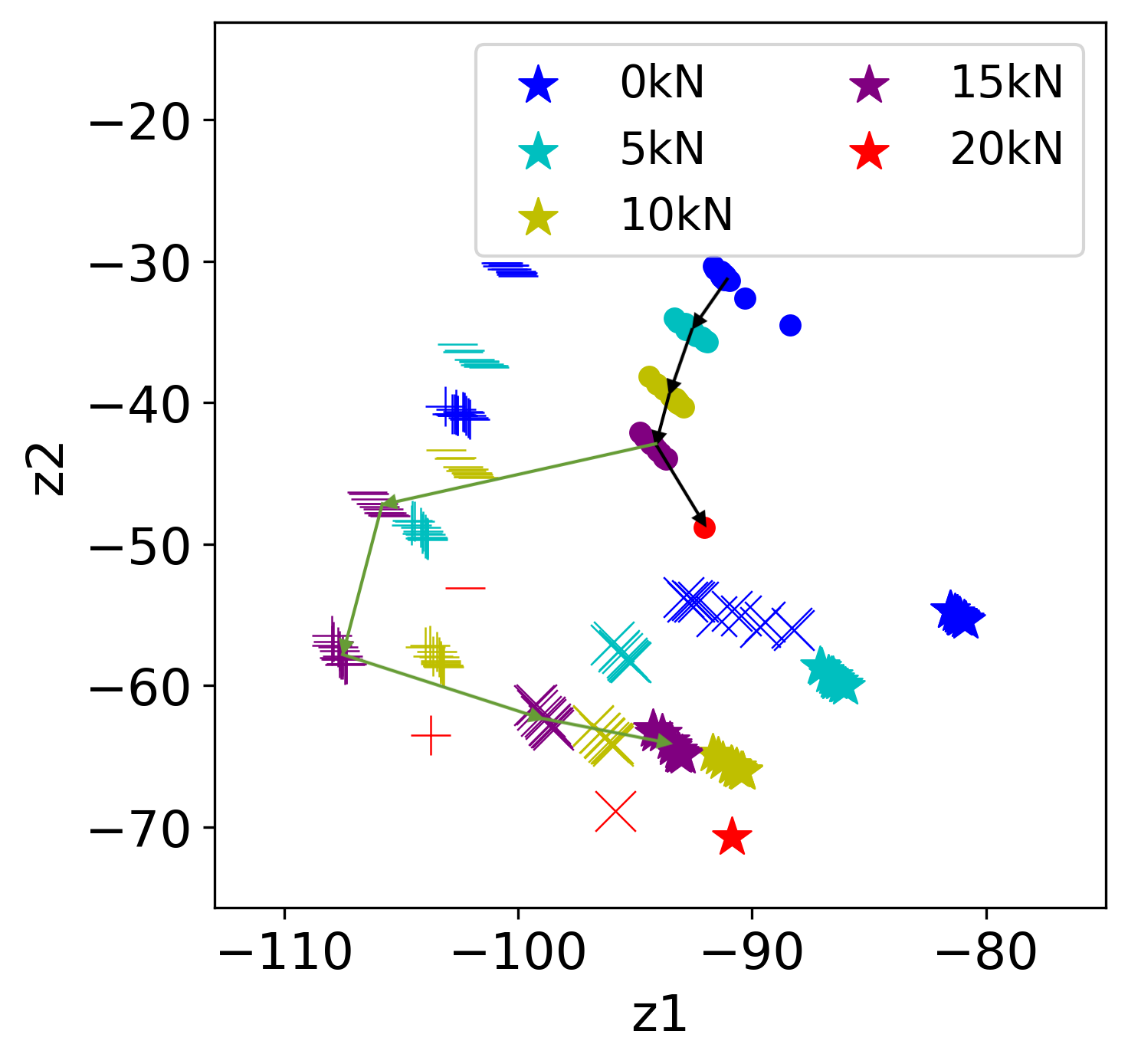}}
    \put(224,202){\includegraphics[width=0.46\textwidth]{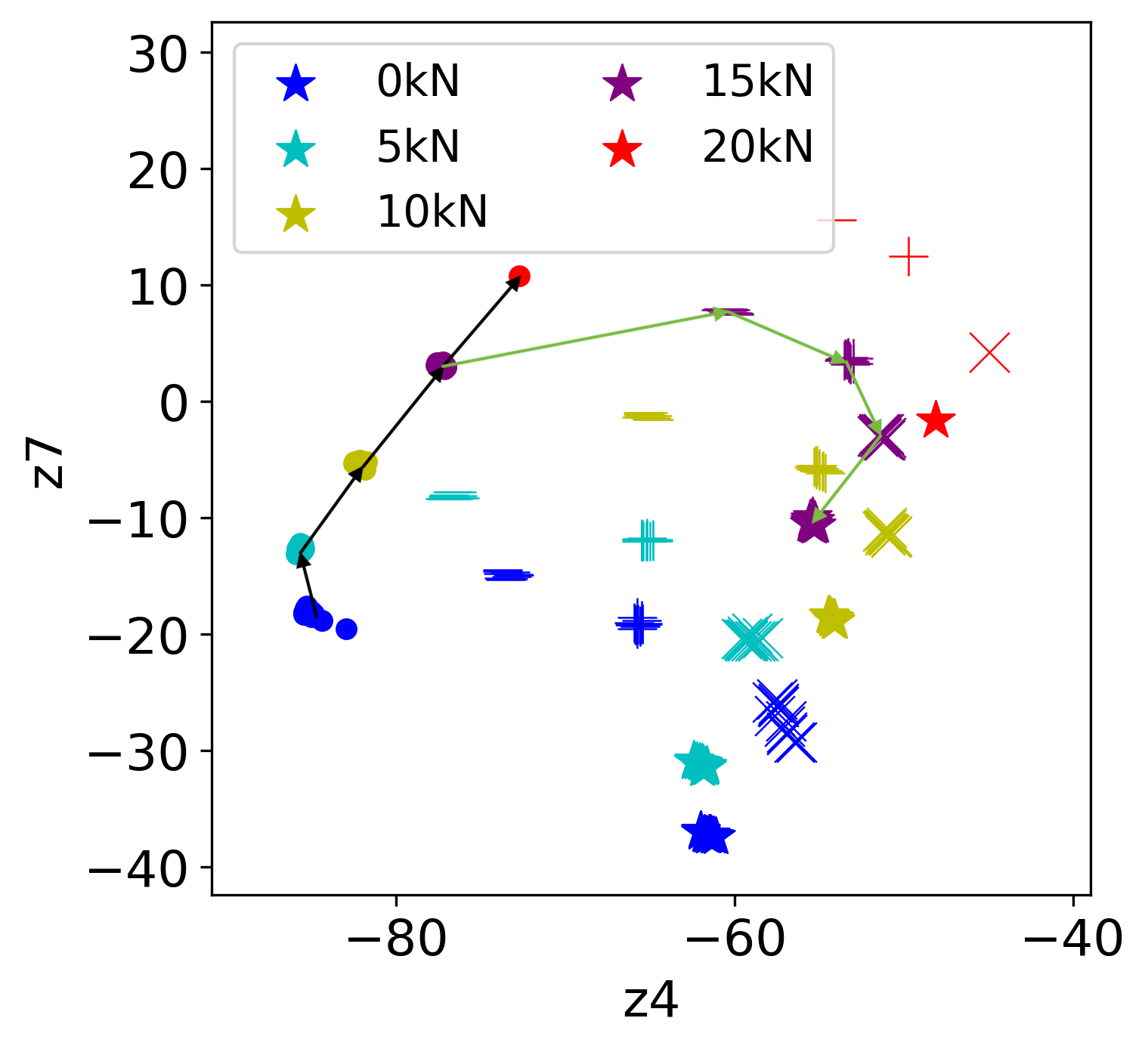}}
    \put(16,1){\includegraphics[width=0.437\textwidth]{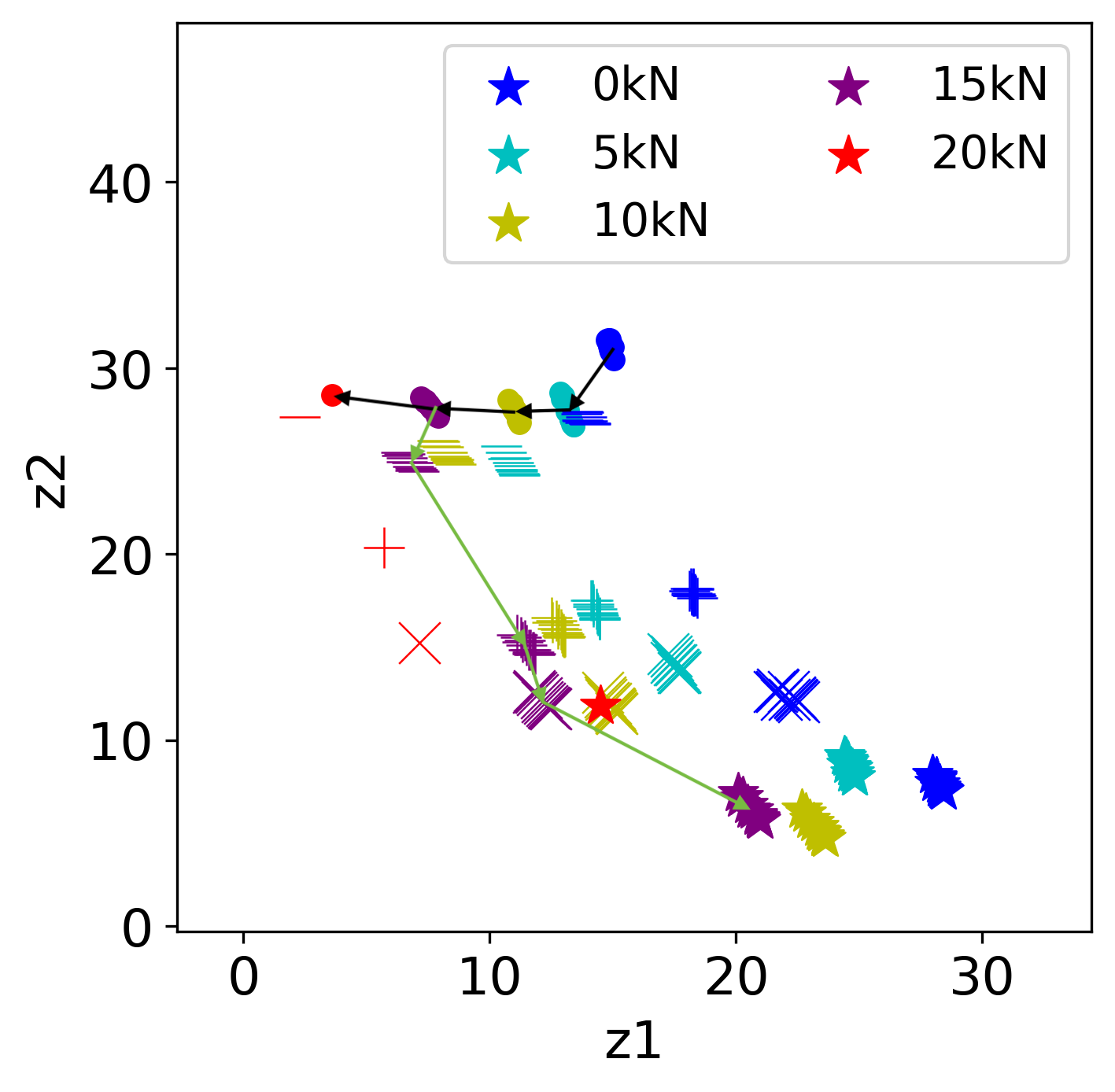}}
    \put(223,2){\includegraphics[width=0.454\textwidth]{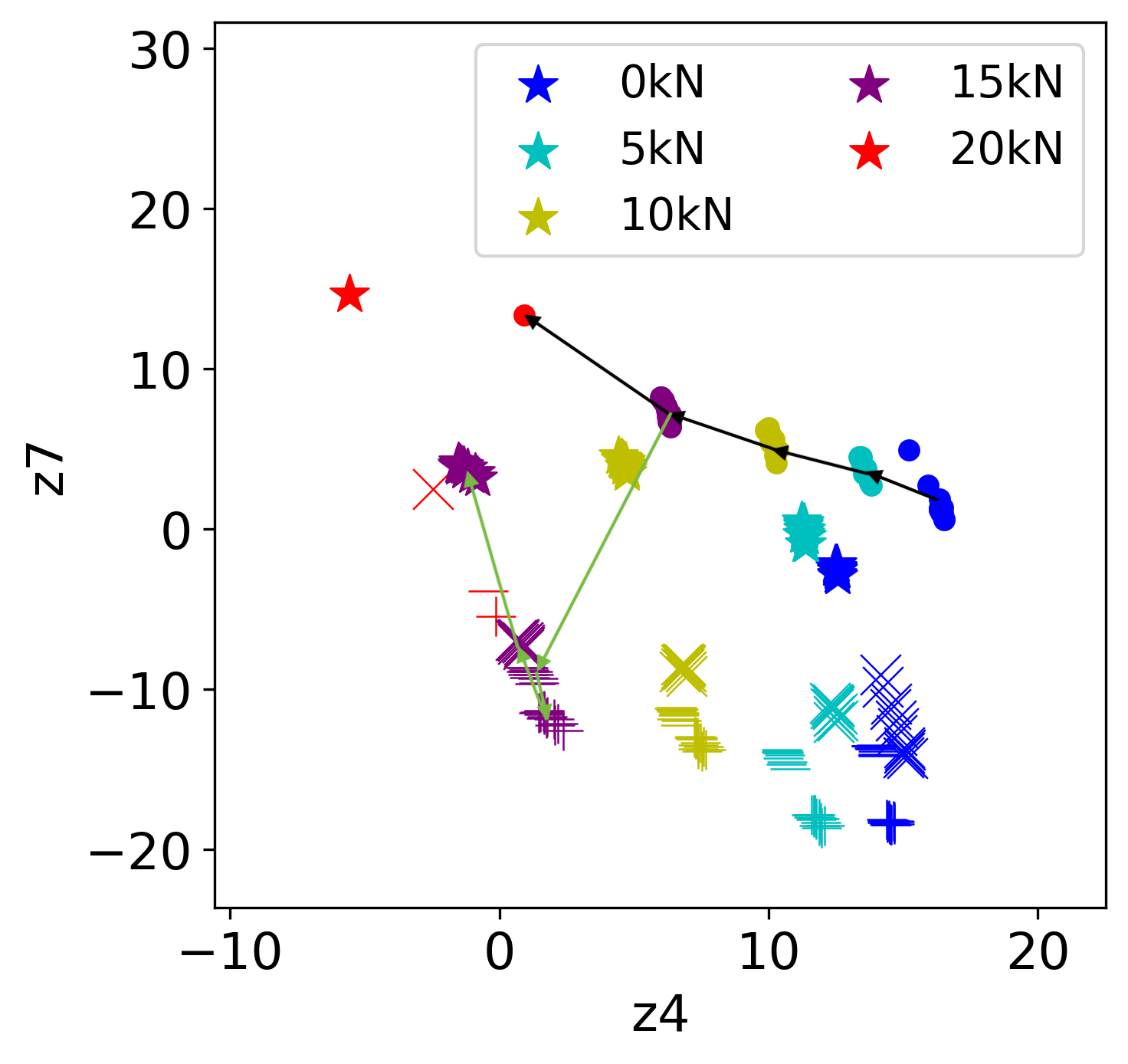}}
    \put(196,242){\color{black} \large {\fontfamily{phv}\selectfont \textbf{a}}}
    \put(410,242){\large {\fontfamily{phv}\selectfont \textbf{b}}}
   \put(196,42){\large {\fontfamily{phv}\selectfont \textbf{c}}} 
   \put(410,42){\large {\fontfamily{phv}\selectfont \textbf{d}}} 
    \end{picture} 
    \caption{Latent space representation under all available states from model Type I. '$\cdot$': level 0; '$-$': level 1; '$+$': level 2; '$\times$': level 3; '$\star$': level 4. Green arrows indicate the trajectory when increasing damage level under 15kN while the black arrows indicate the trajectory when increasing load at healthy state. Panel a: $z_2$ vs $z_1$ for path 2-6; panel b: $z_7$ vs $z_4$ for path 2-6; panel c: $z_2$ vs $z_1$ for path 3-4; panel d: $z_7$ vs $z_4$ for path 3-4.}
\label{fig:lat1} 
\end{figure}

\begin{figure}[t!]
    \begin{picture}(500,185)
    \put(10,0){\includegraphics[width=0.45\textwidth]{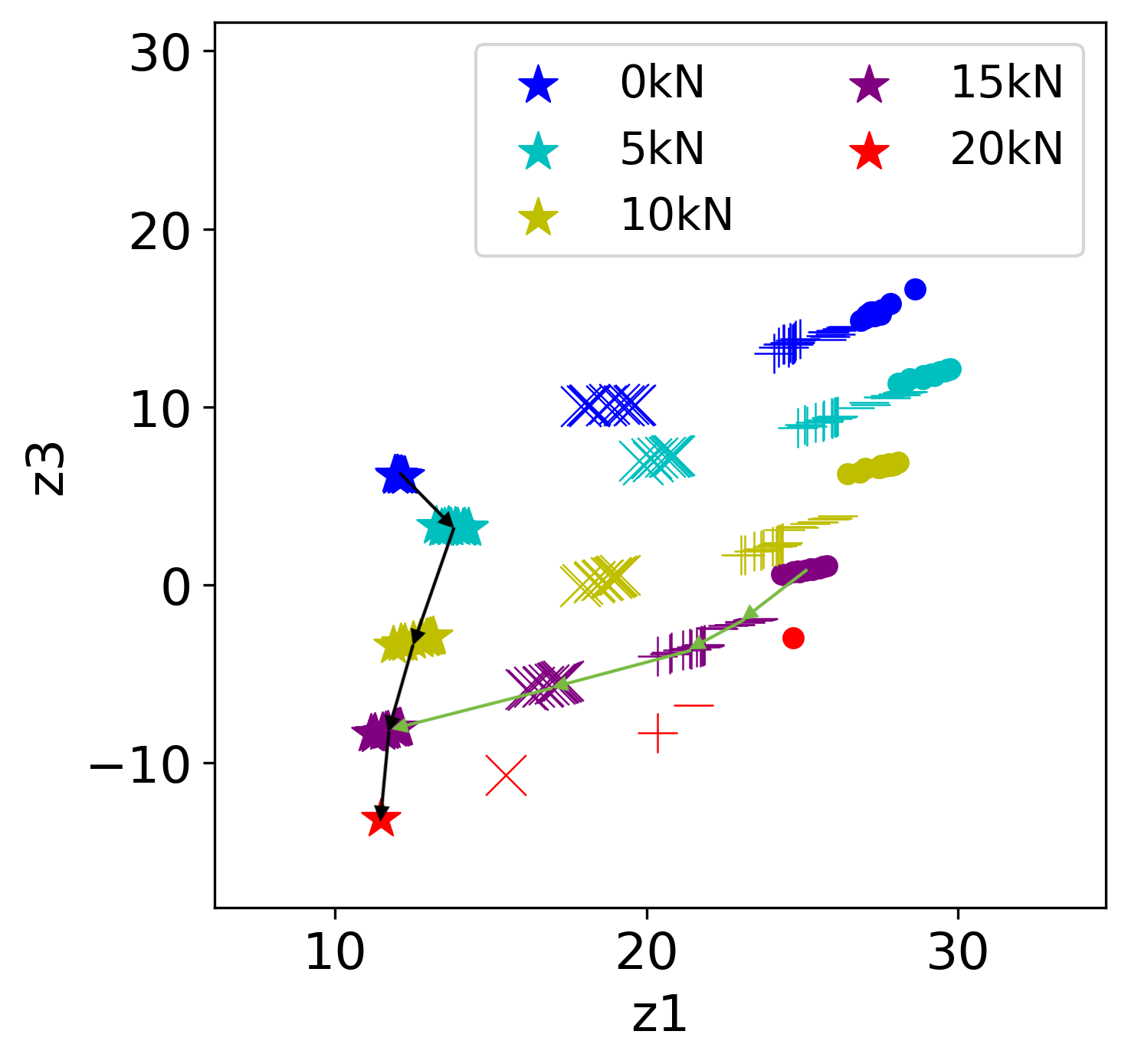}}
    \put(224,0){\includegraphics[width=0.434\textwidth]{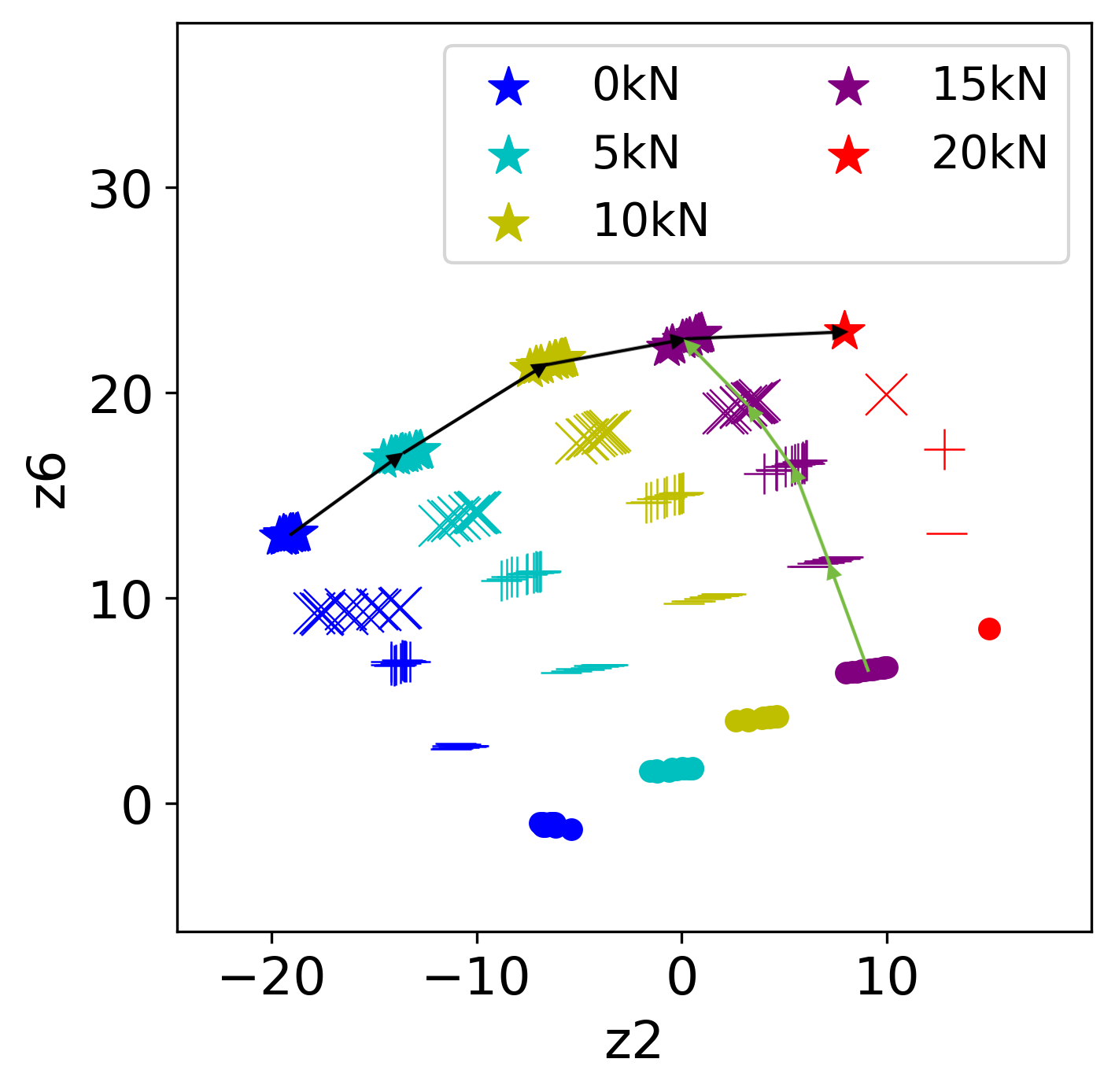}}
    \put(196,40){\color{black} \large {\fontfamily{phv}\selectfont \textbf{a}}}
    \put(404,40){\large {\fontfamily{phv}\selectfont \textbf{b}}}
    \end{picture} 
    \vspace{-20pt}\caption{Latent space representation under all available states from model Type II. '$\cdot$': level 0; '$-$': level 1; '$+$': level 2; '$\times$': level 3; '$\star$': level 4. Green arrows indicate the trajectory when increasing damage level under 15kN while the black arrows indicate the trajectory when increasing load at healthy state. Panel a: $z_3$ vs $z_1$; panel b: $z_6$ vs $z_2$.}
\label{fig:lat2} 
\end{figure}

\subsection{Training Phase}
Under each model type, the three aforementioned NNs-one CAE and two FFNNs—were trained individually. Each network contains several hyperparameters that can significantly influence model performance. To compare the three input construction methods effectively, it is essential to maintain consistency in the hyperparameters across each model type, necessitating a hyperparameter configuration step.

Guided by existing research, two convolutional layers were employed in both the encoder and decoder to achieve high reconstruction accuracy, with the number of filters varying according to different implementation scenarios \cite{sawant2023unsupervised,jana2022cnn}. To identify the optimal CAE configuration, two hyperparameters—the latent space width and the number of filters in the first convolutional layer—were thoroughly analyzed by calculating and comparing the reconstruction errors of the testing data. The number of filters in the second convolutional layer was set to half that of the first layer to reduce the number of parameters. To ensure efficient data compression, the width of the latent space was limited while remaining sufficiently broad to yield low reconstruction errors. In this study, the latent space width was set to range from 2 to 7, and the number of filters was iterated over the values [8, 16, 32, 64, 128]. This setup was motivated by two considerations: 1) a single latent variable may struggle to provide accurate results, particularly when multiple state factors coexist and are difficult to differentiate in the feature space; 2) using filter numbers in powers of two may accelerate the computation process. These steps were repeated for all three types of models derived from the three tensor construction approaches.

To visualize the model performance as the hyperparameters change, the error bar plots of average RSS/SSS($\%$) along with the corresponding 95$\%$ CI are demonstrated in Figure \ref{fig:cae_config}. Note that one hyperparameter remains constant while the other varies, in accordance with the principle of single variable analysis. A decreasing trend is observed for all three models as the latent space width increases. Due to the comparatively larger error associated with the Type I model, the trends for the other two models are less pronounced but still present. Similar conclusions can be drawn when varying the number of filters, despite some slight fluctuations at the beginning. However, it is important to note that the computational burden increases with the number of filters, leading to longer training times per epoch. Taking this into account, a latent width of 7 and a filter number of 64 were selected for the online testing phase, ensuring a fair evaluation across all models. The final detailed CAE configurations for the three CAEs can be found in Tables \ref{tab:cae_compare1} to \ref{tab:cae_compare3}.

Regarding the two FFNNs, their configuration procedures were also discussed. It is noteworthy that both networks aim to establish mappings between states and latent space vectors, but in reverse order. The opposite mapping directions suggest similar yet symmetric structures, meaning the configuration of one network will determine the structure of the other. Starting with the first FFNN, which takes the latent space vector as input and produces state vector estimates as output, the hyperparameters to be tuned include the width and depth of the layers. Following the same approach used for the CAE configuration, the hyperparameters of the FFNN were varied while monitoring the RMSE of the output. As shown in panels c and d of Figure \ref{fig:cae_config}, the RMSE decreases with an increase in either hidden layer width or network depth. However, the rate of decrease diminishes, indicating that significant computational costs need not be sacrificed for incremental performance gains. Accordingly, a hidden layer width of 64 was selected, and the number of hidden layers was set to 5 to achieve a low error while keeping computational costs relatively low.

\subsection{Testing Phase}

With the model configurations finalized, the trained networks are now prepared for integration in testing tasks. As previously introduced, the proposed framework serves two primary functions: estimating states and reconstructing signals. This section will evaluate both functions. Before presenting the results, the latent space corresponding to the input testing signals will be provided to offer readers insight into its characteristics. In each subsection covering latent space representation, state estimation, and signal reconstruction, results from all three models will be presented. Finally, the robustness of the best-performing model will be validated using a sparse dataset.
\subsubsection{Latent Space}
\label{Sec:lat}
In the Type I model, the latent space is time-invariant, with the latent space width set to 7, resulting in 7 latent variables for each time series dataset. To gain a clearer understanding of the underlying details, each pair of latent variables from the test dataset was plotted, as illustrated in Figure \ref{fig:lat1}. In these plots, different colors represent various loads, while distinct marker shapes indicate different damage levels. At each state, multiple markers of the same color and shape correspond to signals collected from multiple trials.

At first glance, it is evident that markers sharing the same shape and color tend to cluster together. This indicates that the divergence in original signals caused by various structural and external conditions is preserved in the latent space, leading to a state-sensitive yet significantly compressed feature space. Furthermore, specific patterns emerge when observing the latent variable pairs as a single state factor changes. For instance, the trajectory indicated by the green arrows connecting the purple markers illustrates how the latent variable values change as the damage level increases under a load of 15 kN. Conversely, the black arrows connecting the dot markers demonstrate the trajectory when the load increases from 0 to 20 kN in the healthy case. These observable patterns across various states confirm that the latent space is sensitive not only to the damage states themselves but also to external EOCs. This finding suggests the feasibility of state estimation based on feature space analysis.

The ability to observe these patterns may prove beneficial in real-world applications, particularly when signal data in certain regions of the target domain is scarce or unavailable. For instance, in this study, signals under 2.5 kN are missing, although data is collected at 5 kN increments. Regression techniques, such as Gaussian Process models, can be utilized in the latent space to estimate latent variable values. Subsequently, the decoder can map these feature vectors back to signals, effectively filling this data gap. This potential will be validated in Section 4.2.4.

For Type II model, the time-invariant latent variable pairs $(z_1,z_3)$ and $(z_2,z_6)$ are shown in Figure \ref{fig:lat2}. Markers of the same color and shape remain clustered in the same region. Following the trajectories indicated by the black and green arrows, similar conclusions can be drawn to those of the Type I model regarding the increase in damage severity or load individually. However, the trend observed in this case is more pronounced and exhibits closer linearity compared to the features extracted by the first model.
%
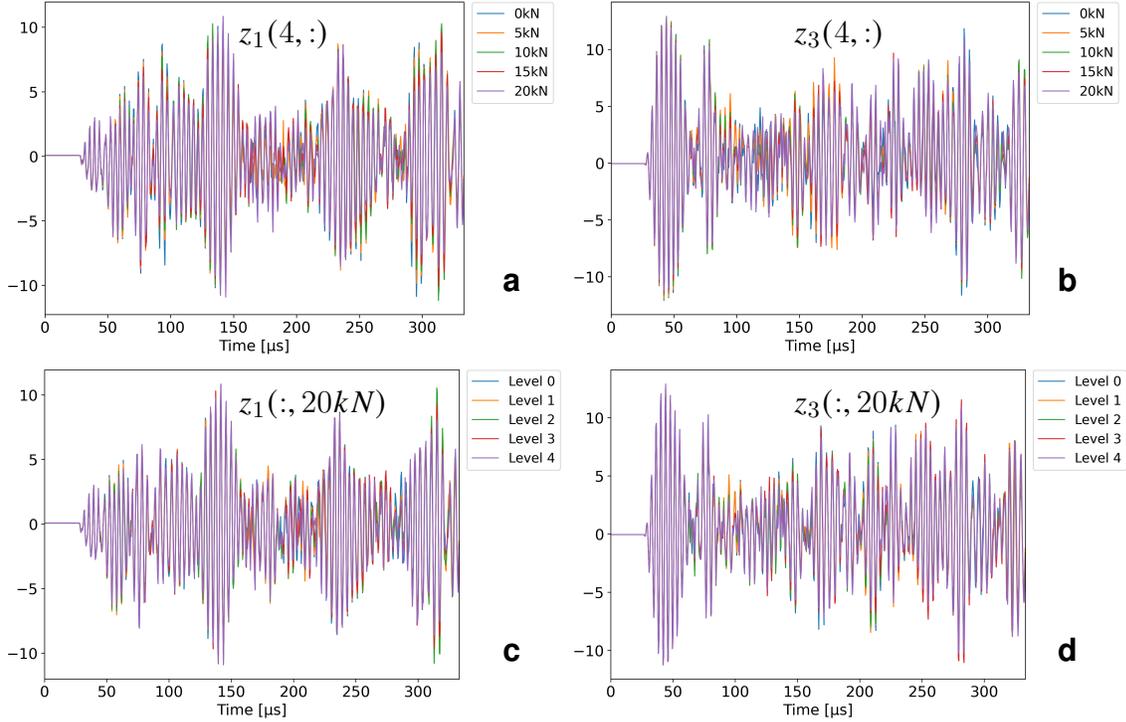
\begin{figure}[t!]
    \begin{picture}(500,410)
    \put(10,250){\includegraphics[width=0.45\textwidth]{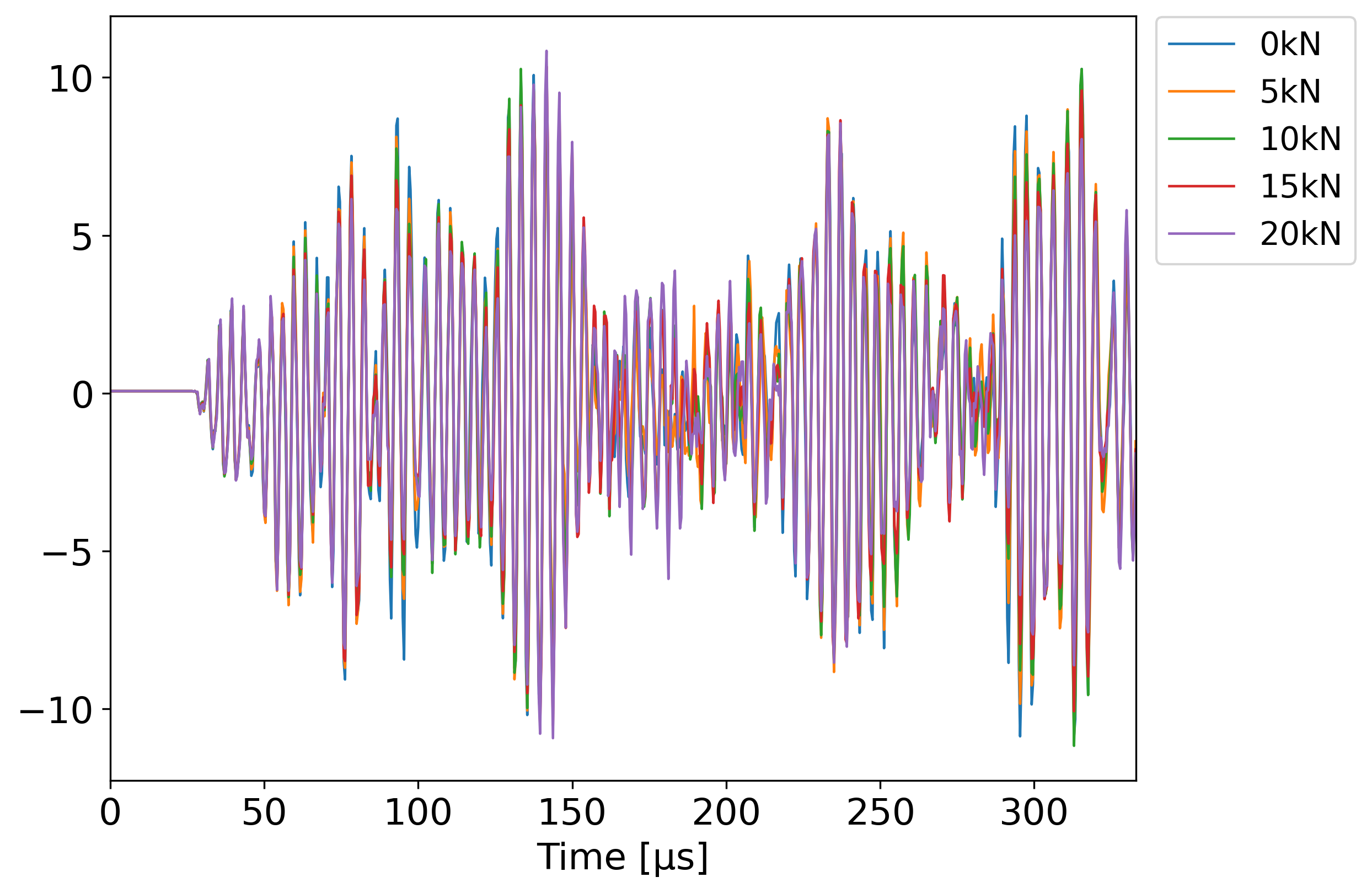}}
    \put(224,250){\includegraphics[width=0.45\textwidth]{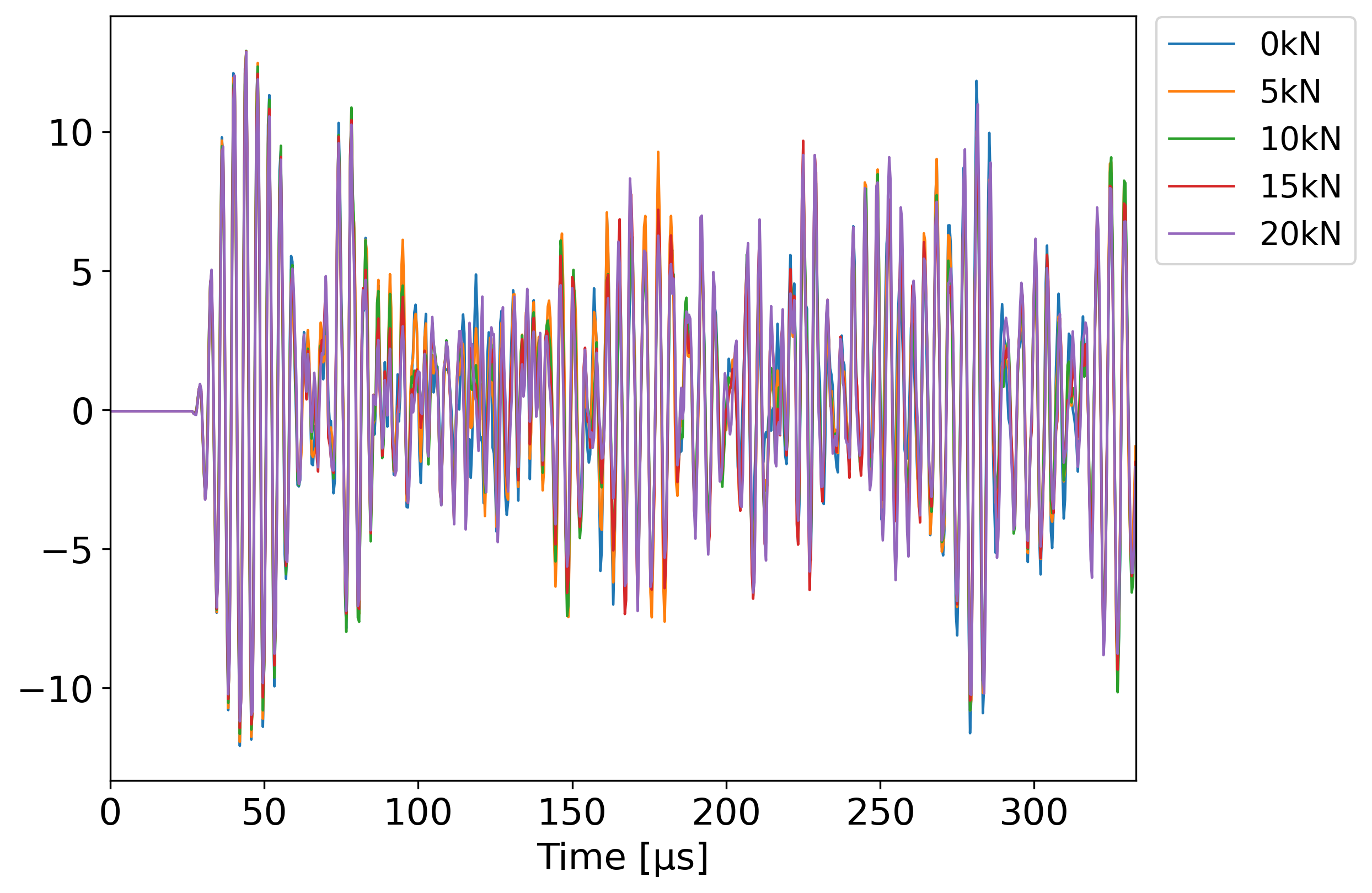}}
    \put(10,112){\includegraphics[width=0.456\textwidth]{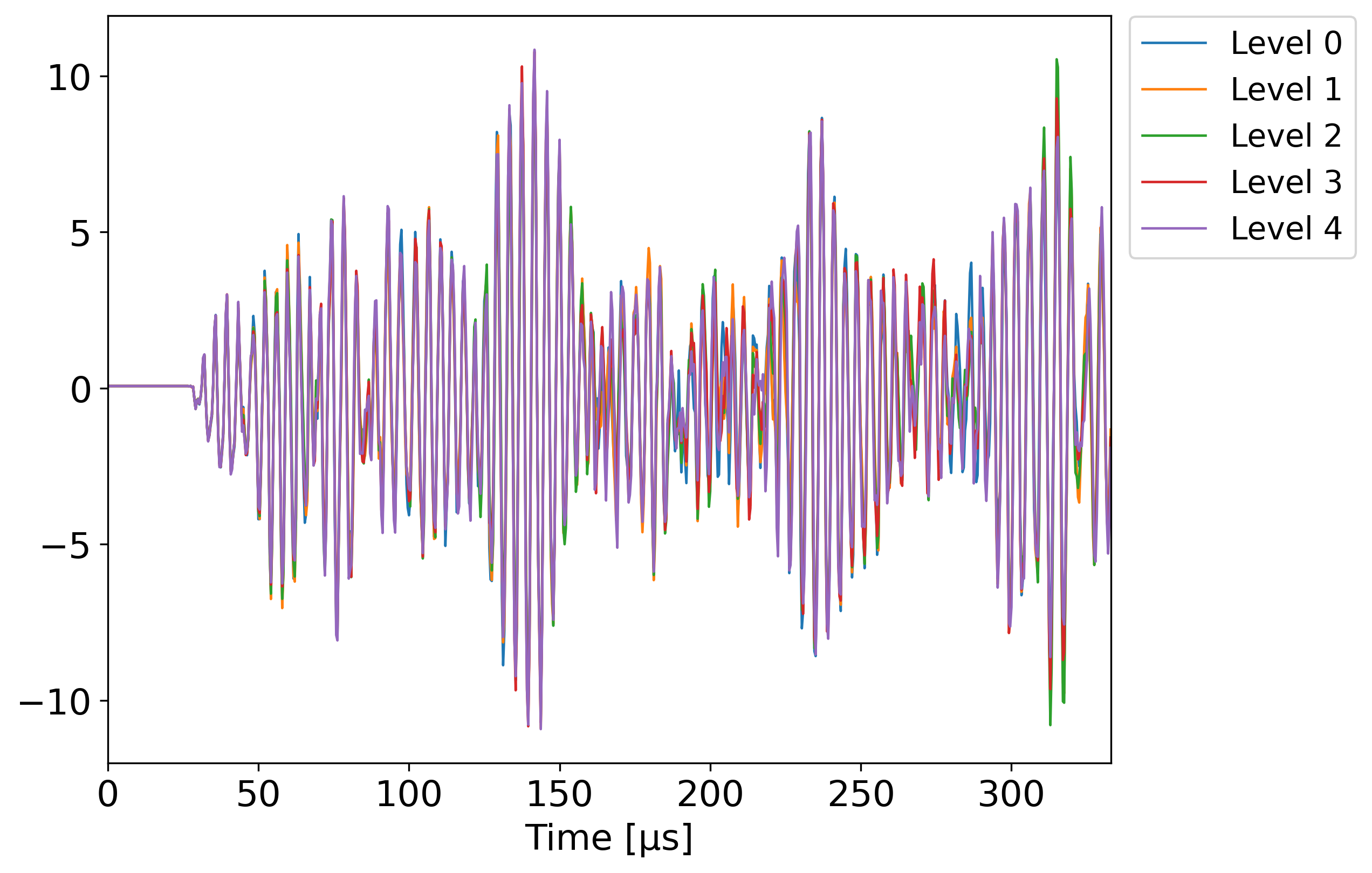}}
    \put(224,112){\includegraphics[width=0.456\textwidth]{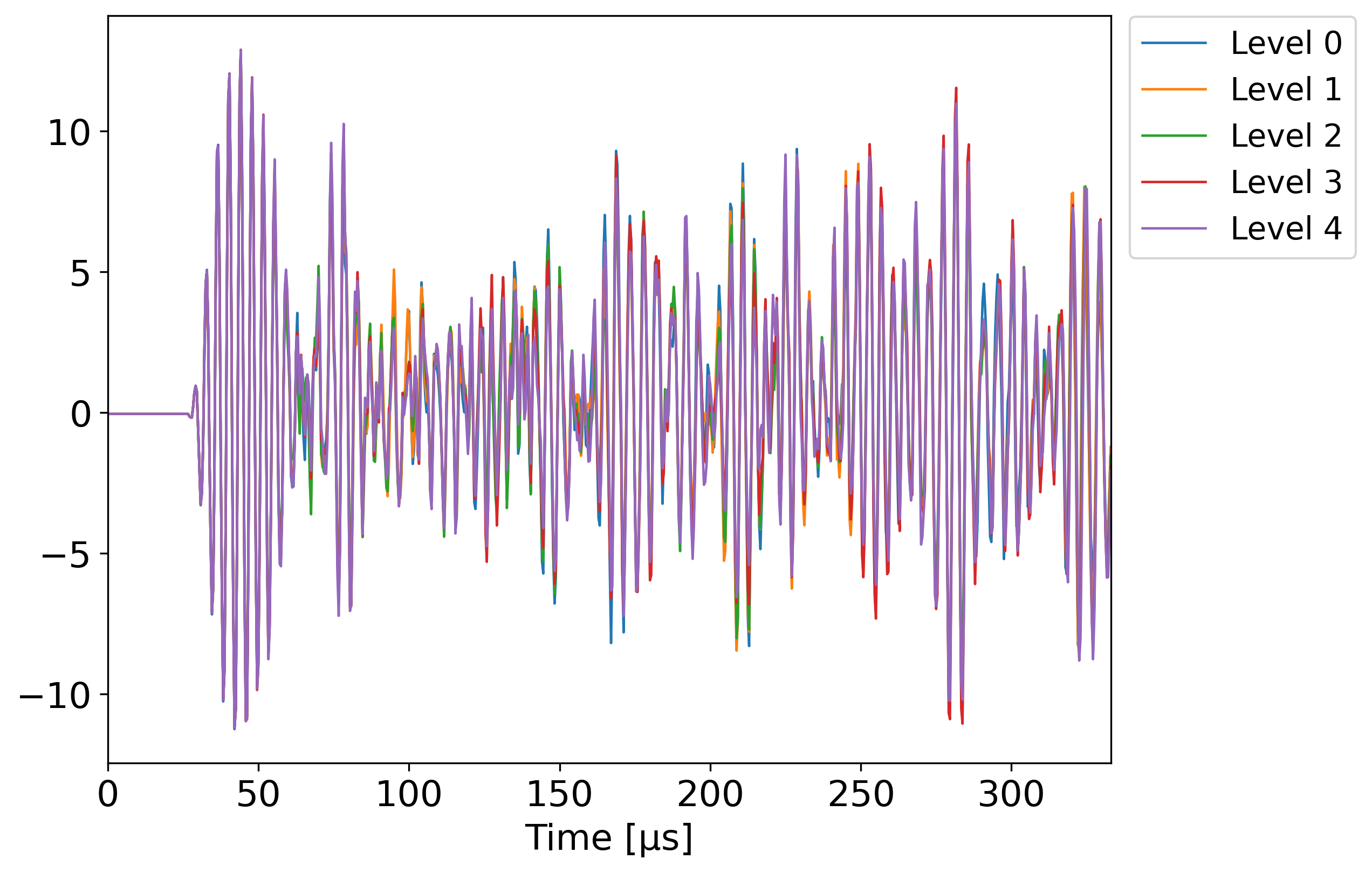}}
    \put(100,370){\color{black} \large {\fontfamily{phv}\selectfont \textbf{$z_1(4,:)$}}}
    \put(310,370){\color{black} \large {\fontfamily{phv}\selectfont \textbf{$z_{3}(4,:)$}}}
    \put(100,230){\color{black} \large {\fontfamily{phv}\selectfont \textbf{$z_{1}(:,20kN)$}}}
    \put(310,230){\color{black} \large {\fontfamily{phv}\selectfont \textbf{$z_{3}(:,20kN)$}}}
    \put(200,276){\color{black} \large {\fontfamily{phv}\selectfont \textbf{a}}}
    \put(410,276){\large {\fontfamily{phv}\selectfont \textbf{b}}}
   \put(200,136){\large {\fontfamily{phv}\selectfont \textbf{c}}} 
   \put(410,136){\large {\fontfamily{phv}\selectfont \textbf{d}}} 
    \end{picture} 
    \vspace{-120pt}
    \caption{Latent space representation from model Type III. Panel a: the 1st variable $z_1$ under damage level 4 and all possible loads; panel b: the 3rd variable $z_3$ under damage level 4 and all possible loads; panel c: the 1st variable $z_1$ under 20kN and all damage levels; panel d: the 3rd variable $z_3$ under 20kN and all damage levels.}
\label{fig:lat3} 
%
\end{figure}
%

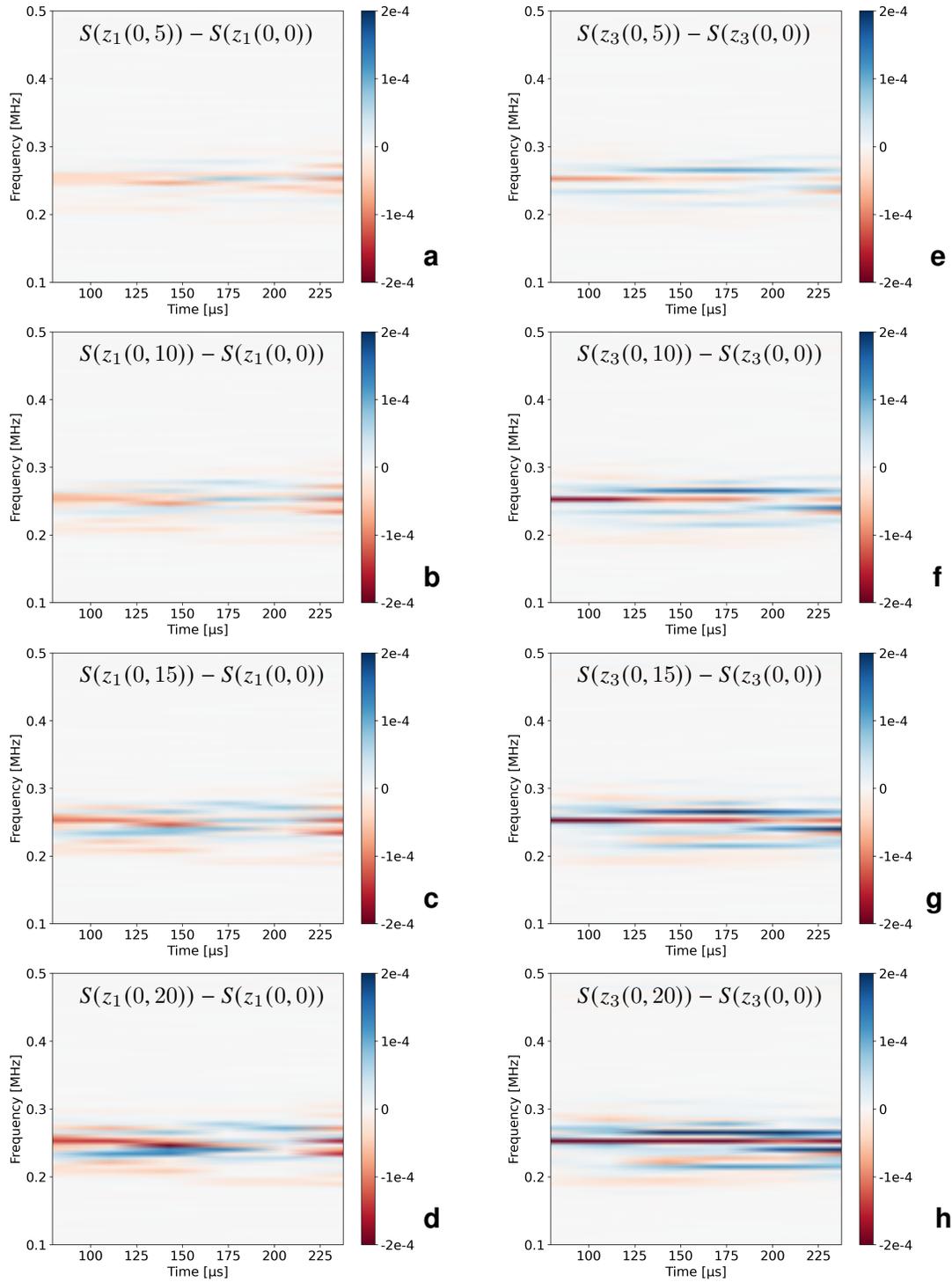
\begin{figure}[t!]
    \begin{picture}(500,530)
    \put(10,414){\includegraphics[width=0.38\textwidth]{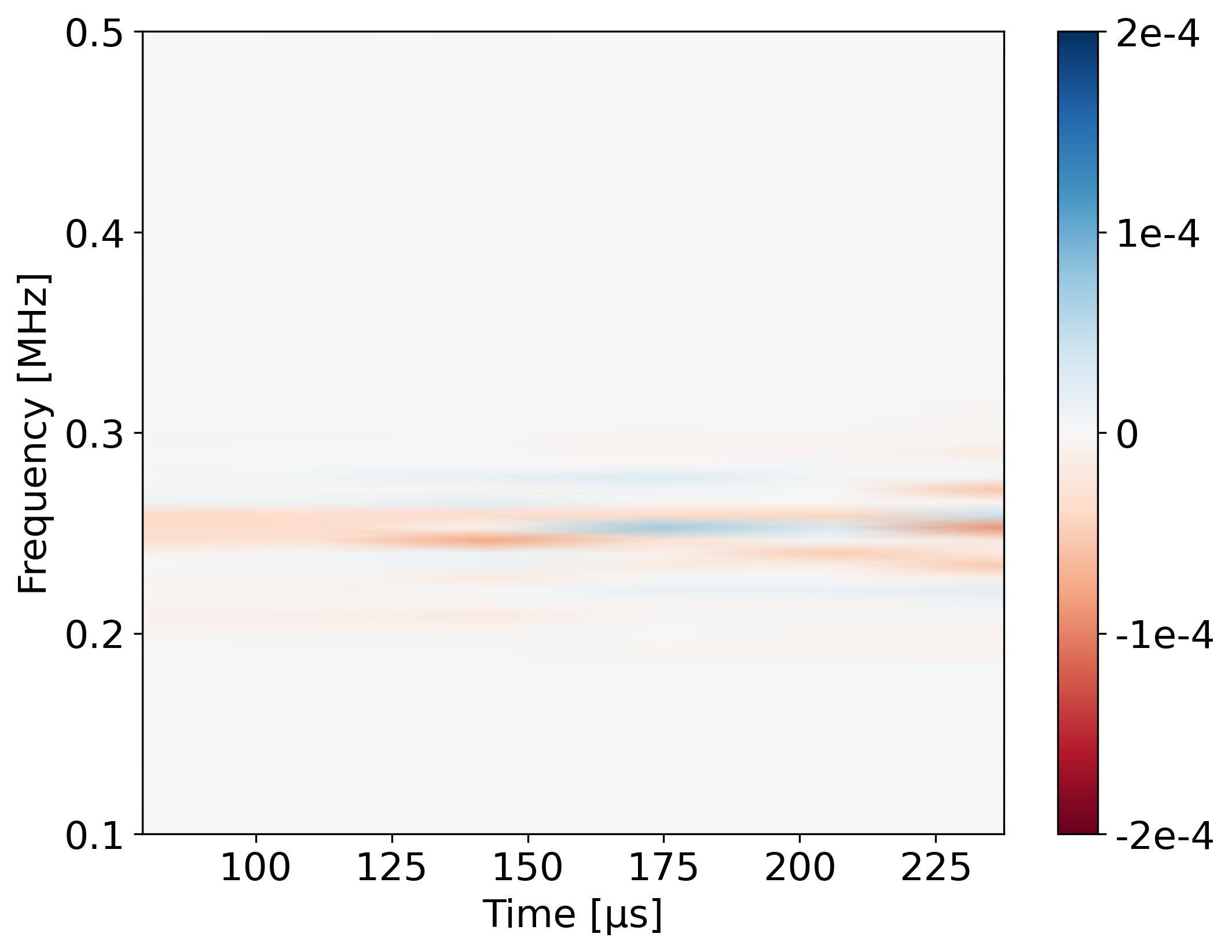}}
    \put(224,414){\includegraphics[width=0.38\textwidth]{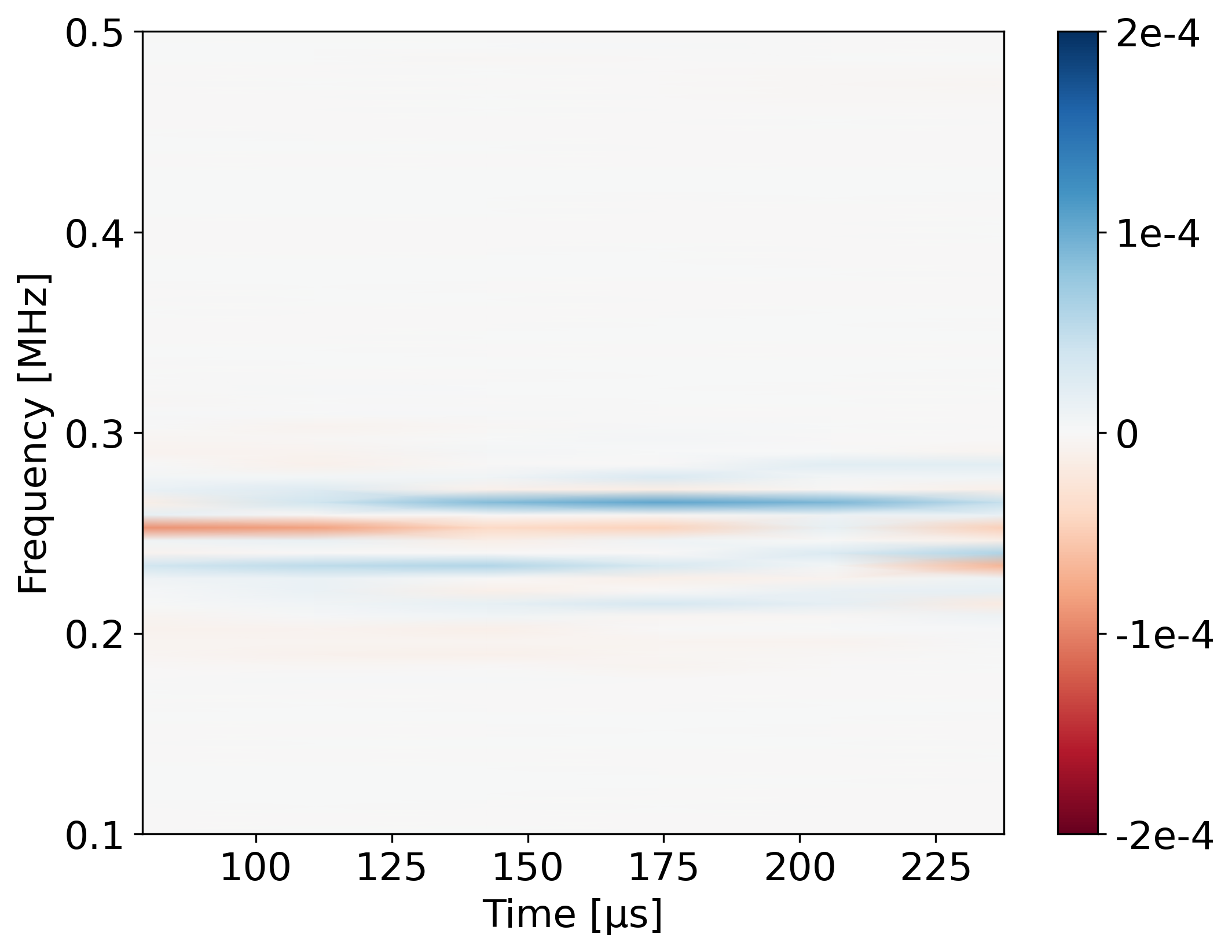}}
    \put(10,276){\includegraphics[width=0.38\textwidth]{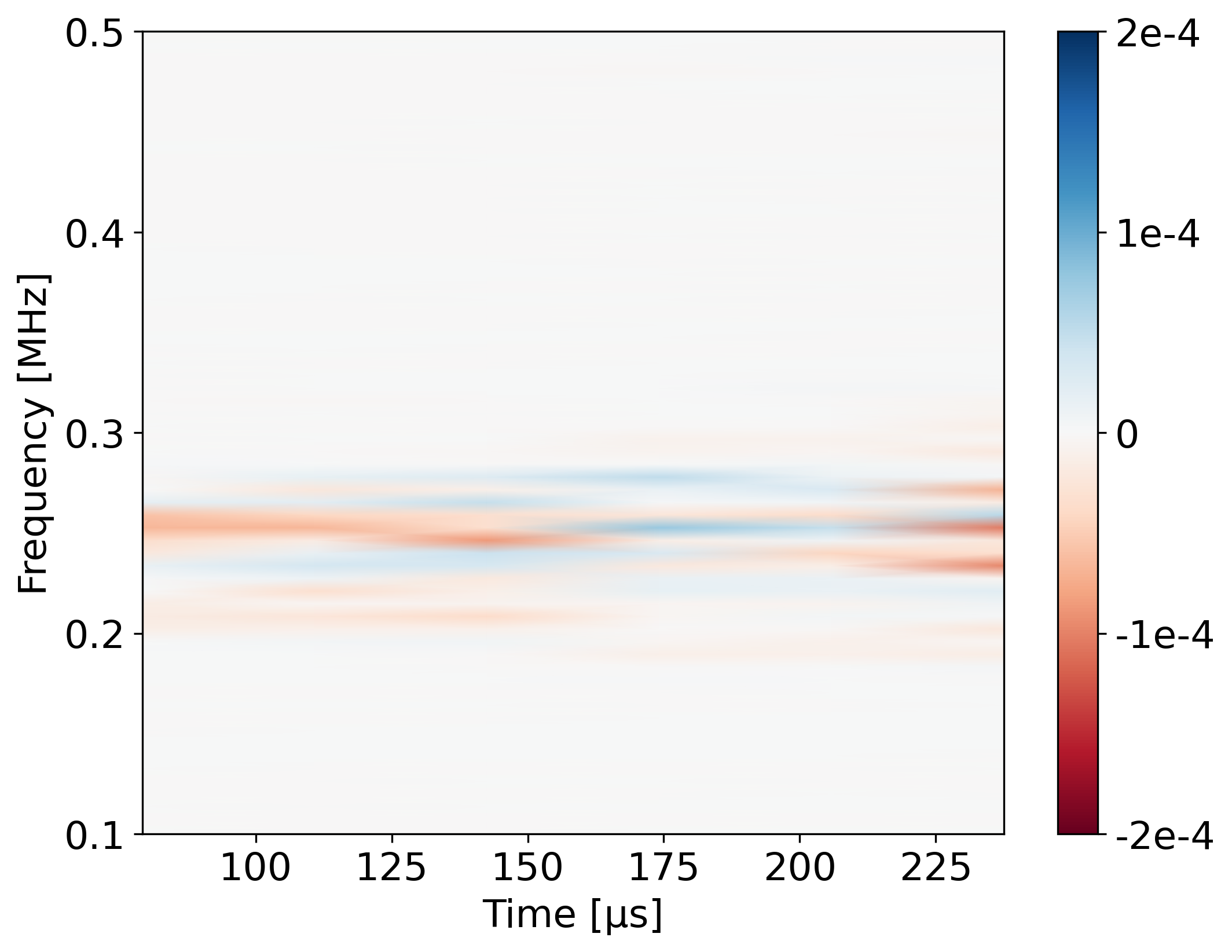}}
    \put(224,276){\includegraphics[width=0.38\textwidth]{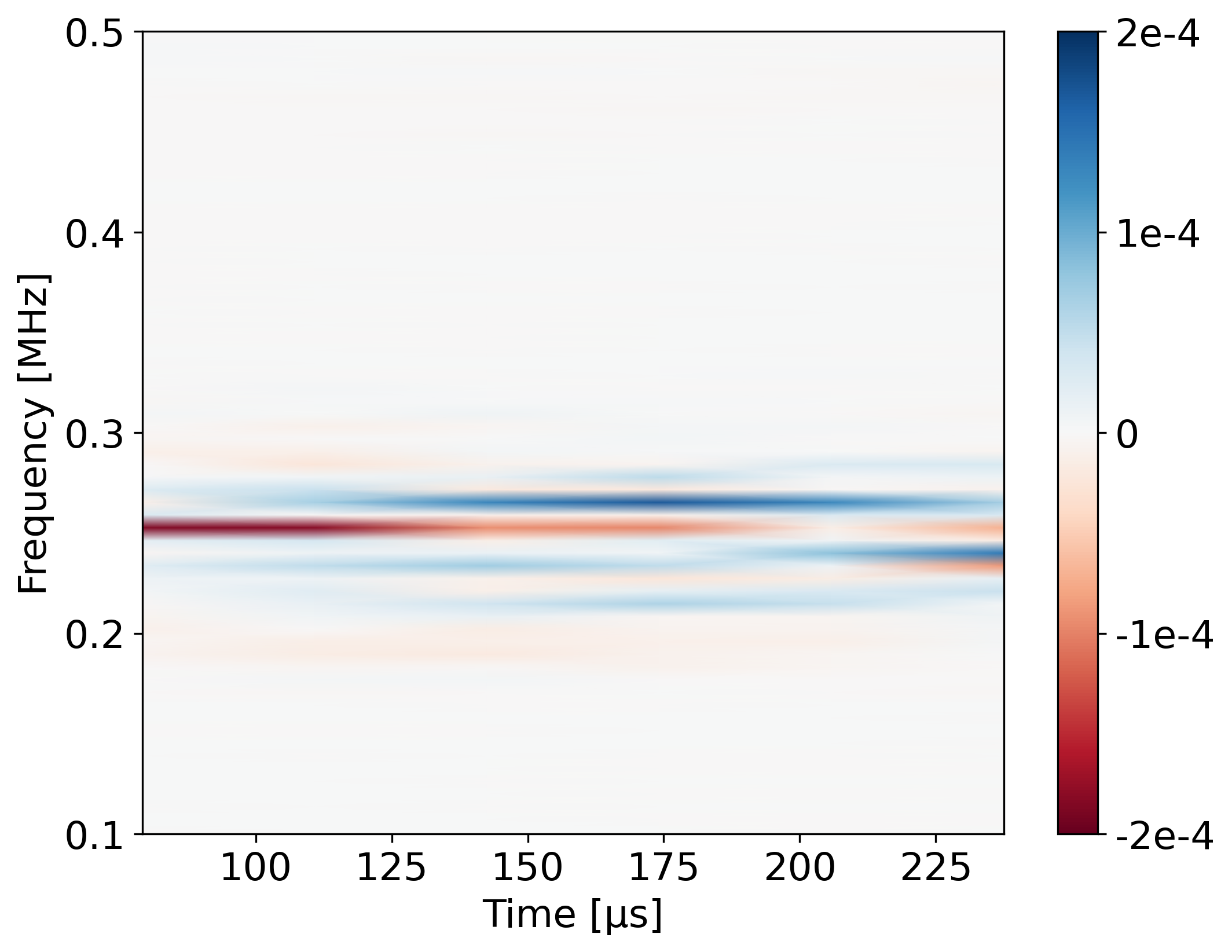}}
    \put(10,138){\includegraphics[width=0.38\textwidth]{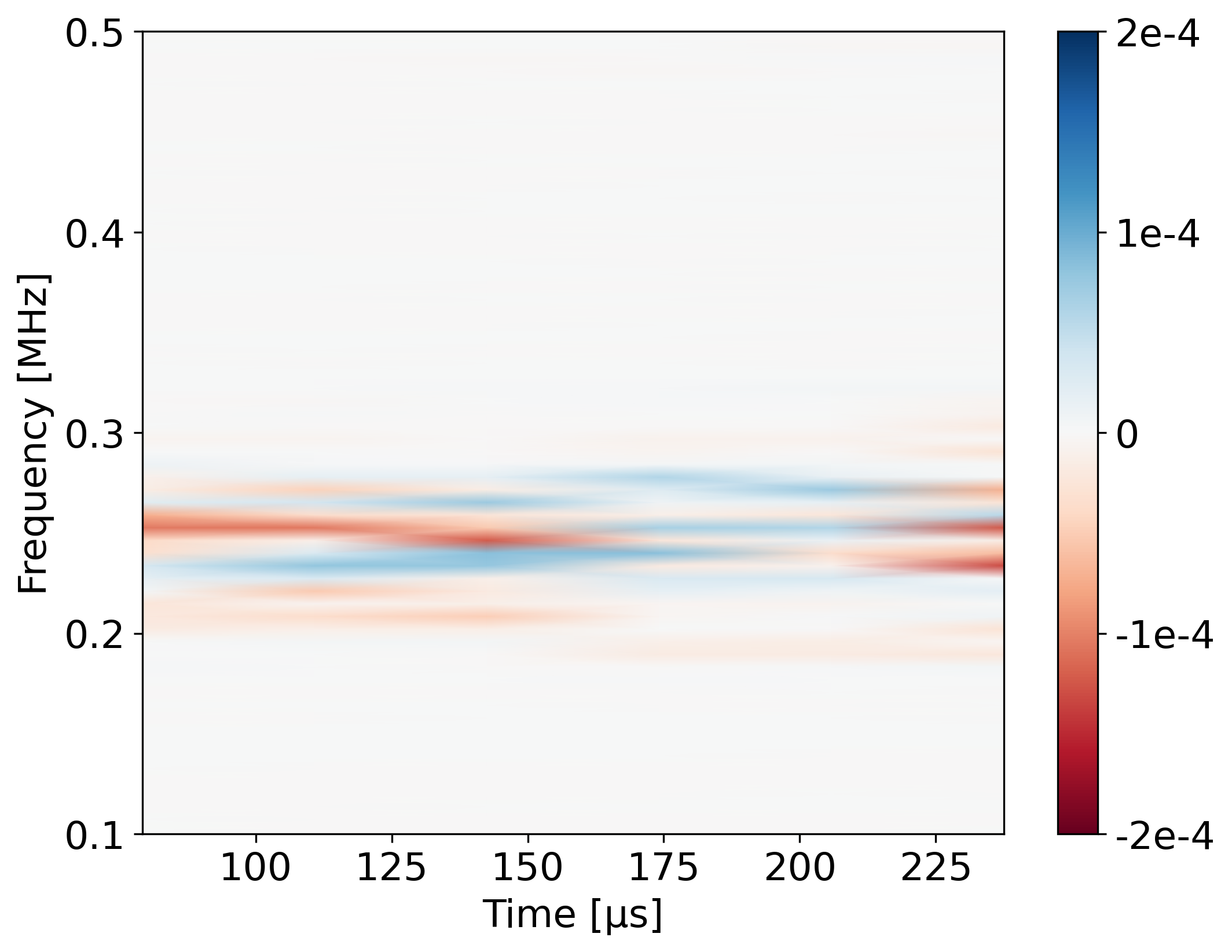}}
    \put(224,138){\includegraphics[width=0.38\textwidth]{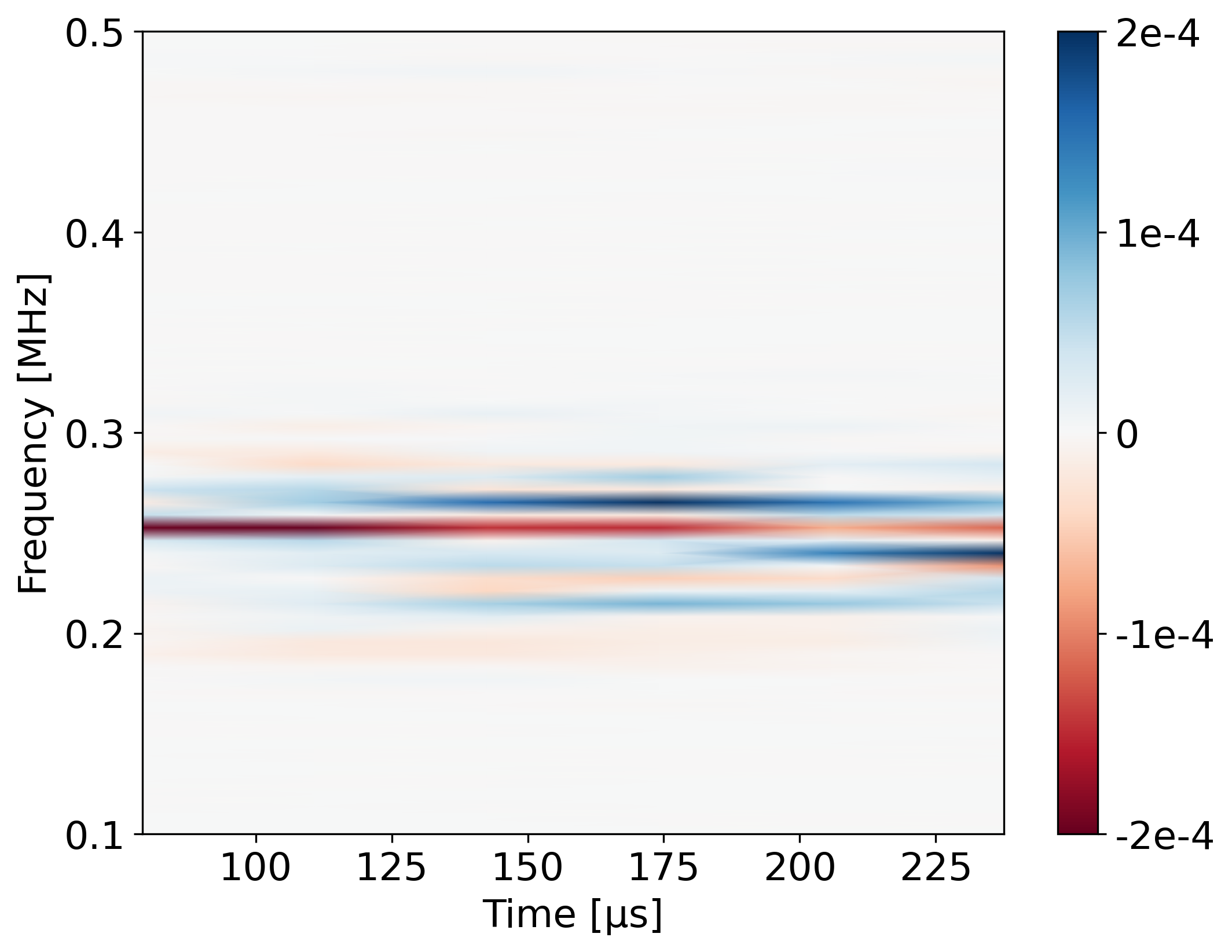}}
    \put(10,0){\includegraphics[width=0.38\textwidth]{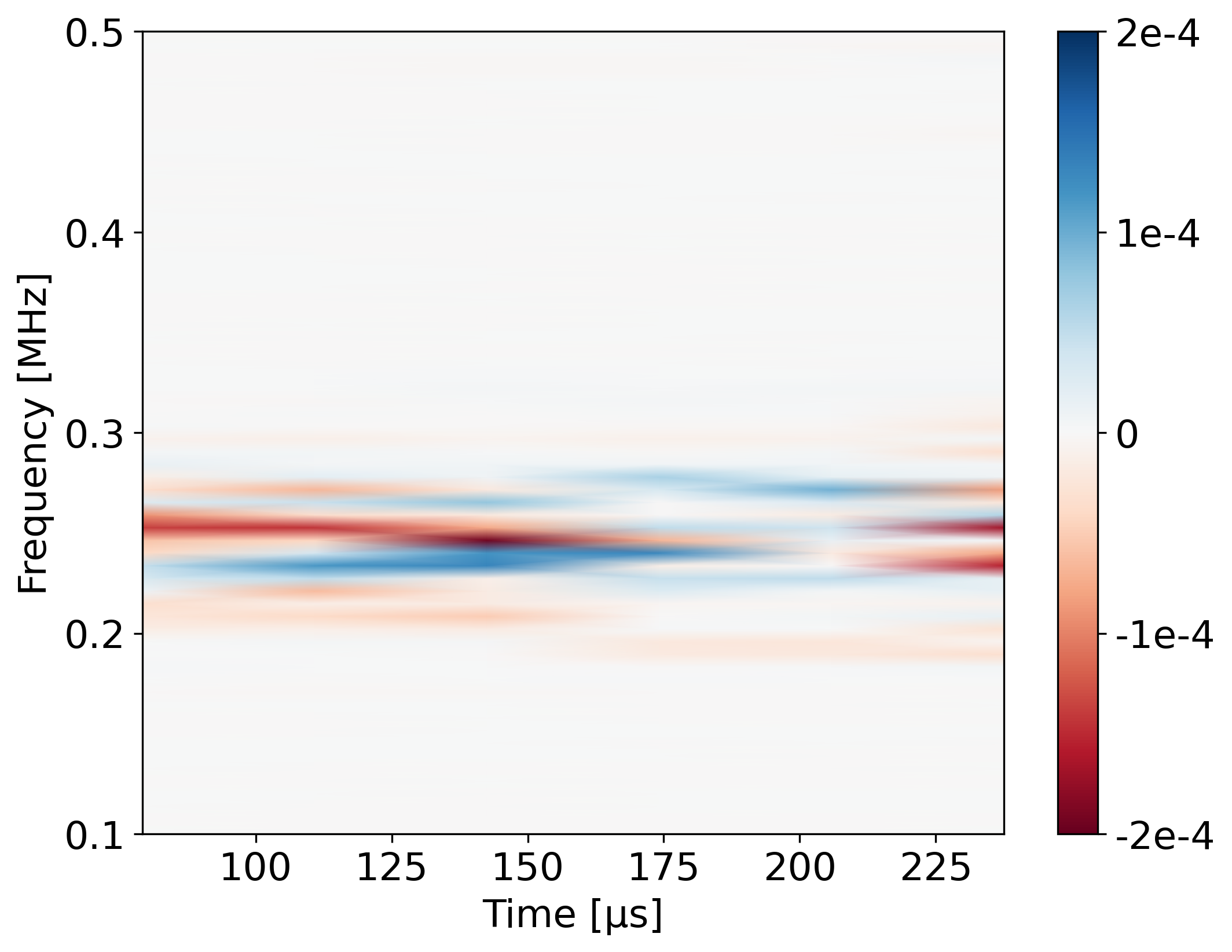}}
    \put(224,0){\includegraphics[width=0.38\textwidth]{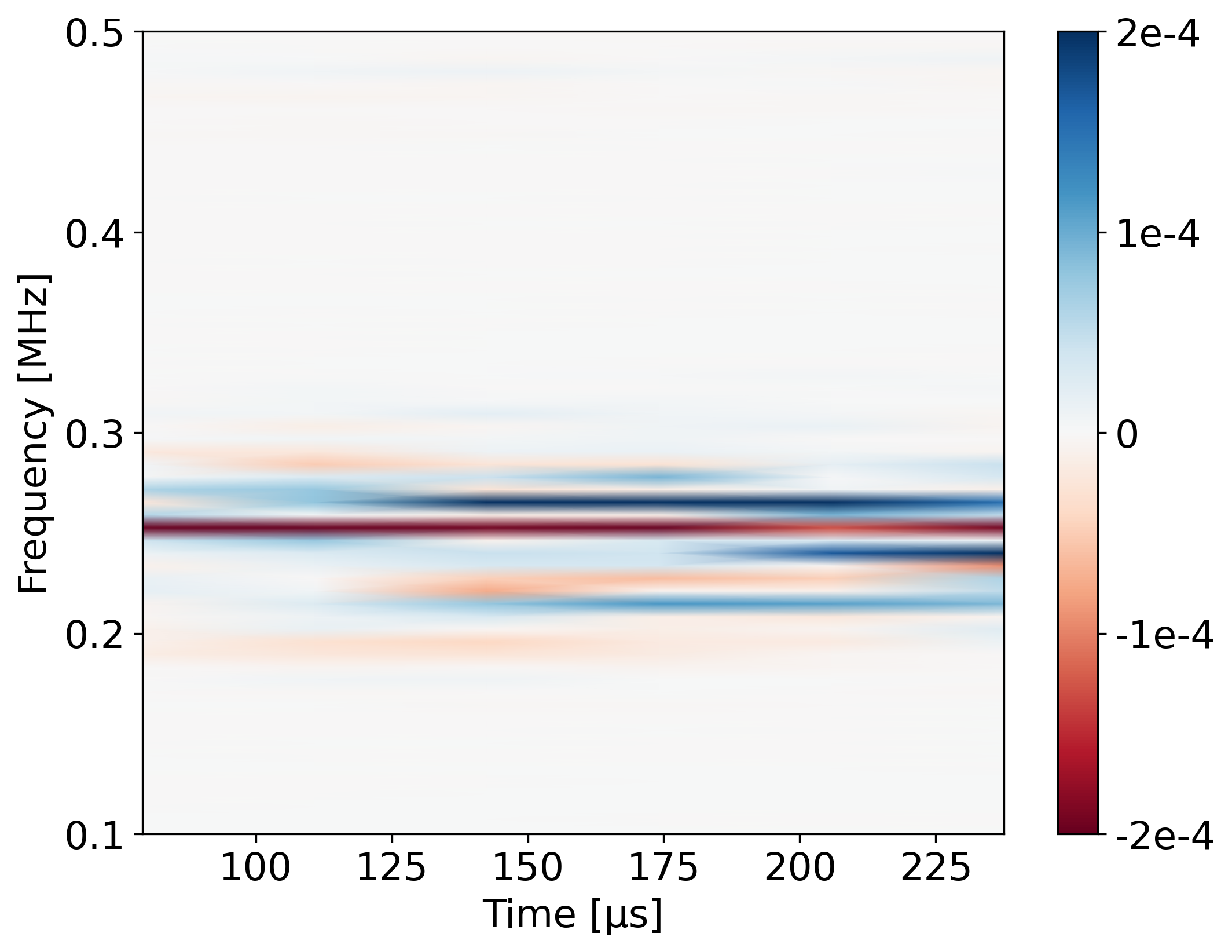}}
    \put(42,535){\color{black} \small {\fontfamily{phv}\selectfont \textbf{$S(z_{1}(0,5))-S(z_{1}(0,0))$}}}
    \put(256,535){\color{black} \small {\fontfamily{phv}\selectfont \textbf{$S(z_{3}(0,5))-S(z_{3}(0,0))$}}}
    \put(42,397){\color{black} \small {\fontfamily{phv}\selectfont \textbf{$S(z_{1}(0,10))-S(z_{1}(0,0))$}}}
    \put(256,397){\color{black} \small
    {\fontfamily{phv}\selectfont \textbf{$S(z_{3}(0,10))-S(z_{3}(0,0))$}}}
    \put(42,259){\color{black} \small {\fontfamily{phv}\selectfont \textbf{$S(z_{1}(0,15))-S(z_{1}(0,0))$}}}
    \put(256,259){\color{black} \small {\fontfamily{phv}\selectfont \textbf{$S(z_{3}(0,15))-S(z_{3}(0,0))$}}}
    \put(42,121){\color{black} \small {\fontfamily{phv}\selectfont \textbf{$S(z_{1}(0,20))-S(z_{1}(0,0))$}}}
    \put(256,121){\color{black} \small {\fontfamily{phv}\selectfont \textbf{$S(z_{3}(0,20))-S(z_{3}(0,0))$}}}
    \put(190,438){\color{black} \large {\fontfamily{phv}\selectfont \textbf{a}}}
    \put(190,300){\large {\fontfamily{phv}\selectfont \textbf{b}}} 
    \put(190,162){\large {\fontfamily{phv}\selectfont \textbf{c}}}
    \put(190,24){\large {\fontfamily{phv}\selectfont \textbf{d}}}
   
   \put(408,438){\large {\fontfamily{phv}\selectfont \textbf{e}}} 
   \put(409,300){\large {\fontfamily{phv}\selectfont \textbf{f}}} 
   \put(406,162){\large {\fontfamily{phv}\selectfont \textbf{g}}} 
   \put(410,24){\large {\fontfamily{phv}\selectfont \textbf{h}}} 
    \end{picture} 
    \caption{Scatter plots of the spectrogram of latent space subtracting by the spectrogram of the baseline latent space  representation from model Type III. $S(\cdot)$ indicates the function of spectrogram. Panels a-d: spectrogram of $z_1$ at healthy case under 5, 10, 15, 20kN subtracting the spectrogram of $z_1$ at healthy case under 0kN; panels e-h: spectrogram of $z_3$ at healthy case under 5, 10, 15, 20kN subtracting the spectrogram  of $z_3$ at healthy case under 0kN.}
\label{fig:lat3_1} 
\end{figure}

\begin{figure}[t!]
    \begin{picture}(500,530)
    \put(10,414){\includegraphics[width=0.38\textwidth]{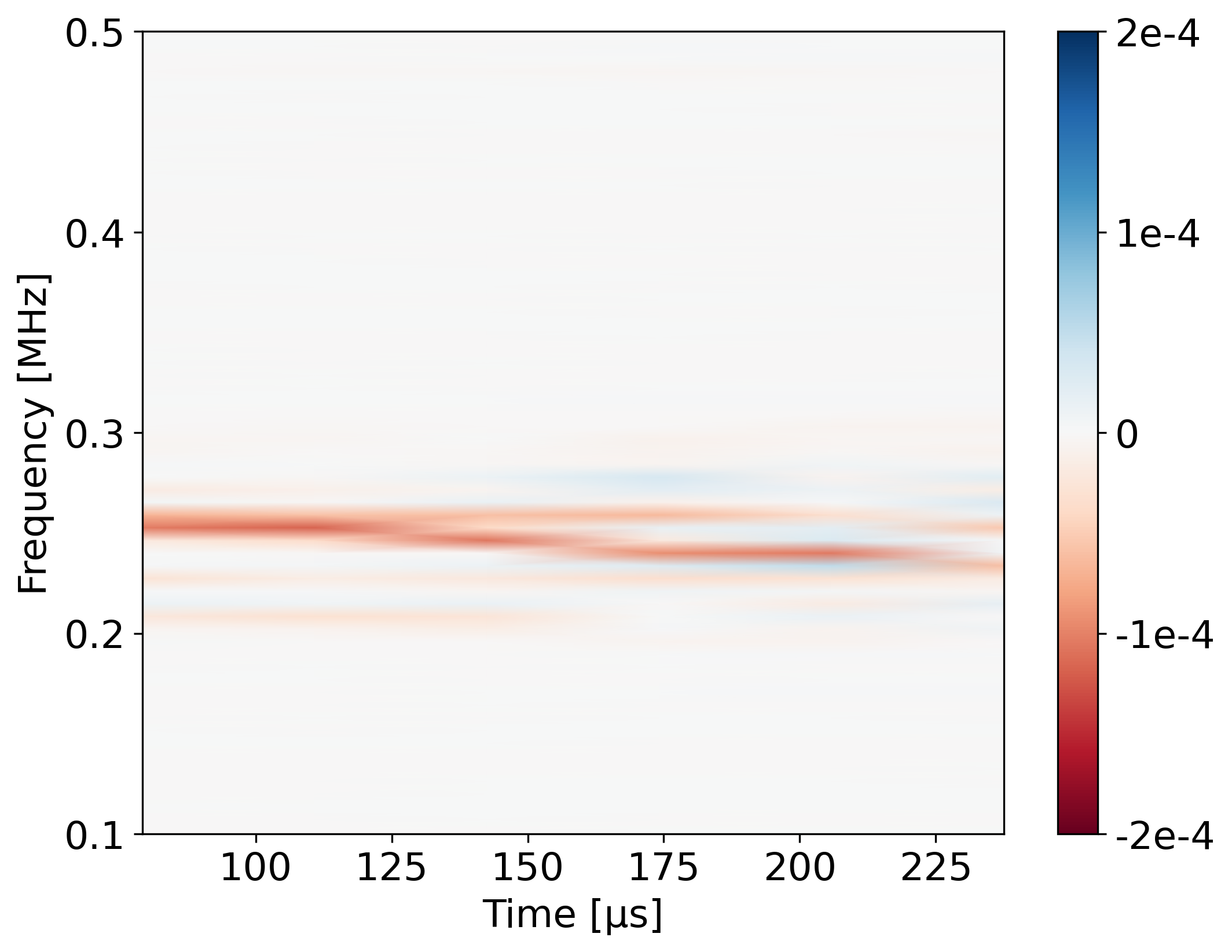}}
    \put(224,414){\includegraphics[width=0.38\textwidth]{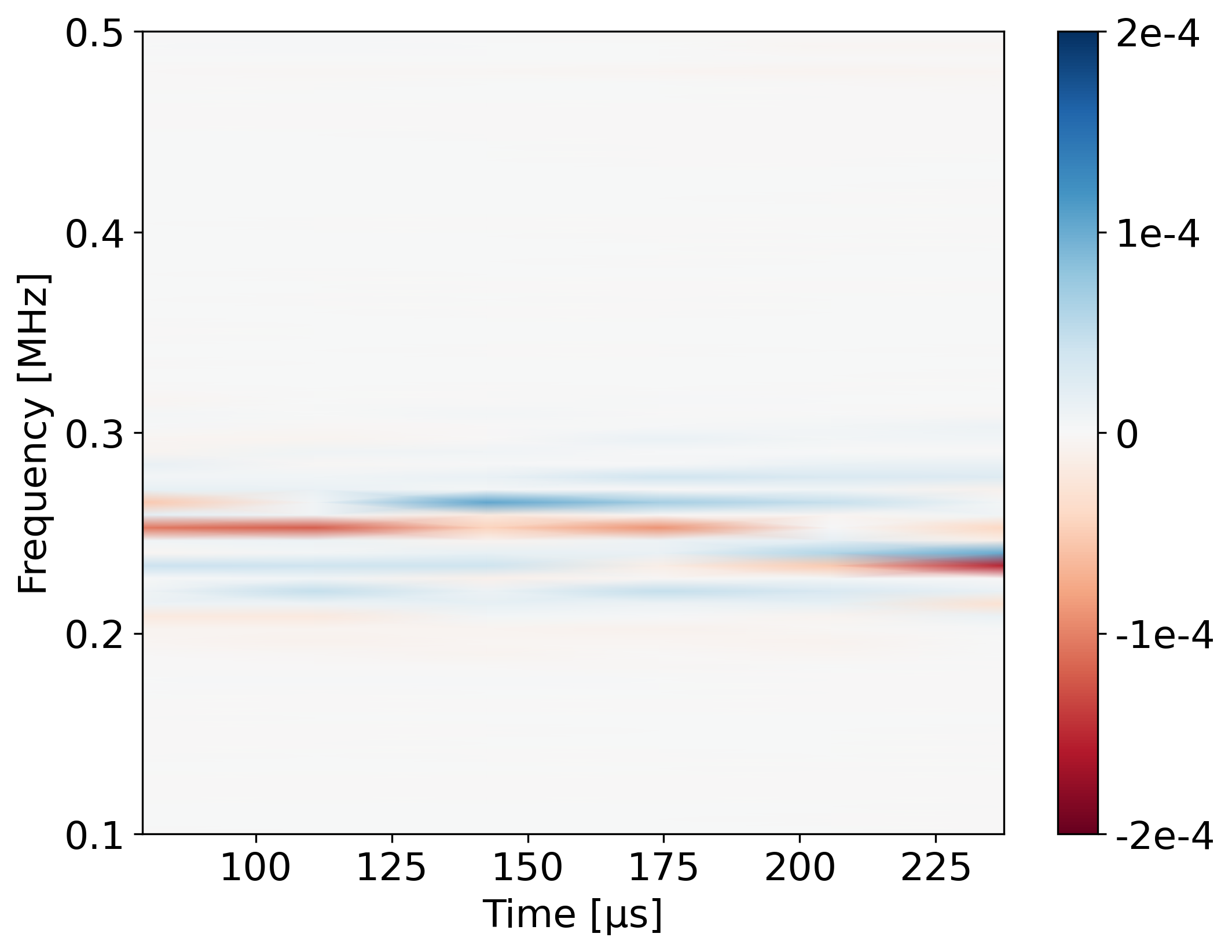}}
    \put(10,276){\includegraphics[width=0.38\textwidth]{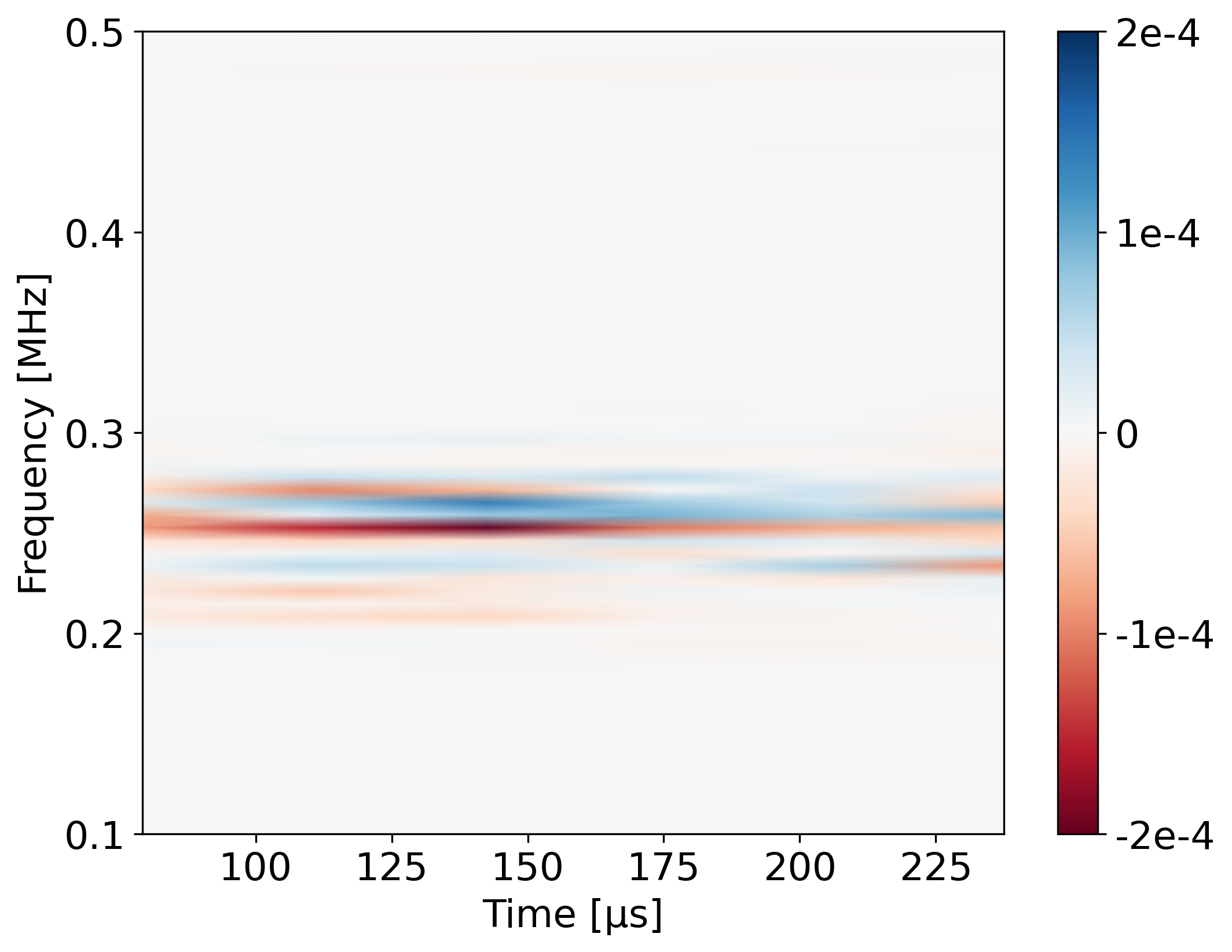}}
    \put(224,276){\includegraphics[width=0.38\textwidth]{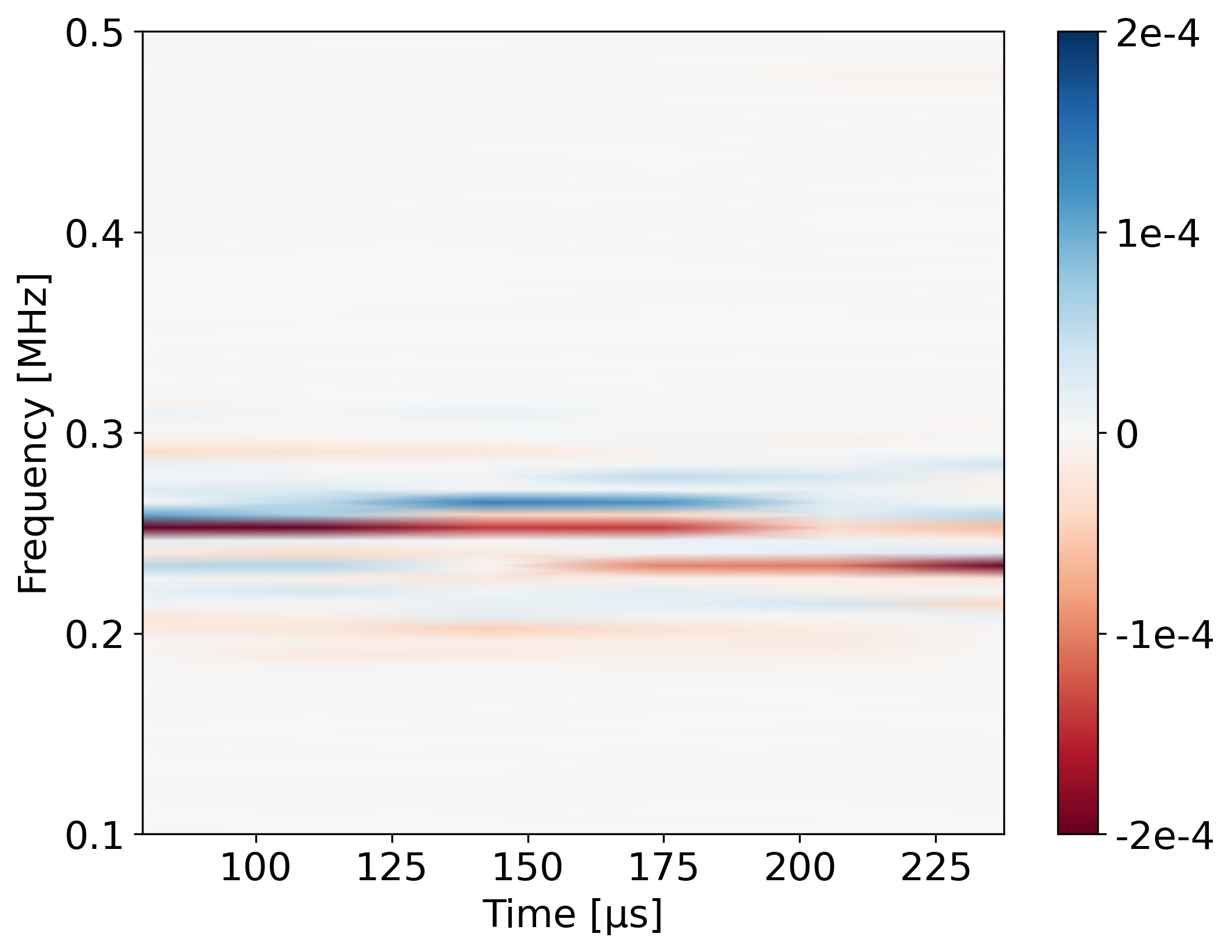}}
    \put(10,138){\includegraphics[width=0.38\textwidth]{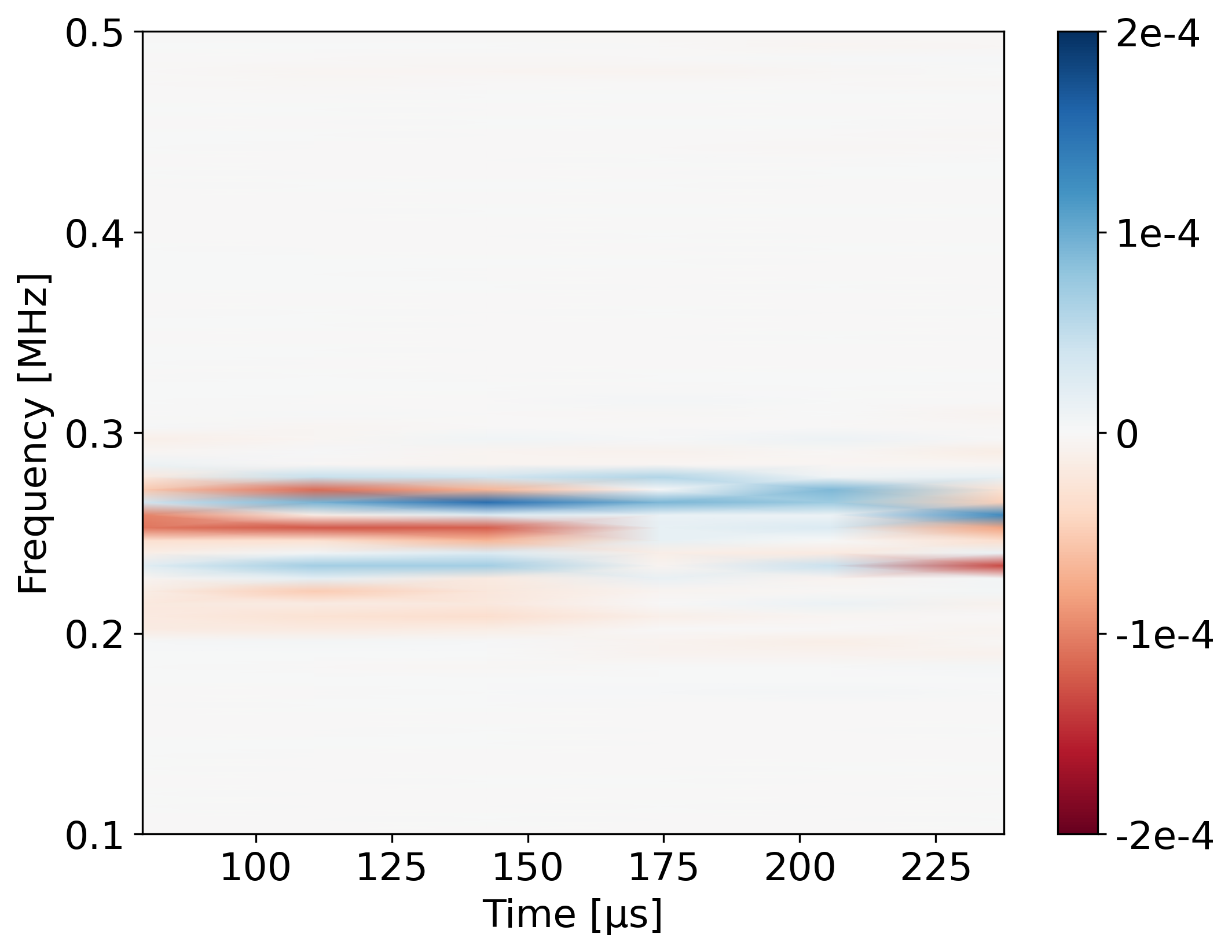}}
    \put(224,138){\includegraphics[width=0.38\textwidth]{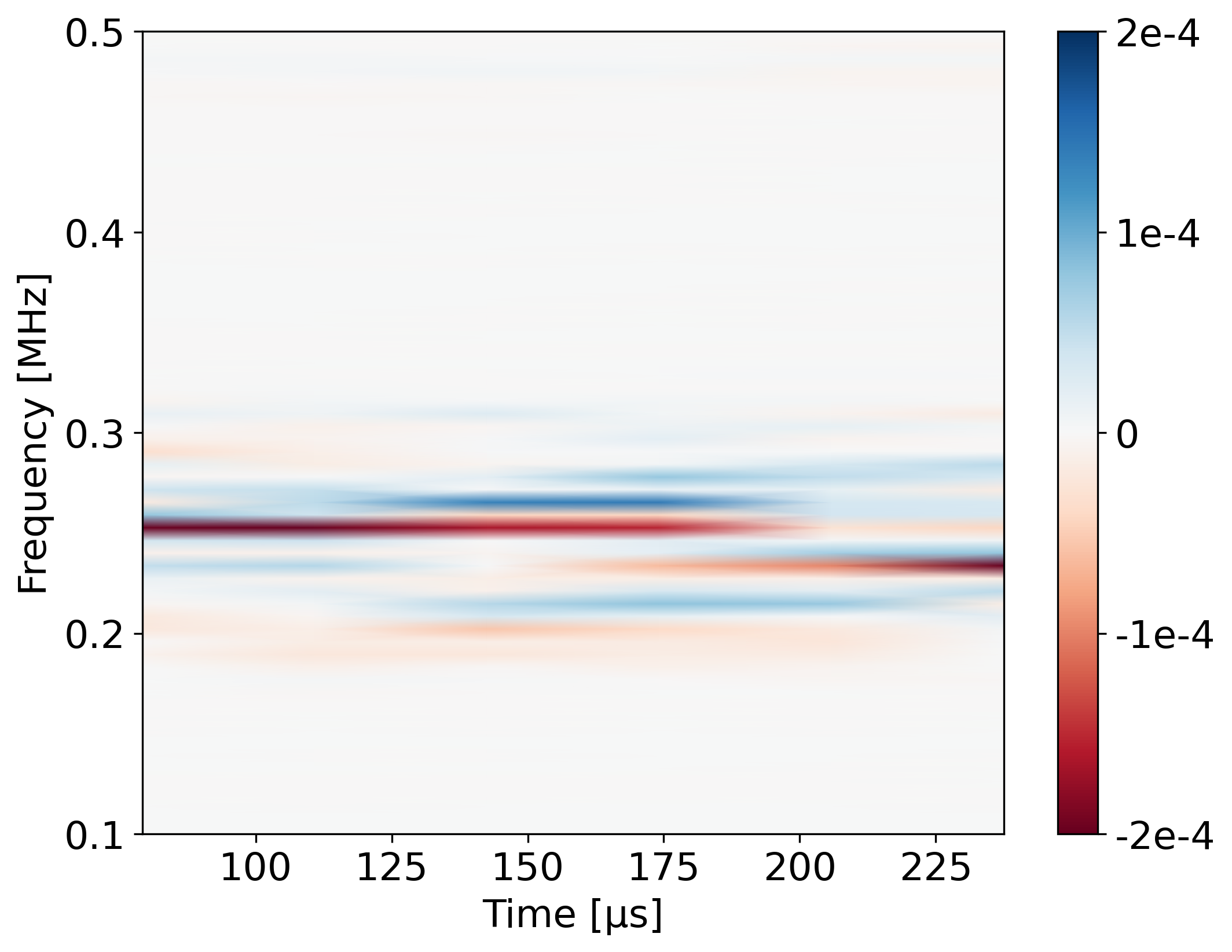}}
    \put(10,0){\includegraphics[width=0.38\textwidth]{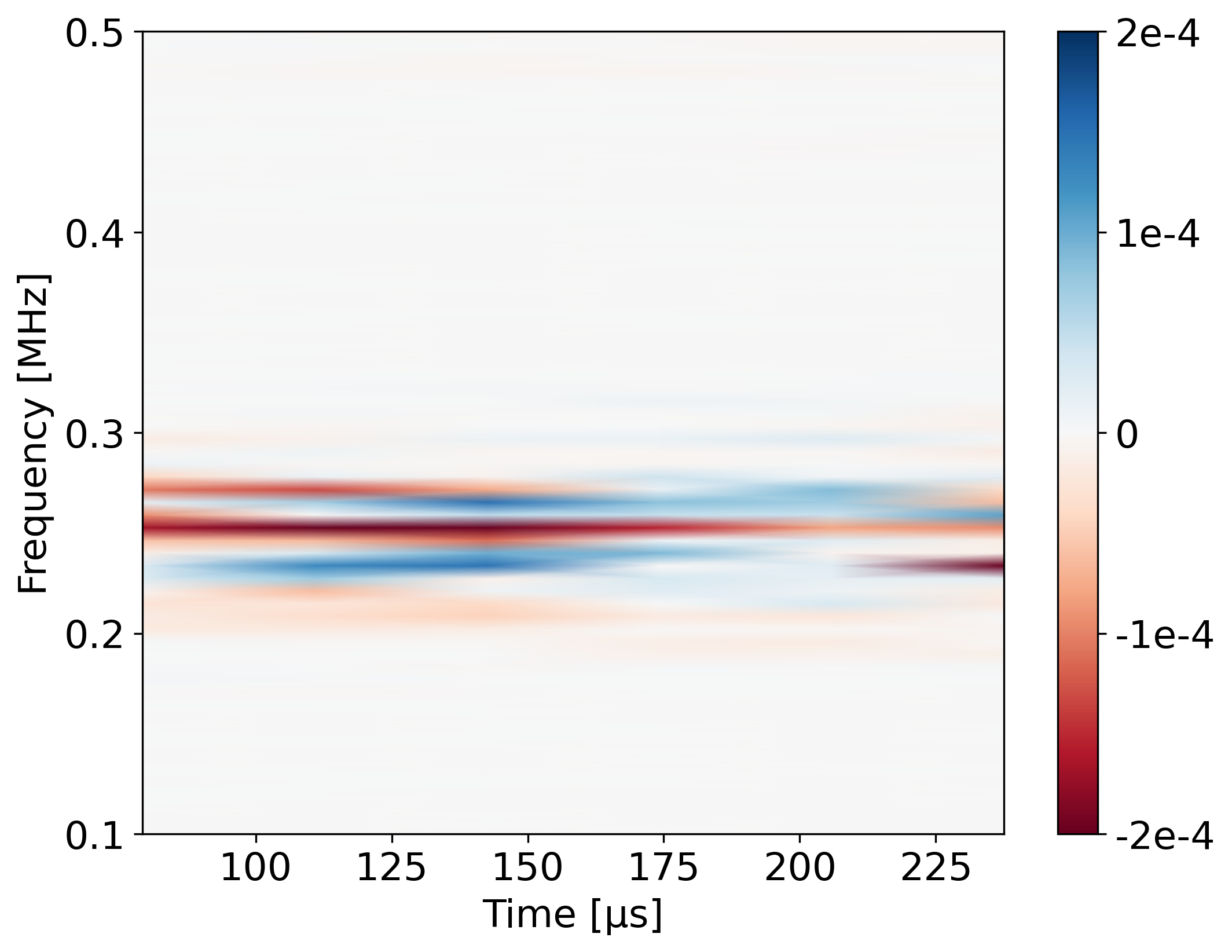}}
    \put(224,0){\includegraphics[width=0.38\textwidth]{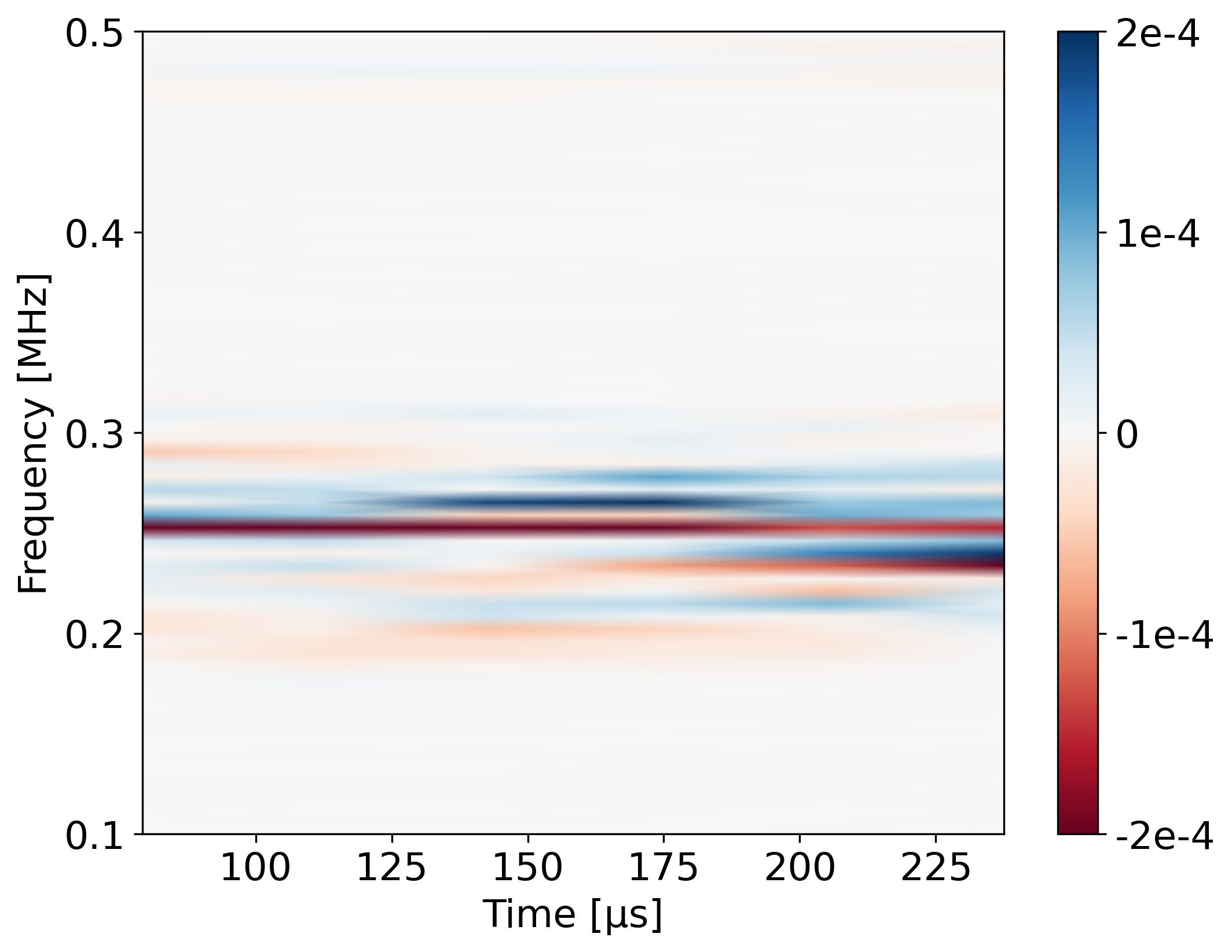}}
    \put(42,535){\color{black} \small {\fontfamily{phv}\selectfont \textbf{$S(z_{1}(4,5))-S(z_{1}(4,0))$}}}
    \put(256,535){\color{black} \small {\fontfamily{phv}\selectfont \textbf{$S(z_{3}(4,5))-S(z_{3}(4,0))$}}}
    \put(42,397){\color{black} \small {\fontfamily{phv}\selectfont \textbf{$S(z_{1}(4,10))-S(z_{1}(4,0))$}}}
    \put(256,397){\color{black} \small
    {\fontfamily{phv}\selectfont \textbf{$S(z_{3}(4,10))-S(z_{3}(4,0))$}}}
    \put(42,259){\color{black} \small {\fontfamily{phv}\selectfont \textbf{$S(z_{1}(4,15))-S(z_{1}(4,0))$}}}
    \put(256,259){\color{black} \small {\fontfamily{phv}\selectfont \textbf{$S(z_{3}(4,15))-S(z_{3}(4,0))$}}}
    \put(42,121){\color{black} \small {\fontfamily{phv}\selectfont \textbf{$S(z_{1}(4,20))-S(z_{1}(4,0))$}}}
    \put(256,121){\color{black} \small {\fontfamily{phv}\selectfont \textbf{$S(z_{3}(4,20))-S(z_{3}(4,0))$}}}
    \put(190,438){\color{black} \large {\fontfamily{phv}\selectfont \textbf{a}}}
    \put(190,300){\large {\fontfamily{phv}\selectfont \textbf{b}}} 
    \put(190,162){\large {\fontfamily{phv}\selectfont \textbf{c}}}
    \put(190,24){\large {\fontfamily{phv}\selectfont \textbf{d}}}
   
   \put(408,438){\large {\fontfamily{phv}\selectfont \textbf{e}}} 
   \put(409,300){\large {\fontfamily{phv}\selectfont \textbf{f}}} 
   \put(406,162){\large {\fontfamily{phv}\selectfont \textbf{g}}} 
   \put(410,24){\large {\fontfamily{phv}\selectfont \textbf{h}}} 
    \end{picture} 
    \caption{Scatter plots of the spectrogram of latent space subtracting by the spectrogram of the baseline latent space  representation from model Type III. $S(\cdot)$ indicates the function of spectrogram. Panels a-d: spectrogram of $z_1$ at damage level 4 under 5, 10, 15, 20kN subtracting the spectrogram  of $z_1$ at damage level 4 under 0kN; panels e-h: spectrogram of $z_3$ at damage level 4 under 5, 10, 15, 20kN subtracting the spectrogram of $z_3$ at damage level 4 under 0kN.}
\label{fig:lat3_2} 
\end{figure}
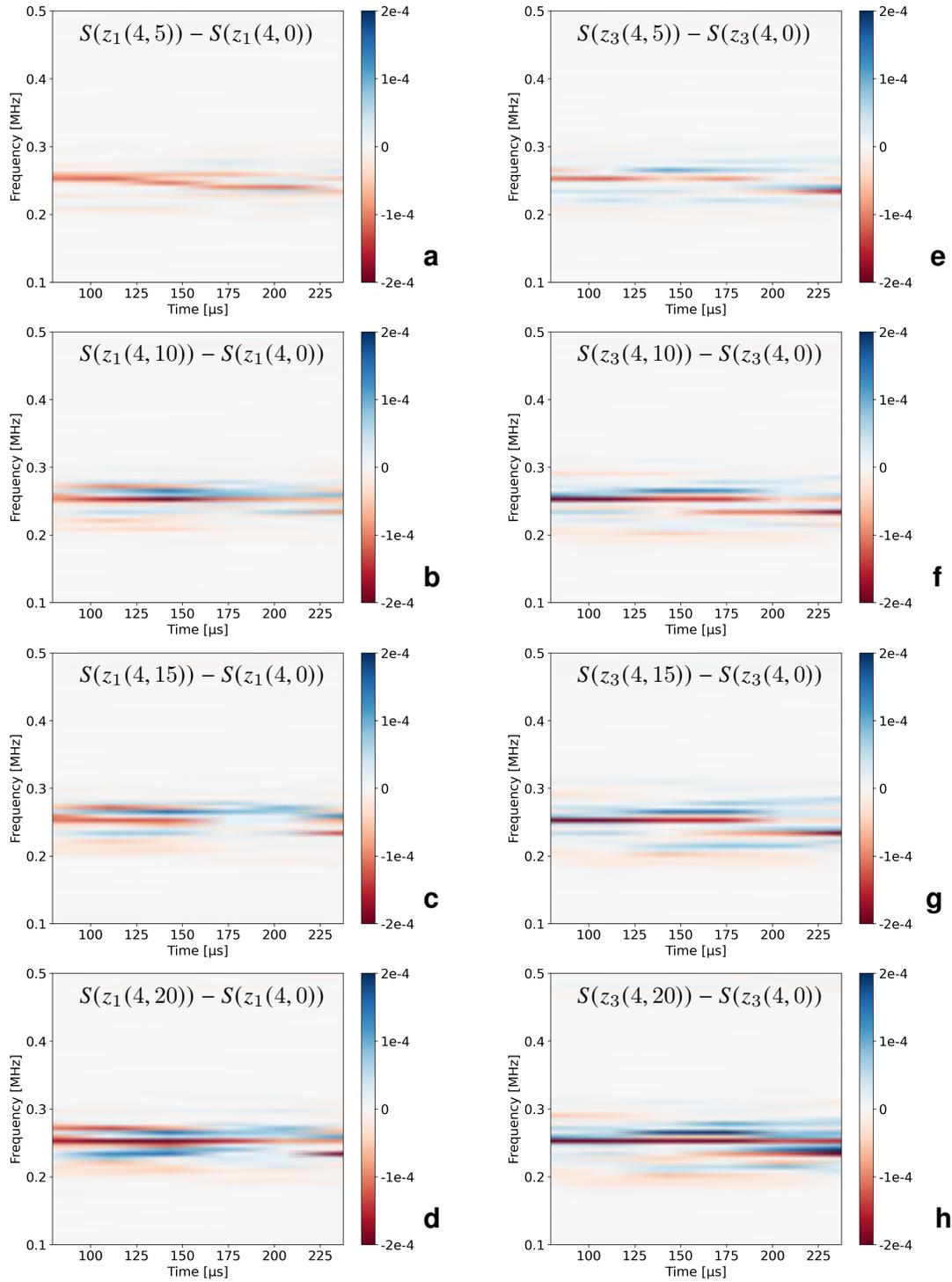

\begin{figure}[t!]
    \begin{picture}(500,530)
    \put(10,414){\includegraphics[width=0.38\textwidth]{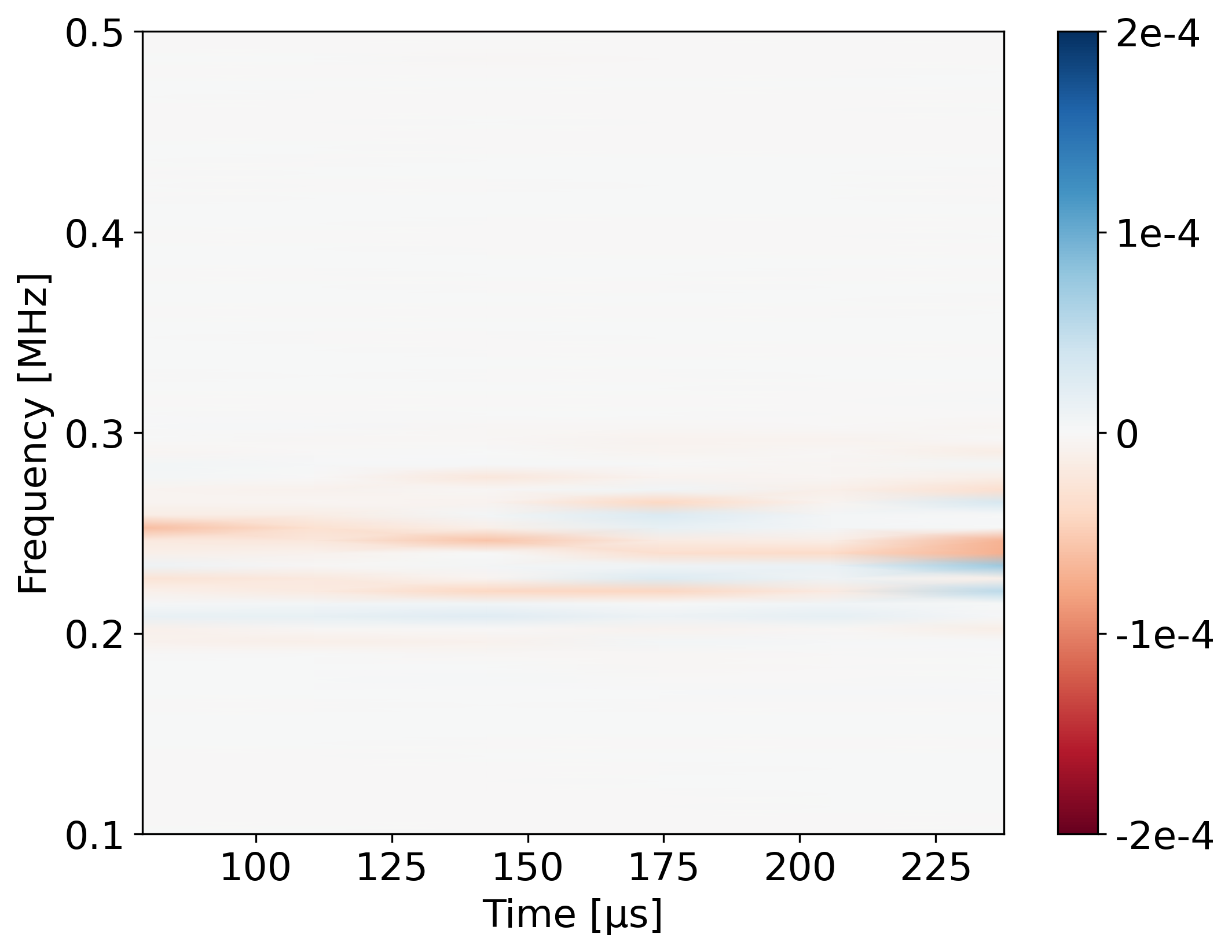}}
    \put(224,414){\includegraphics[width=0.38\textwidth]{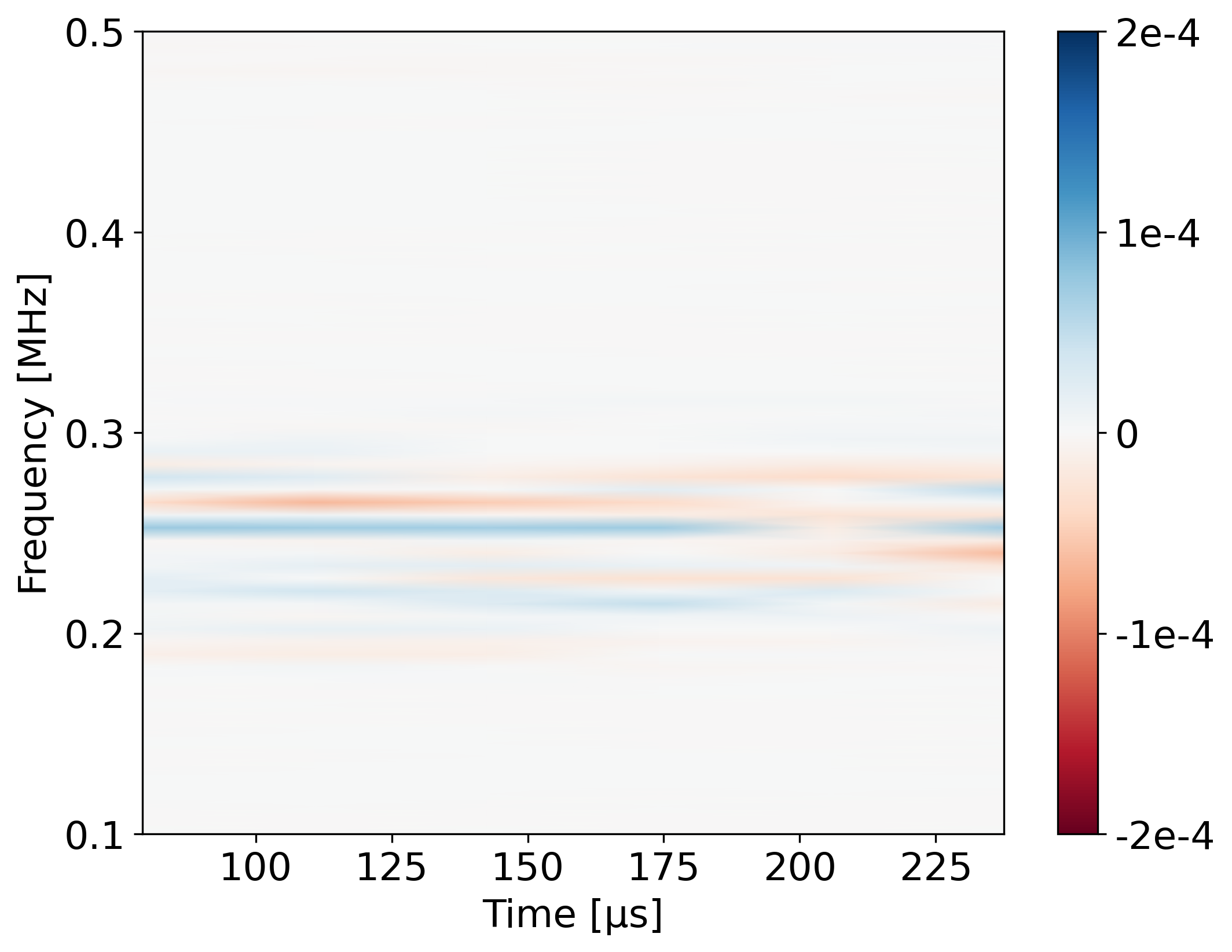}}
    \put(10,276){\includegraphics[width=0.38\textwidth]{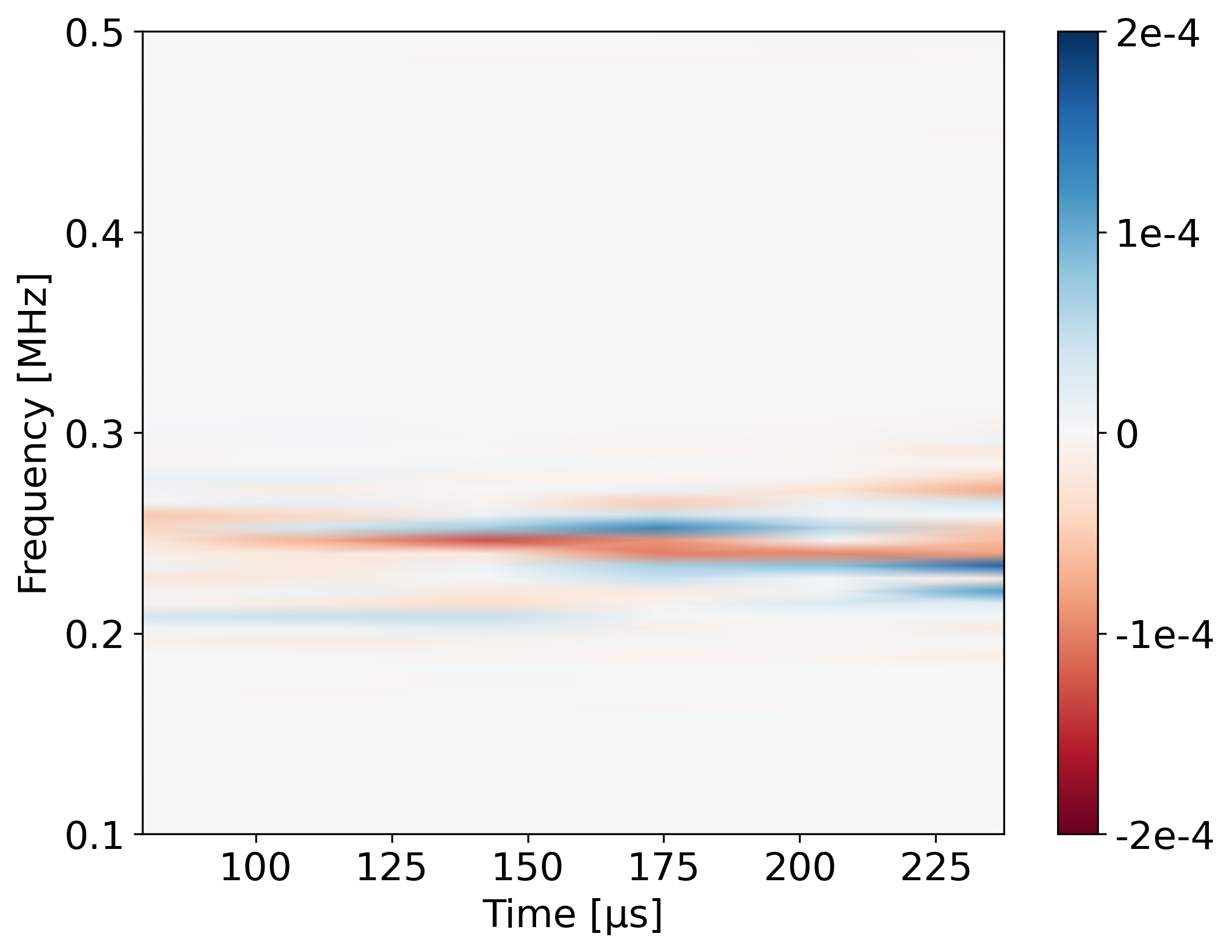}}
    \put(224,276){\includegraphics[width=0.38\textwidth]{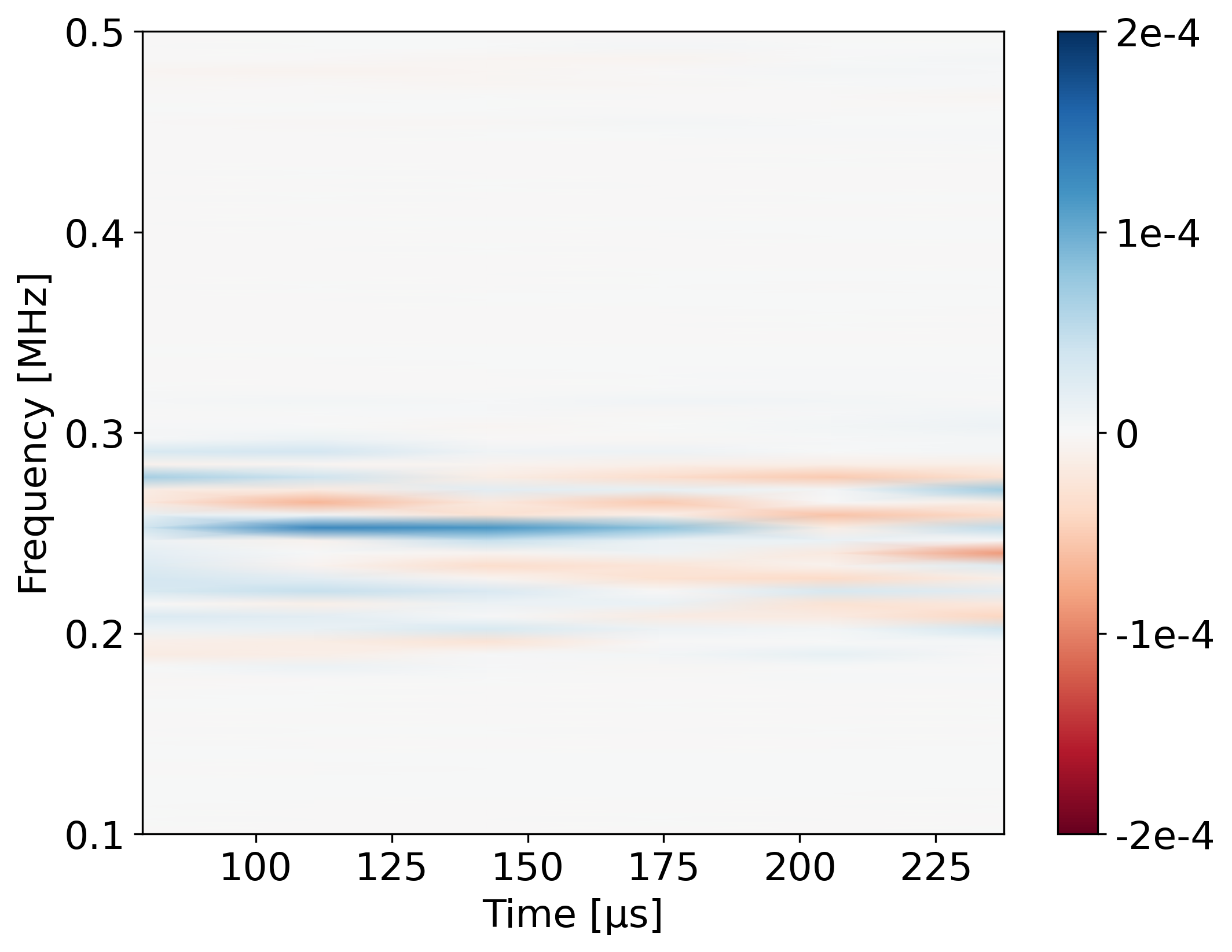}}
    \put(10,138){\includegraphics[width=0.38\textwidth]{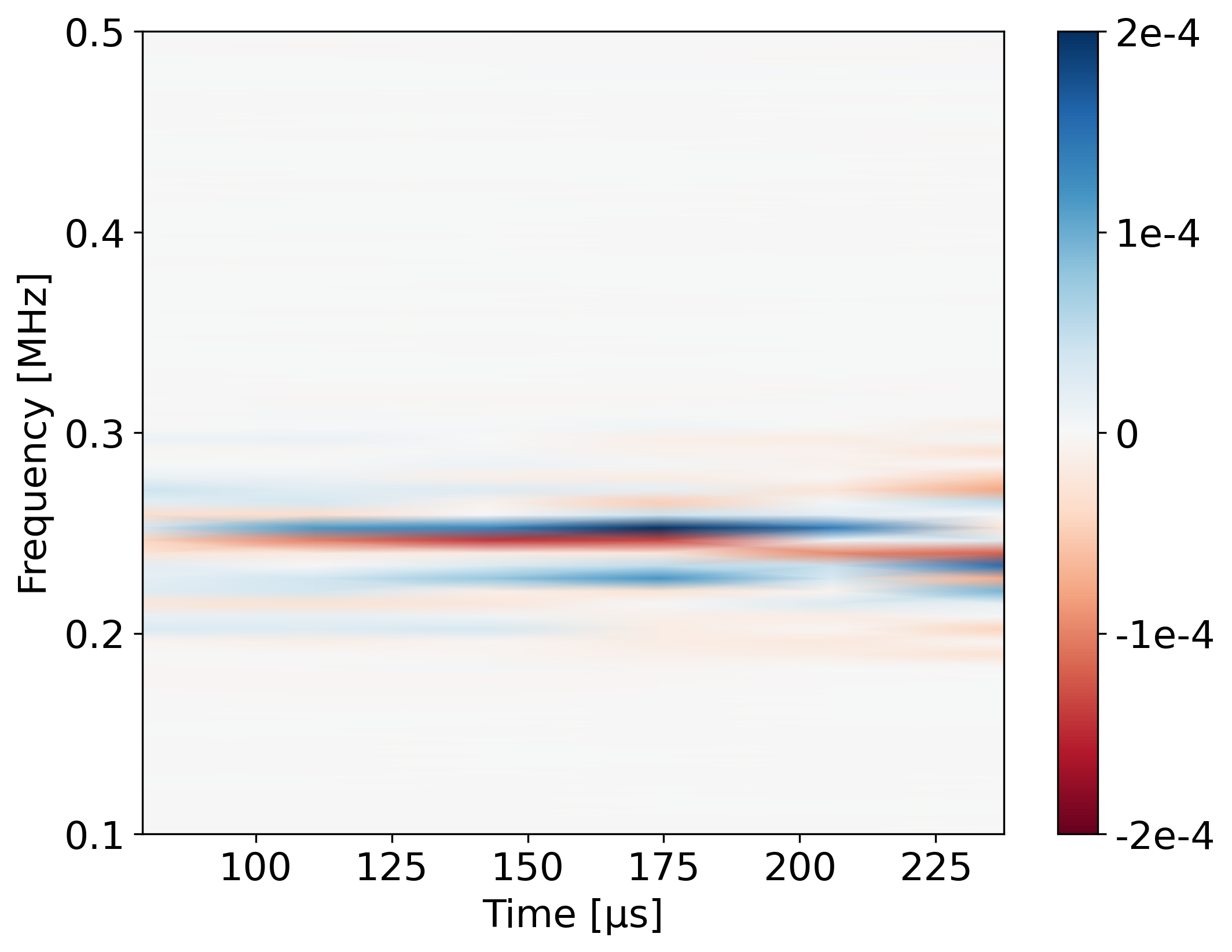}}
    \put(224,138){\includegraphics[width=0.38\textwidth]{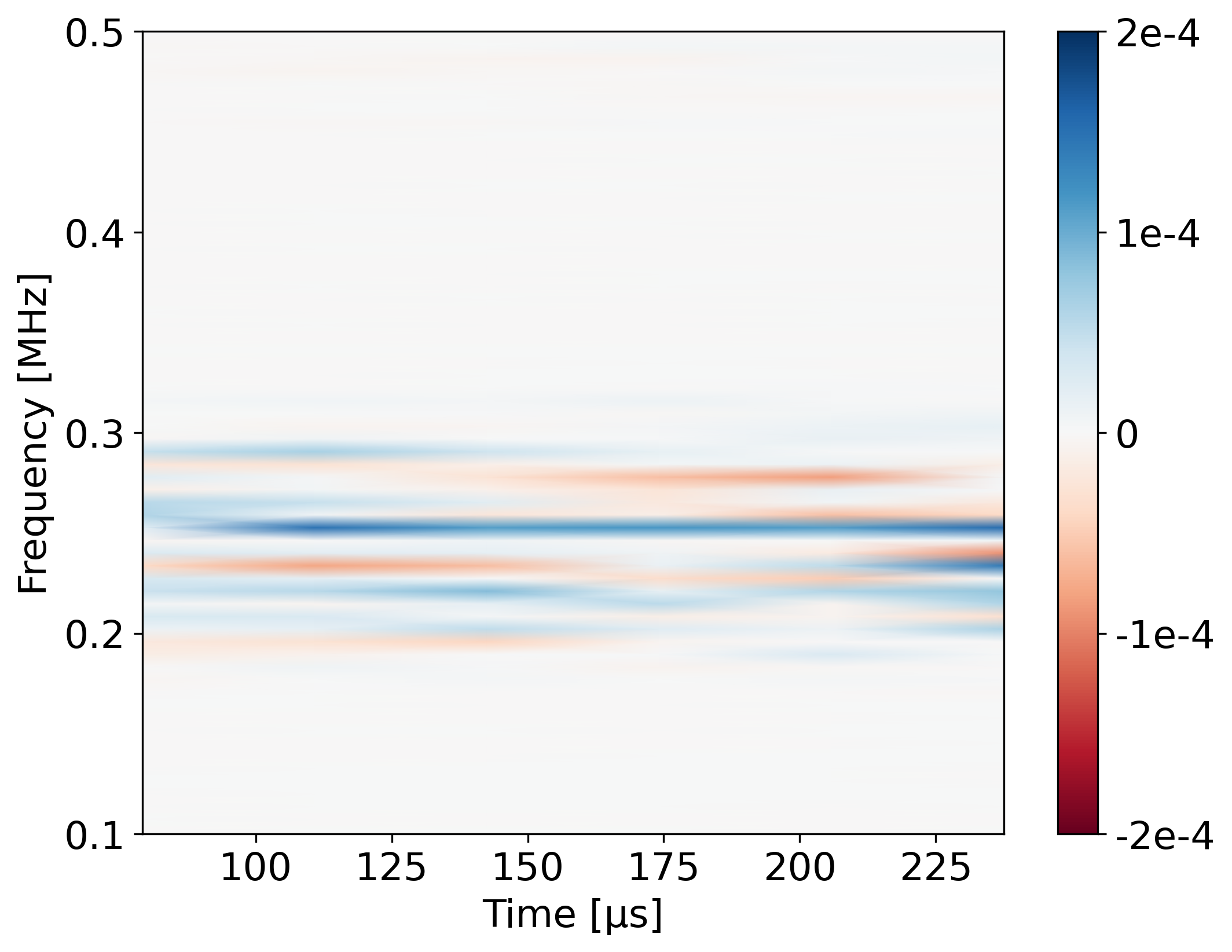}}
    \put(10,0){\includegraphics[width=0.38\textwidth]{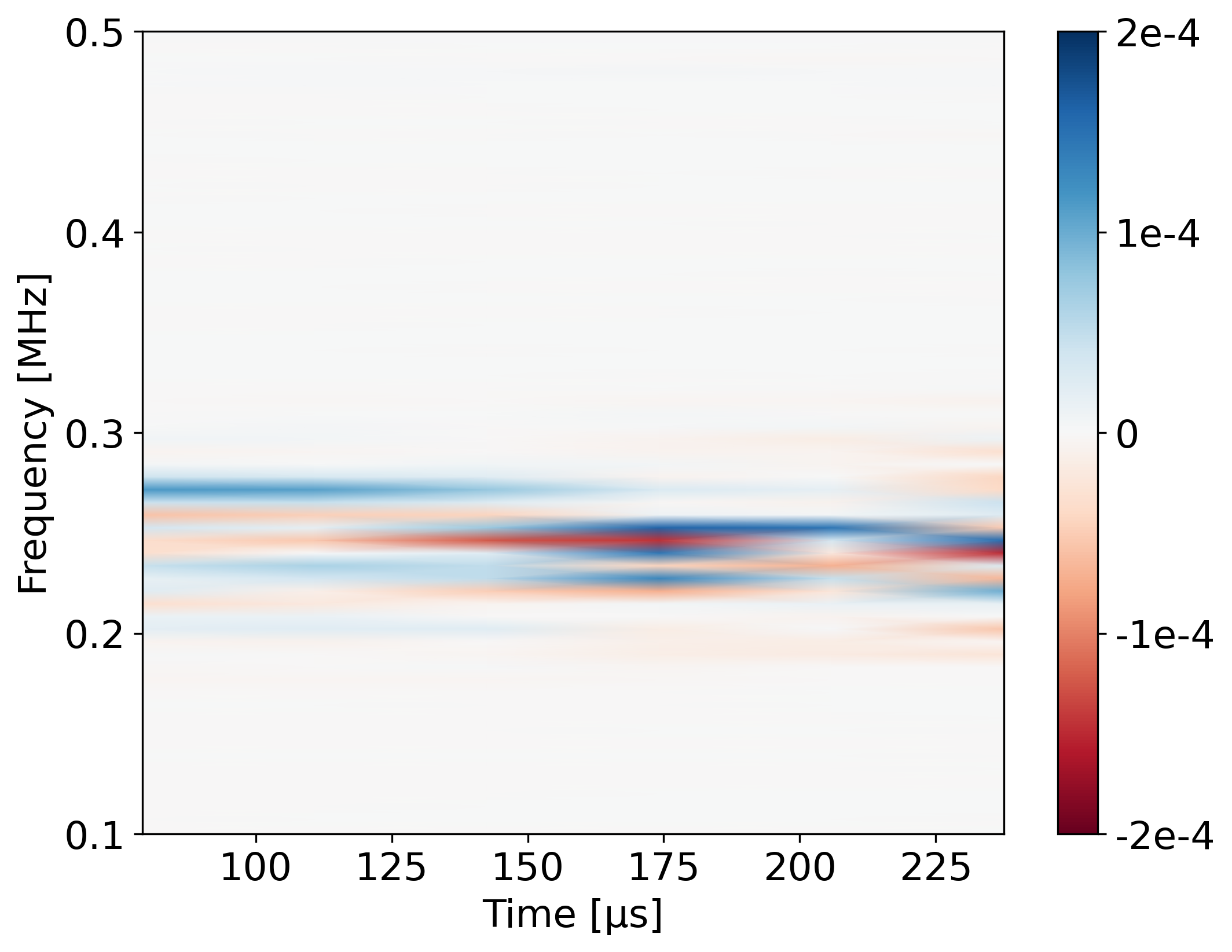}}
    \put(224,0){\includegraphics[width=0.38\textwidth]{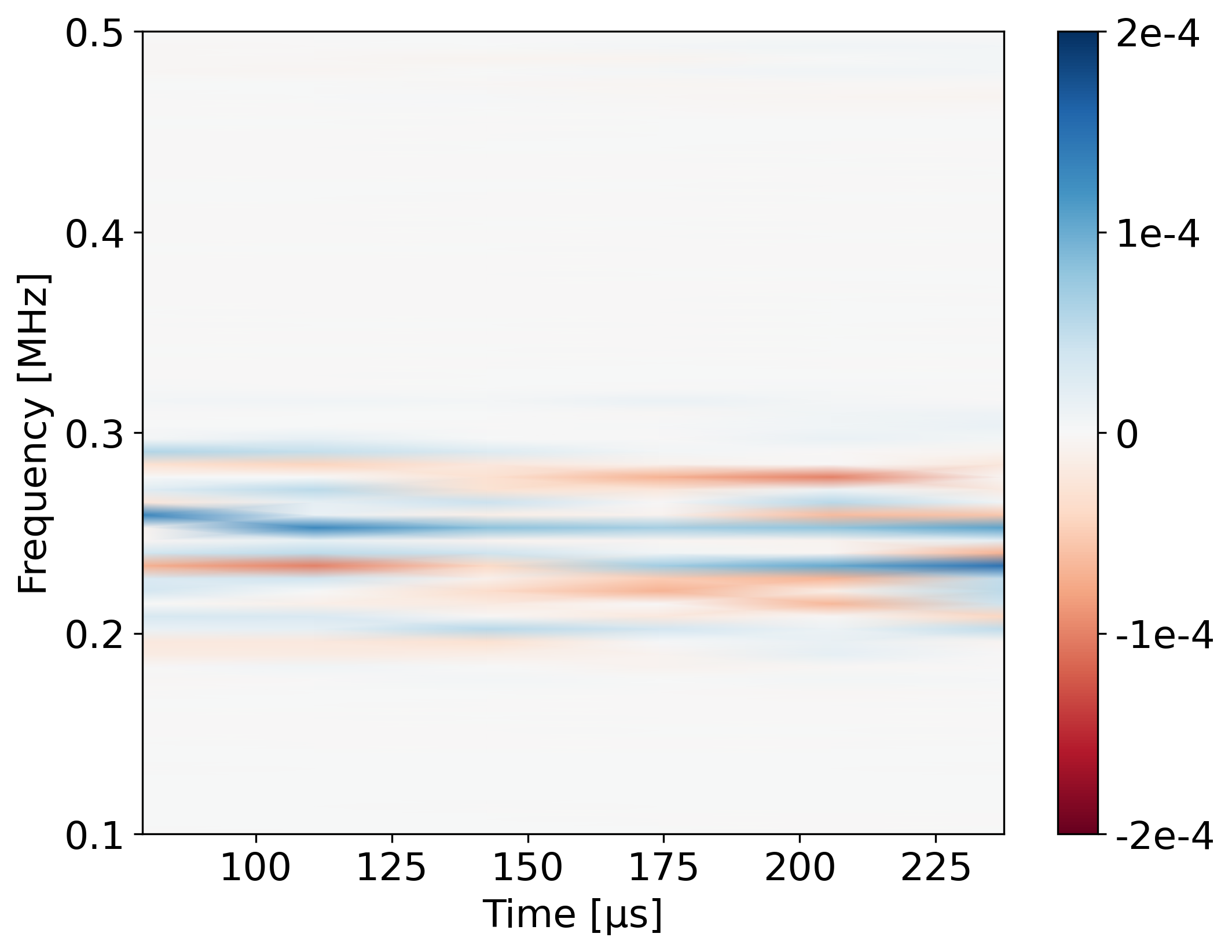}}
    \put(42,535){\color{black} \small {\fontfamily{phv}\selectfont \textbf{$S(z_{1}(1,0))-S(z_{1}(0,0))$}}}
    \put(256,535){\color{black} \small {\fontfamily{phv}\selectfont \textbf{$S(z_{3}(1,0))-S(z_{3}(0,0))$}}}
    \put(42,397){\color{black} \small {\fontfamily{phv}\selectfont \textbf{$S(z_{1}(2,0))-S(z_{1}(0,0))$}}}
    \put(256,397){\color{black} \small
    {\fontfamily{phv}\selectfont \textbf{$S(z_{3}(2,0))-S(z_{3}(0,0))$}}}
    \put(42,259){\color{black} \small {\fontfamily{phv}\selectfont \textbf{$S(z_{1}(3,0))-S(z_{1}(0,0))$}}}
    \put(256,259){\color{black} \small {\fontfamily{phv}\selectfont \textbf{$S(z_{3}(3,0))-S(z_{3}(0,0))$}}}
    \put(42,121){\color{black} \small {\fontfamily{phv}\selectfont \textbf{$S(z_{1}(4,0))-S(z_{1}(0,0))$}}}
    \put(256,121){\color{black} \small {\fontfamily{phv}\selectfont \textbf{$S(z_{3}(4,0))-S(z_{3}(0,0))$}}}
    \put(190,438){\color{black} \large {\fontfamily{phv}\selectfont \textbf{a}}}
    \put(190,300){\large {\fontfamily{phv}\selectfont \textbf{b}}} 
    \put(190,162){\large {\fontfamily{phv}\selectfont \textbf{c}}}
    \put(190,24){\large {\fontfamily{phv}\selectfont \textbf{d}}}
   
   \put(408,438){\large {\fontfamily{phv}\selectfont \textbf{e}}} 
   \put(409,300){\large {\fontfamily{phv}\selectfont \textbf{f}}} 
   \put(406,162){\large {\fontfamily{phv}\selectfont \textbf{g}}} 
   \put(410,24){\large {\fontfamily{phv}\selectfont \textbf{h}}} 
    \end{picture} 
    \caption{Scatter plots of the spectrogram of latent space subtracting by the spectrogram of the baseline latent space  representation from model Type III. $S(\cdot)$ indicates the function of spectrogram. Panels a-d: spectrogram of $z_1$ under 0kN at damage level 1-4 subtracting the spectrogram of $z_1$ under 0kN at healthy case; panels e-h: spectrogram of $z_3$ under 0kN at damage level 1-4 subtracting the spectrogram of $z_3$ under 0kN at healthy case.}
\label{fig:lat3_3} 
\end{figure}

\begin{figure}[t!]
    \begin{picture}(500,530)
    \put(10,414){\includegraphics[width=0.38\textwidth]{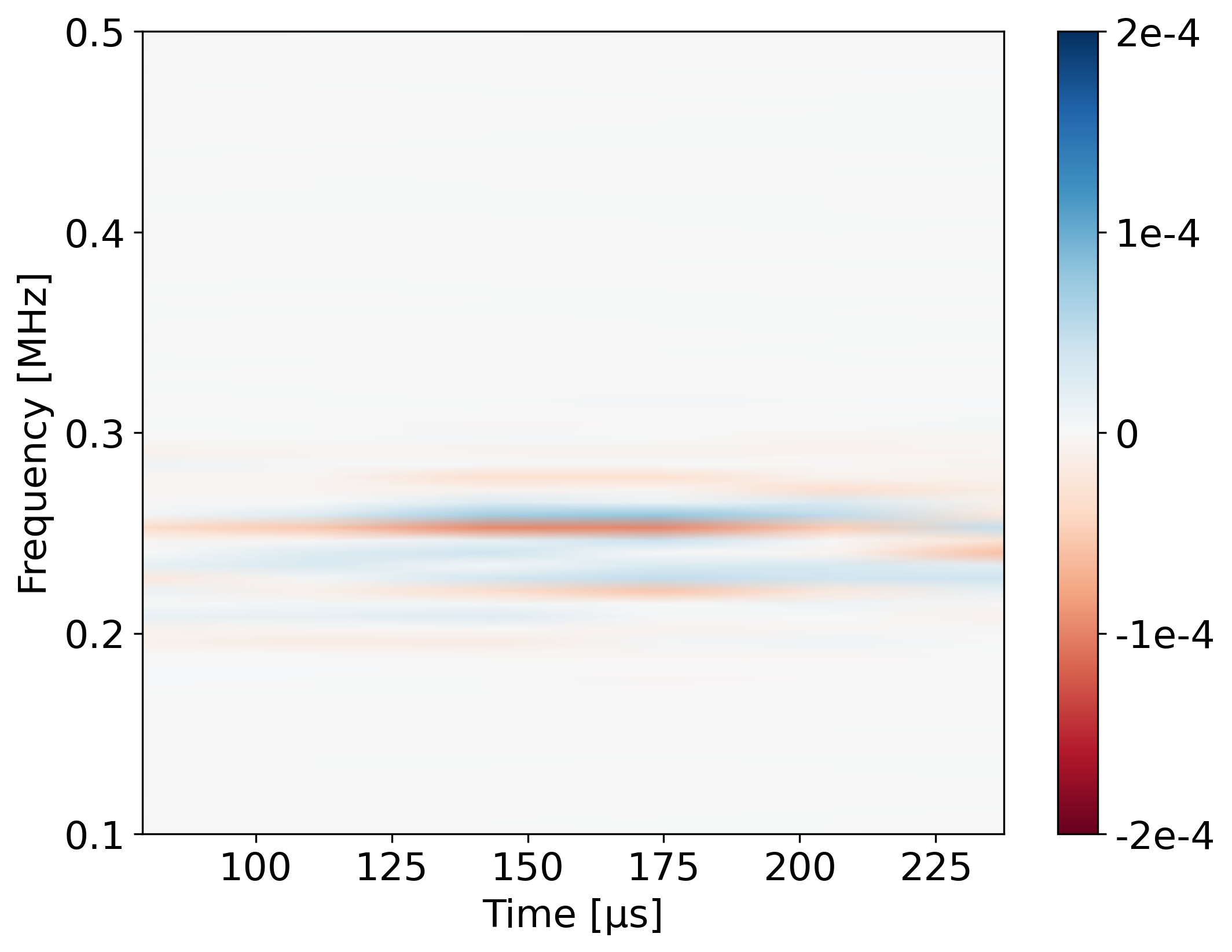}}
    \put(224,414){\includegraphics[width=0.38\textwidth]{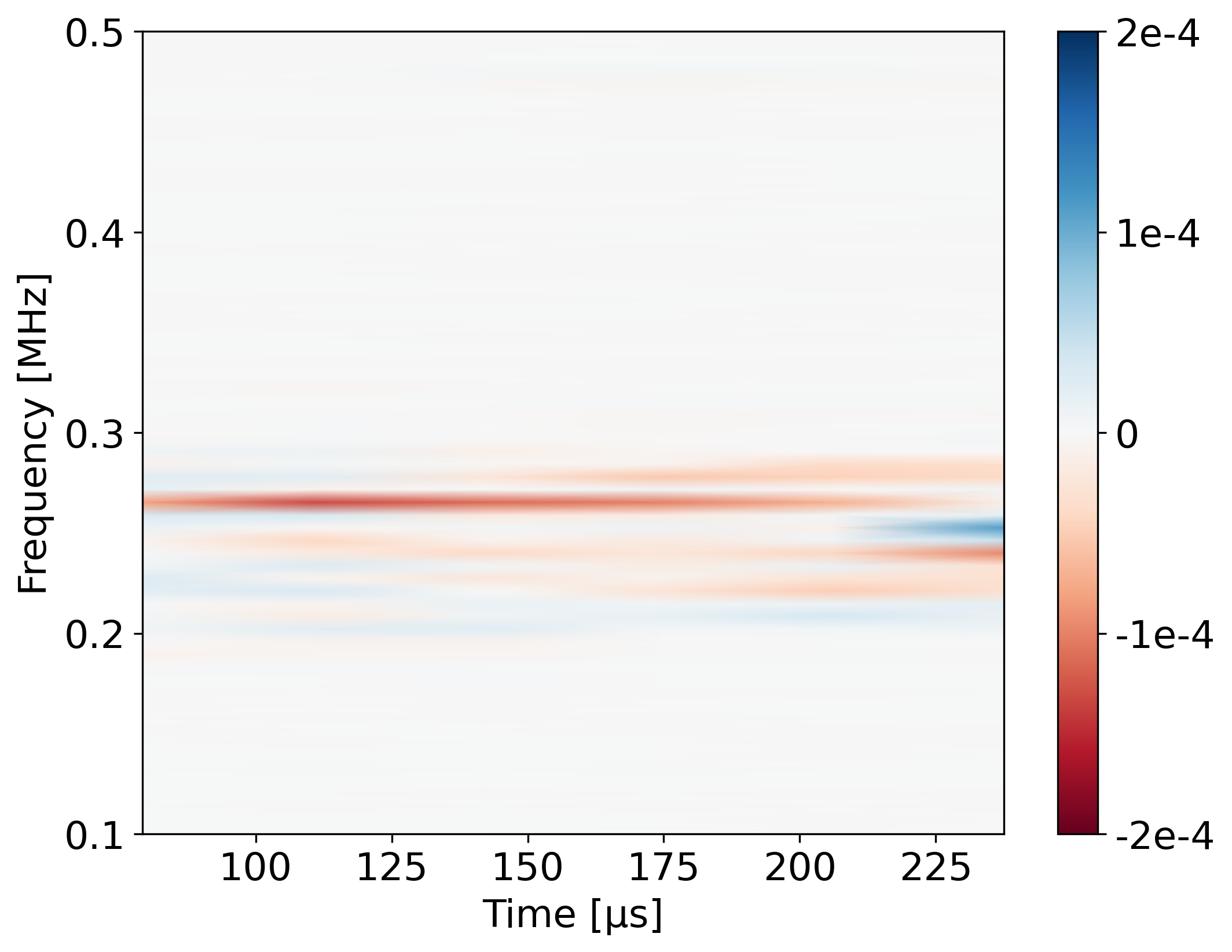}}
    \put(10,276){\includegraphics[width=0.38\textwidth]{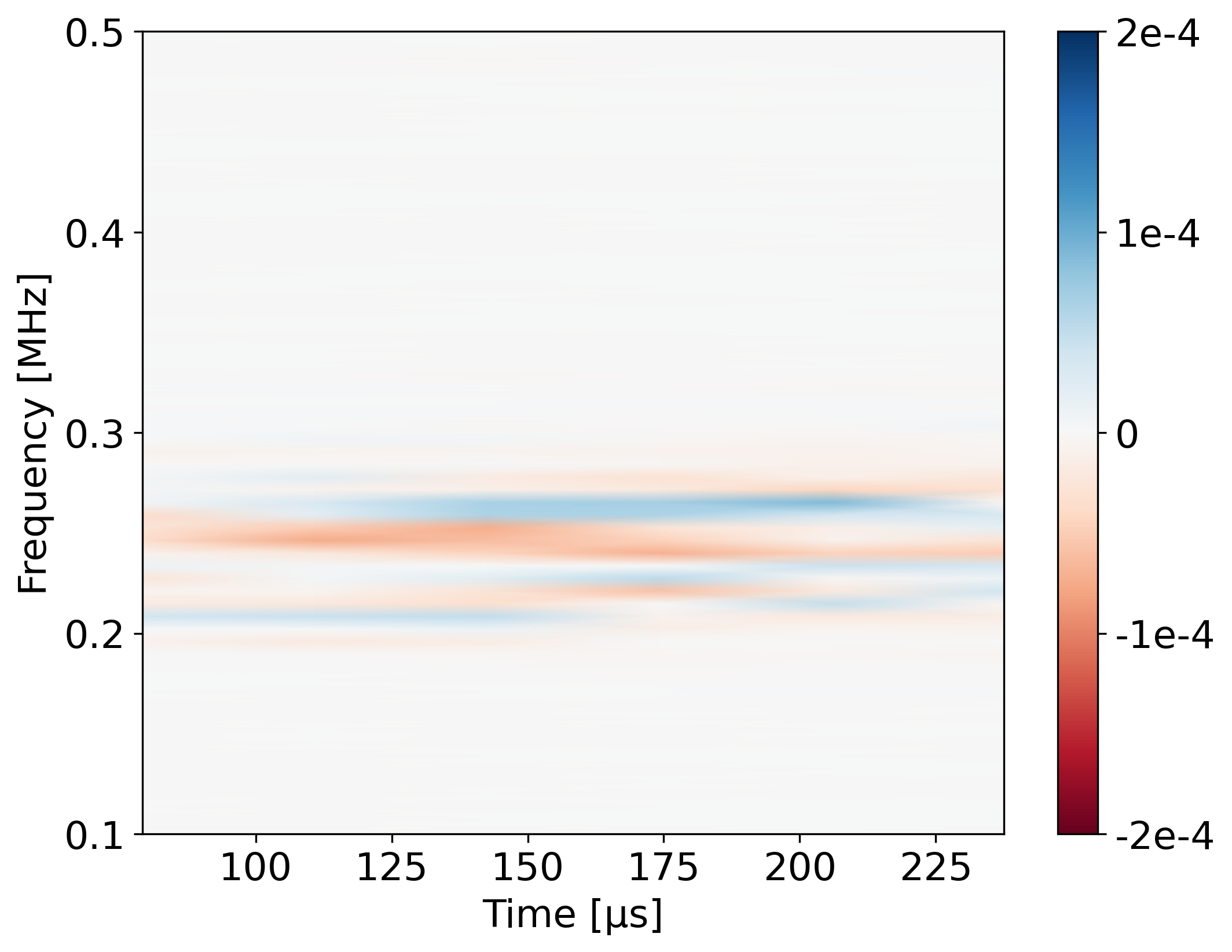}}
    \put(224,276){\includegraphics[width=0.38\textwidth]{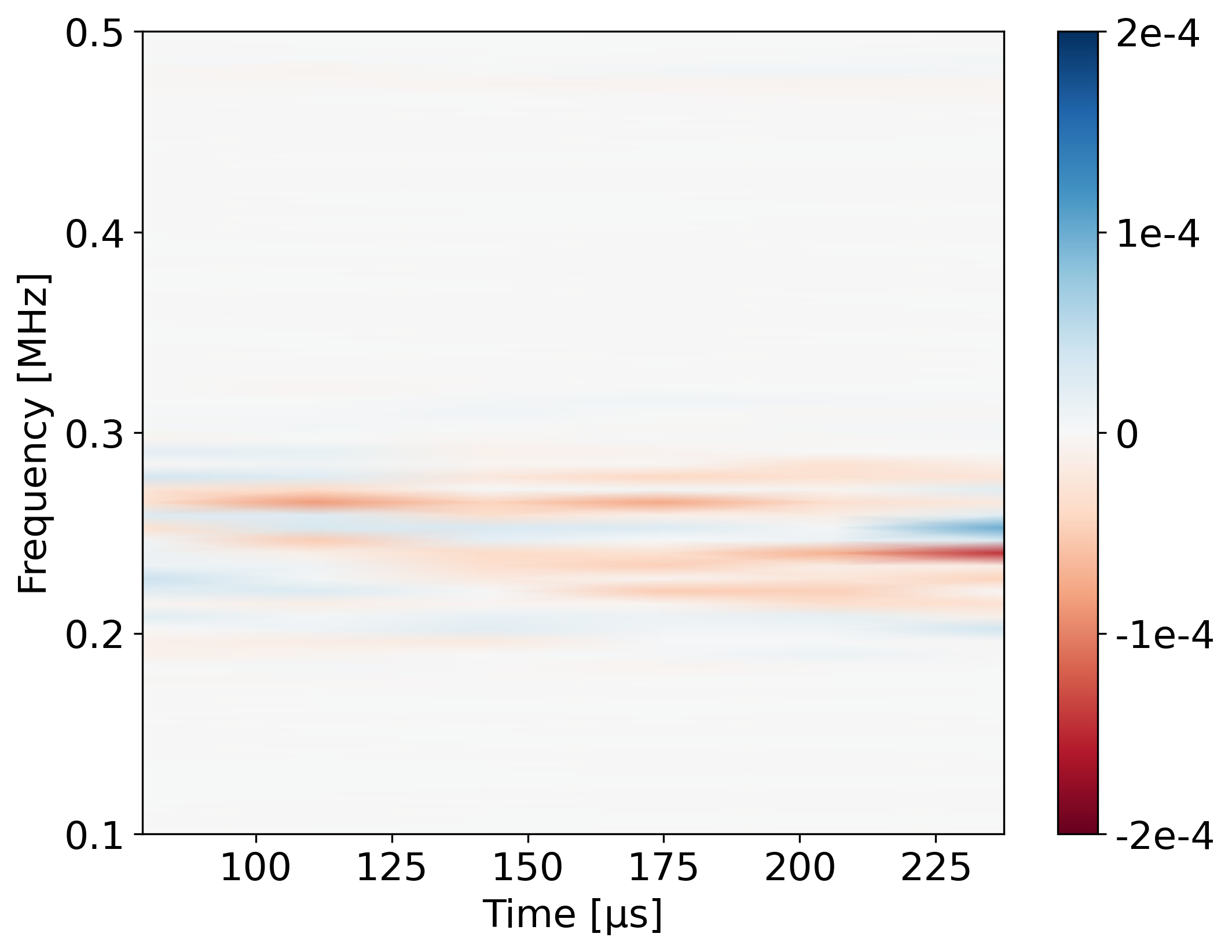}}
    \put(10,138){\includegraphics[width=0.38\textwidth]{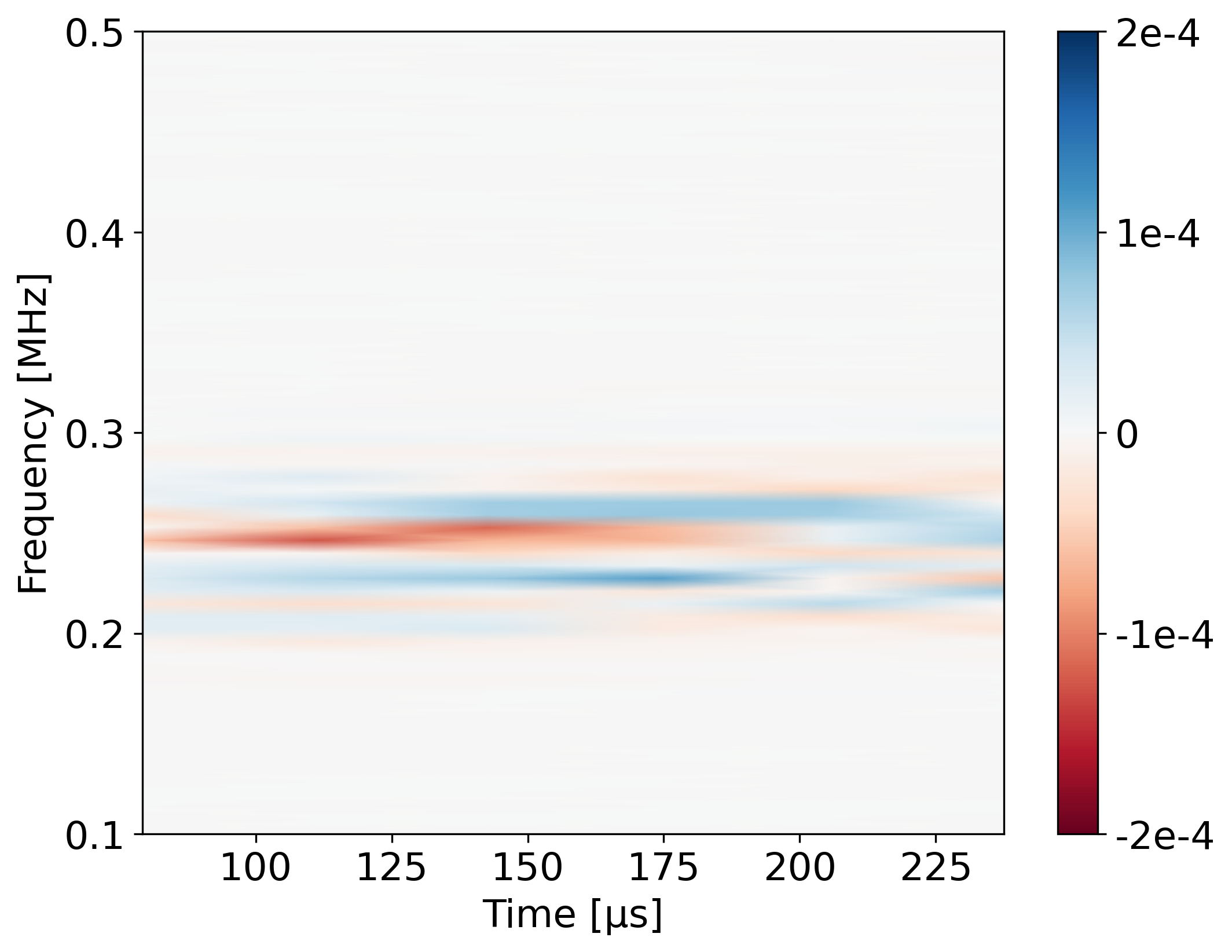}}
    \put(224,138){\includegraphics[width=0.38\textwidth]{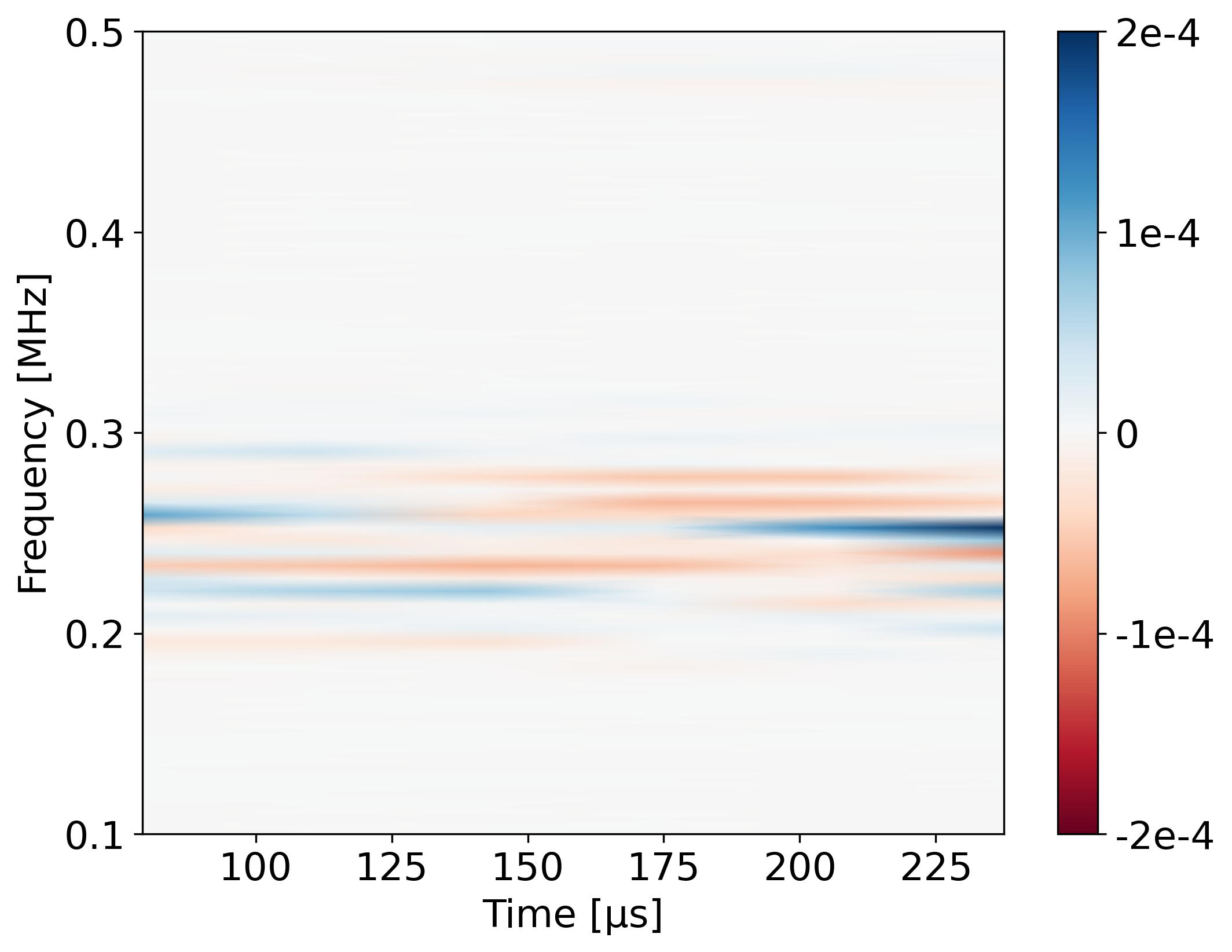}}
    \put(10,0){\includegraphics[width=0.38\textwidth]{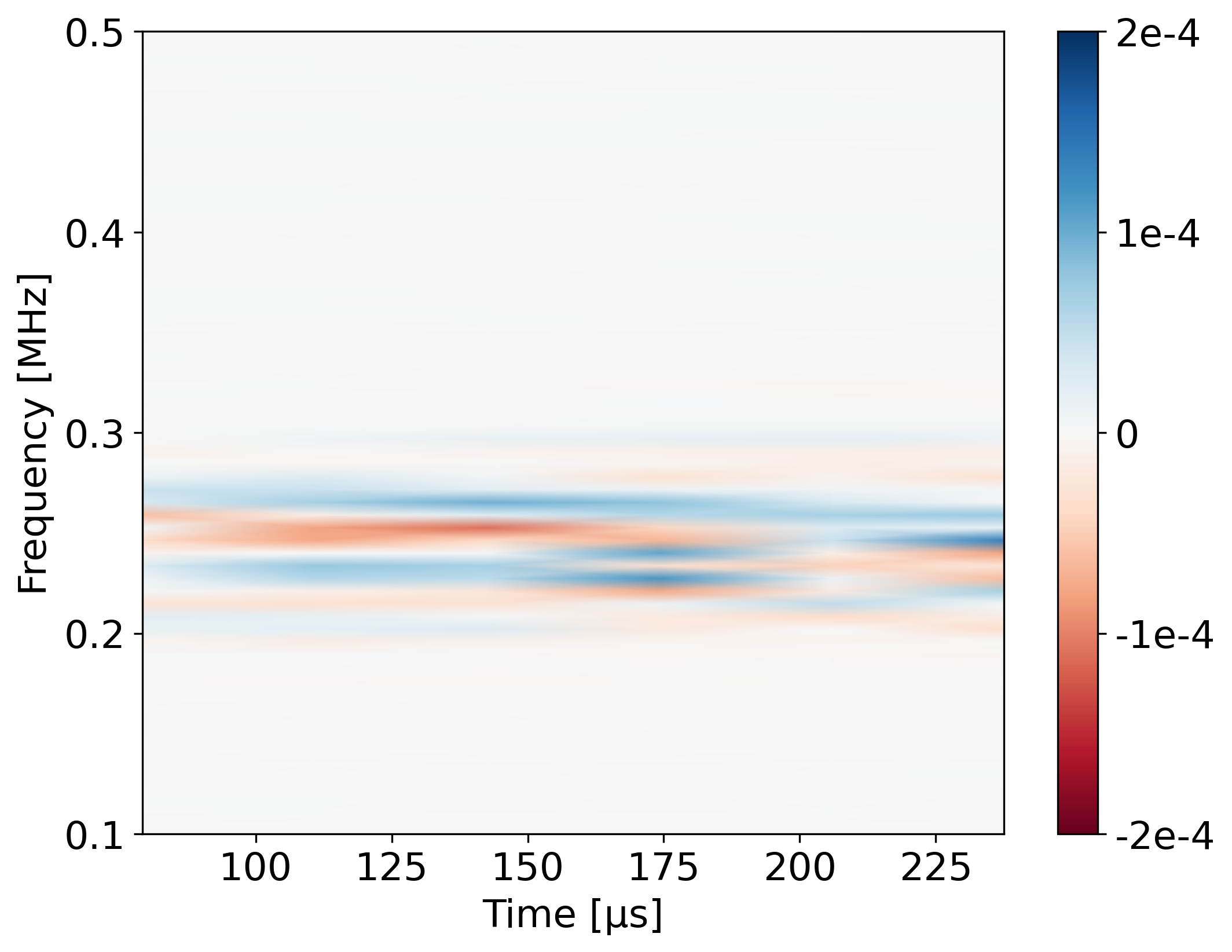}}
    \put(224,0){\includegraphics[width=0.38\textwidth]{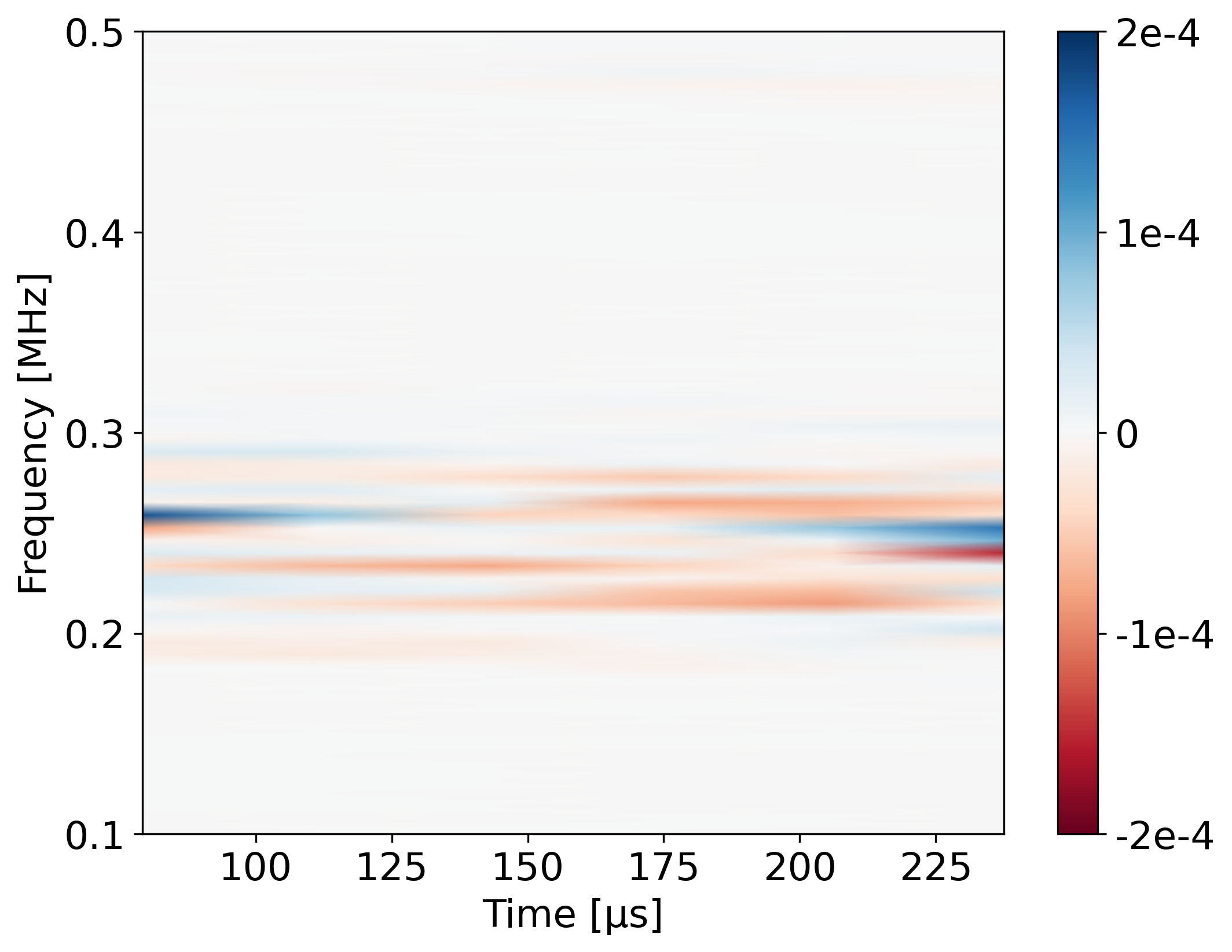}}
    \put(42,535){\color{black} \small {\fontfamily{phv}\selectfont \textbf{$S(z_{1}(1,20))-S(z_{1}(0,20))$}}}
    \put(256,535){\color{black} \small {\fontfamily{phv}\selectfont \textbf{$S(z_{3}(1,20))-S(z_{3}(0,20))$}}}
    \put(42,397){\color{black} \small {\fontfamily{phv}\selectfont \textbf{$S(z_{1}(2,20))-S(z_{1}(0,20))$}}}
    \put(256,397){\color{black} \small
    {\fontfamily{phv}\selectfont \textbf{$S(z_{3}(2,20))-S(z_{3}(0,20))$}}}
    \put(42,259){\color{black} \small {\fontfamily{phv}\selectfont \textbf{$S(z_{1}(3,20))-S(z_{1}(0,20))$}}}
    \put(256,259){\color{black} \small {\fontfamily{phv}\selectfont \textbf{$S(z_{3}(3,20))-S(z_{3}(0,20))$}}}
    \put(42,121){\color{black} \small {\fontfamily{phv}\selectfont \textbf{$S(z_{1}(4,20))-S(z_{1}(0,20))$}}}
    \put(256,121){\color{black} \small {\fontfamily{phv}\selectfont \textbf{$S(z_{3}(4,20))-S(z_{3}(0,20))$}}}
    \put(190,438){\color{black} \large {\fontfamily{phv}\selectfont \textbf{a}}}
    \put(190,300){\large {\fontfamily{phv}\selectfont \textbf{b}}} 
    \put(190,162){\large {\fontfamily{phv}\selectfont \textbf{c}}}
    \put(190,24){\large {\fontfamily{phv}\selectfont \textbf{d}}}
   
   \put(408,438){\large {\fontfamily{phv}\selectfont \textbf{e}}} 
   \put(409,300){\large {\fontfamily{phv}\selectfont \textbf{f}}} 
   \put(406,162){\large {\fontfamily{phv}\selectfont \textbf{g}}} 
   \put(410,24){\large {\fontfamily{phv}\selectfont \textbf{h}}} 
    \end{picture} 
    \caption{Scatter plots of the spectrogram of latent space subtracting by the spectrogram of the baseline latent space  representation from model Type III. $S(\cdot)$ indicates the function of spectrogram. Panels a-d: spectrogram of $z_1$ under 20kN at damage level 1-4 subtracting the spectrogram of $z_1$ under 20kN at healthy case; panels e-h: spectrogram of $z_3$ under 20kN at damage level 1-4 subtracting the spectrogram of $z_3$ under 20kN at healthy case.}
\label{fig:lat3_4} 
\end{figure}

The feature space derived from the Type III model exhibits distinct characteristics compared to the previous two models, primarily due to its time-varying nature, as illustrated in Figure \ref{fig:lat3}. Panels (a) and (b) display the two latent variables, $z_1$ and $z_3$, under different loads at damage level 4. Panels (c) and (d) present the same two variables across varying damage levels under a load of 20 kN. Each latent space variable maintains the same length as the original data, changing its value at each sampling instant and can be interpreted as a “latent signal.” However, it lacks similar amplitude characteristics when compared to the input signals. Given its irregularity in the temporal domain, the feature space is transformed into the frequency domain for further analysis.

From the spectrograms, the peak frequency of the time-varying latent space closely aligns with the original central frequency of 250 kHz. Each data segment is 256 samples long, with overlap points rounding to 95$\%$ of the segment length. To analyze the evolution of latent space variables as states change, scatter plots were generated using the spectrograms of target states and baseline states. As shown in Figure \ref{fig:lat3_1}, the spectrograms of $z_1$ at damage level 0 (healthy case) under loads from 5 to 20 kN, with 5 kN increments, were subtracted from the spectrogram at damage level 0 under 0 kN, presented in panels (a-d). It is evident that as the load increases, the scatter plot patterns become more complex. Not only does the number of crests and troughs increase, but the maximum absolute extreme values also trend upward, indicated by darker colors. Panels (e-h) depict the spectrogram scatter plots of $z_3$ under the same states as panels (a-d), yielding similar conclusions. This trend is attributed to the fact that larger loads lead to greater deformation of the plate, which in turn affects the wave speed. Consequently, the compressed signals exhibit more variations compared to the baseline signal.

Instead of illustrating patterns under the healthy case, Figure \ref{fig:lat3_2} highlights the evolution of the spectrogram scatter plots of $z_1$ and $z_3$ at damage level 4. In panels (a-d), the load increases from 5 to 20 kN while the damage severity remains at level 4, with the spectrogram of $z_1$ under 0 kN load serving as the baseline. Panels (e-h) display the case for $z_3$. The deepened color in both latent variables indicates heightened sensitivity to external load. A comparison between Figure \ref{fig:lat3_1} and Figure \ref{fig:lat3_2} reveals that the sensitivity of the time-varying latent space to loading conditions is apparent across various damage levels. 

In addition to external loading conditions, increasing damage severity also results in larger extreme values and more complex scatter plot patterns. As depicted in Figure 
\ref{fig:lat3_3}, the color contrast of $z_1$ scatters at a more severe damage level, specifically level 4 in panel (d) under 0 kN load, is significantly more pronounced than in panels (a), (b), and (c). A similar observation is made for in panels (e-h). 

To reinforce these findings, results under 20 kN are presented in Figure \ref{fig:lat3_4}. A brief conclusion can be drawn here: more severe damages lead to greater disparity in the time-varying latent space. A comparison of Figure \ref{fig:lat3_3} and Figure \ref{fig:lat3_4} shows that this pattern persists even with varying EOCs. The simultaneous observation and capture of changes in damage state and EOCs imply the effectiveness of the latent space-based classifier.

\subsubsection{State Estimation}
\begin{table}[b!]
\centering
\caption{\uppercase{Summary of the state estimation performance of Model I.}}\label{tab:type1}
\renewcommand{\arraystretch}{1.2}
{\footnotesize
\tabcolsep=0.11cm
\begin{tabular}{|c|c|c|c|c|c|} 
\hline
 Level & 0 &1&2&3&4 \\

\cline{1-6}

 $\mu$ & 0.0472
 & 1.0119
 & 2.0357 & 2.9912 & 4.0026 \\
\hline
STD & 0.0780
 & 0.0916
 & 0.0658 & 0.0805 & 0.1122 \\
\hline
Load(kN) & 0 &5&10&15&20 \\
\hline
$\mu$ & 0.0178
 & 5.003
 & 0.9840 & 14.9013 & 19.5305 \\
\hline
STD & 0.0755
 & 0.0616
 & 0.0671 & 0.0892 & 0.1195 \\
\hline

\end{tabular}} 
\end{table}
%
\begin{table}[t!]
\centering
\caption{\uppercase{Summary of the state estimation performance of Model II.}}\label{tab:type2}
\renewcommand{\arraystretch}{1.2}
{\footnotesize
\tabcolsep=0.11cm
\begin{tabular}{|c|c|c|c|c|c|} 
\hline
 Level & 0 &1&2&3&4 \\

\cline{1-6}

 $\mu$ & 0.0092
 & 1.0123
 & 2.0186 & 2.9977 & 4.0059 \\
\hline
STD & 0.0119
 & 0.0093
 & 0.0106 & 0.0528 & 0.0138 \\
\hline
Load(kN) & 0 &5&10&15&20 \\
\hline
$\mu$ & 0.0044
 & 5.0725
 & 10.0052 & 14.9901 & 19.8665 \\
\hline
STD & 0.0188
 & 0.0239
 & 0.0111 & 0.0126 & 0.0194 \\
\hline

\end{tabular}} 
\end{table}
%
\begin{table}[t!]
\centering
\caption{\uppercase{Summary of the state estimation performance of Model III.}}\label{tab:type3}
\renewcommand{\arraystretch}{1.2}
{\footnotesize
\tabcolsep=0.11cm
\begin{tabular}{|c|c|c|c|c|c|} 
\hline
 Level & 0 &1&2&3&4 \\

\cline{1-6}

 $\mu$ & -0.0091
 & 1.0032
 & 2.0184 & 3.0068 & 3.9933 \\
\hline
STD & 0.0062
 & 0.0108
 & 0.0249 & 0.0176 & 0.0136 \\
\hline
Load(kN) & 0 &5&10&15&20 \\
\hline
$\mu$ & 0.0032
 & 4.9930
 & 9.9395 & 14.9452 & 19.9527 \\
\hline
STD & 0.0066
 & 0.0104
 & 0.0155 & 0.0134 & 0.0213 \\
\hline

\end{tabular}} 
\end{table}
Recall that the first function of the proposed framework is state estimation based on signals collected from unknown states. After the signals are compressed by the encoder, the data is forwarded to the first FFNN for state prediction. Damage levels are represented by indices 0-4, while load levels are represented by indices 0, 5, 10, 15, and 20. Since the predictions are generated as float data, they are rounded to the nearest integer to approximate the states with the highest probability. Nearly perfect results are observed during this process. For the Type I model, only one misprediction occurs throughout the entire test set, while for the other two model types, all predicted results align with the ground truth states. This outcome can be viewed as a mutual validation of the previous latent space analysis presented in Section \ref{Sec:lat}, where latent variables for different states could be effectively differentiated.

At this point, the conclusion is that the model's estimation capability is directly influenced by the classification quality in the latent space. Tables \ref{tab:type1}-\ref{tab:type3} provide the average float results, including the mean and standard deviation of predictions across different states using the full test dataset. As shown, the mean values closely match the indices representing actual damage or load levels, and the standard deviations are small, indicating the model's excellent state prediction ability.


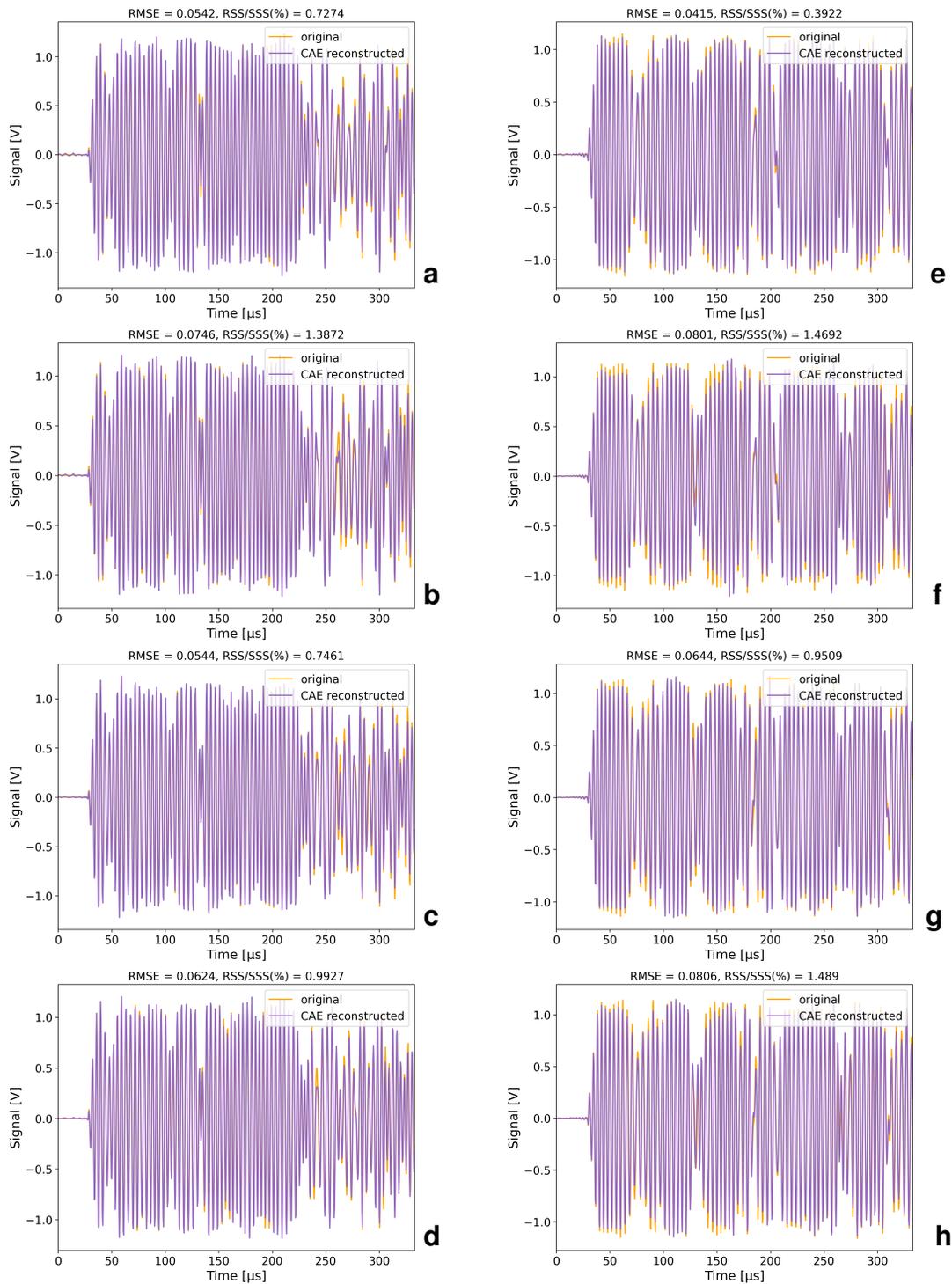
\begin{figure}[t!]
    \begin{picture}(500,530)
    \put(10,414){\includegraphics[width=0.38\textwidth]{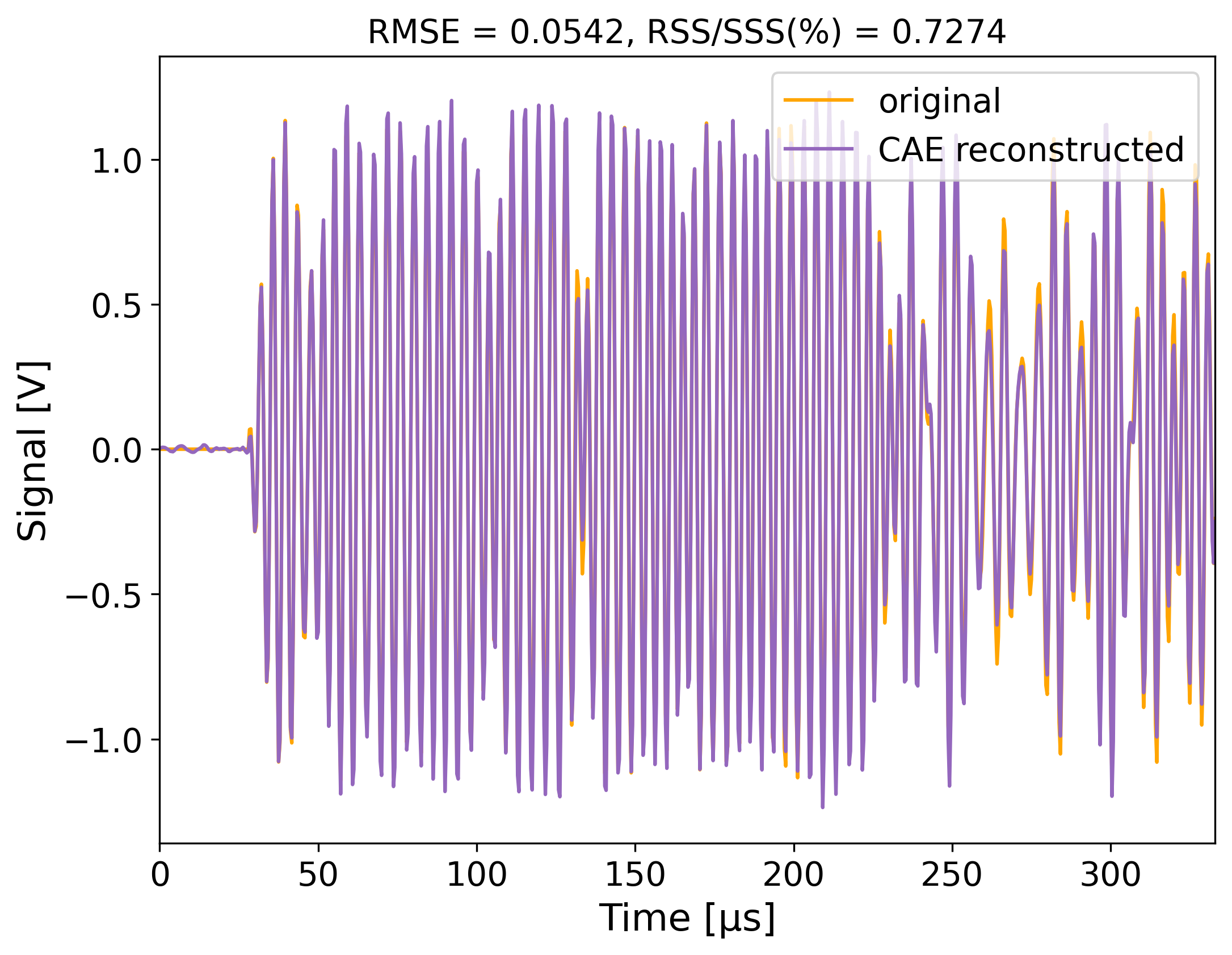}}
    \put(224,414){\includegraphics[width=0.38\textwidth]{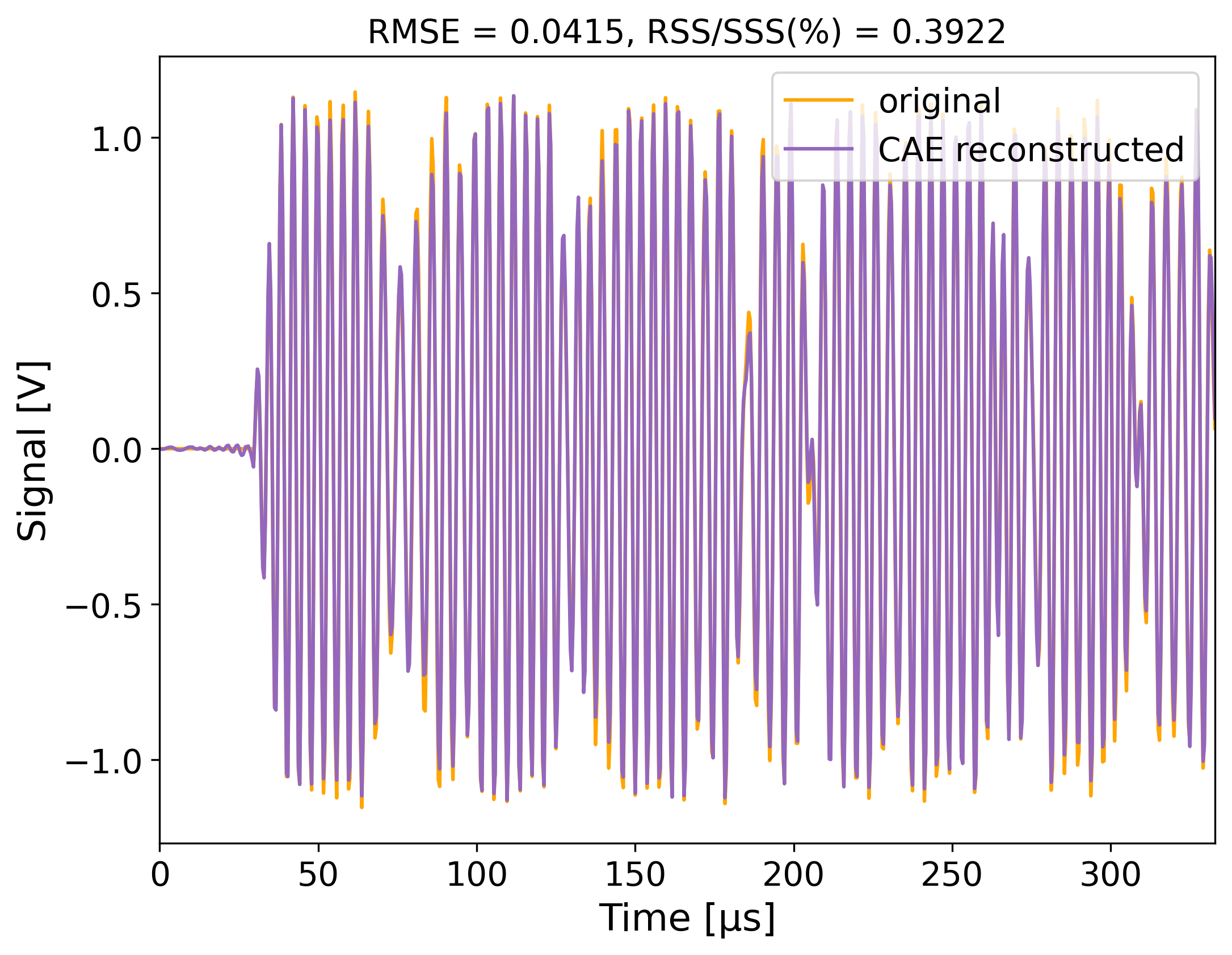}}
    \put(10,276){\includegraphics[width=0.38\textwidth]{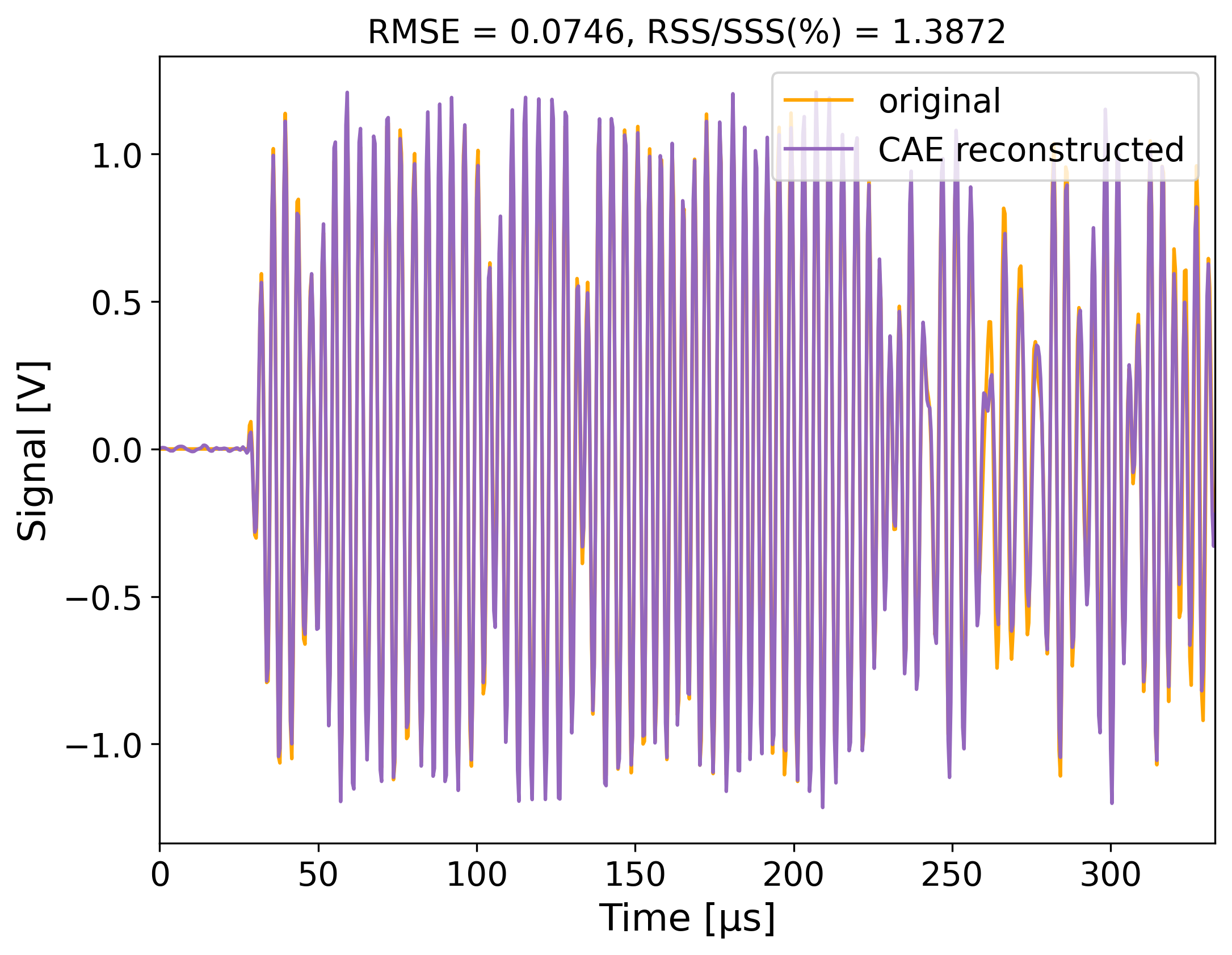}}
    \put(224,276){\includegraphics[width=0.38\textwidth]{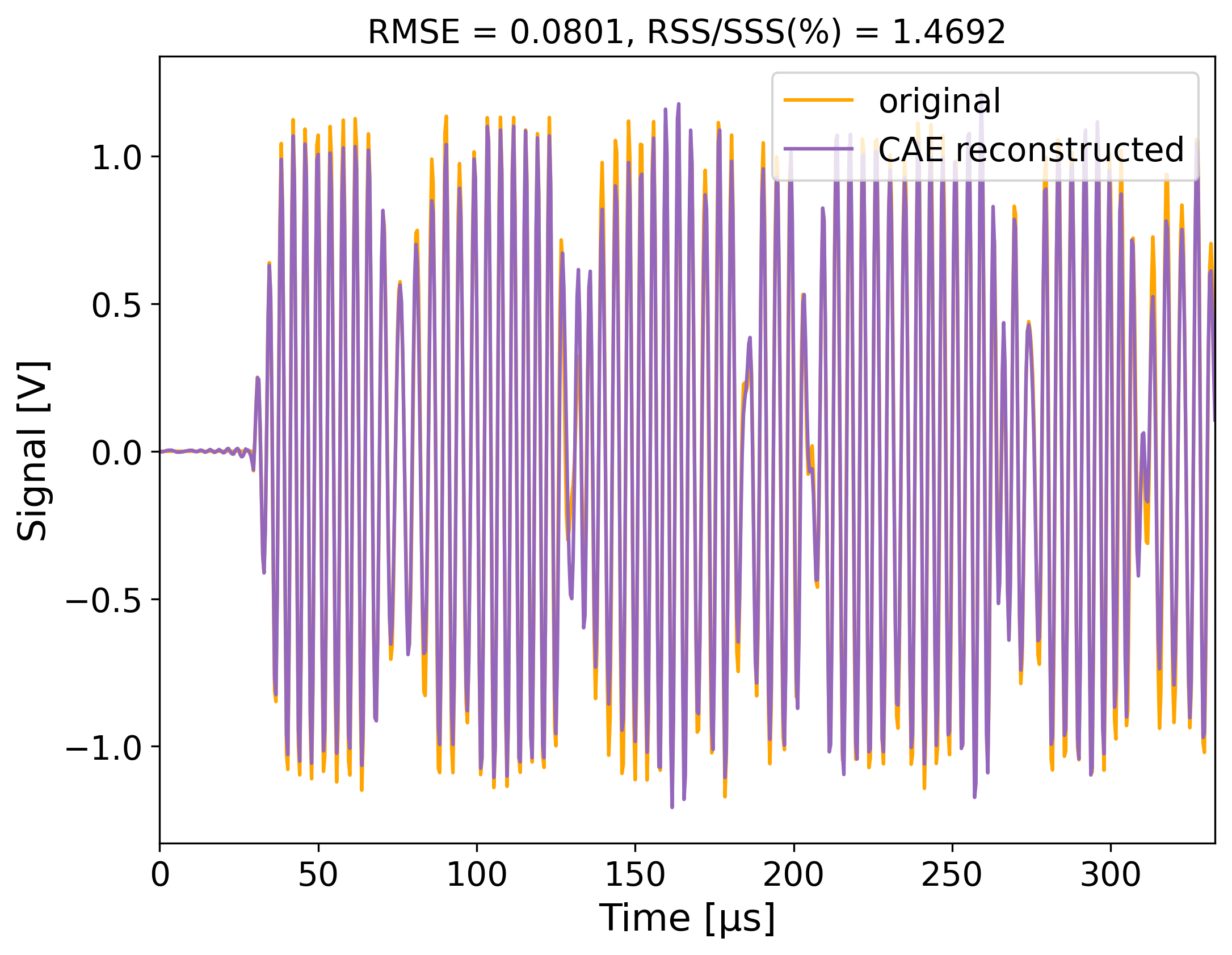}}
    \put(10,138){\includegraphics[width=0.38\textwidth]{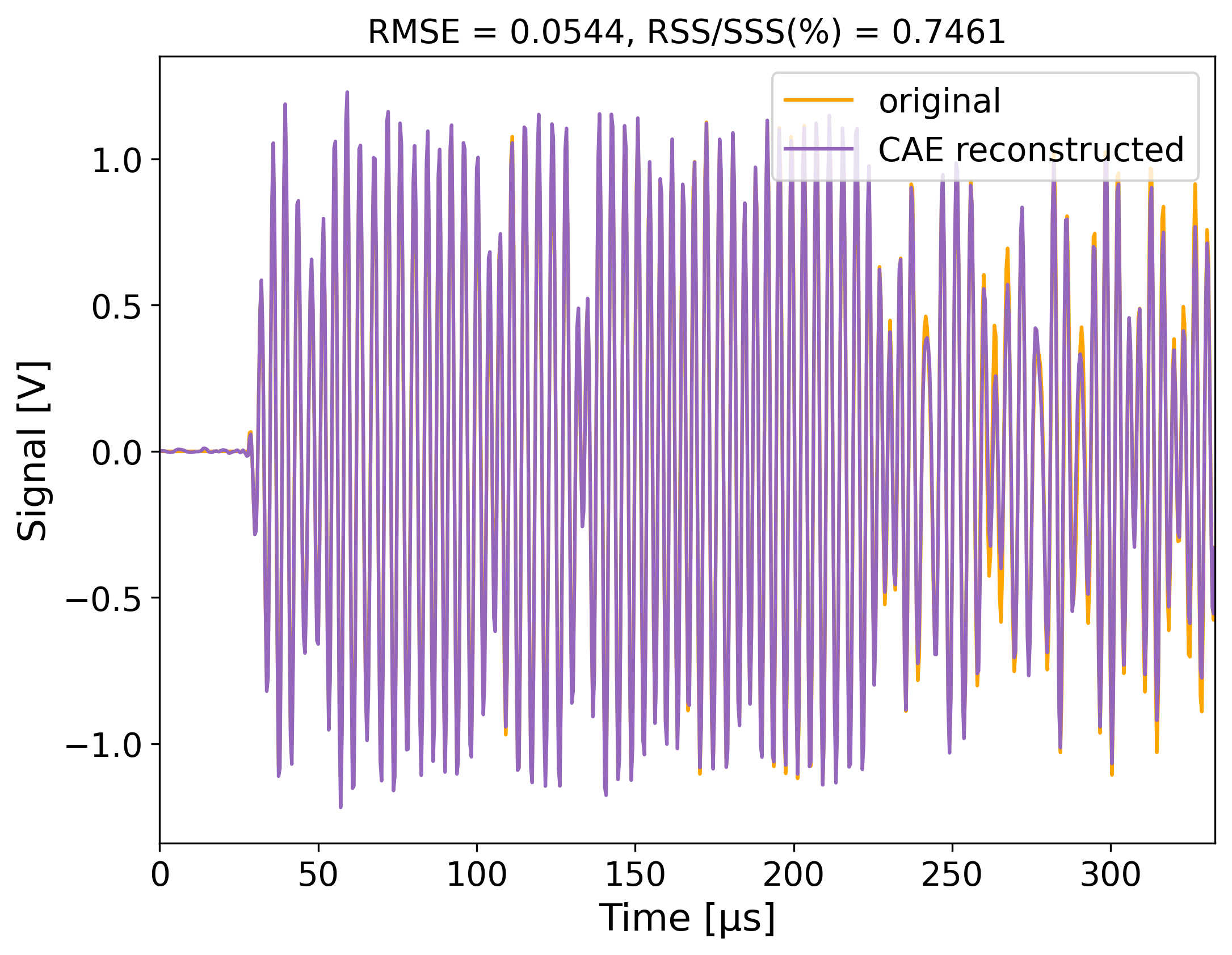}}
    \put(224,138){\includegraphics[width=0.38\textwidth]{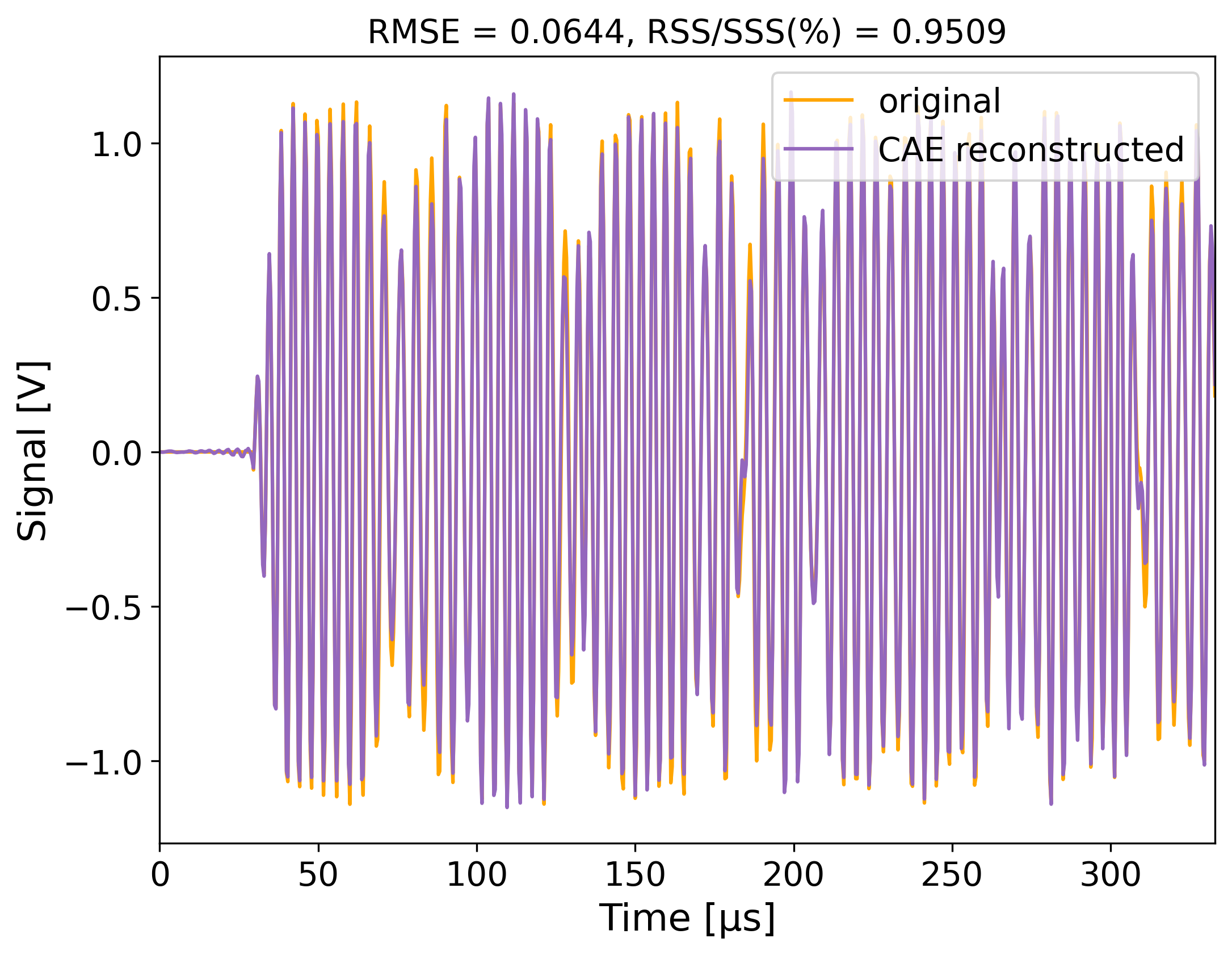}}
    \put(10,0){\includegraphics[width=0.38\textwidth]{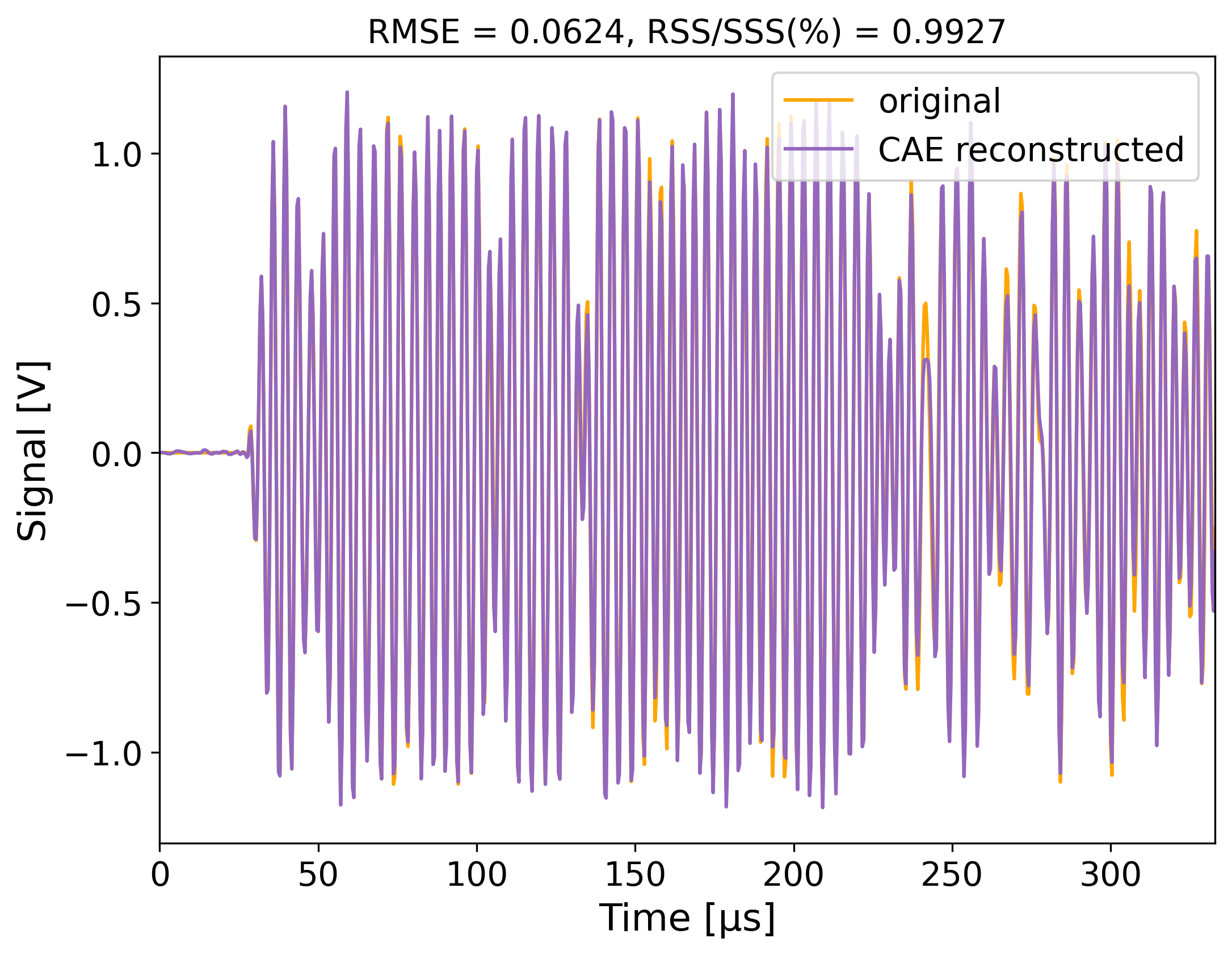}}
    \put(224,0){\includegraphics[width=0.38\textwidth]{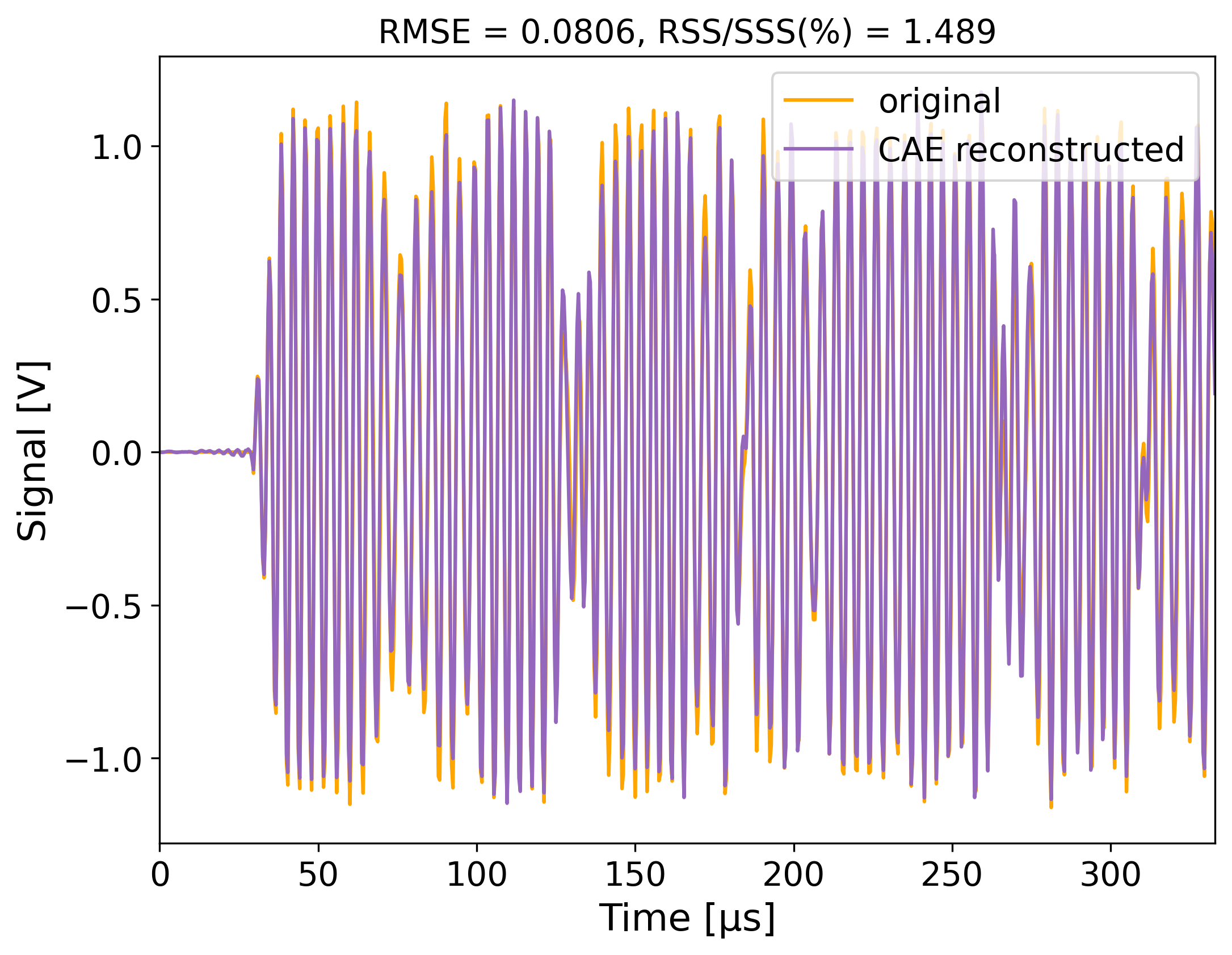}}
    \put(190,432){\color{black} \large {\fontfamily{phv}\selectfont \textbf{a}}}
    \put(190,294){\large {\fontfamily{phv}\selectfont \textbf{b}}} 
    \put(190,156){\large {\fontfamily{phv}\selectfont \textbf{c}}}
    \put(190,18){\large {\fontfamily{phv}\selectfont \textbf{d}}}
   
   \put(408,432){\large {\fontfamily{phv}\selectfont \textbf{e}}} 
   \put(409,294){\large {\fontfamily{phv}\selectfont \textbf{f}}} 
   \put(406,156){\large {\fontfamily{phv}\selectfont \textbf{g}}} 
   \put(410,18){\large {\fontfamily{phv}\selectfont \textbf{h}}} 
    \end{picture} 
    \caption{Reconstructed signals from Model I. Panel a: under healthy case with 10kN from path 2-6; panel b: under healthy case with 20kN from path 2-6; panel c: under damage level 2 with 10kN from path 2-6; panel d: under damage level 4 with 20kN from path 2-6; panel e: under healthy case with 10kN from path 3-4; panel f: under healthy case with 20kN from path 3-4; panel g: under damage level 2 with 10kN from path 3-4; panel h: under damage level 4 with 20kN from path 3-4.}
\label{fig:recon_1} 
\end{figure}

\begin{figure}[t!]
    \begin{picture}(500,314)
    \put(10,204){\includegraphics[width=0.48\textwidth]{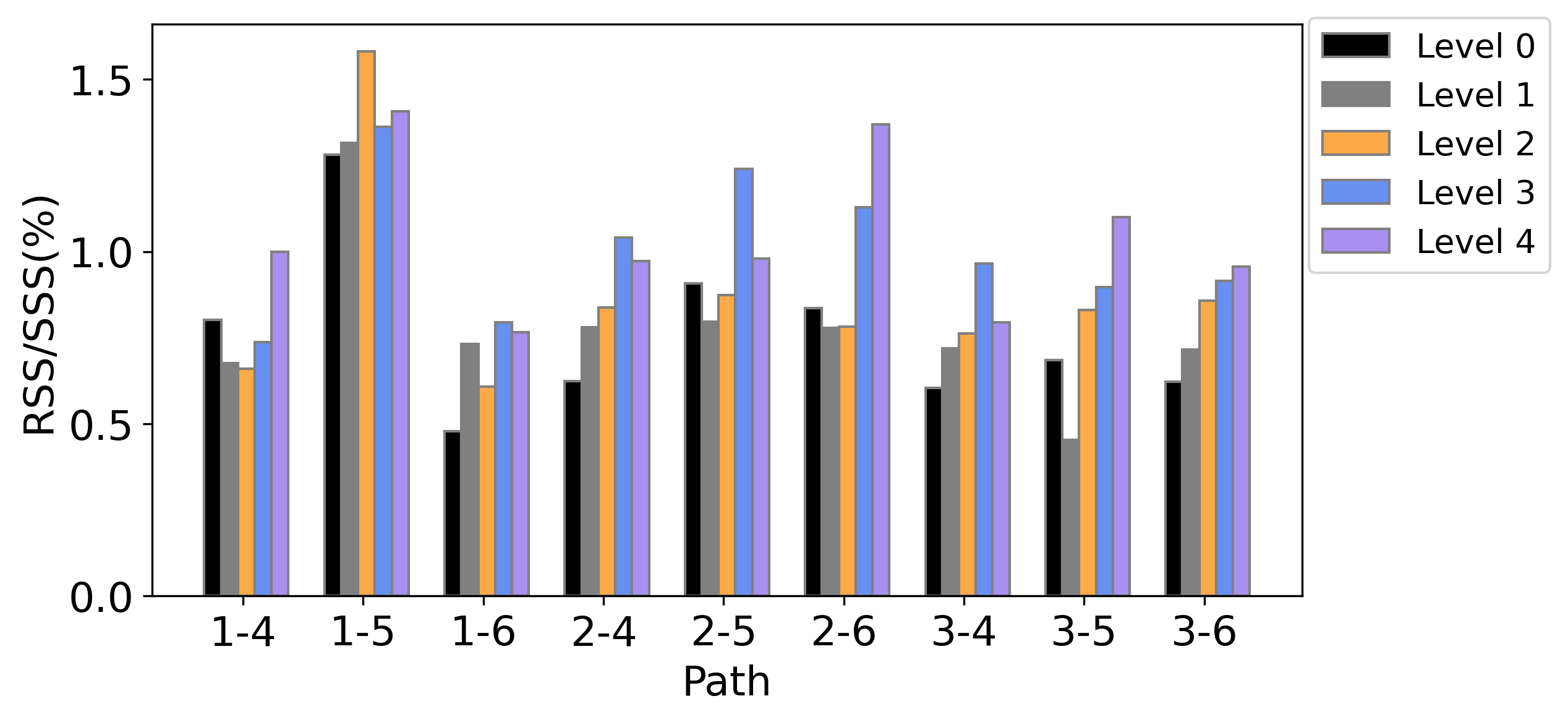}}
    \put(240,204){\includegraphics[width=0.48\textwidth]{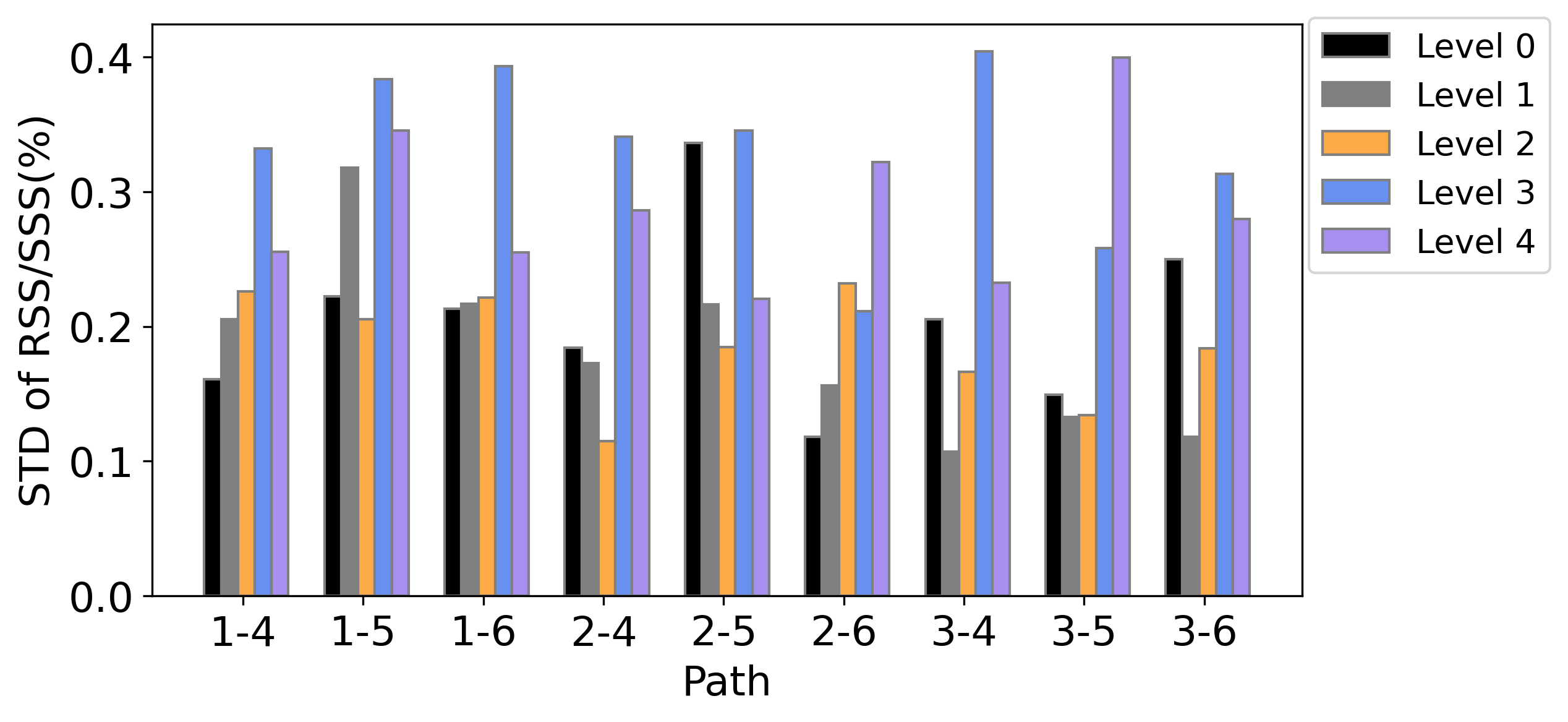}}
    \put(6,102){\includegraphics[width=0.486\textwidth]{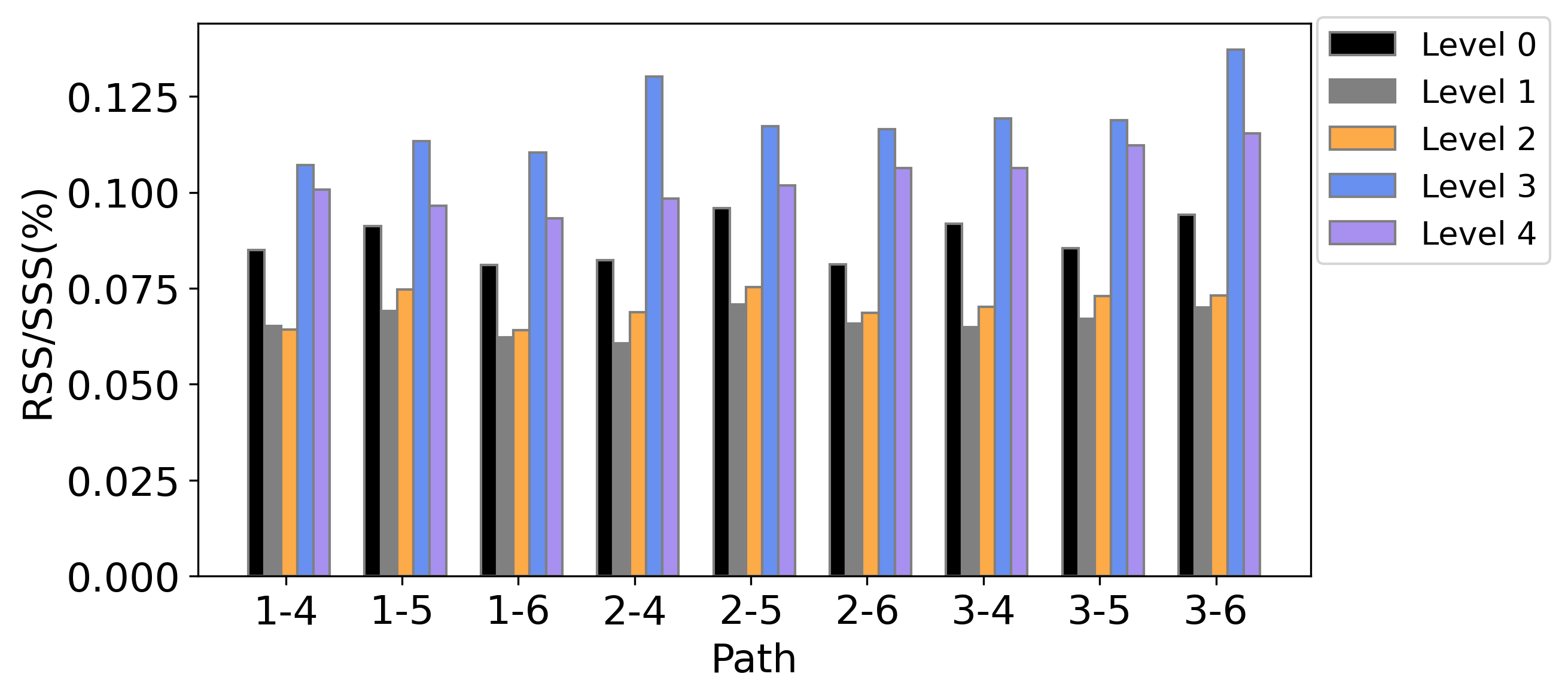}}
    \put(238,102){\includegraphics[width=0.48\textwidth]{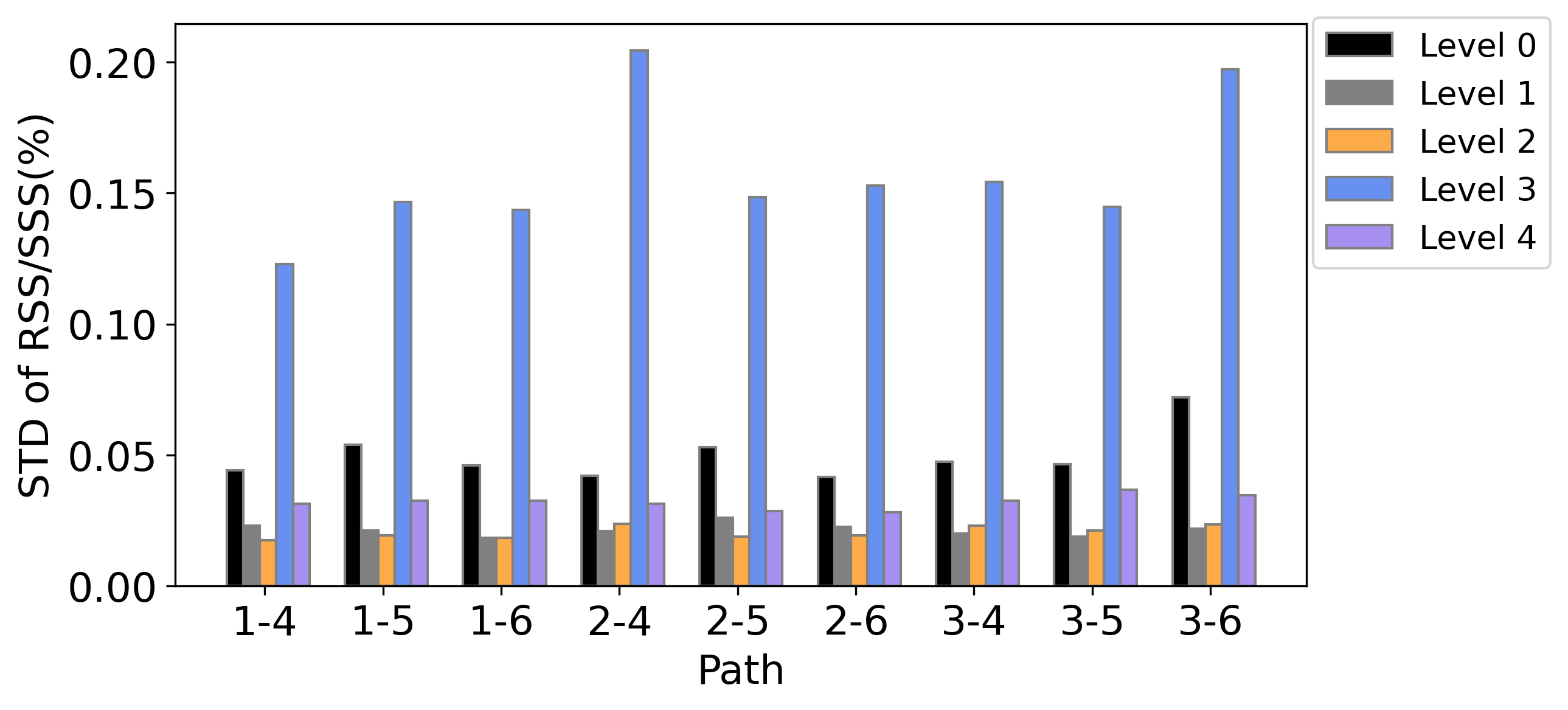}}
    \put(10,0){\includegraphics[width=0.48\textwidth]{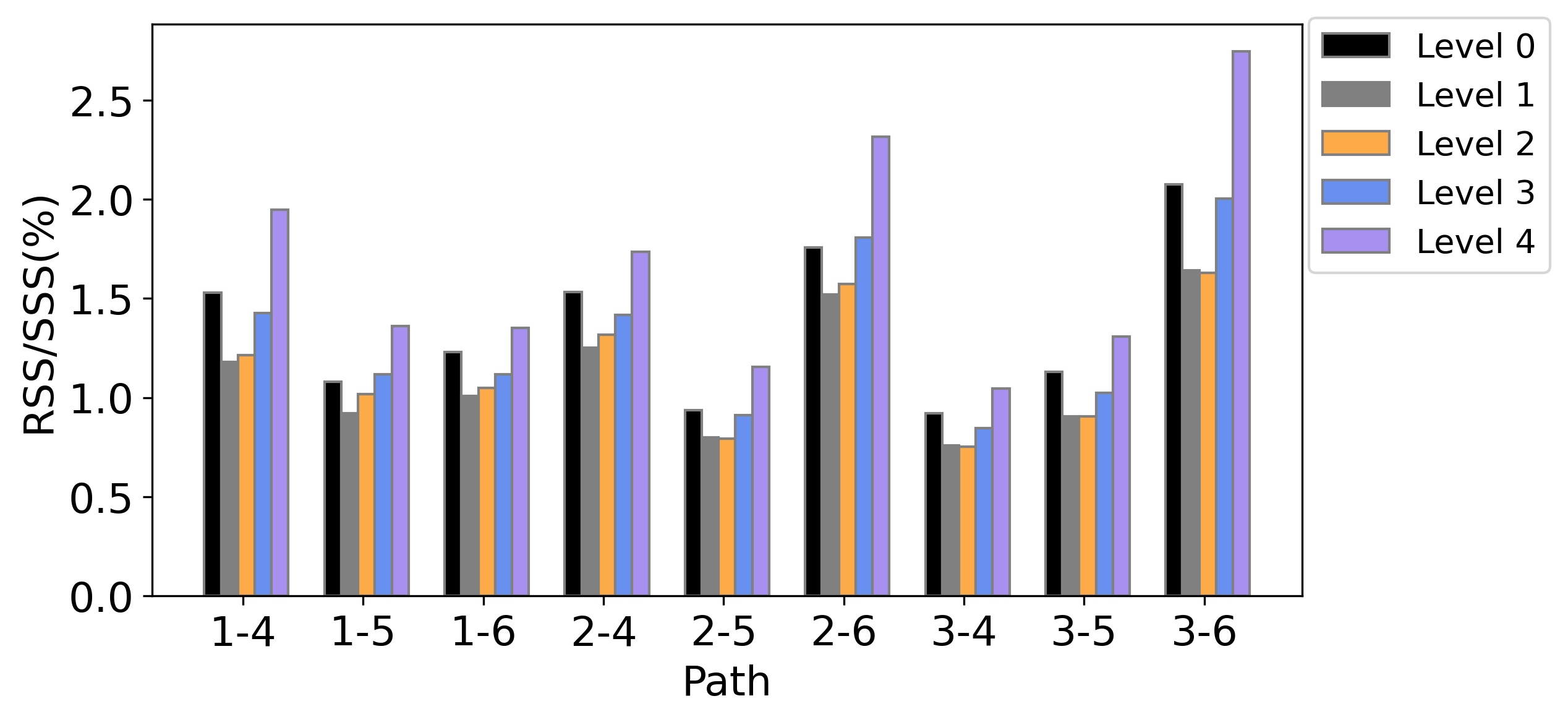}}
    \put(240,0){\includegraphics[width=0.48\textwidth]{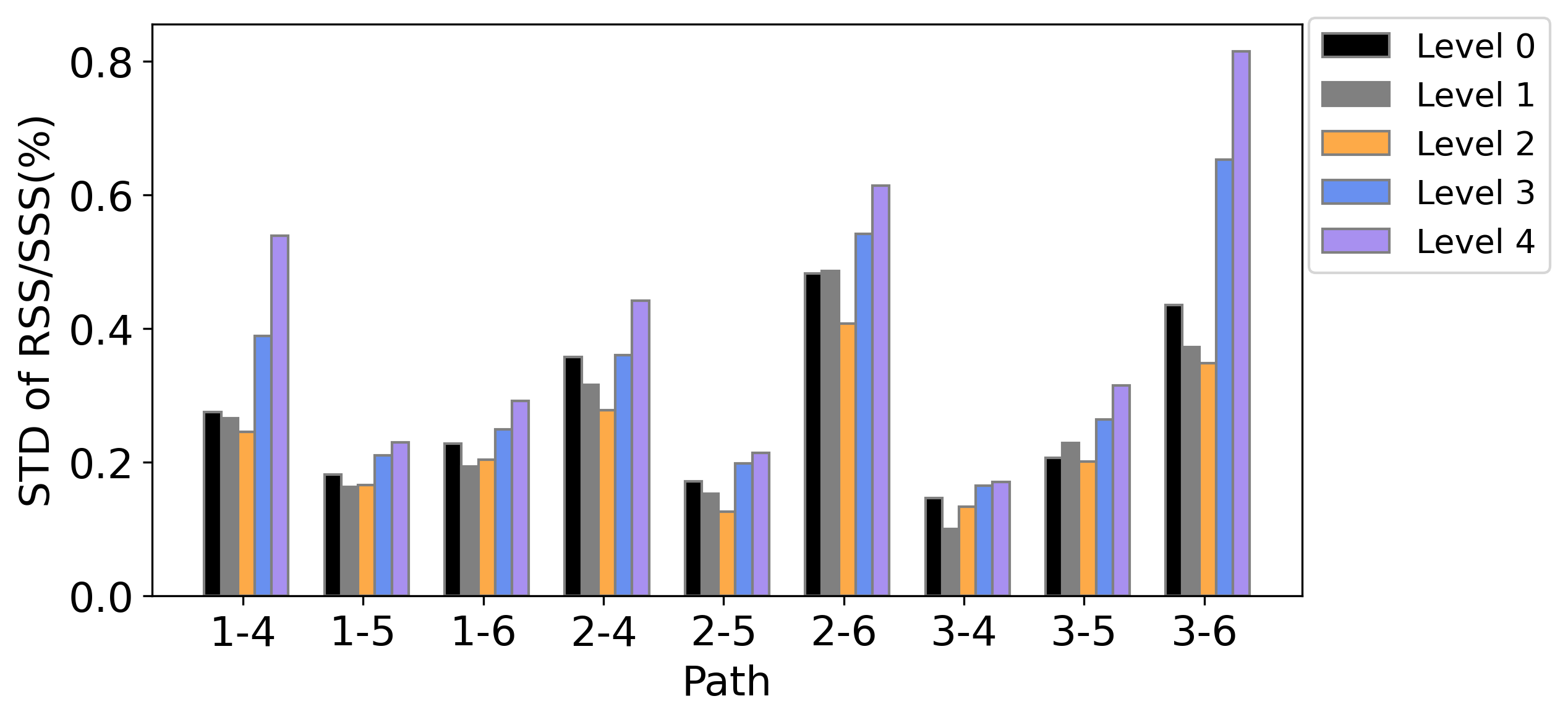}}
    \put(200,212){\color{black} \large {\fontfamily{phv}\selectfont \textbf{a}}}
    \put(200,114){\color{black} \large {\fontfamily{phv}\selectfont \textbf{c}}}
    \put(200,16){\large {\fontfamily{phv}\selectfont \textbf{e}}} 
    \put(430,212){\large {\fontfamily{phv}\selectfont \textbf{b}}}
   \put(430,114){\large {\fontfamily{phv}\selectfont \textbf{d}}} 
   \put(430,16){\large {\fontfamily{phv}\selectfont \textbf{f}}} 
    \end{picture} 
    \caption{Summary of the signal reconstruction performance of the three types of models. Panel a and b: Mean and standard deviation of signal reconstruction error of Type I model w.r.t each damage level under each path, respectively; Panel c and d: Mean and standard deviation of signal reconstruction error of Type II model w.r.t each damage level under each path, respectively; Panel e and f: Mean and standard deviation of signal reconstruction error of Type III model w.r.t each damage level under each path, respectively.}
\label{fig:recon_2} 
\end{figure}
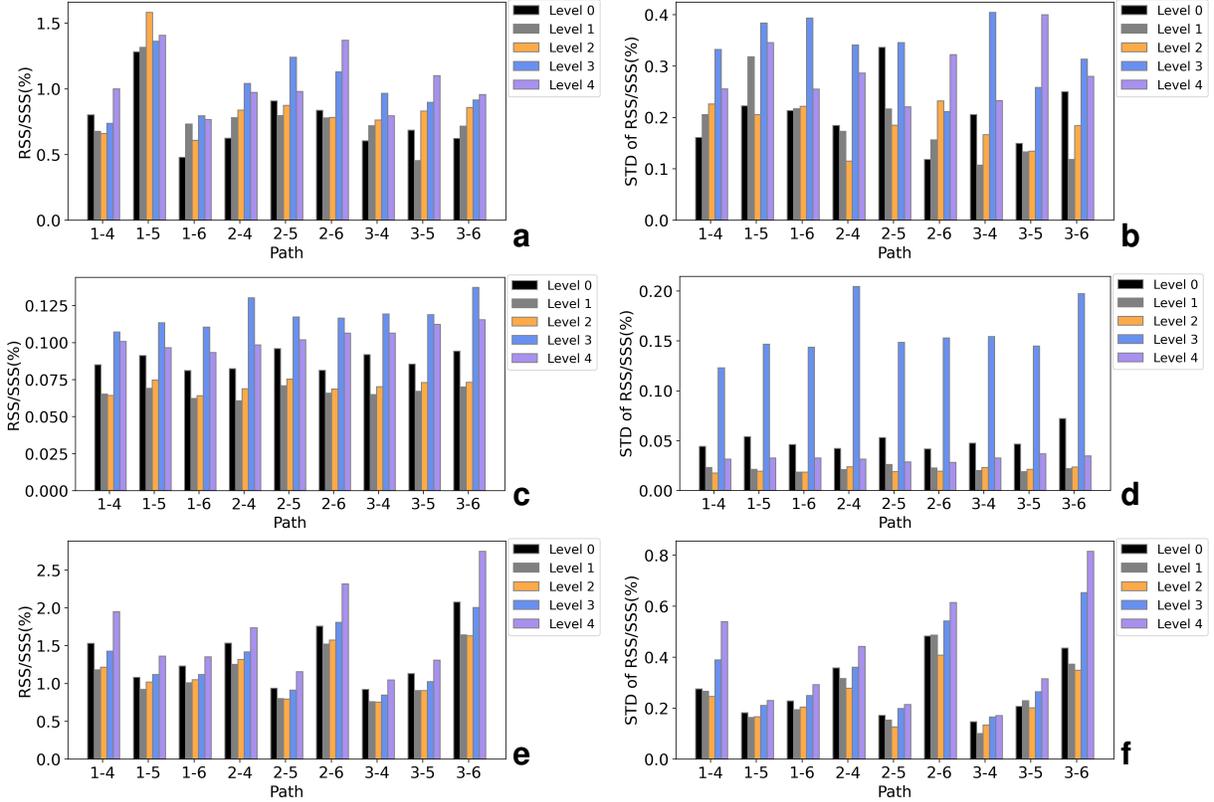
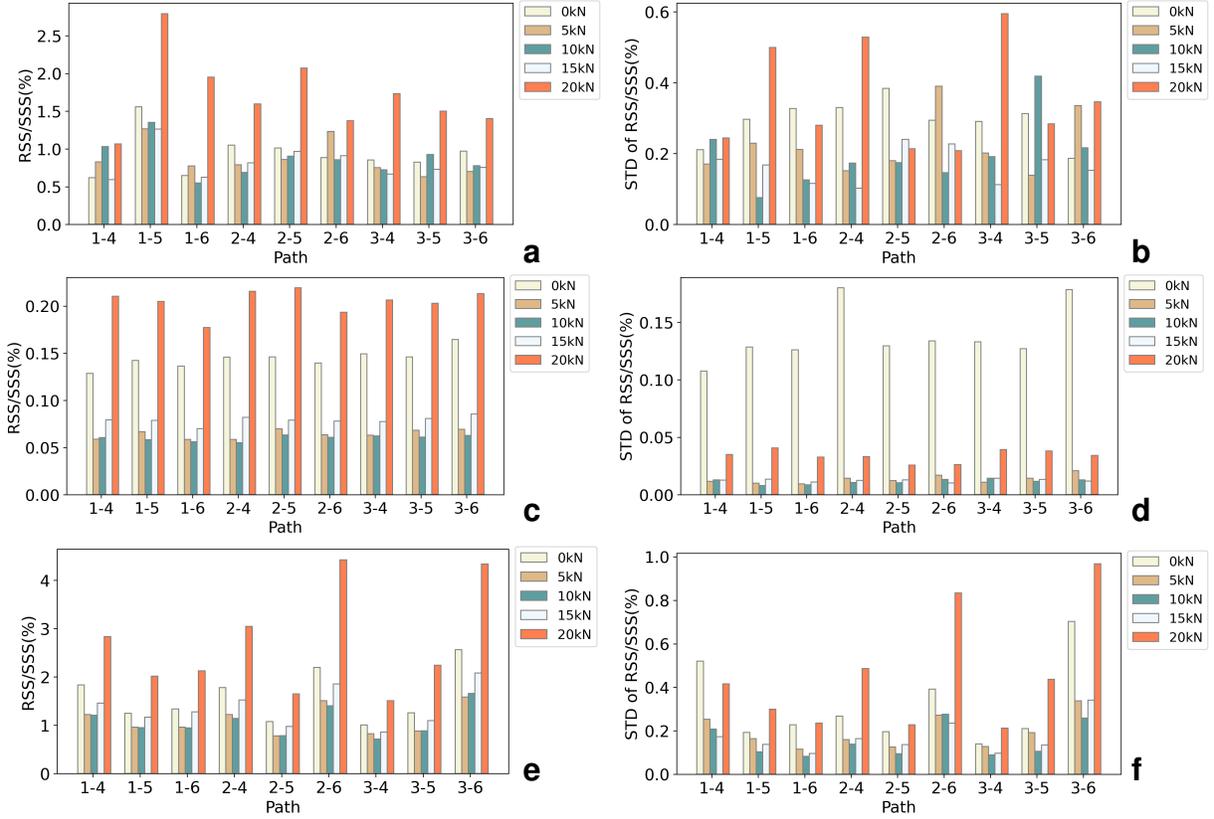
\begin{figure}[t!]
    \begin{picture}(500,320)
    \put(10,208){\includegraphics[width=0.48\textwidth]{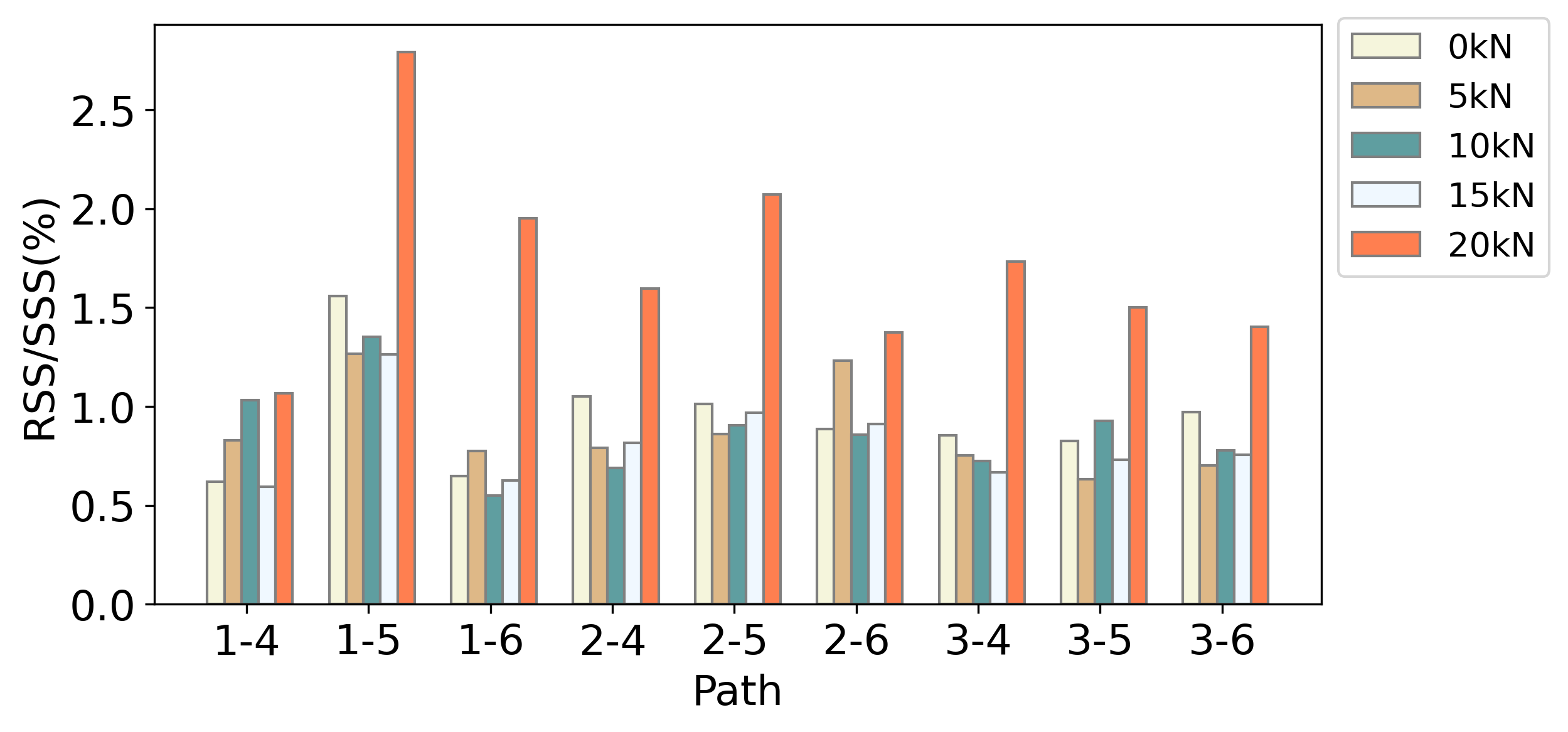}}
    \put(240,208){\includegraphics[width=0.48\textwidth]{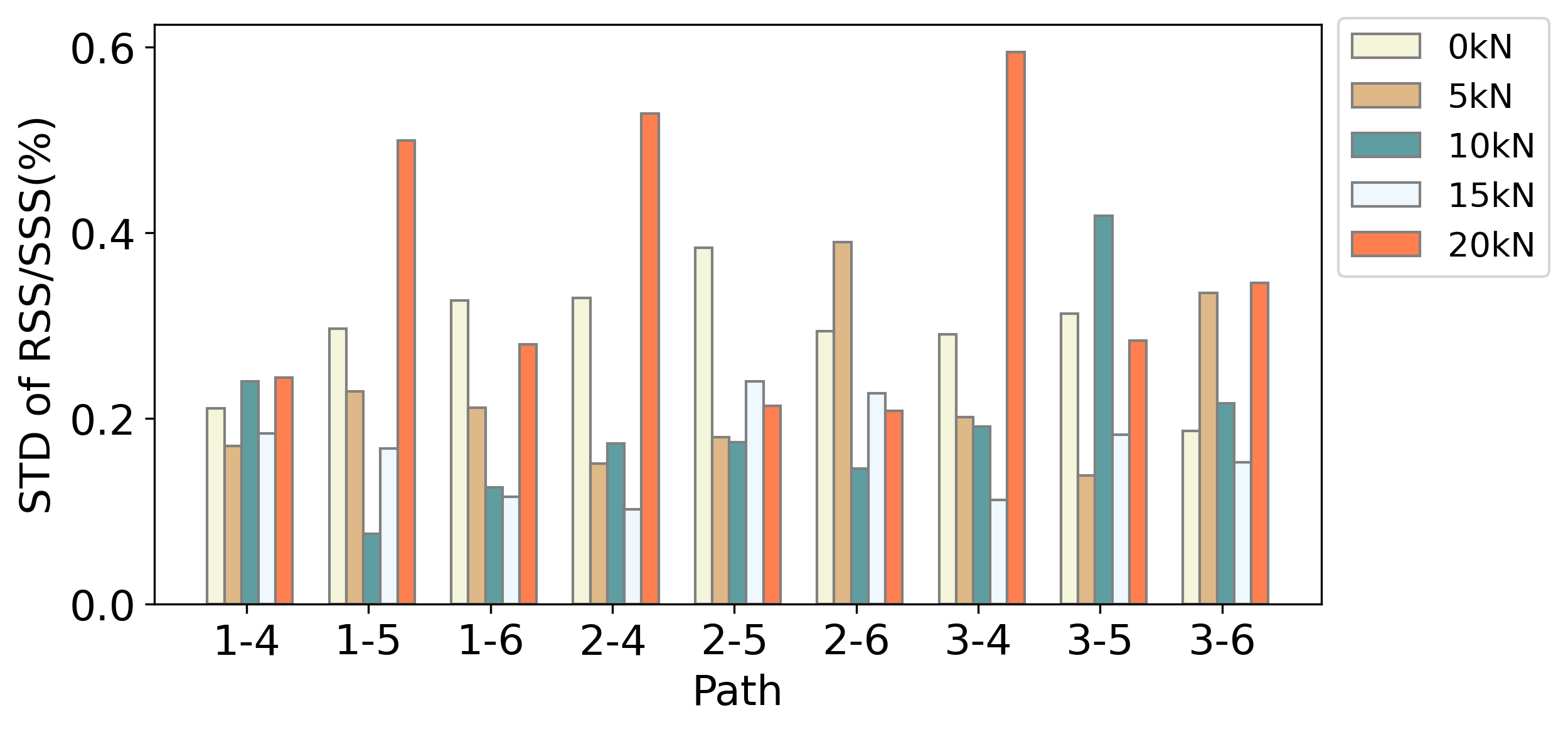}}
    \put(6,106){\includegraphics[width=0.48\textwidth]{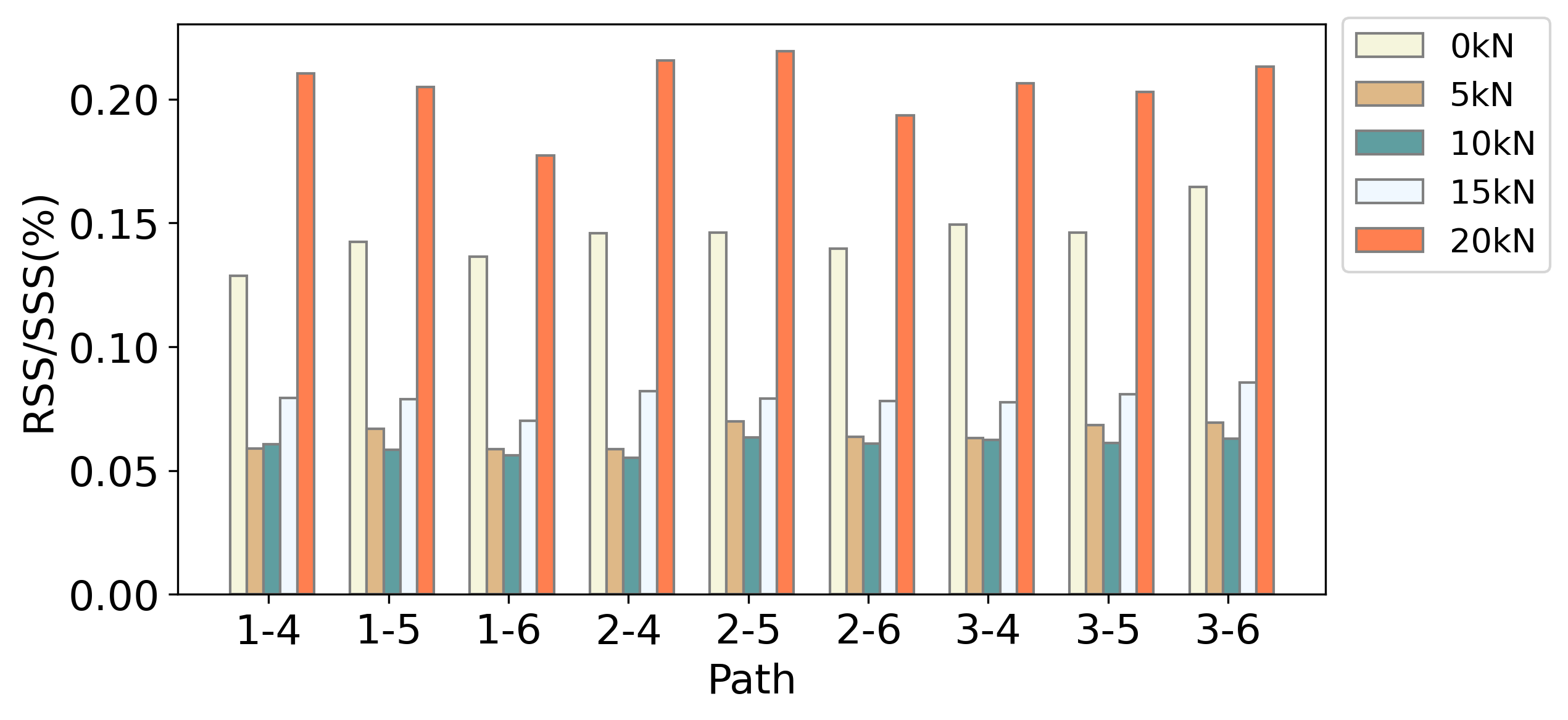}}
    \put(238,106){\includegraphics[width=0.48\textwidth]{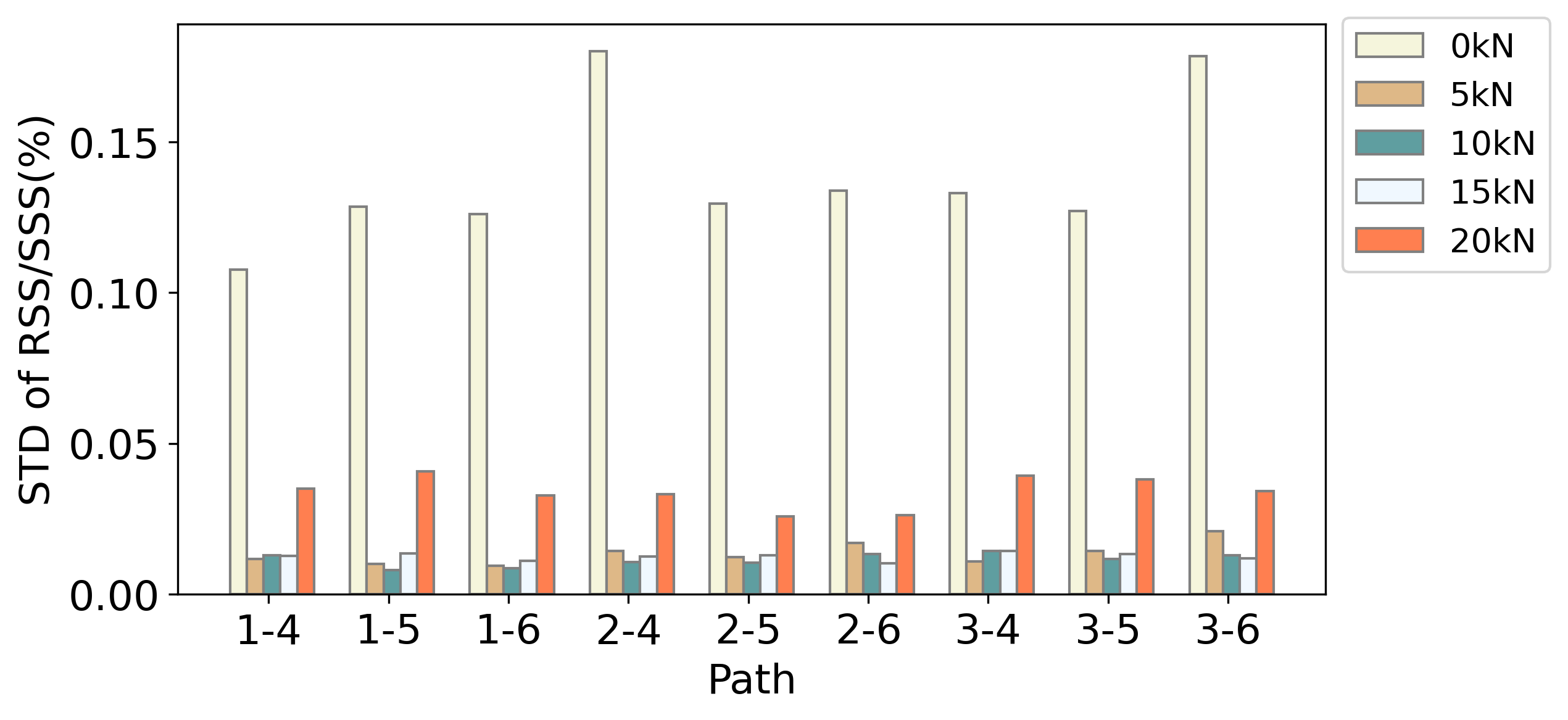}}
    \put(11,0){\includegraphics[width=0.475\textwidth]{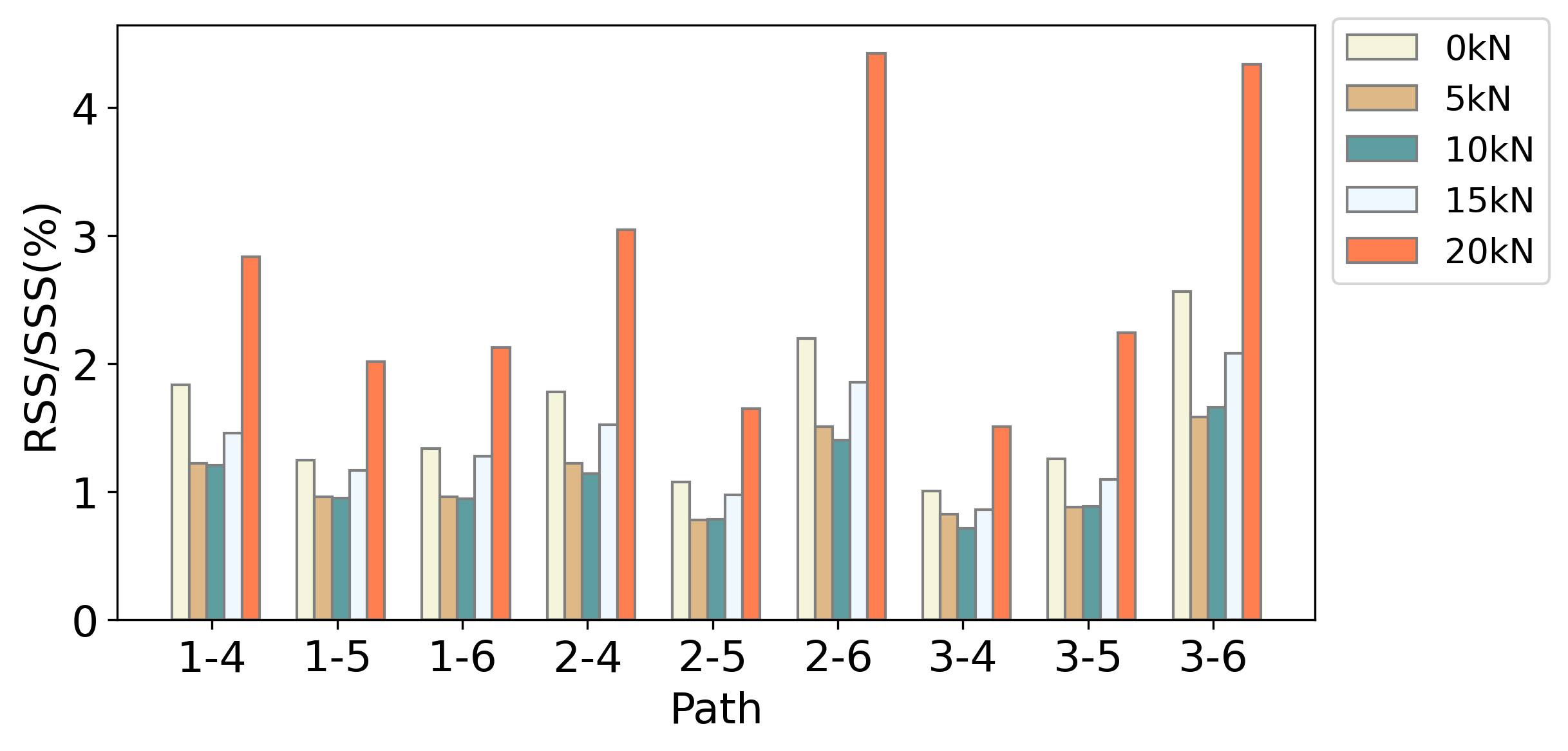}}
    \put(240,0){\includegraphics[width=0.48\textwidth]{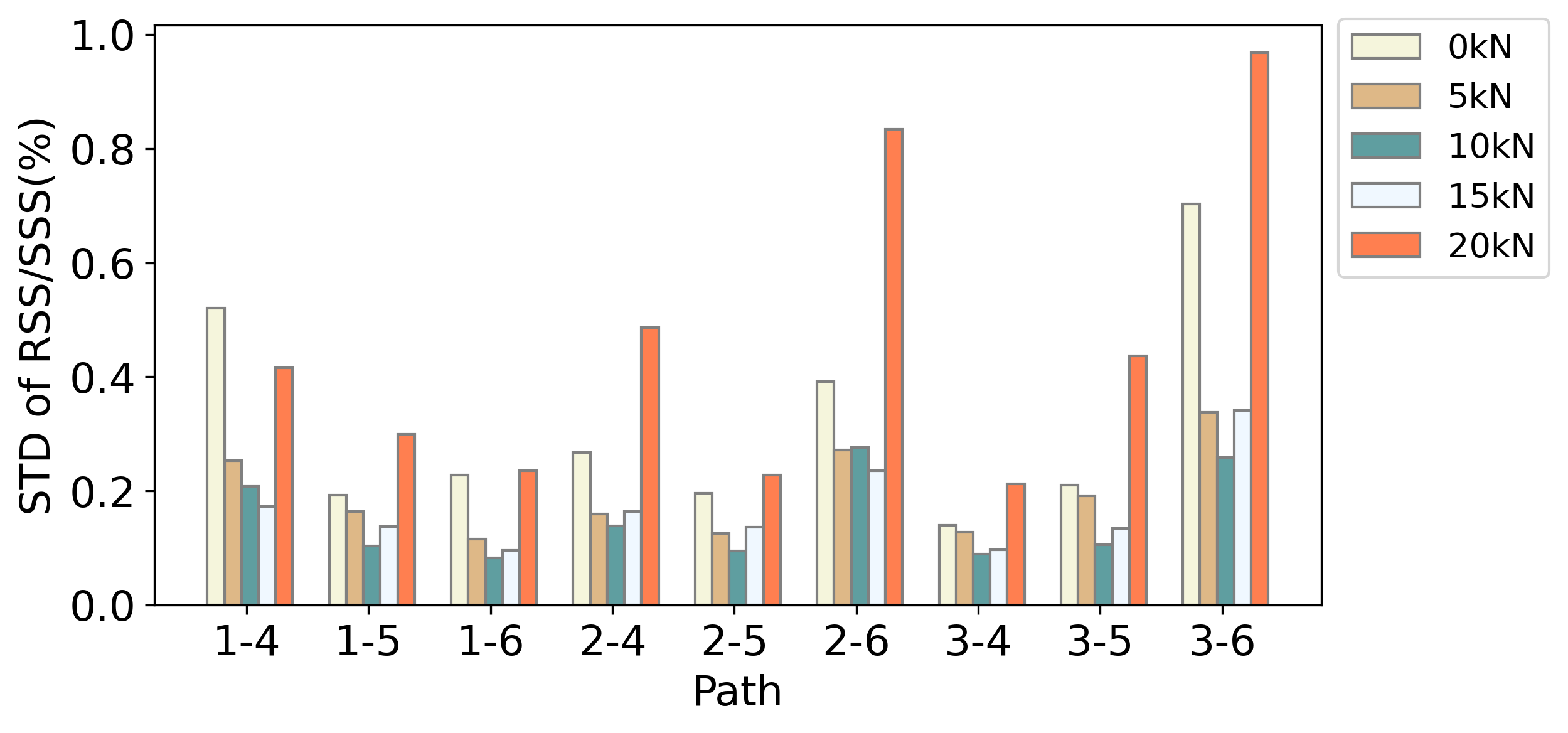}}
    \put(204,212){\color{black} \large {\fontfamily{phv}\selectfont \textbf{a}}}
    \put(204,114){\color{black} \large {\fontfamily{phv}\selectfont \textbf{c}}}
    \put(204,16){\large {\fontfamily{phv}\selectfont \textbf{e}}} 
    \put(434,212){\large {\fontfamily{phv}\selectfont \textbf{b}}}
   \put(434,114){\large {\fontfamily{phv}\selectfont \textbf{d}}} 
   \put(434,16){\large {\fontfamily{phv}\selectfont \textbf{f}}} 
    \end{picture} 
    \caption{Summary of the signal reconstruction performance of the three types of models. Panel a and b: Mean and standard deviation of signal reconstruction error of Type I model w.r.t each load under each path, respectively; Panel c and d: Mean and standard deviation of signal reconstruction error of Type II model w.r.t each load under each path, respectively; Panel e and f: Mean and standard deviation of signal reconstruction error of Type III model w.r.t each load under each path, respectively.}
\label{fig:recon_3} 
\end{figure}

\begin{figure}[t!]
    \begin{picture}(500,530)
    \put(10,414){\includegraphics[width=0.38\textwidth]{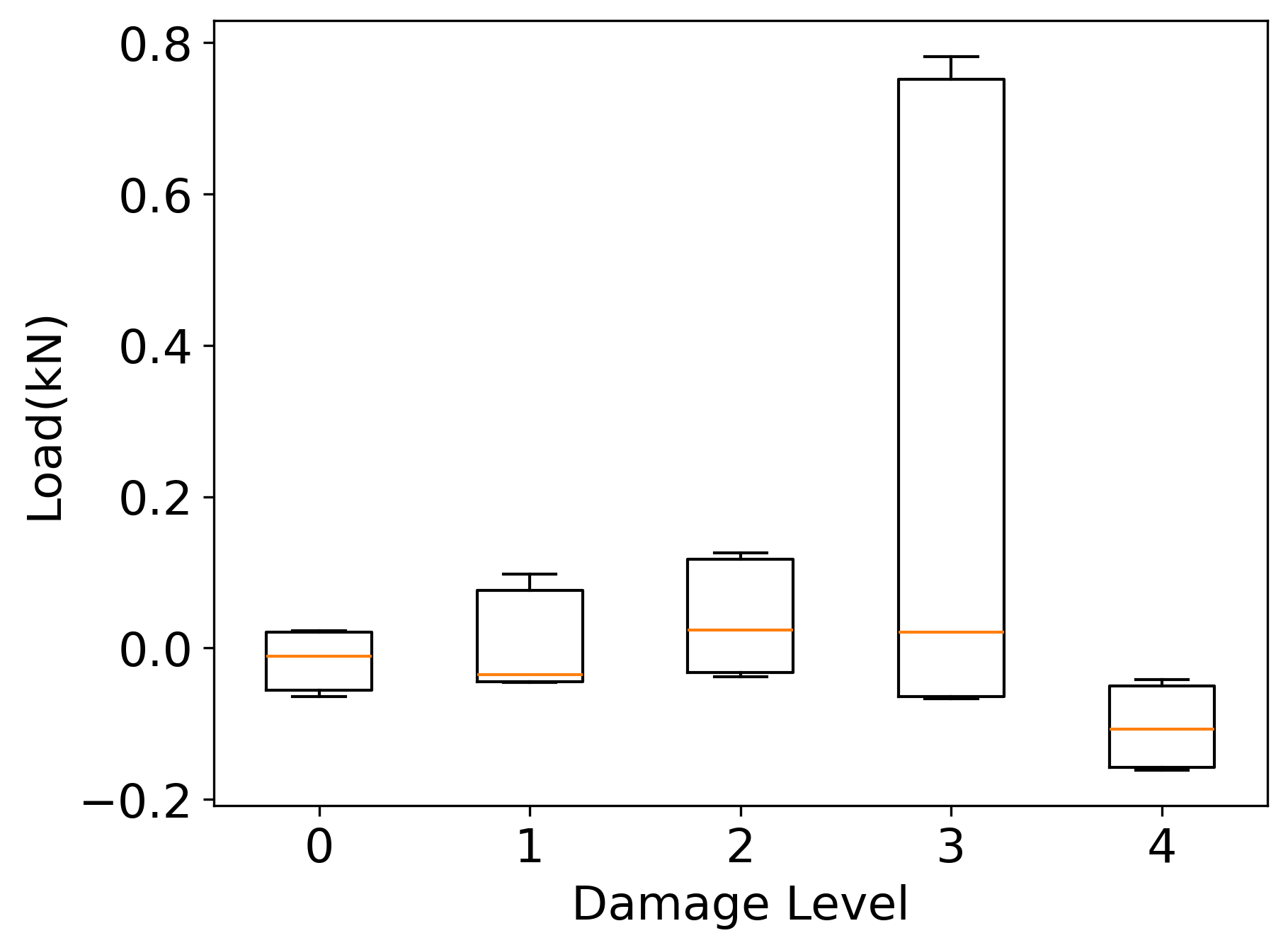}}
    \put(224,414){\includegraphics[width=0.38\textwidth]{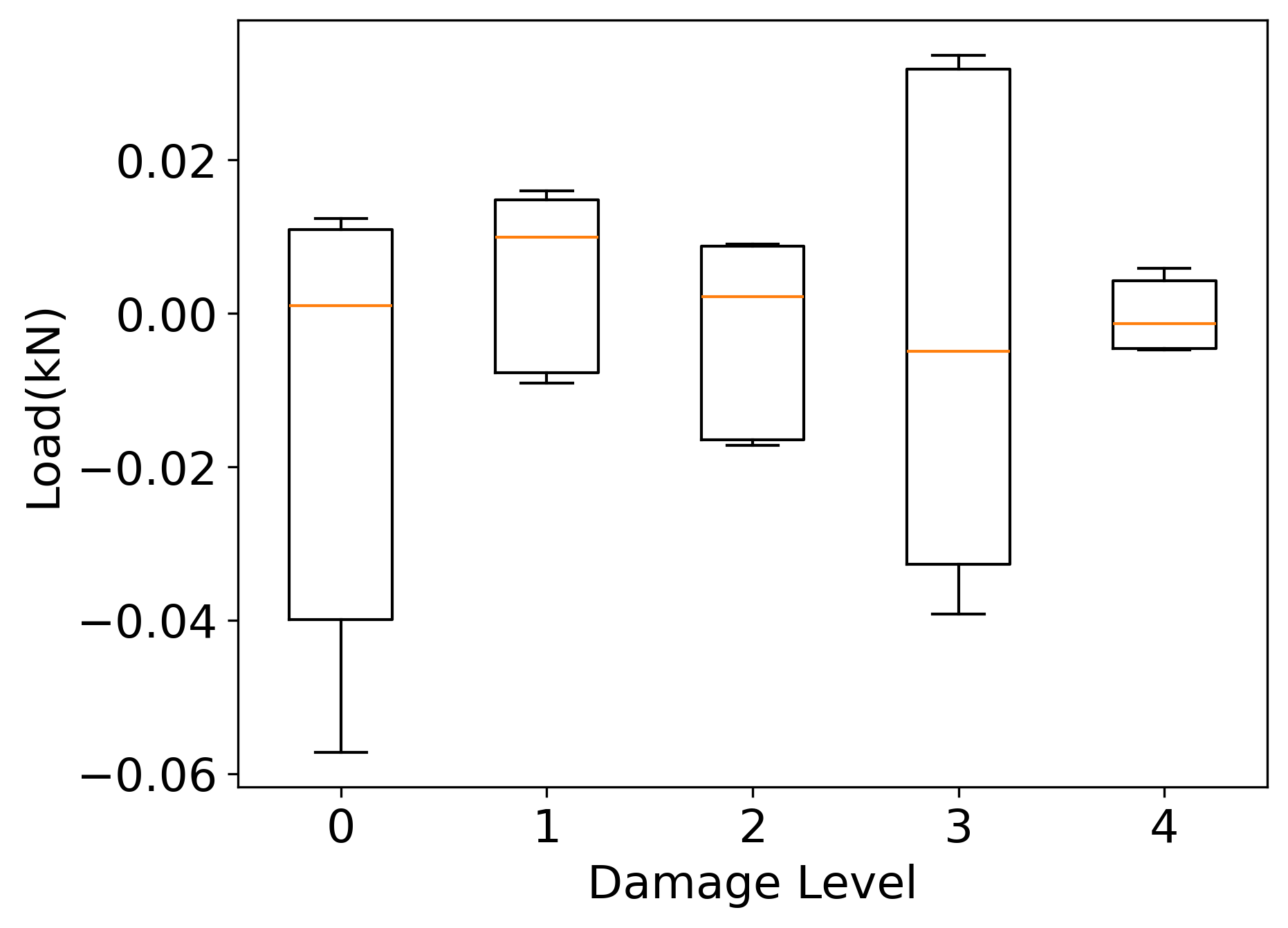}}
    \put(10,276){\includegraphics[width=0.38\textwidth]{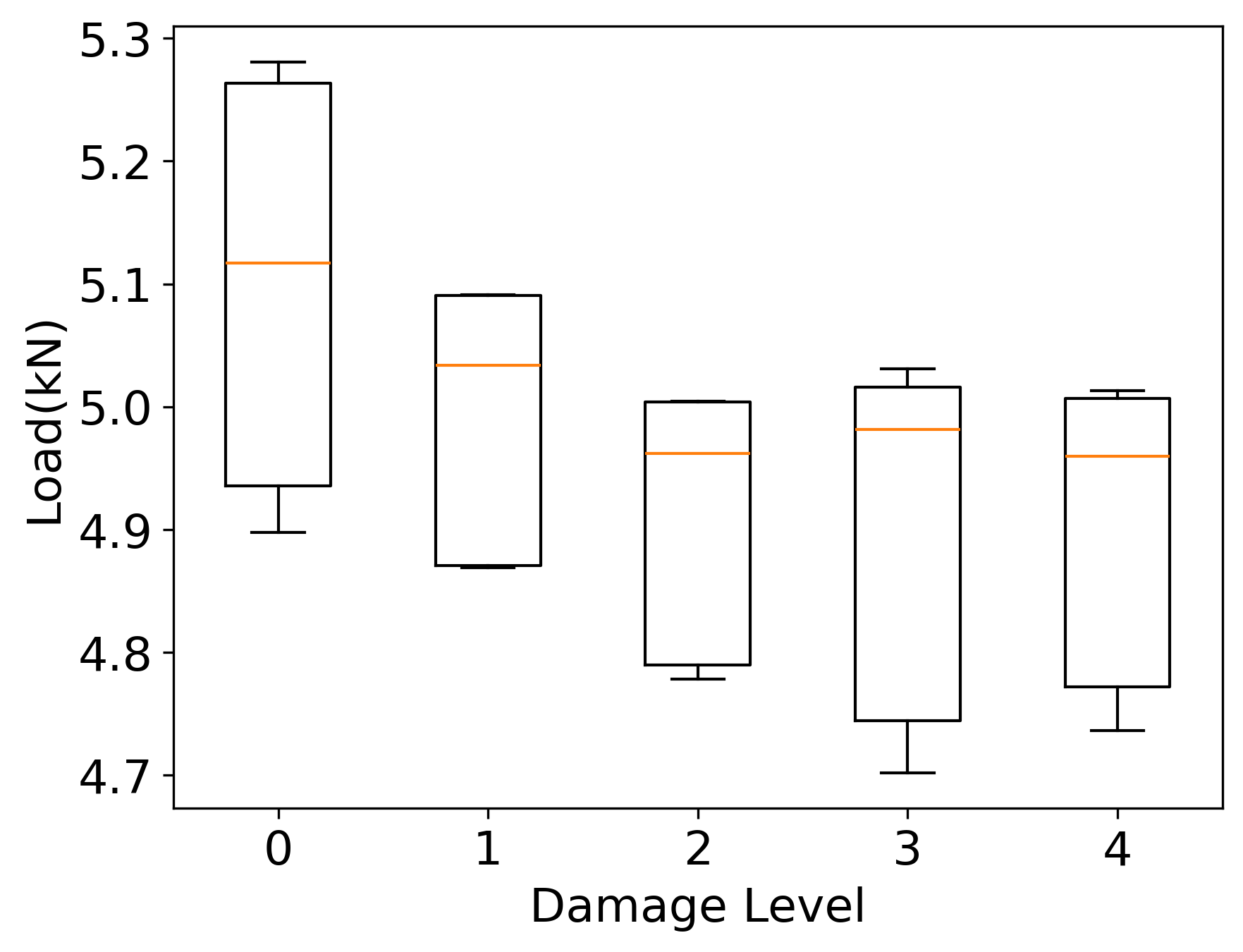}}
    \put(224,276){\includegraphics[width=0.38\textwidth]{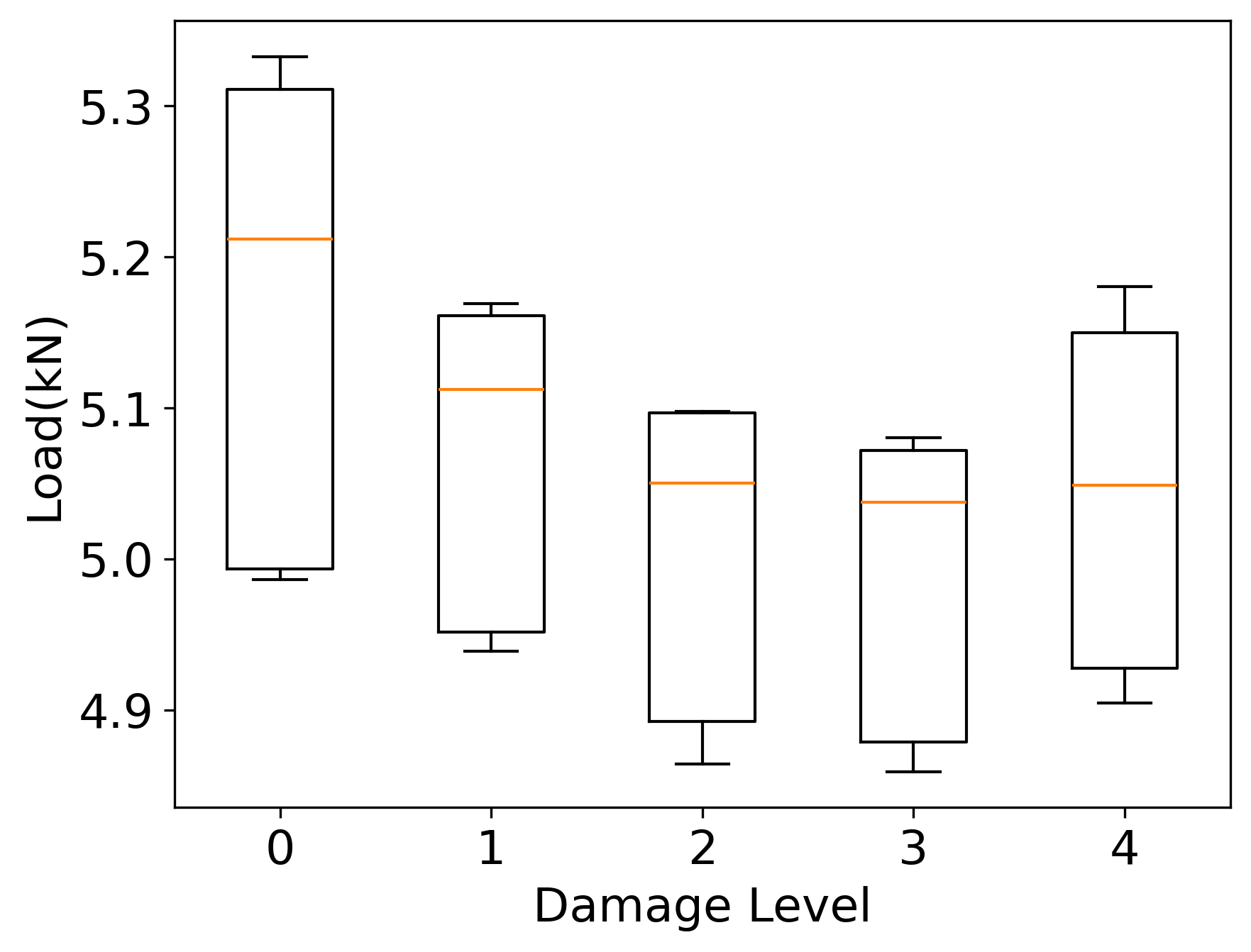}}
    \put(10,138){\includegraphics[width=0.38\textwidth]{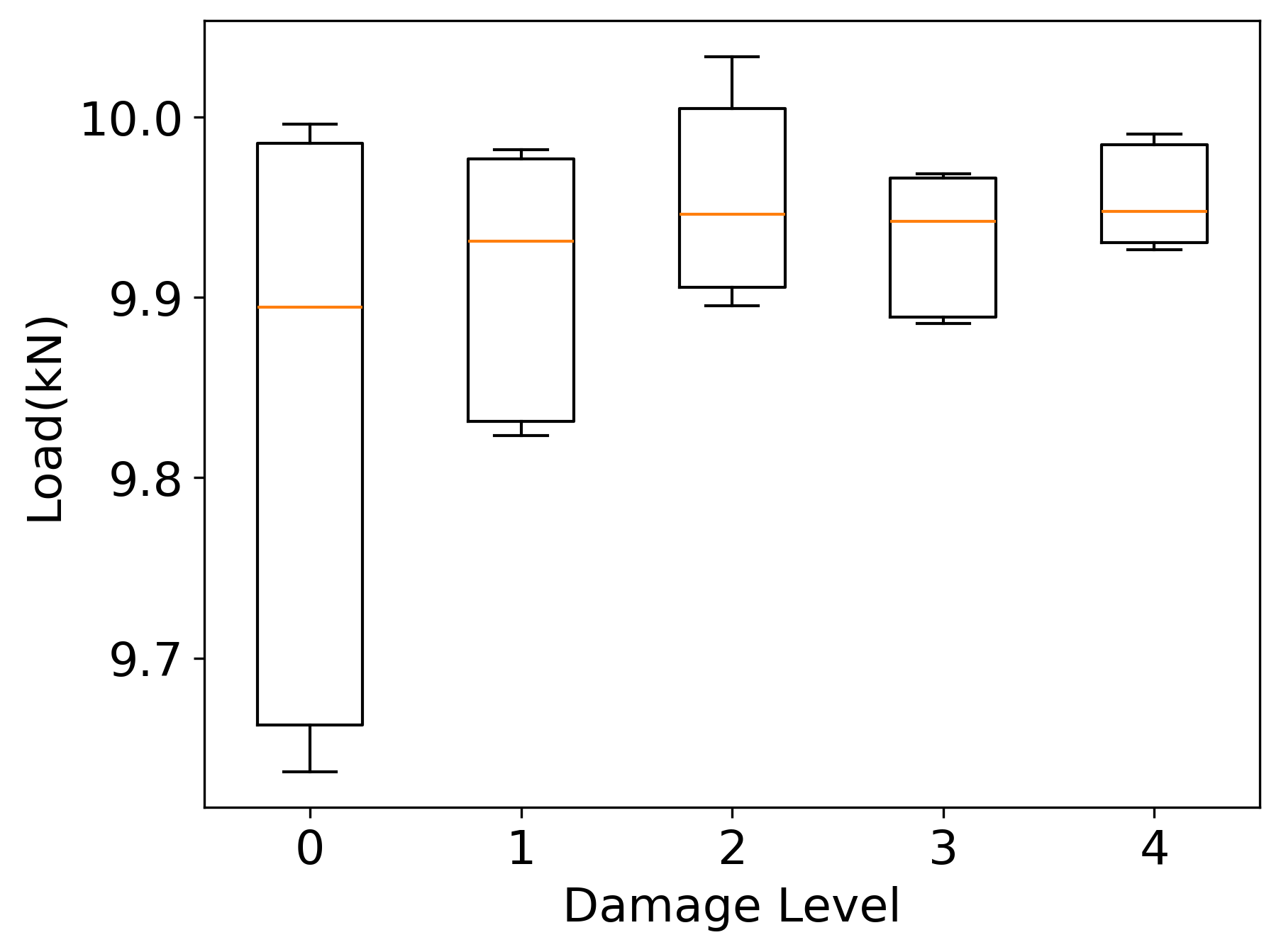}}
    \put(224,138){\includegraphics[width=0.38\textwidth]{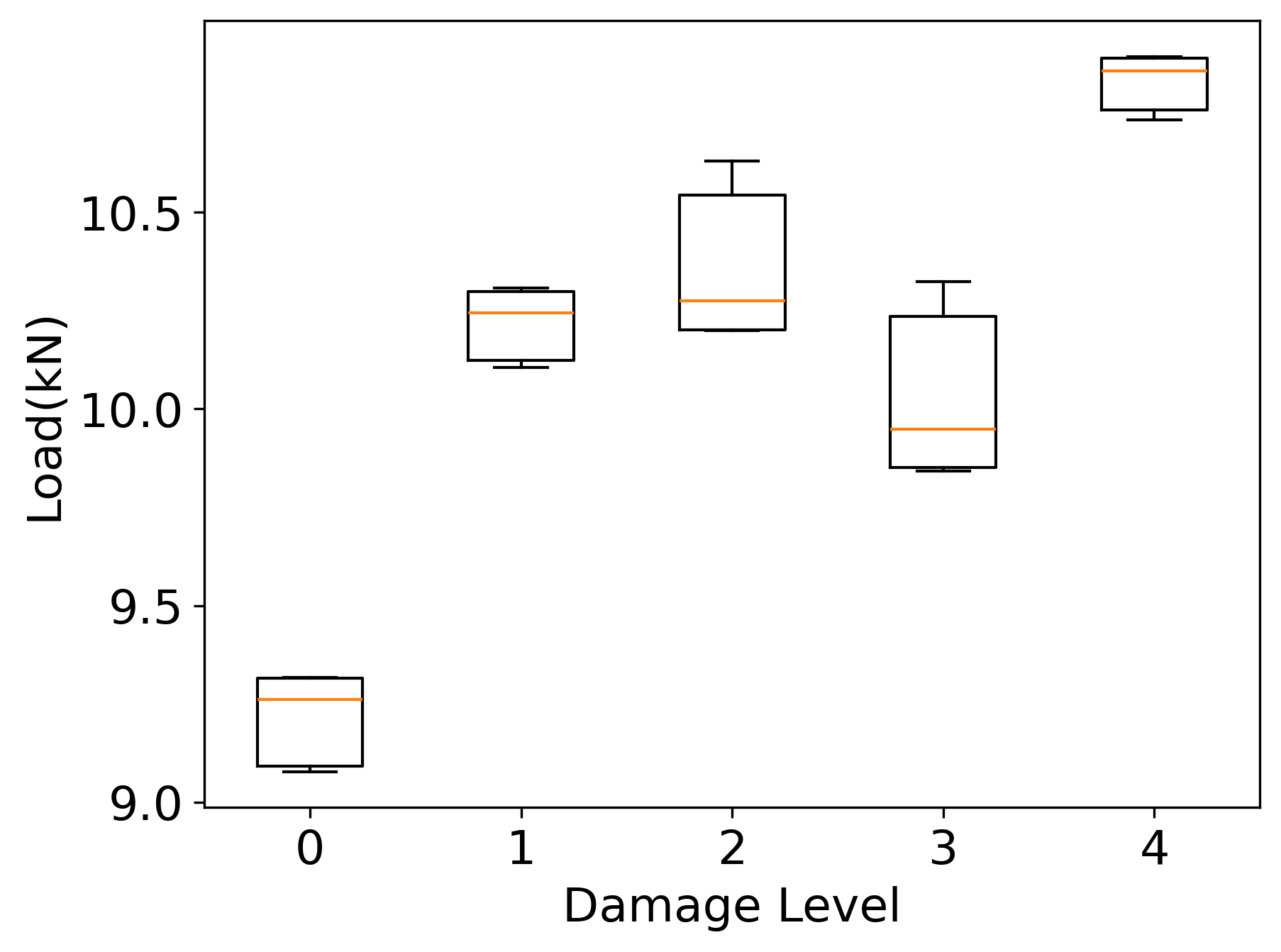}}
    \put(10,0){\includegraphics[width=0.38\textwidth]{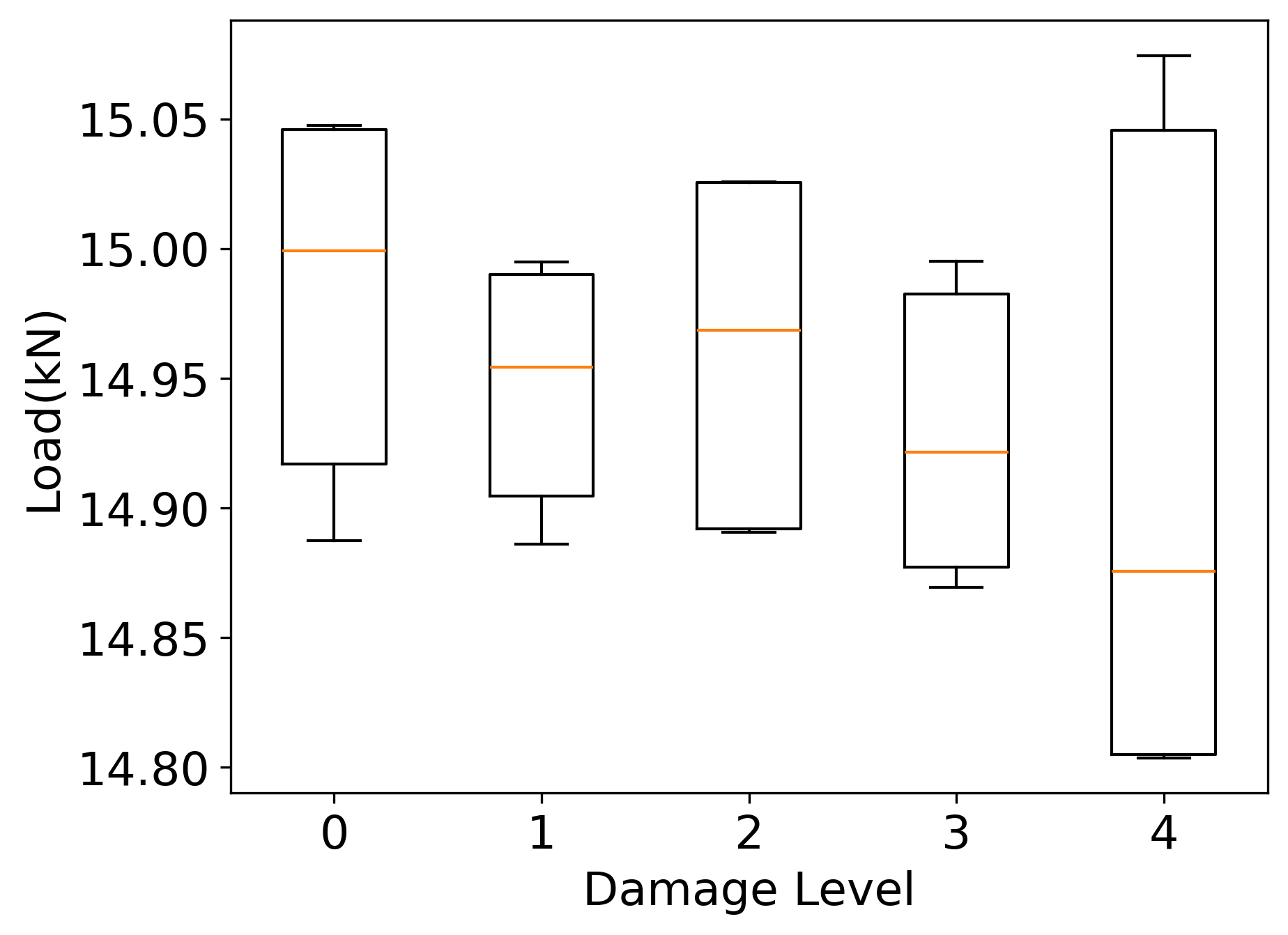}}
    \put(224,0){\includegraphics[width=0.38\textwidth]{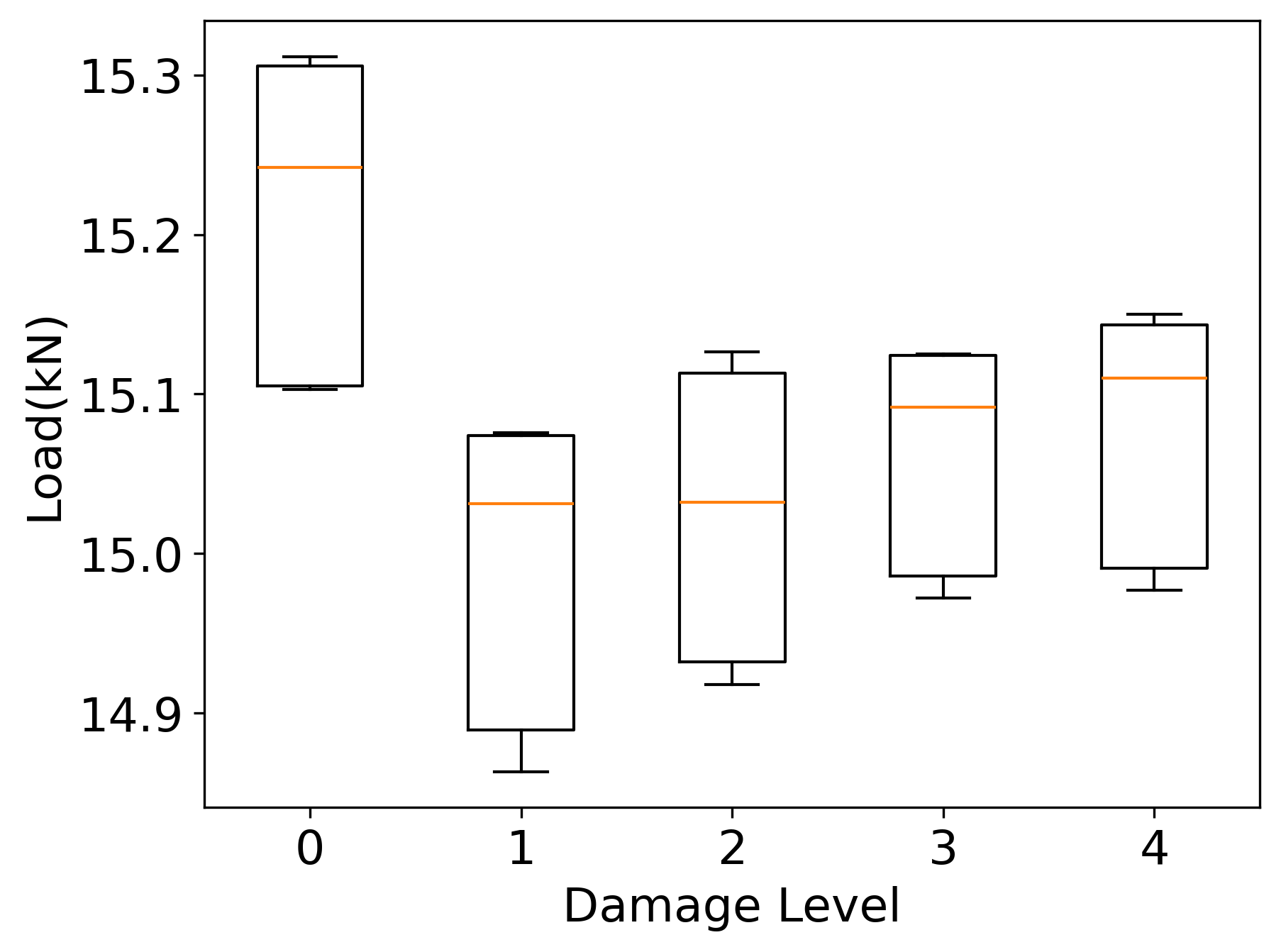}}
    \put(190,432){\color{black} \large {\fontfamily{phv}\selectfont \textbf{a}}}
    \put(190,294){\large {\fontfamily{phv}\selectfont \textbf{b}}} 
    \put(190,156){\large {\fontfamily{phv}\selectfont \textbf{c}}}
    \put(190,18){\large {\fontfamily{phv}\selectfont \textbf{d}}}
   
   \put(408,432){\large {\fontfamily{phv}\selectfont \textbf{e}}} 
   \put(409,294){\large {\fontfamily{phv}\selectfont \textbf{f}}} 
   \put(406,156){\large {\fontfamily{phv}\selectfont \textbf{g}}} 
   \put(410,18){\large {\fontfamily{phv}\selectfont \textbf{h}}} 
    \end{picture} 
    \caption{Box plots of force estimation under all available damage conditions from models trained with the full set and the sparse set, respectively. The box covers 95$\%$ CI and the orange line indicates the mean value. Panels a-d use the full dataset while panels e-h use the incomplete one without 10kN case.}
\label{fig:robust} 
\end{figure}
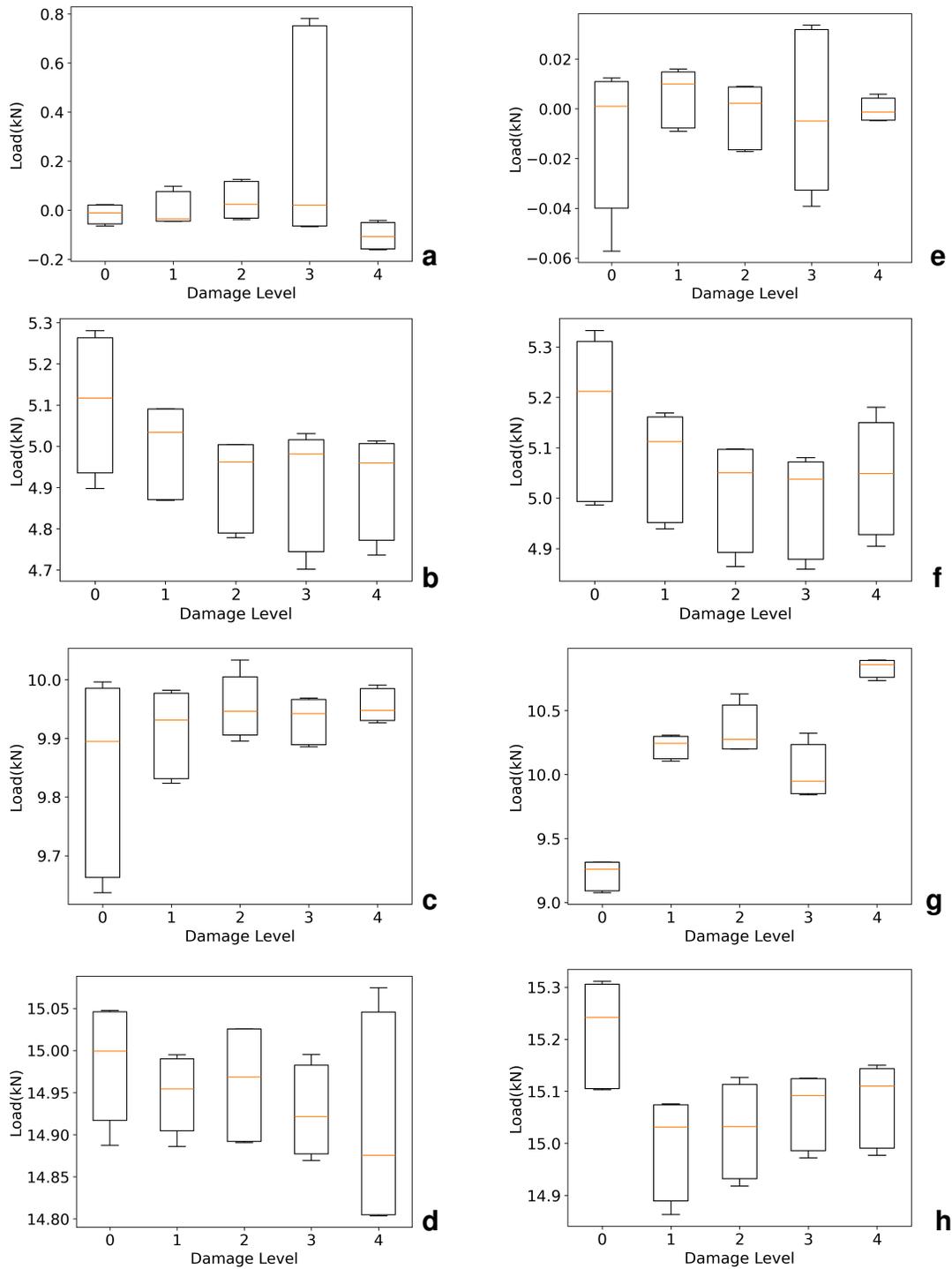

\begin{figure}[t!]
    \begin{picture}(500,200)
    \put(10,108){\includegraphics[width=0.48\textwidth]{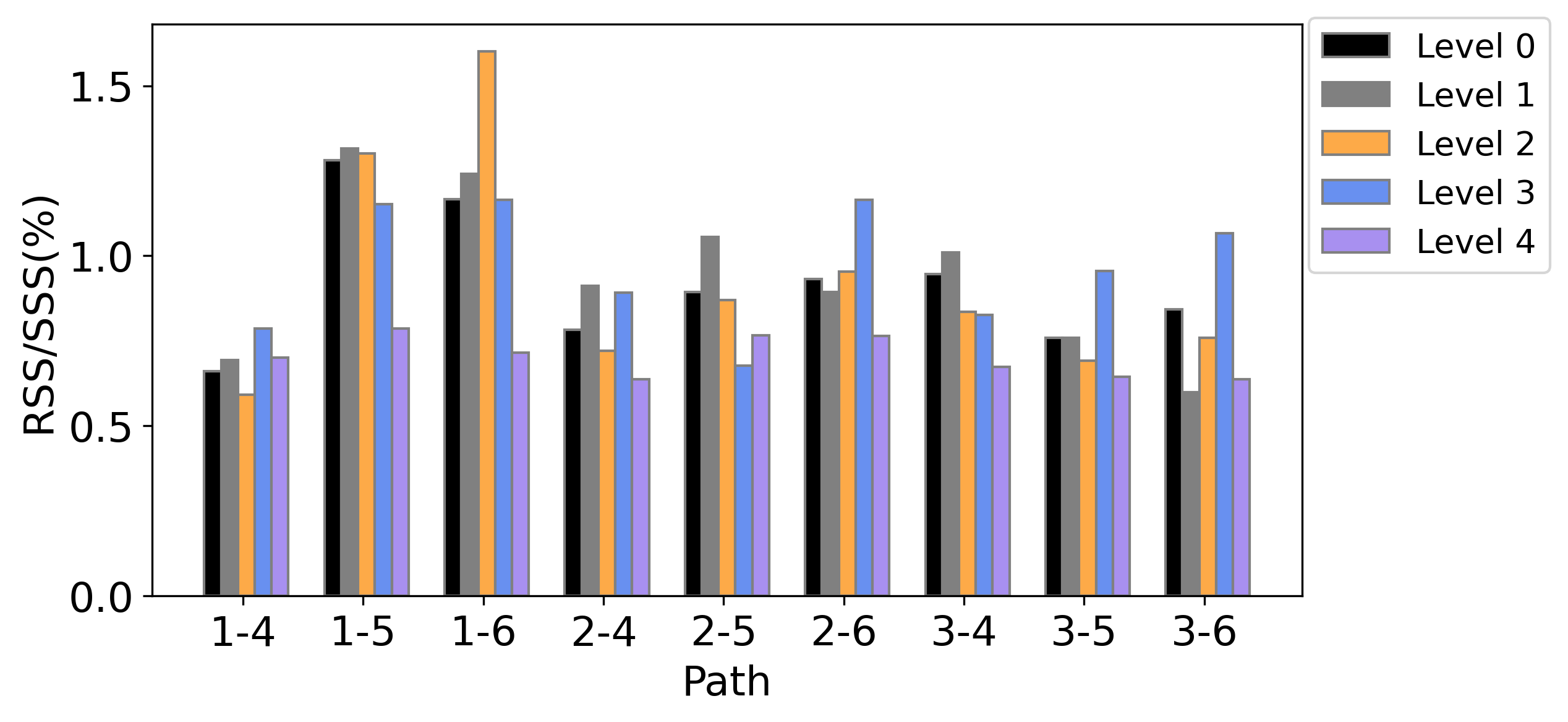}}
    \put(244,108){\includegraphics[width=0.48\textwidth]{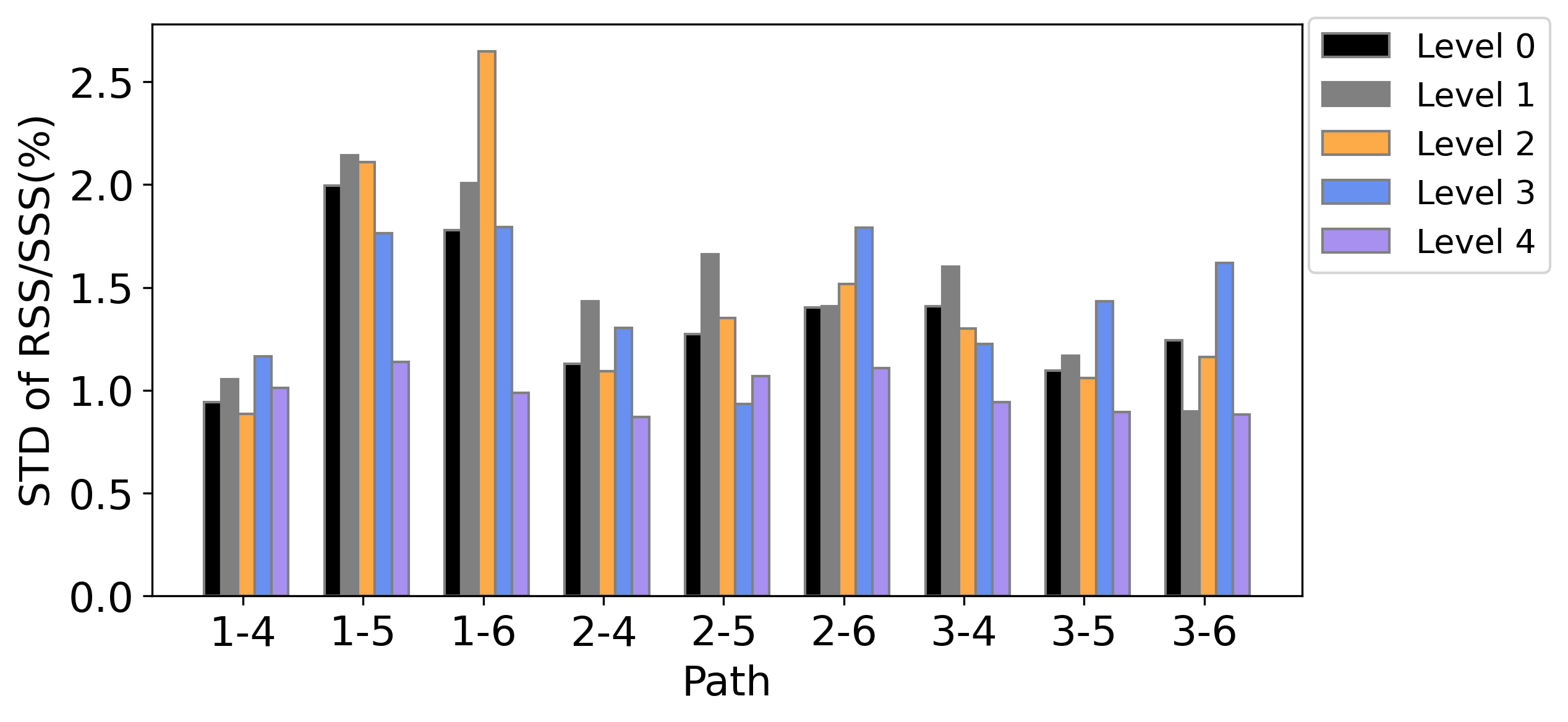}}
    \put(11,0){\includegraphics[width=0.474\textwidth]{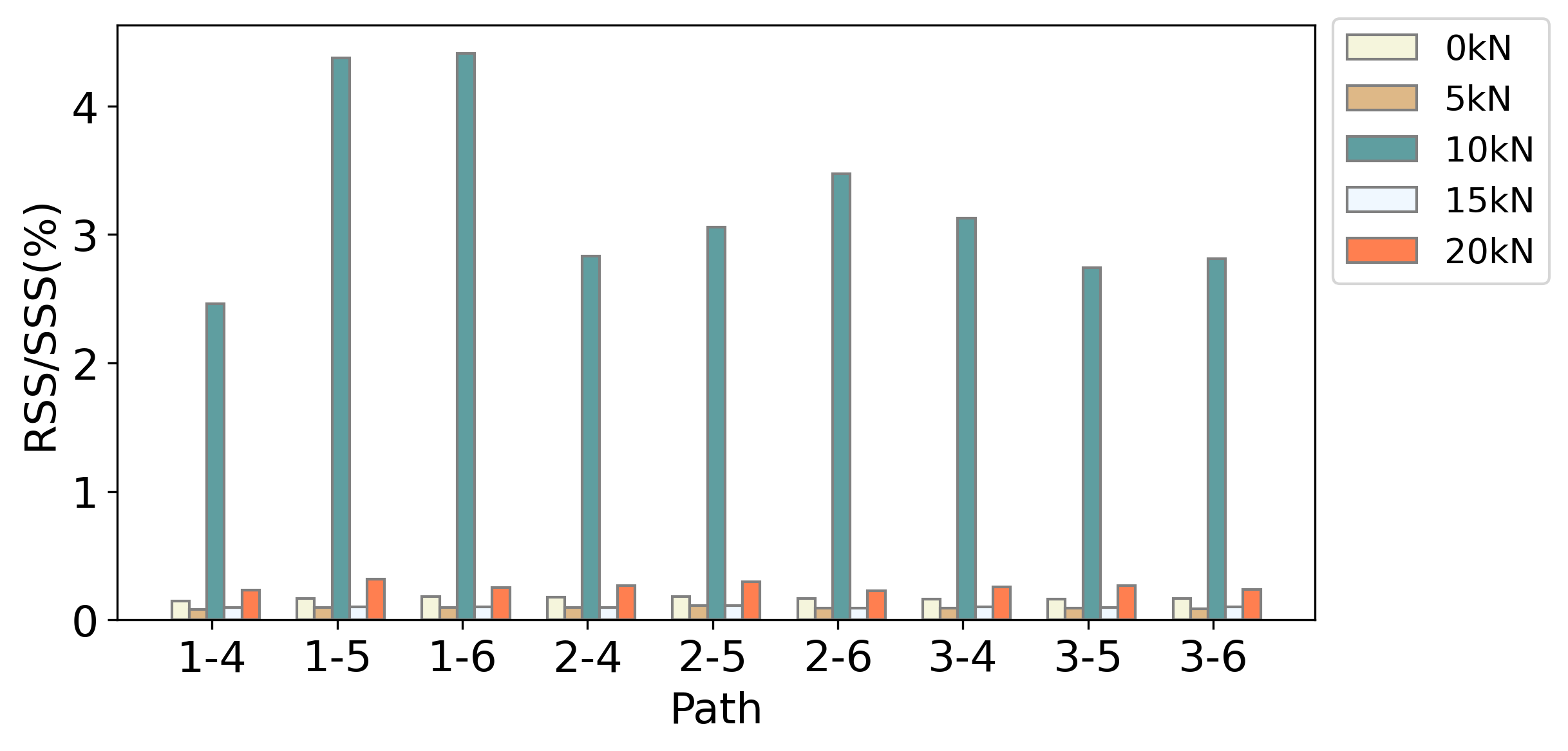}}
    \put(242,0){\includegraphics[width=0.48\textwidth]{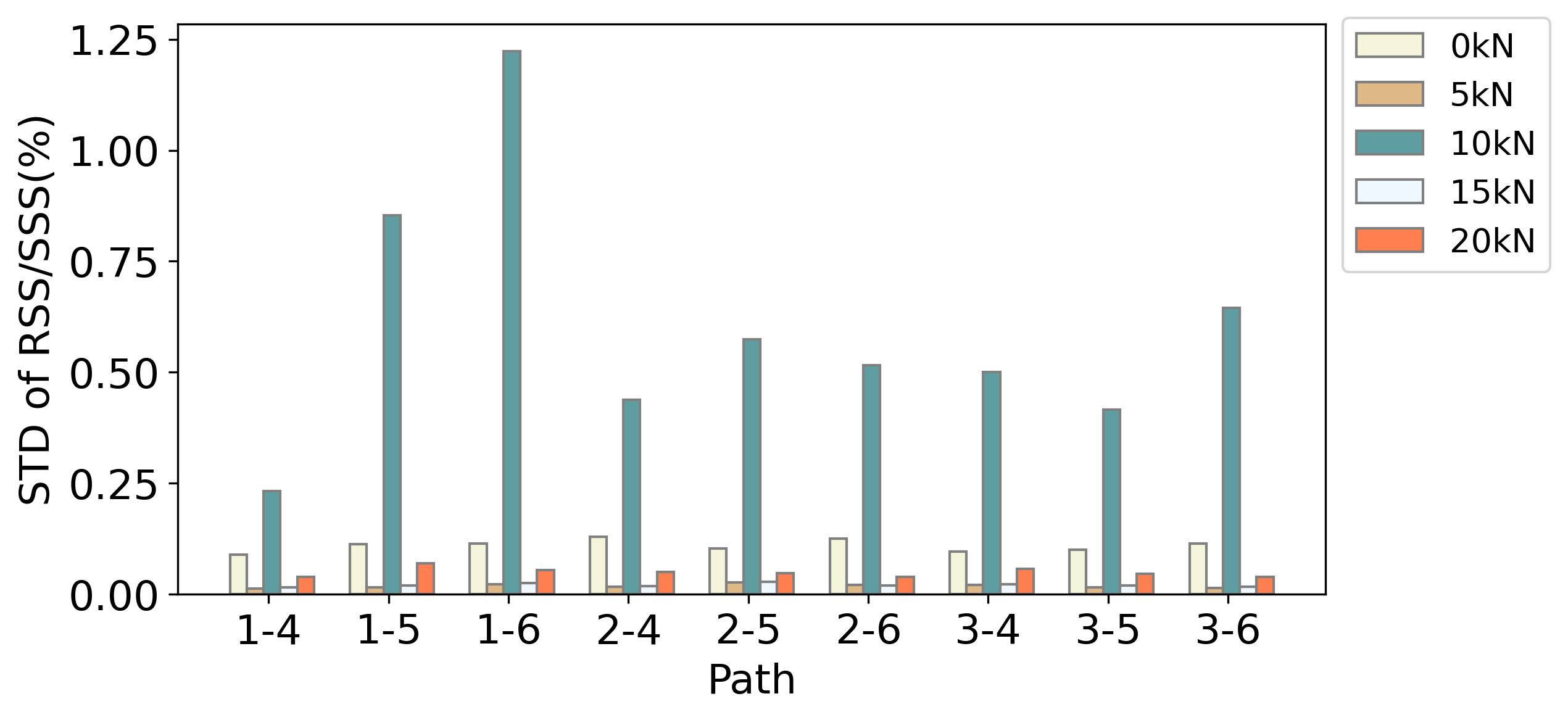}}
    \put(200,114){\color{black} \large {\fontfamily{phv}\selectfont \textbf{a}}}
    \put(200,12){\large {\fontfamily{phv}\selectfont \textbf{c}}} 
   \put(434,114){\large {\fontfamily{phv}\selectfont \textbf{b}}} 
   \put(434,12){\large {\fontfamily{phv}\selectfont \textbf{d}}} 
    \end{picture} 
    \caption{Summary of the signal reconstruction performance of Type II model using sparse set. Panel a and b: Mean and standard deviation of signal reconstruction error w.r.t each damage level under each path, respectively; Panel c and d: Mean and standard deviation of signal reconstruction error w.r.t each load under each path.}
\label{fig:rob_2} 
\end{figure}

\subsubsection{Signal Reconstruction}

Examples of the reconstructed signals from the Type I model are presented in Figure 
\ref{fig:recon_1}. Panels (a-d) show the original and reconstructed signals in a healthy state under varying loads from sensor paths 2-6, while panels (e-h) display the corresponding signals from paths 3-4 under the same conditions. The Type I model used here employs the hyperparameters determined from earlier steps. In these plots, the CAE-reconstructed signals nearly overlap with the original signals, with negligible RMSE and RSS/SSS($\%$).

The overall signal reconstruction performance from the three methods is summarized in Figure \ref{fig:recon_2}, where
the bar plots indicate the mean and standard deviation of RSS/SSS($\%$) values for each damage level and sensor path, using all test data. When comparing panels (a), (c), and (e), it is evident that the Type II model yields the lowest average reconstruction error, with the maximum average value below approximately $0.15\%$. The standard deviation of the Type II model is also the smallest, as seen in panels (b), (d), and (f), with a maximum value around $0.2\%$. In contrast, the Type I model's highest average error occurs in path 1-5, with a value just over 1.5$\%$ at damage level 2, as indicated by the orange bar in panel (a). For the Type III model, the largest error—over 2.5$\%$—is observed at damage level 4 in path 3-6, as shown in panel (c), which is the highest error among all three models.

A similar observation can be made in Figure \ref{fig:recon_3}, where the bar plots represent the reconstruction errors with respect to varying loads, using data across all damage states. Once again, Model II demonstrates superior performance, while Model III exhibits the largest error. The reasons behind these results are as follows: 1) Due to the CAE structure, the Type II model requires data from all sensor paths under each state to be provided simultaneously during both training and testing phases. This contrasts with the Type I model, where signals from different paths can be processed independently. As a result, the Type II model benefits from additional information when compressing and reconstructing the data, leading to potentially better performance than the Type I model; 2) In the Type III model, where the sensor paths are further decomposed into two dimensions, the data from all paths at each state is also required due to its structure. While this ensures that the inputs contain the same amount of information as in the Type II model, the decomposition approach does not fully show its advantages in this application, as only a limited number of paths are included. However, the author considers the decomposition method worth exploring further. In cases where a large sensor array is employed and numerous paths are available, the Type III model could offer significant computational cost savings.

\subsubsection{Model Robustness Verification}

To verify the robustness of the model, the best-performing model, Type II, was tested with a sparse dataset. Specifically, signals collected under 10kN were removed from the original dataset, and the remaining data was used to train the CAE and FFNN, keeping the model configurations unchanged. The results are compared with those obtained from a complete dataset for validation.

Figure \ref{fig:robust} presents the comparison of box plots for force estimation across all available damage conditions, using Type II models trained on the full and sparse datasets, respectively. Panels (a)-(d) display results from the full dataset, while panels (e)-(h) represent the incomplete dataset. The boxes cover the $95\%$ CI, with the minimum and maximum values denoted by the horizontal lines extending from the boxes. Despite the differences in training datasets, both models are evaluated on the same test set, which includes data from all states. 

As introduced in Section 4.2.2, the model trained on the complete dataset achieved near-perfect state predictions, with mean state index estimates close to the true values and negligible standard deviations. This is confirmed by the yellow lines of the box plots indicating the mean estimates are close to the true values as well as the small variance. Comparing panels a\&e, b\&f and d\&h, the load predictions for 0, 5 and 15kN remain accurate and are not significantly impacted by the absence of 10kN data. However, in panels c\&g, a noticeable difference is observed in the predicted load ranges. Specifically, the range of the load axis expands from about 0.3kN to over 1.5kN. This is as expected, since the model did not have access to 10kN data during training. Nonetheless, this broader confidence interval does not severely affect the final prediction accuracy because the box length remains much smaller than the data resolution (5kN).

After evaluating the state estimation capability, the signal reconstruction function of the Type II model was also tested using the sparse dataset. From Figure \ref{fig:rob_2}, panels (a) and (b) show that the mean and standard deviation of the reconstruction error, with respect to damage levels, are higher compared to the model trained on the full dataset as shown in Figure \ref{fig:recon_2}, but still remain relatively low. Panels (c) and (d) illustrate the reconstruction error with respect to external loads, where the error at 10kN stands out as expected. Although the complete dataset yields more precise reconstructed signals, the model trained on the incomplete data still achieves a respectable level of accuracy, demonstrating the robustness of the proposed framework.
\section{Conclusion} \label{Sec:conc}

In this study, a CAE- and FFNN-based active-sensing SHM framework has been developed for damage quantification and signal reconstruction tasks across various damage states and EOCs. The proposed approach is flexible and can be easily generalized to more complex EOCs. Time series data were directly used as input, while both time-invariant and time-varying latent spaces were obtained by constructing input tensors in different ways. During training, model configurations were optimized for performance. The trained networks were then restructured and combined to fulfill different tasks. The framework, utilizing three distinct model types, was validated and compared using PZT readings from an aluminum plate. Both time-invariant and time-varying latent spaces demonstrated sensitivity to changes in structural conditions and EOCs. The framework successfully executed two main functions: state estimation and signal reconstruction. Additionally, the robustness of the models was tested under sparse data conditions. The proposed method achieved nearly perfect state estimation and low signal reconstruction errors, demonstrating strong robustness even with limited data.

\vspace{-6pt}

\bibliographystyle{aiaa} 

\bibliography{References} 

\vspace{-6pt}

\clearpage
\appendix
\appendixpage
\addappheadtotoc

\setcounter{table}{0}
\renewcommand\thetable{\Alph{section}.\arabic{table}}

\setcounter{figure}{0}
\renewcommand\thefigure{\Alph{section}.\arabic{figure}}
\section{Additional Results}
\begin{figure}[h!]
    \begin{picture}(500,410)
    \put(10,250){\includegraphics[width=0.45\textwidth]{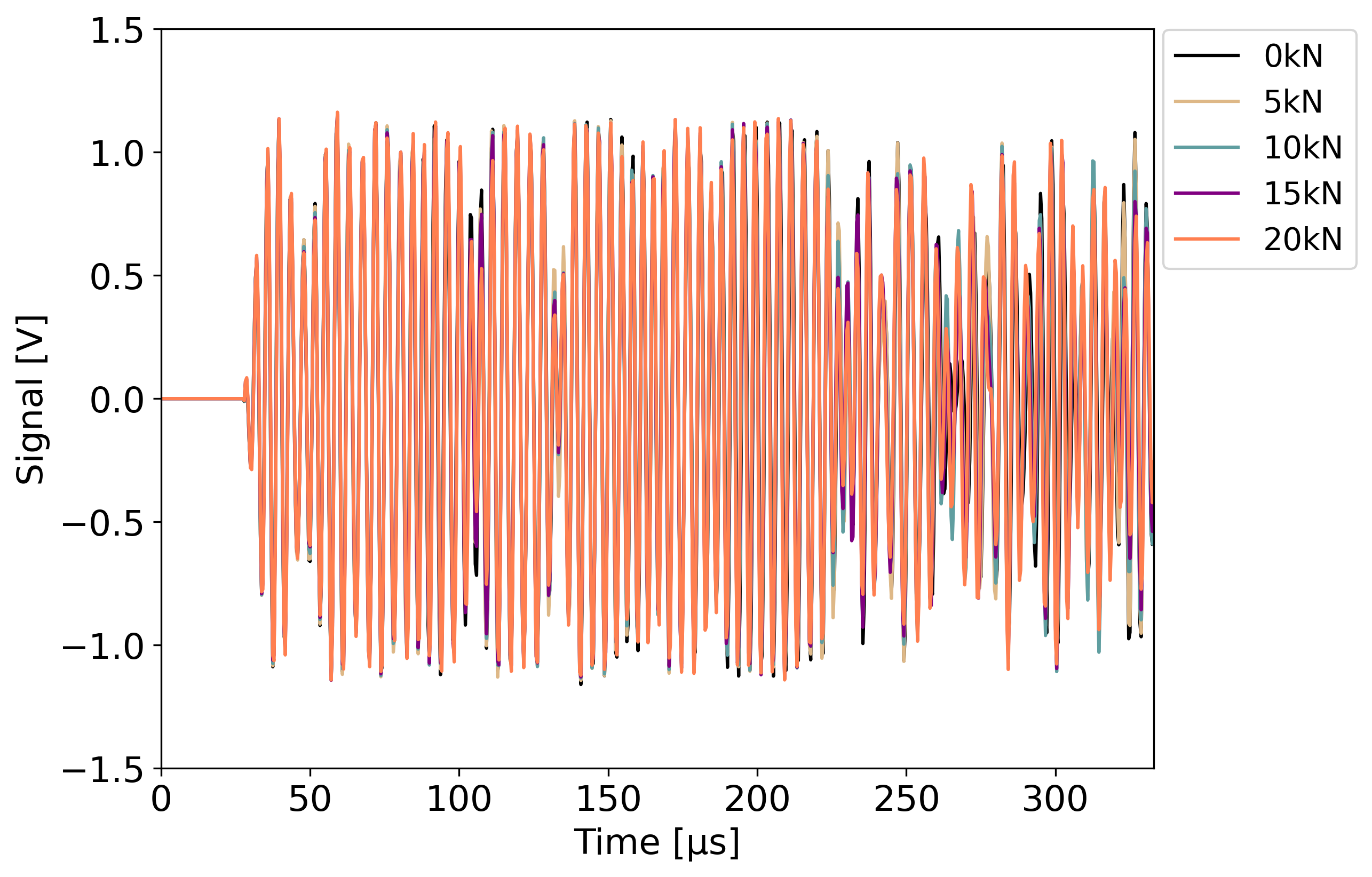}}
    \put(224,250){\includegraphics[width=0.45\textwidth]{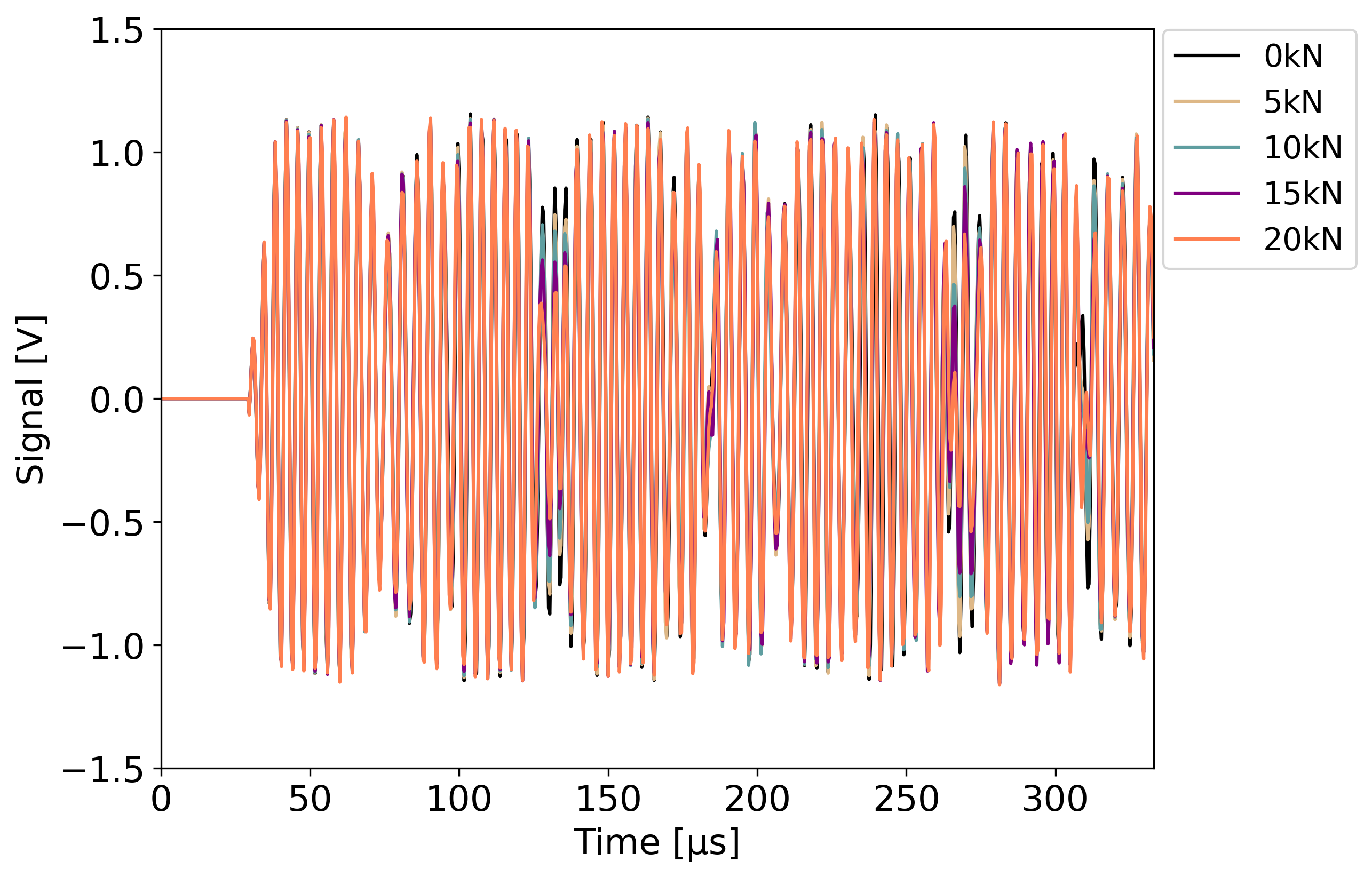}}
    \put(10,112){\includegraphics[width=0.45\textwidth]{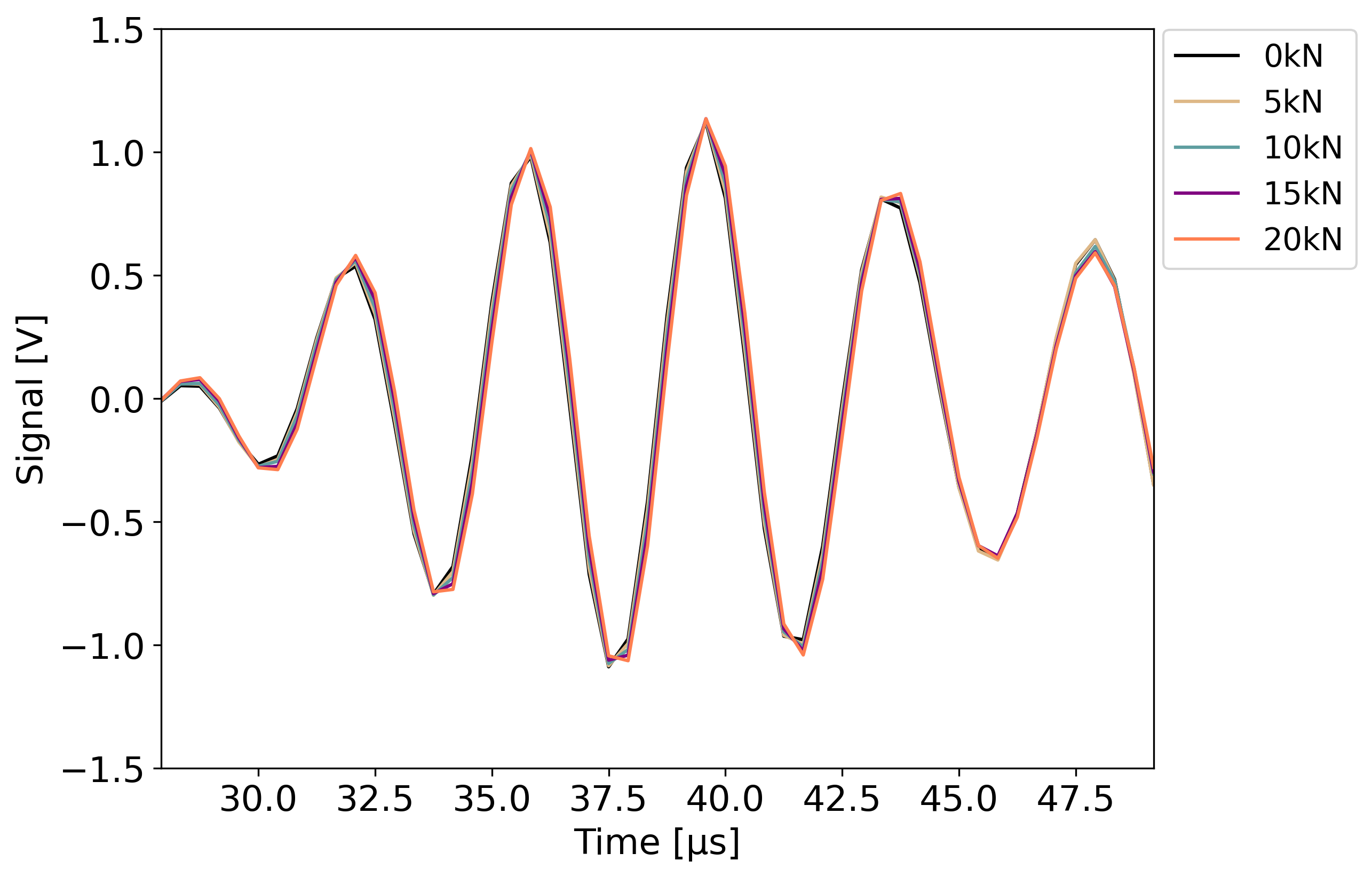}}
    \put(224,112){\includegraphics[width=0.45\textwidth]{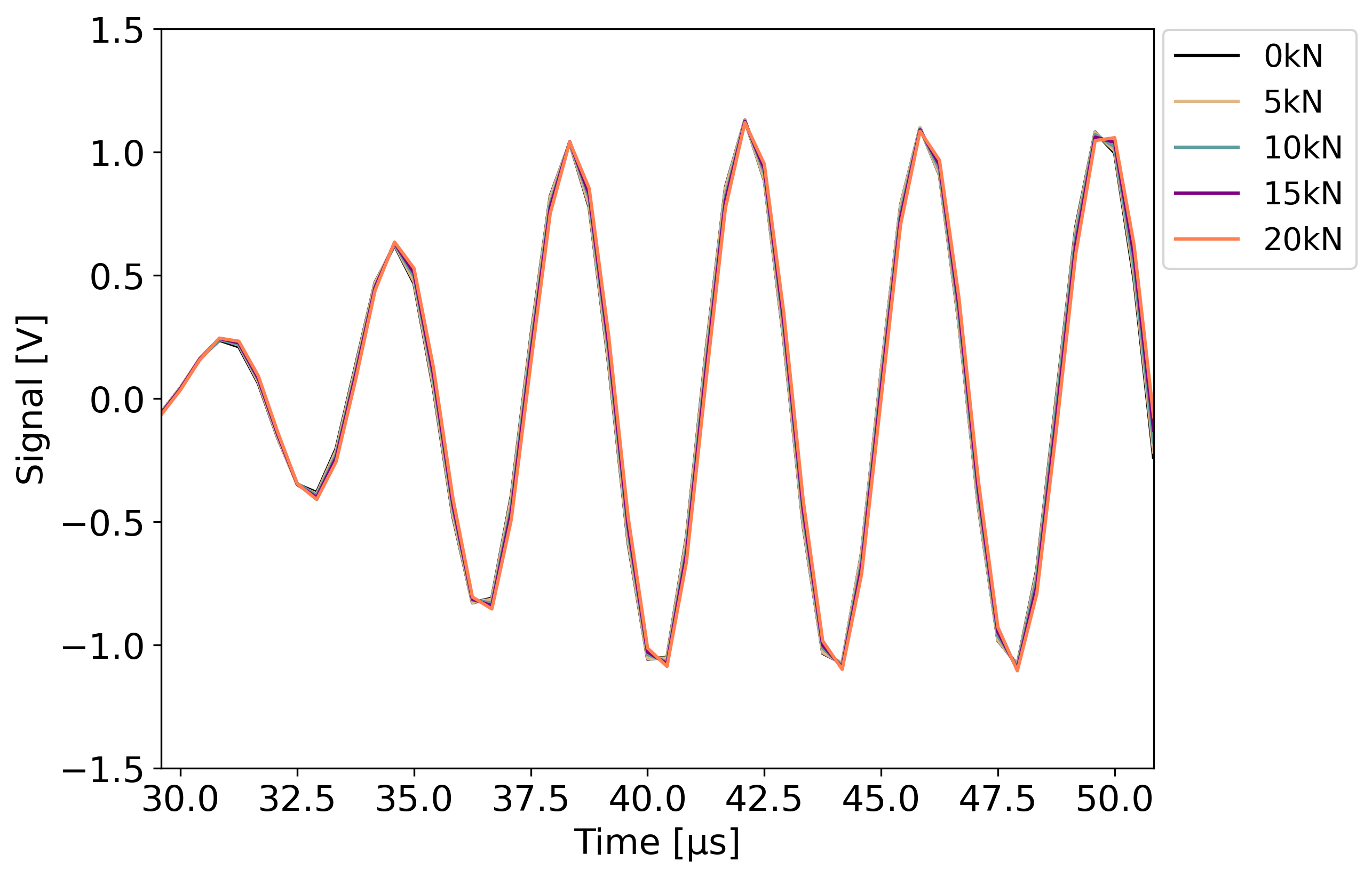}}
    \put(10,-26){\includegraphics[width=0.45\textwidth]{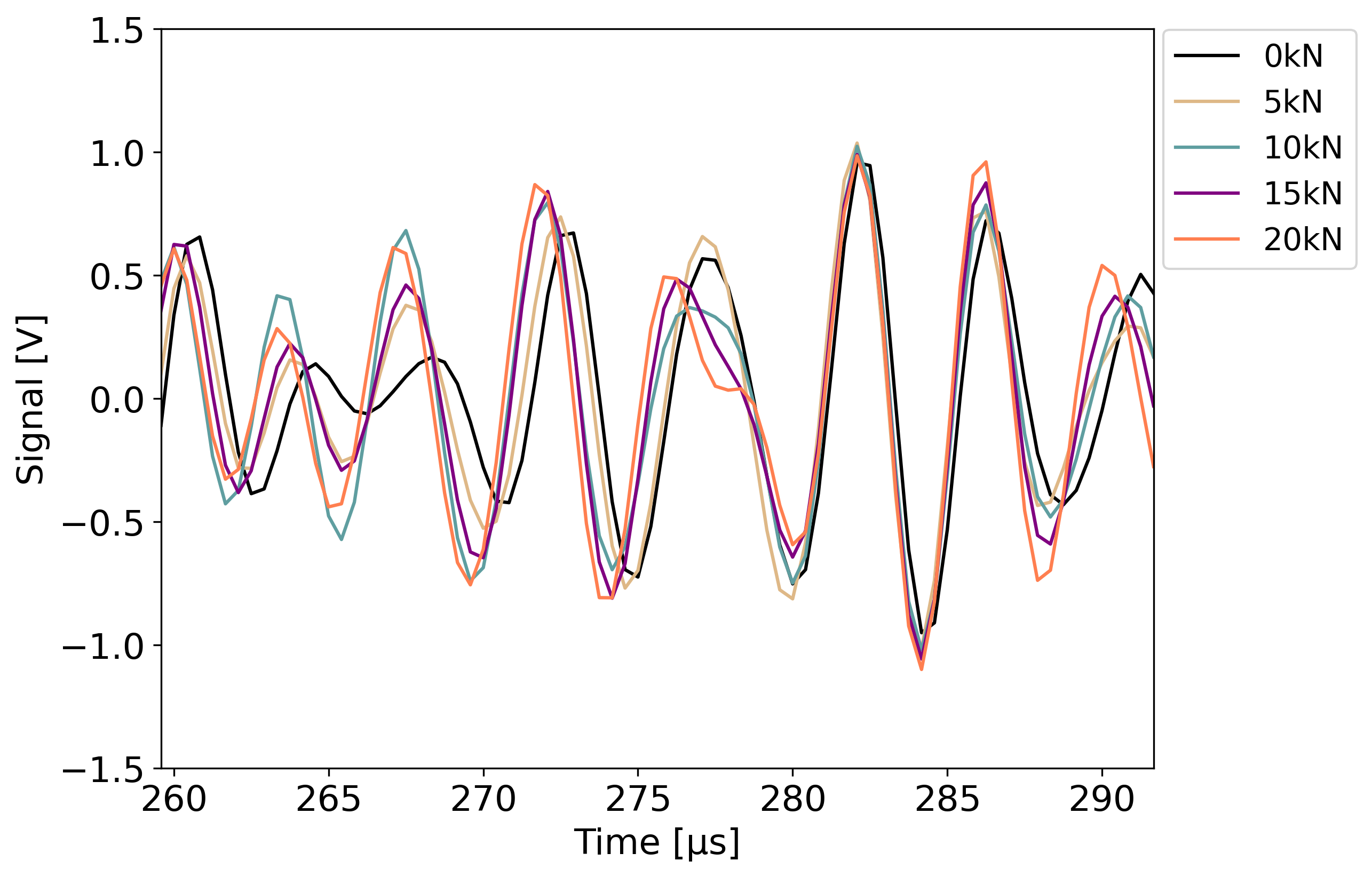}}
    \put(224,-26){\includegraphics[width=0.45\textwidth]{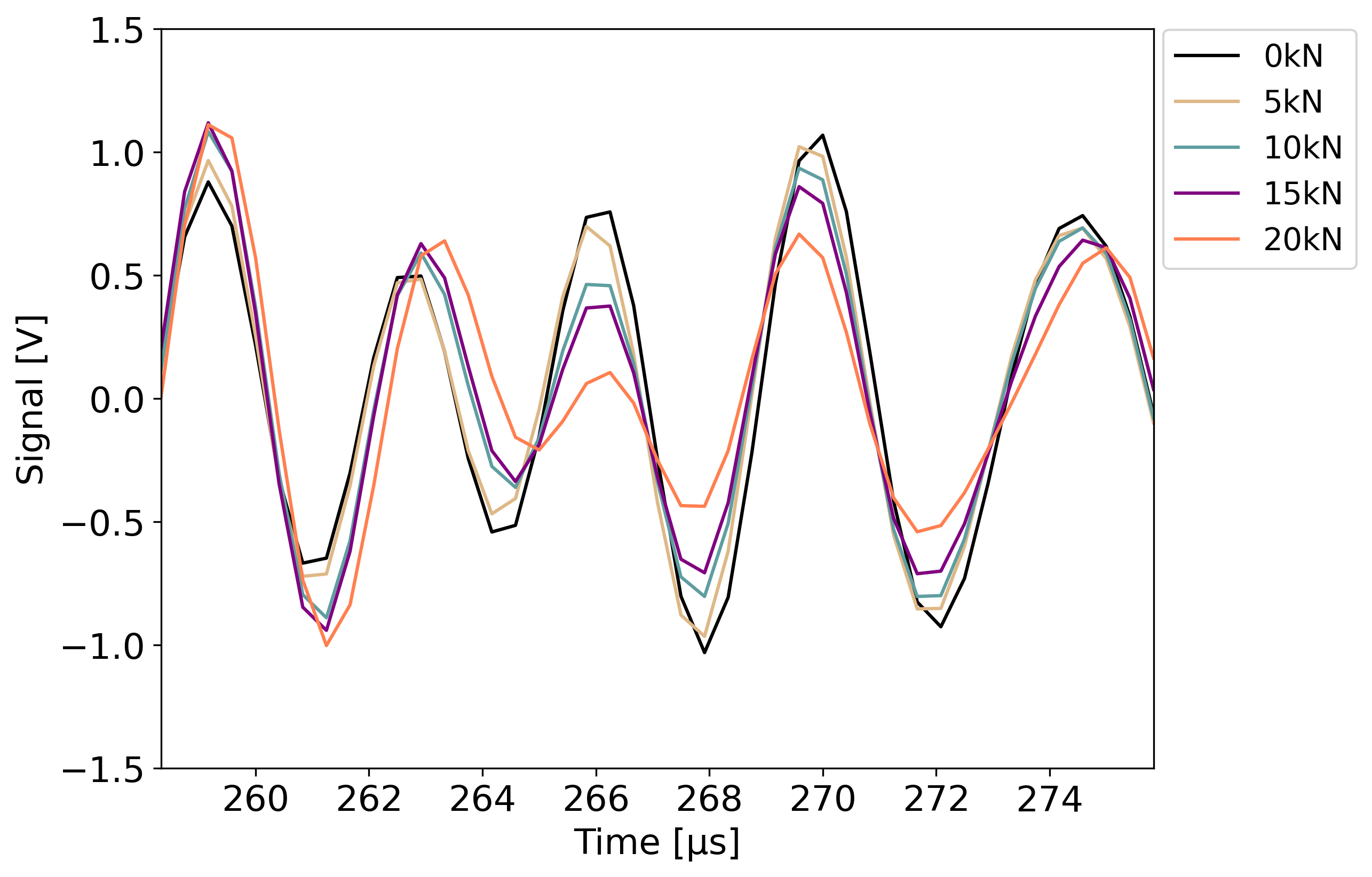}}
    \put(200,280){\color{black} \large {\fontfamily{phv}\selectfont \textbf{a}}}
    \put(410,280){\large {\fontfamily{phv}\selectfont \textbf{d}}}
   \put(200,140){\large {\fontfamily{phv}\selectfont \textbf{b}}} 
   \put(410,140){\large {\fontfamily{phv}\selectfont \textbf{e}}}
   \put(200,0){\large {\fontfamily{phv}\selectfont \textbf{c}}} 
   \put(410,0){\large {\fontfamily{phv}\selectfont \textbf{f}}} 
    \end{picture} 
    \vspace{-0pt}
    \caption{Sample signals in this study after downsampling. Panel a: full signals of damage level 2 under all loads from path 2-6; panel b: the first wave packet of damage level 2 under all loads from path 2-6; panel c: partial signals of damage level 2 under all loads from path 2-6; panel d: full signals of damage level 2 under all loads from path 3-4; panel b: the first wave packet of damage level 2 under all loads from path 3-4; panel c: partial signals of damage level 2 under all loads from path 3-4.}
\label{fig:signals_DIs} 
\end{figure}
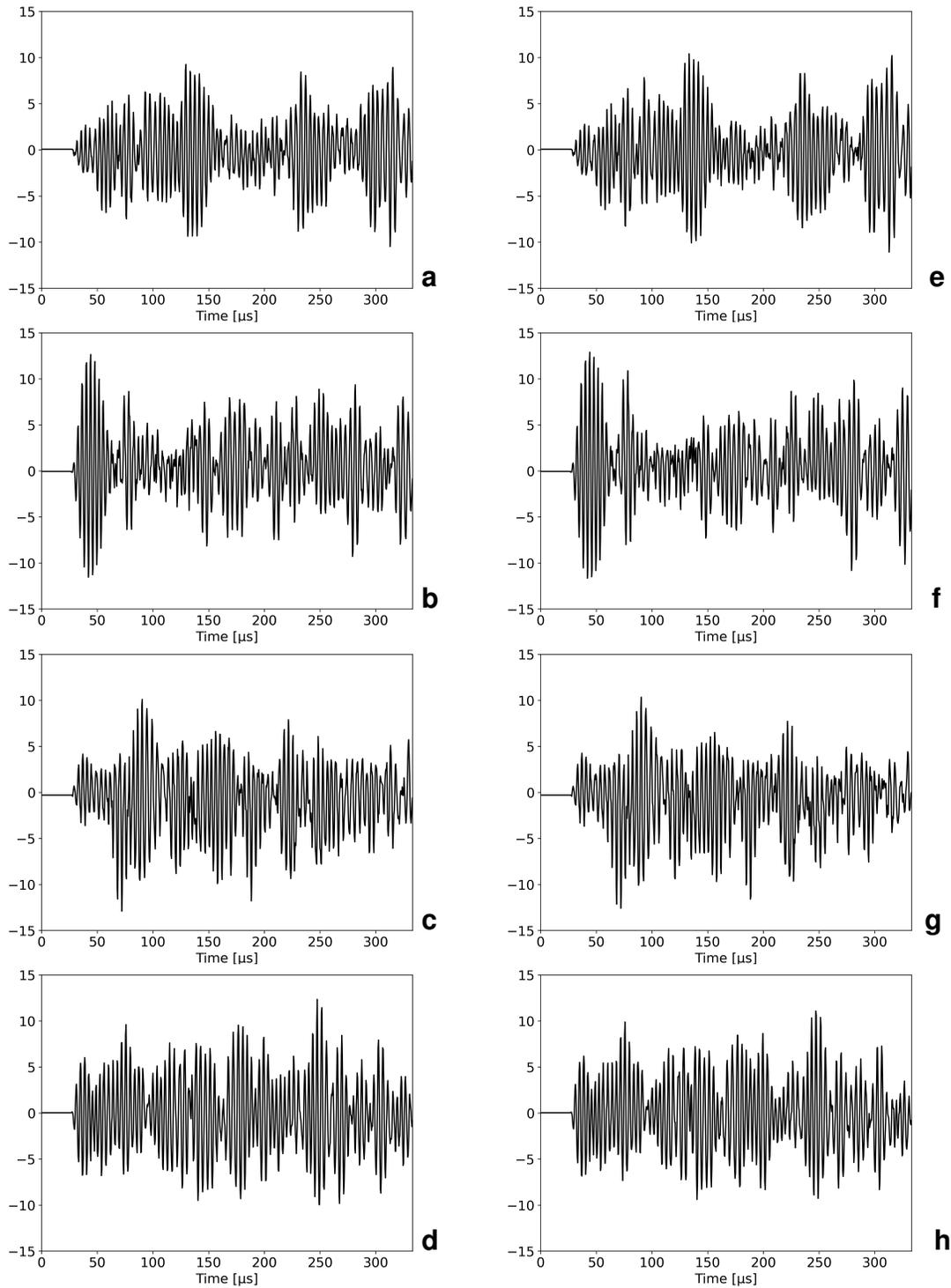
\begin{figure}[t!]
    \begin{picture}(500,410)
    \put(10,250){\includegraphics[width=0.38\textwidth]{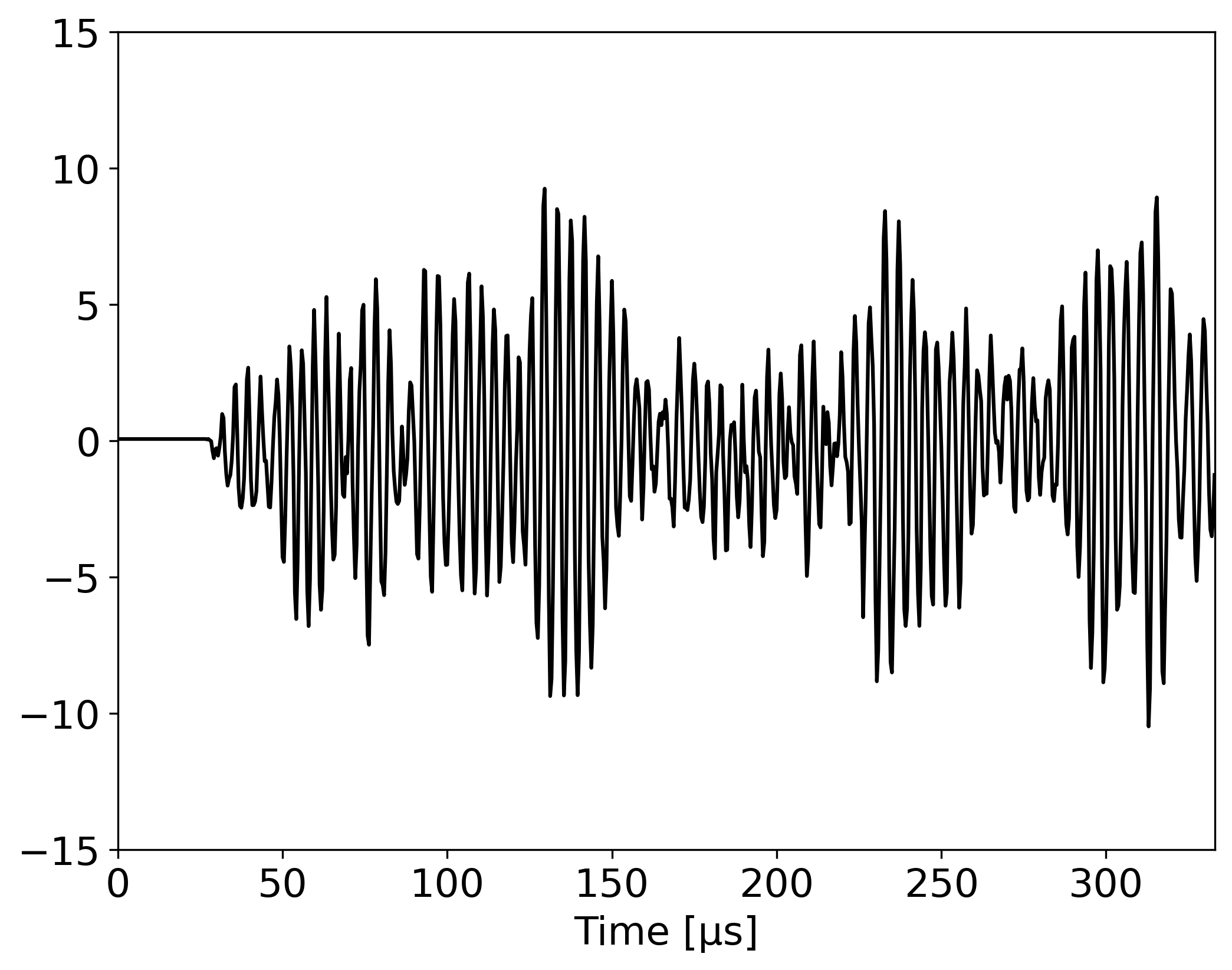}}
    \put(224,250){\includegraphics[width=0.38\textwidth]{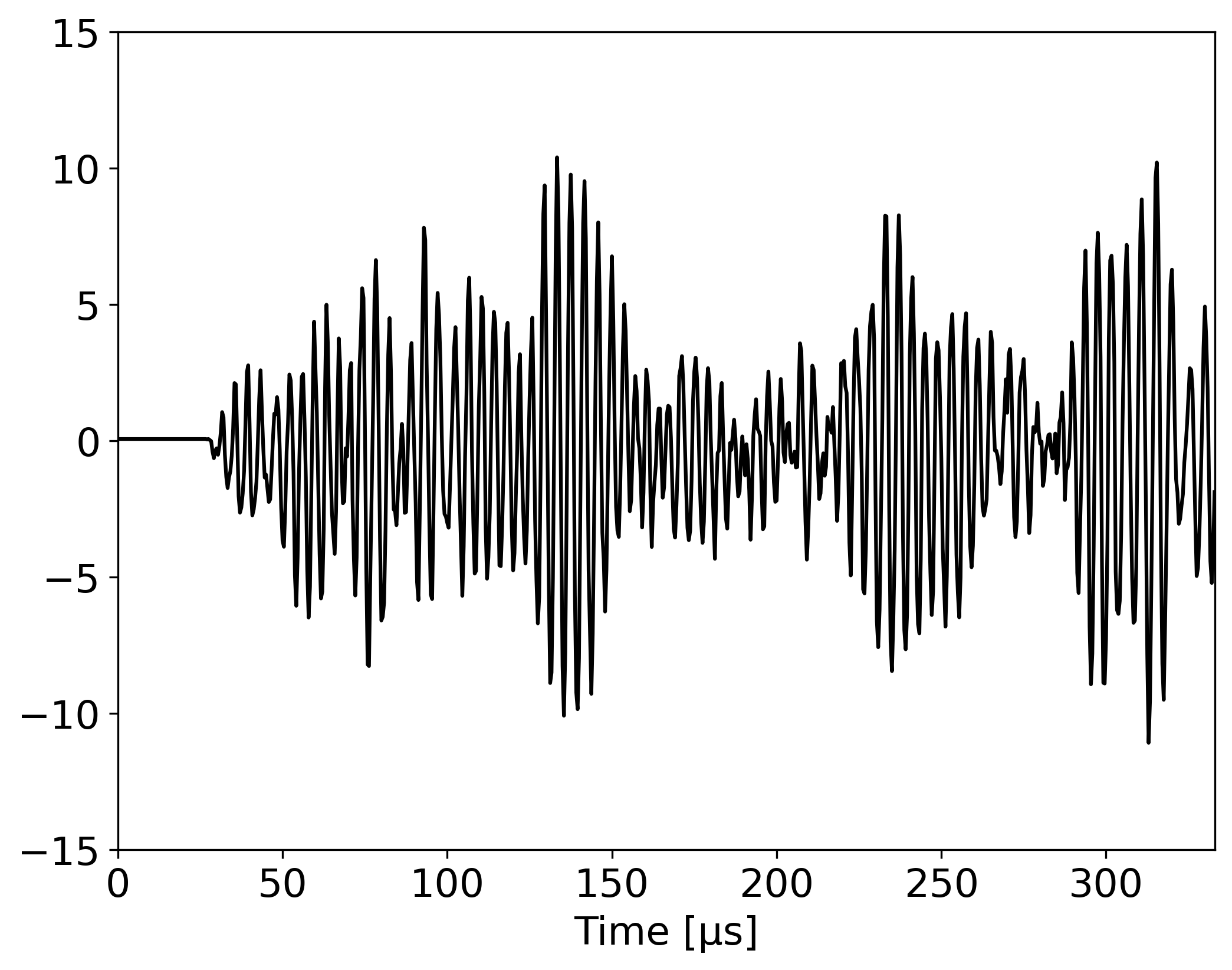}}
    \put(10,112){\includegraphics[width=0.38\textwidth]{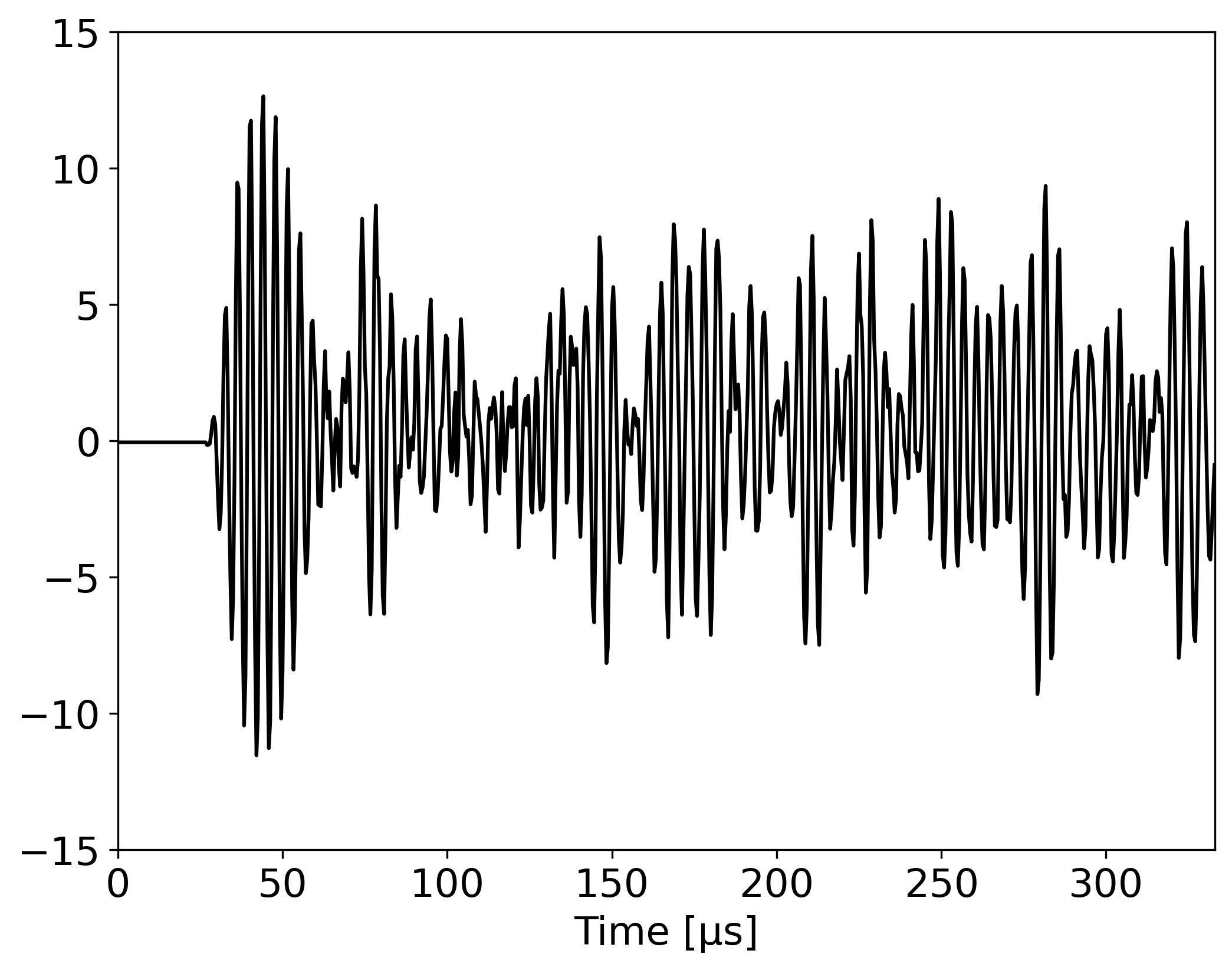}}
    \put(224,112){\includegraphics[width=0.38\textwidth]{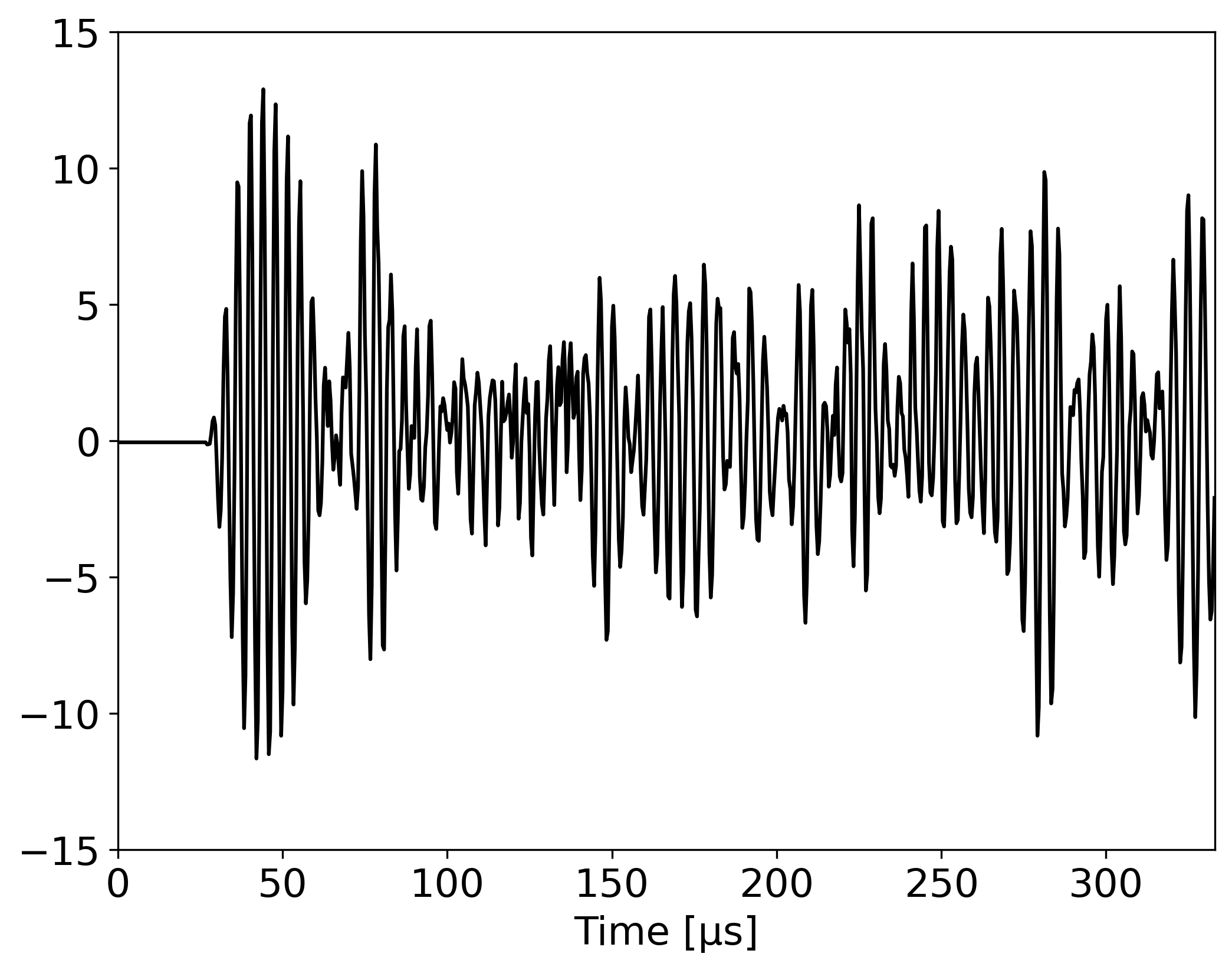}}
    \put(10,-26){\includegraphics[width=0.38\textwidth]{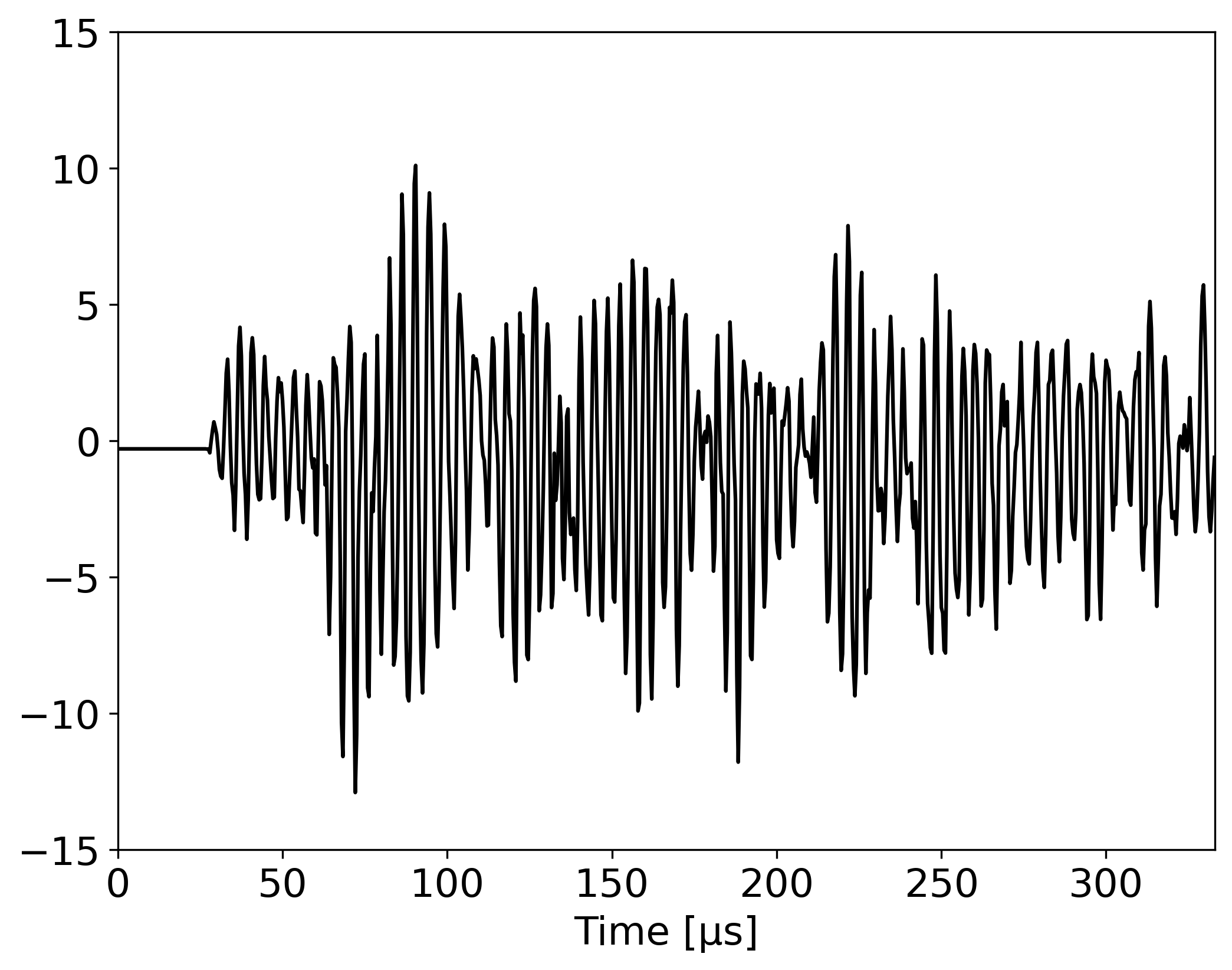}}
    \put(224,-26){\includegraphics[width=0.38\textwidth]{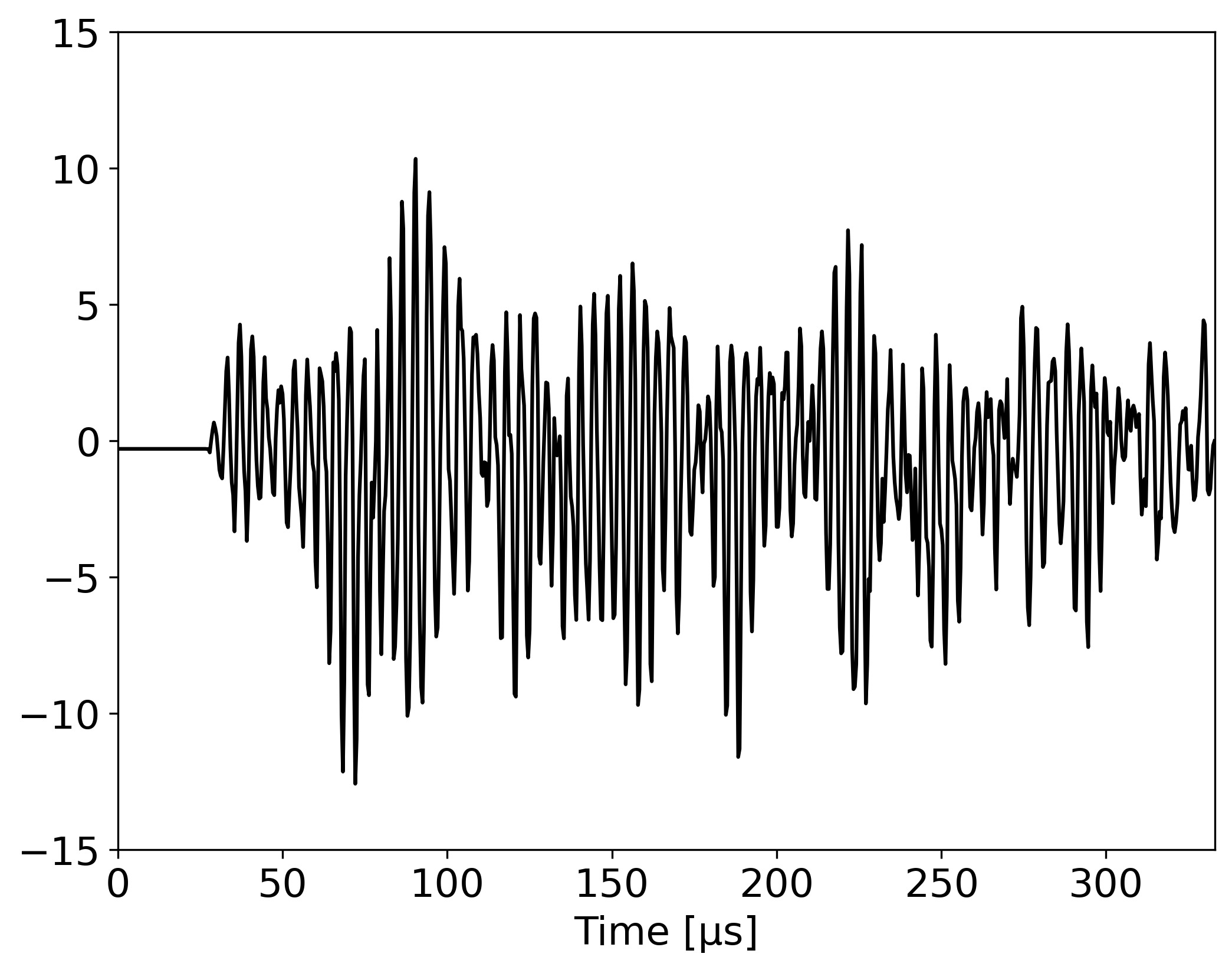}}
    \put(10,-164){\includegraphics[width=0.38\textwidth]{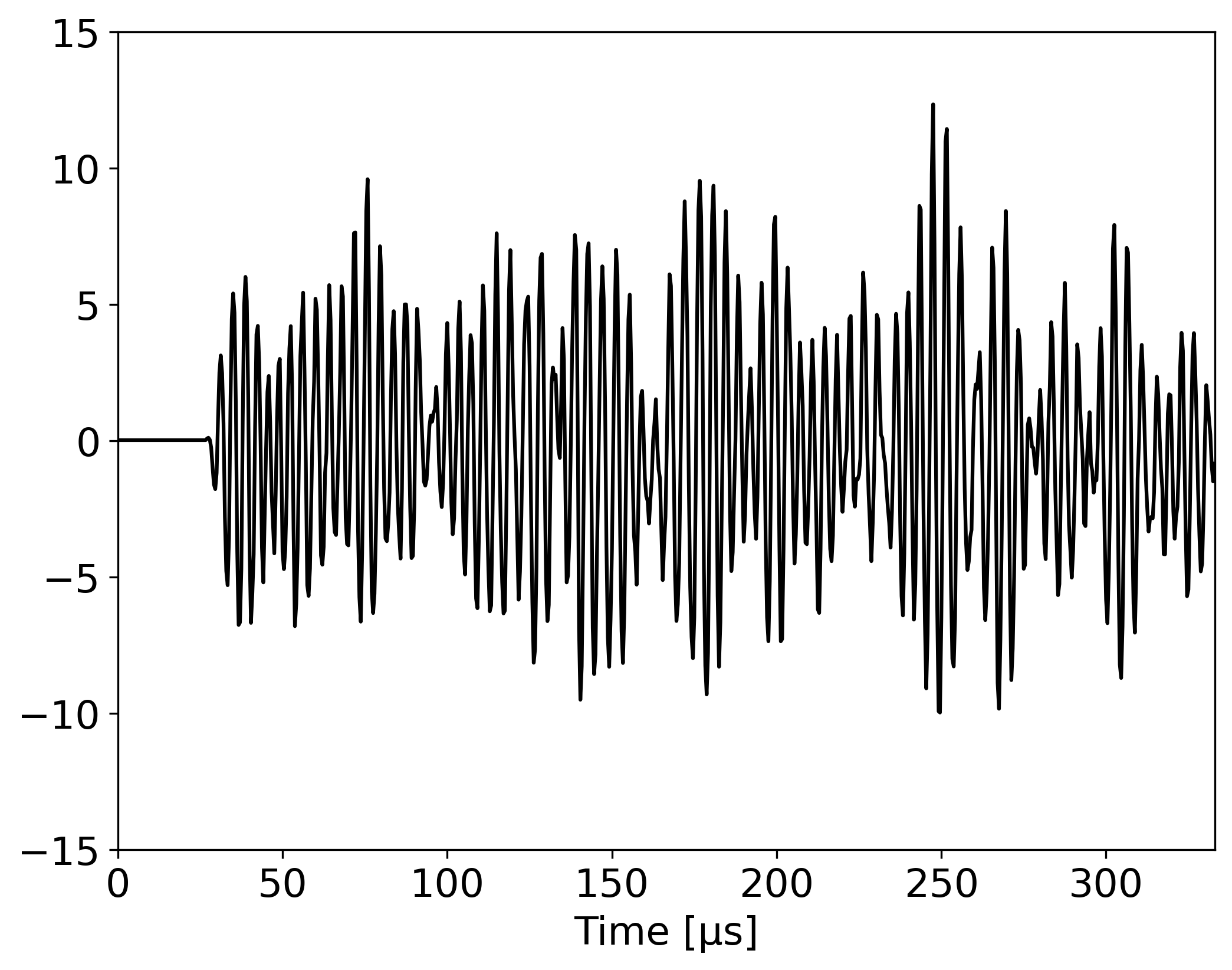}}
    \put(224,-164){\includegraphics[width=0.38\textwidth]{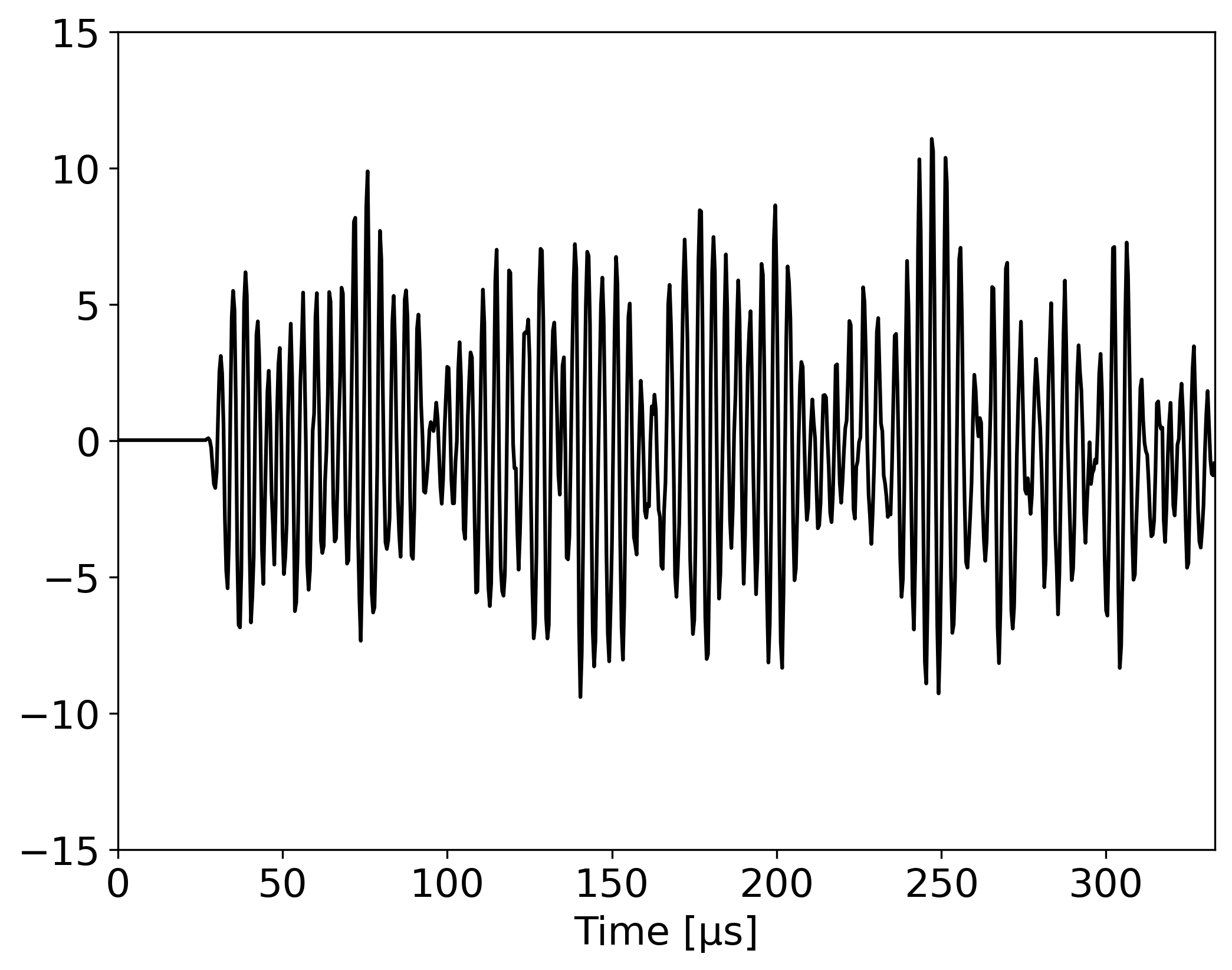}}
    \put(190,268){\color{black} \large {\fontfamily{phv}\selectfont \textbf{a}}}
    \put(190,130){\large {\fontfamily{phv}\selectfont \textbf{b}}} 
    \put(190,-8){\large {\fontfamily{phv}\selectfont \textbf{c}}}
    \put(190,-146){\large {\fontfamily{phv}\selectfont \textbf{d}}}
   
   \put(408,268){\large {\fontfamily{phv}\selectfont \textbf{e}}} 
   \put(409,130){\large {\fontfamily{phv}\selectfont \textbf{f}}} 
   \put(406,-8){\large {\fontfamily{phv}\selectfont \textbf{g}}} 
   \put(410,-146){\large {\fontfamily{phv}\selectfont \textbf{h}}} 
    \end{picture} 
    \vspace{+140pt}
    \caption{Latent space representations from model Type III. Panel a-d: the 1st, 3rd, 5th and 7th latent variables under healthy case with 10kN; Panel a-d: the 1st, 3rd, 5th and 7th latent variables under damage level 4 with 10kN.}
\label{fig:lat2_appendix} 
\end{figure}
%
\begin{figure}[t!]
    \begin{picture}(500,410)
    \put(10,250){\includegraphics[width=0.38\textwidth]{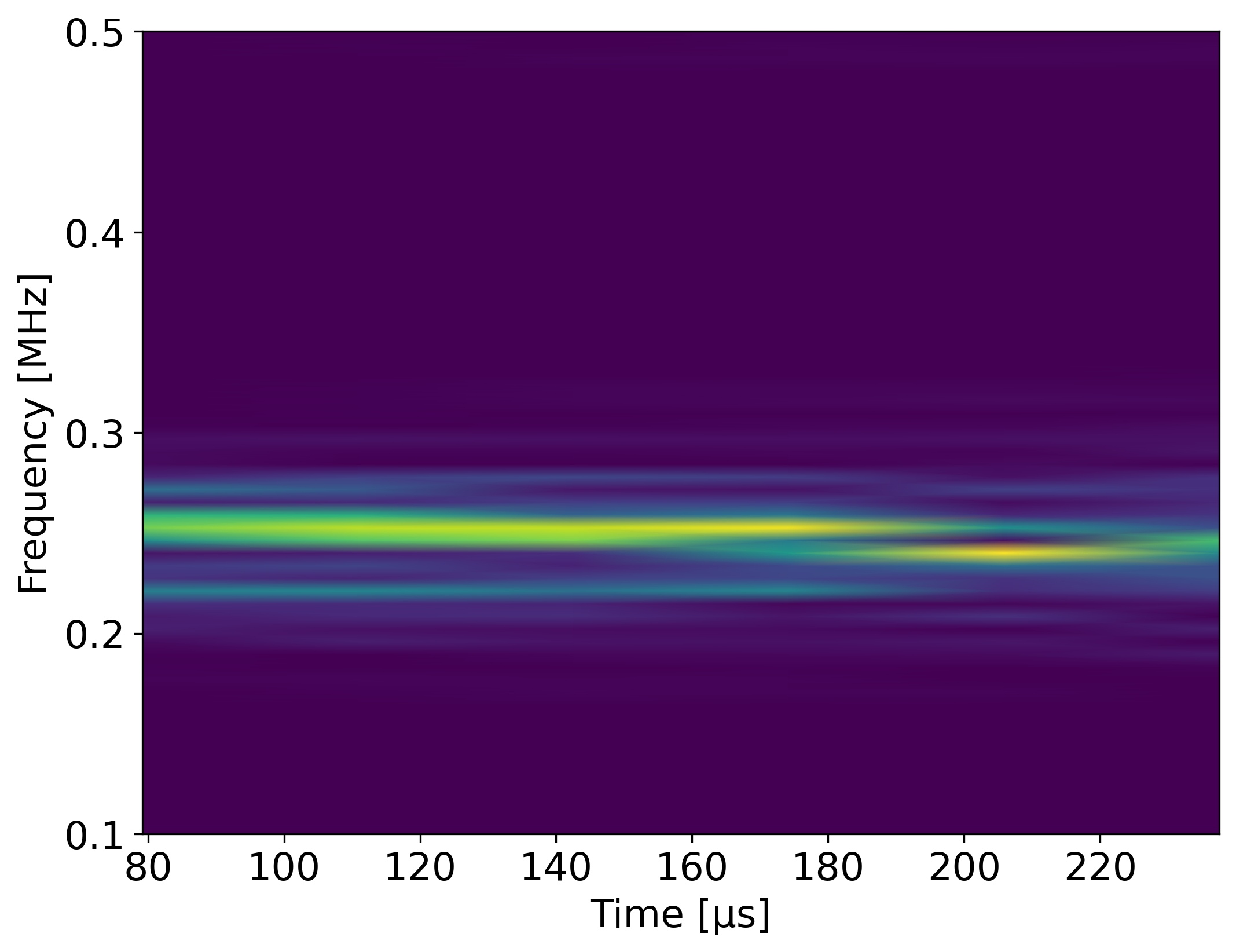}}
    \put(224,250){\includegraphics[width=0.38\textwidth]{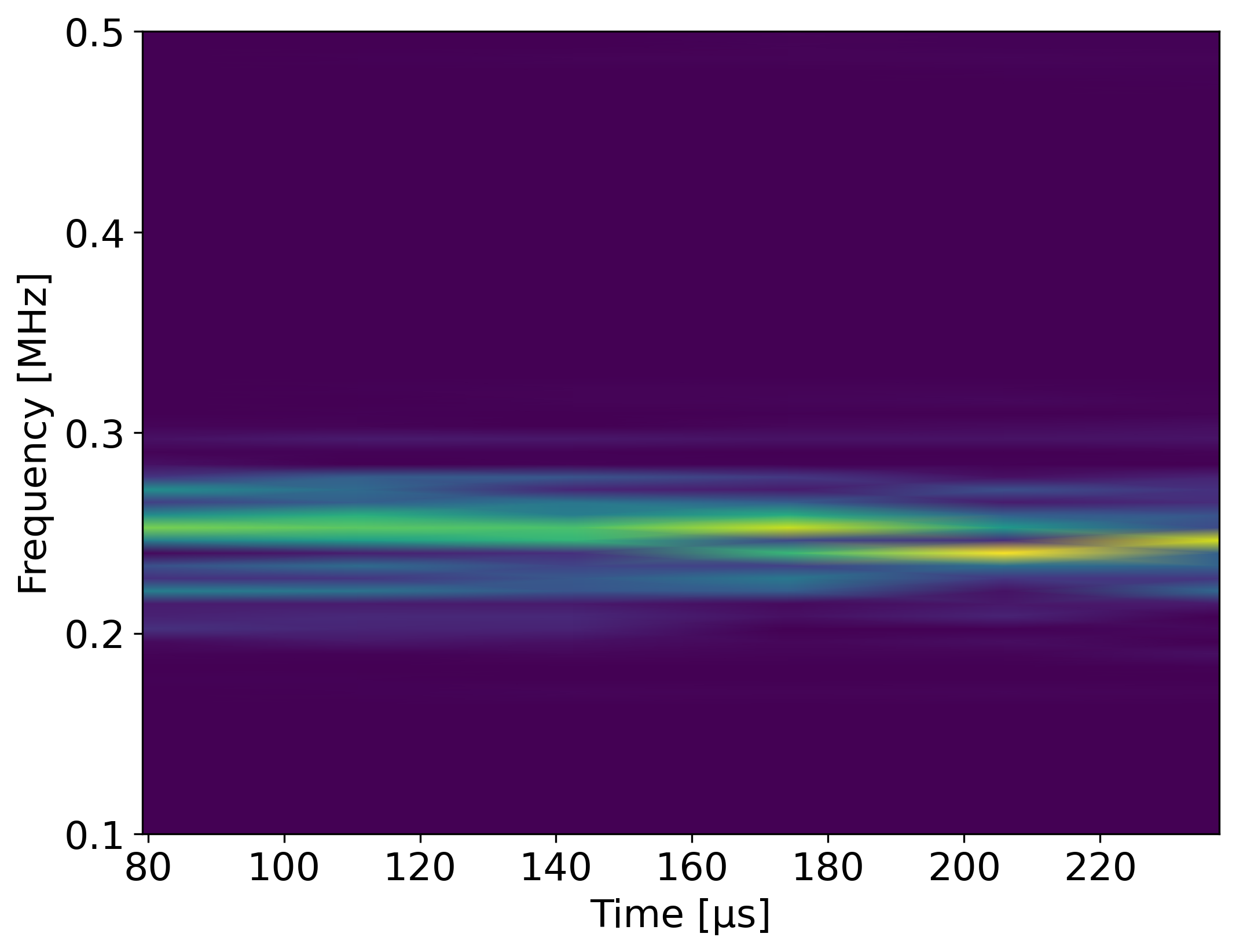}}
    \put(10,112){\includegraphics[width=0.38\textwidth]{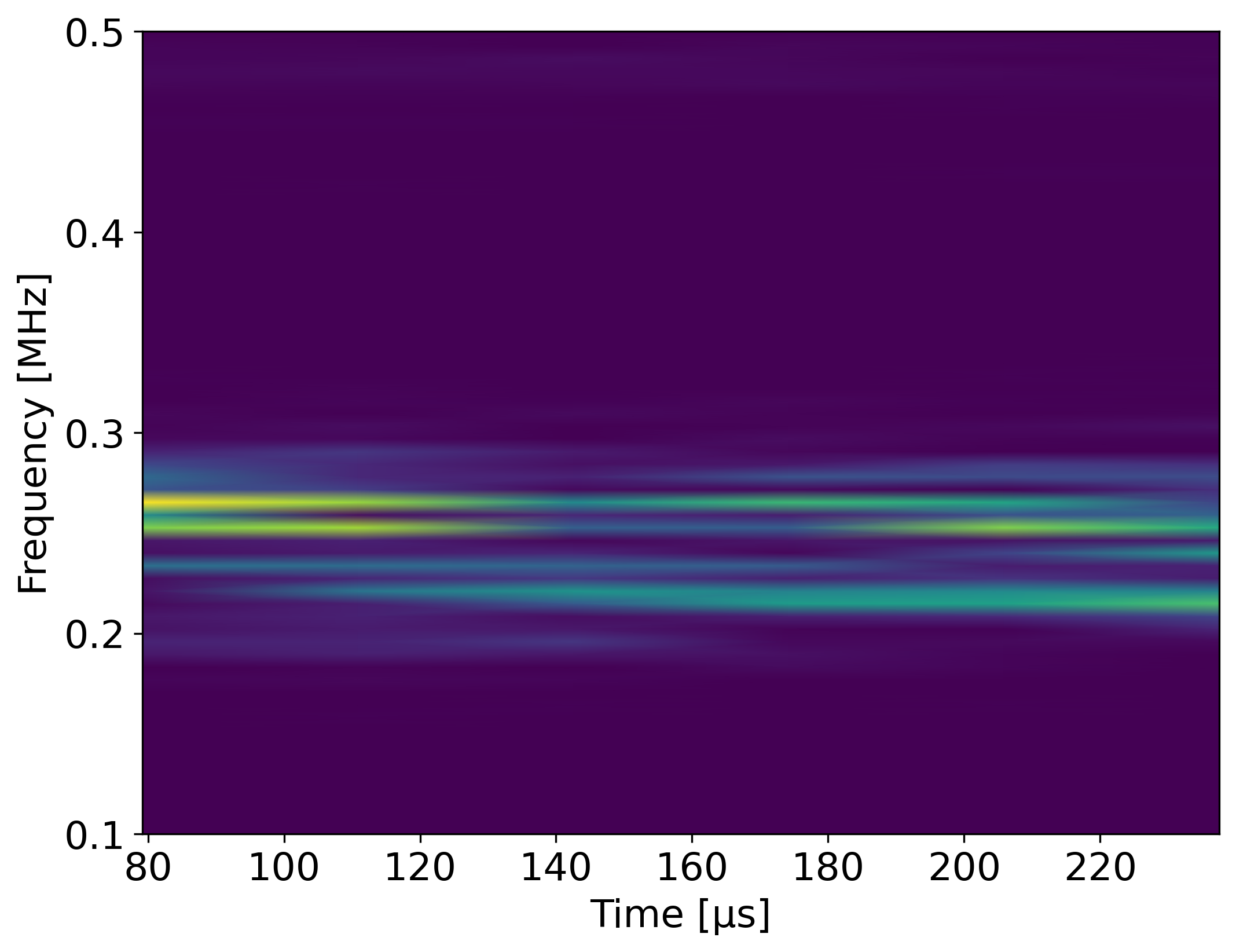}}
    \put(224,112){\includegraphics[width=0.38\textwidth]{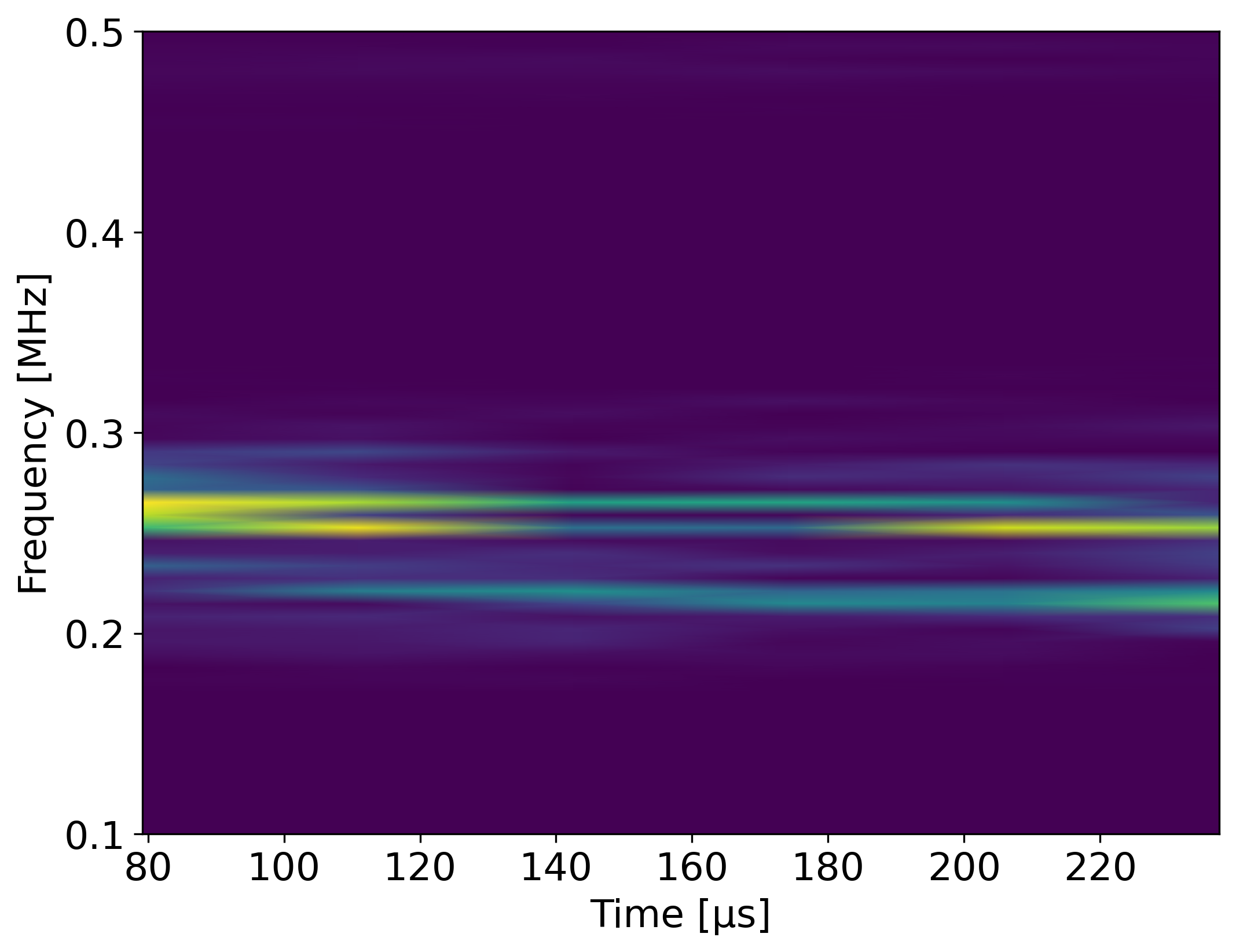}}
    \put(10,-26){\includegraphics[width=0.38\textwidth]{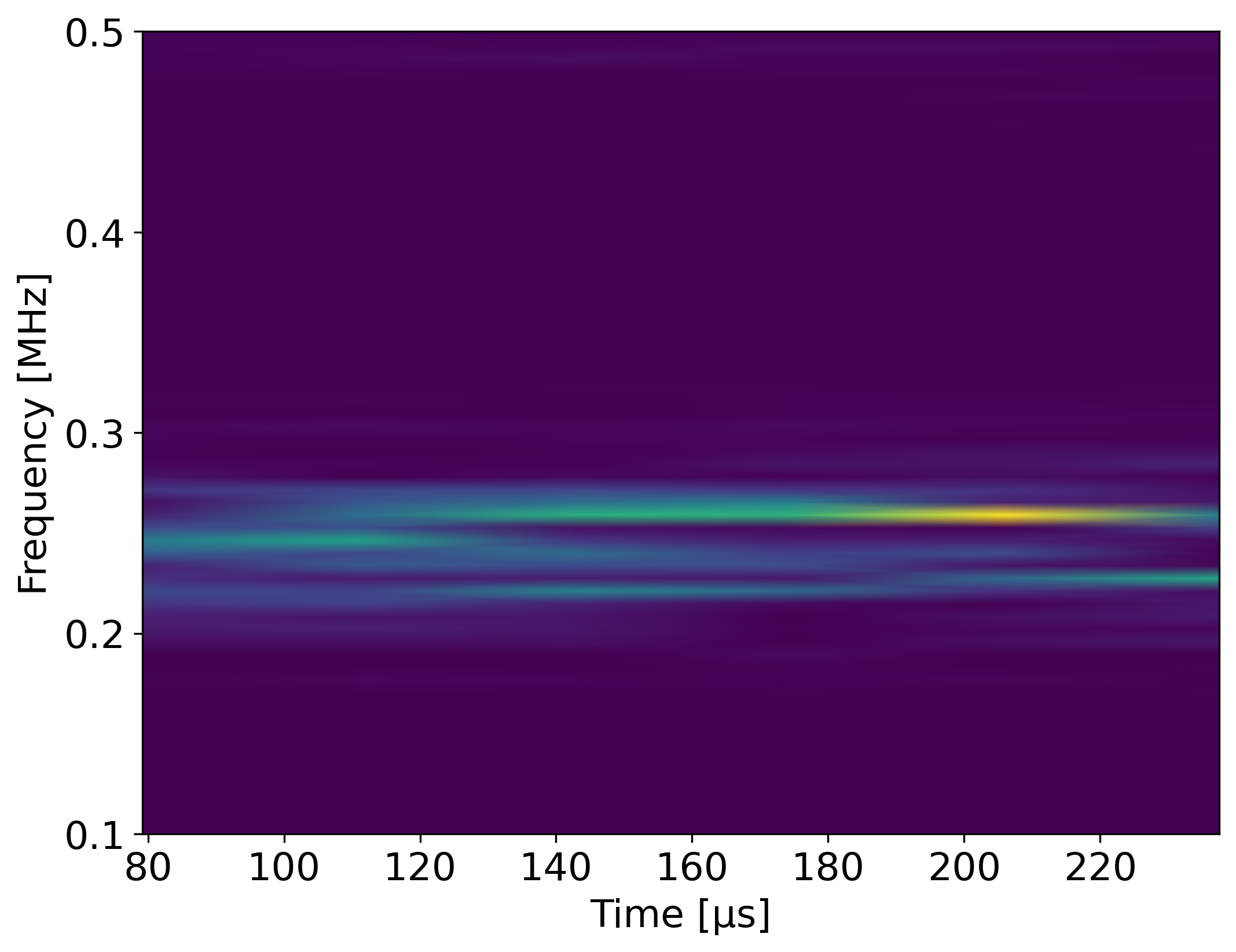}}
    \put(224,-26){\includegraphics[width=0.38\textwidth]{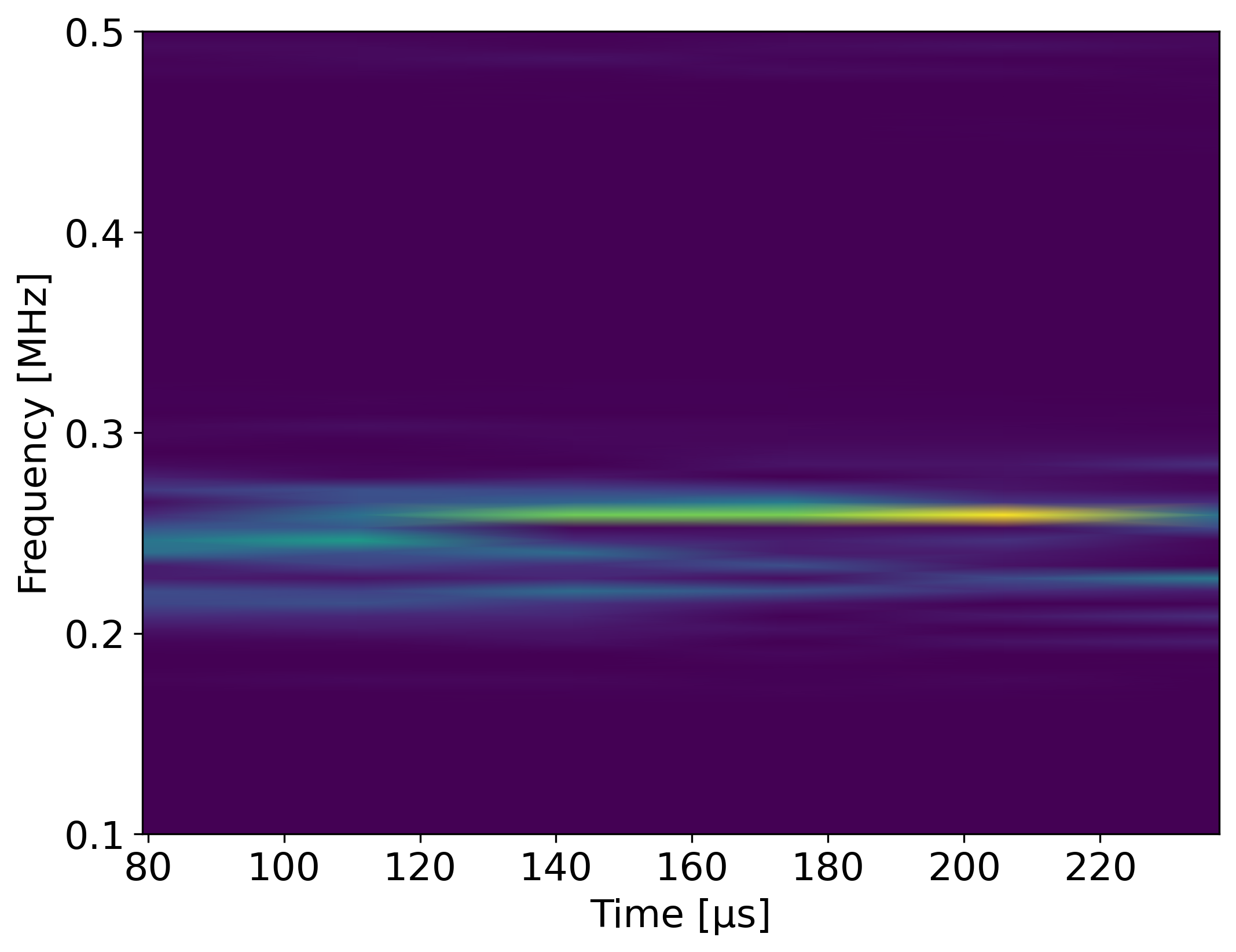}}
    \put(10,-164){\includegraphics[width=0.38\textwidth]{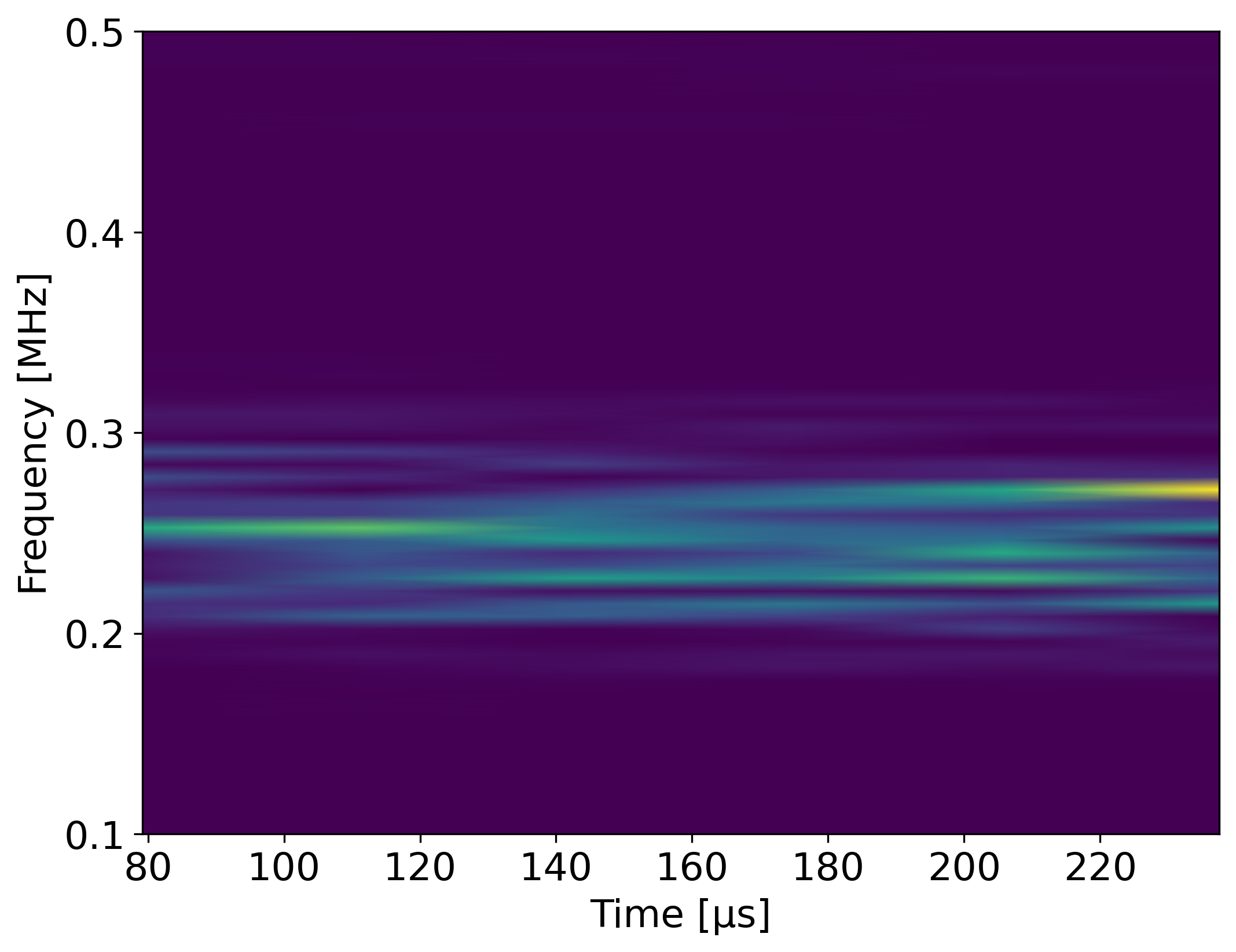}}
    \put(224,-164){\includegraphics[width=0.38\textwidth]{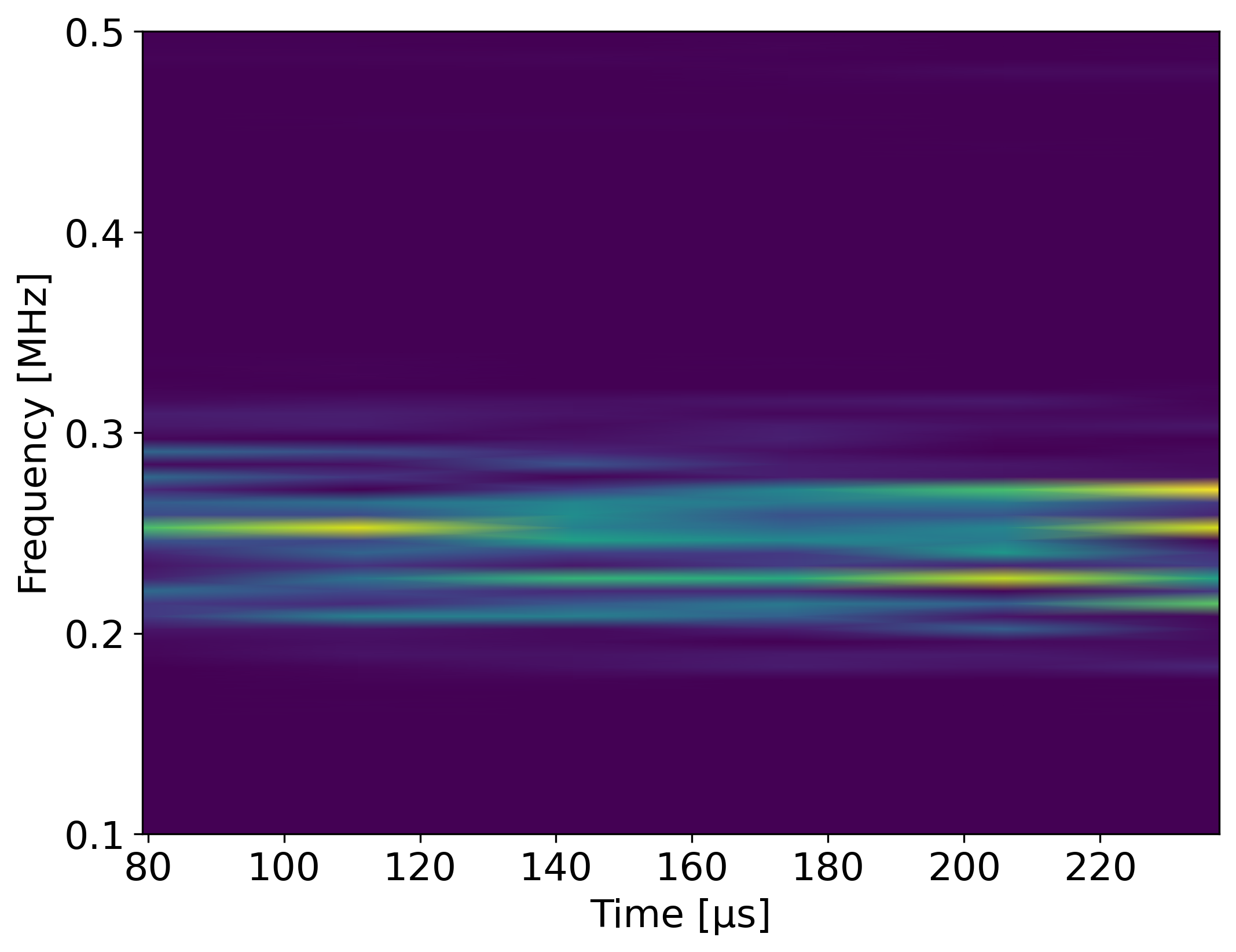}}
    \put(190,268){\color{black} \large {\fontfamily{phv}\selectfont \textbf{a}}}
    \put(190,130){\large {\fontfamily{phv}\selectfont \textbf{b}}} 
    \put(190,-8){\large {\fontfamily{phv}\selectfont \textbf{c}}}
    \put(190,-146){\large {\fontfamily{phv}\selectfont \textbf{d}}}
   
   \put(408,268){\large {\fontfamily{phv}\selectfont \textbf{e}}} 
   \put(409,130){\large {\fontfamily{phv}\selectfont \textbf{f}}} 
   \put(406,-8){\large {\fontfamily{phv}\selectfont \textbf{g}}} 
   \put(410,-146){\large {\fontfamily{phv}\selectfont \textbf{h}}} 
    \end{picture} 
    \vspace{+140pt}
    \caption{PSD of the latent space representations from model Type III. Panel a-d: PSD of the 1st, 3rd, 5th and 7th latent variables under healthy case with 10kN; Panel a-d: PSD of the 1st, 3rd, 5th and 7th latent variables under damage level 4 with 10kN. Window length is 380 sample length and number of overlap is about 80\% of window length.}
\label{fig:lat2_appendix1} 
\end{figure}
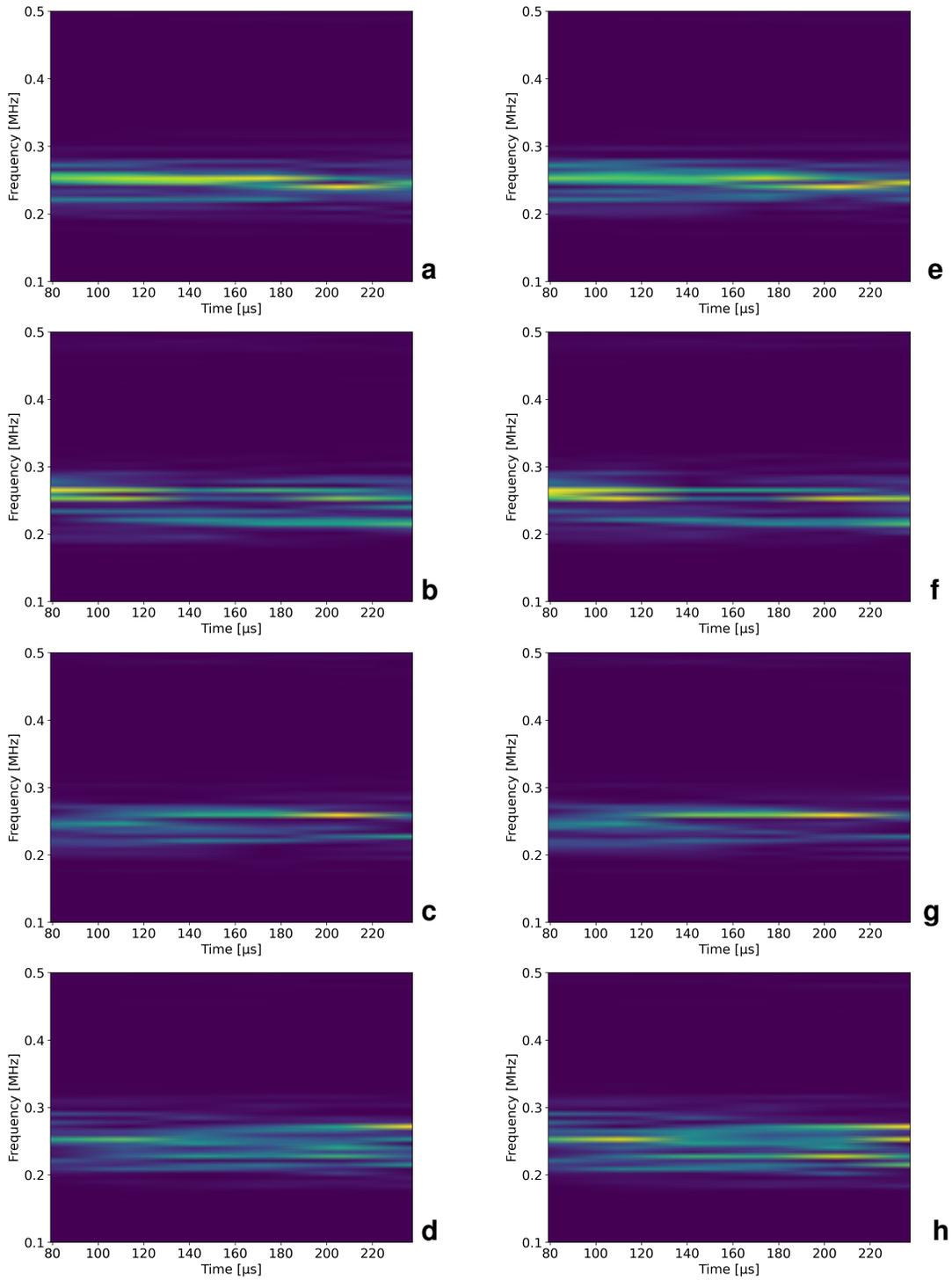







\end{document}